\documentclass[preprint,12pt, authoryear]{aastex} 

\usepackage{helvet}

\bibpunct{(}{)}{;}{a}{}{,}   

\usepackage{amssymb}
\usepackage{appendix}
\usepackage{placeins}
\usepackage{lscape}
\usepackage{rotating}
\usepackage{amsmath}
\usepackage{multicol}
\usepackage{mathtools}
\usepackage{graphicx}

\usepackage{booktabs}
\usepackage{alltt} 

\usepackage{standalone}
\usepackage{longtable}

\newcount\Comments  
\usepackage[usenames,dvipsnames]{xcolor}
\definecolor{darkgreen}{rgb}{0,0.5,0}
\definecolor{purple}{rgb}{1,0,1}
\definecolor{darkpurple}{rgb}{0.5,0,0.5}
\definecolor{lightgreen}{RGB}{135,220,0}

\newcommand{\g}{$\gamma$}

\newcommand{\kms}{km s$^{-1}$}
\newcommand{\st}{$^{\mathrm{st}}$}        
\newcommand{\Fermi}{\emph{Fermi}}  
\newcommand{\FermiLat}{\emph{Fermi} LAT}     
\newcommand{\ptlike}{{\tt pointlike}}
\newcommand{\gtlike}{{\tt gtlike}}
\newcommand{\logL}{$\log \mathcal{L}$}
\newcommand{\hi}{$\mathrm{H\,\scriptstyle{I}}$}
\newcommand{\Hii}{$\mathrm{H\,\scriptstyle{II}}$}
\newcommand{\overlaploc}{\mathrm{Overlap}_\mathrm{loc}}
\newcommand{\overlapext}{\mathrm{Overlap}_\mathrm{ext}}

\newcommand{\nGalSNRs}{$279$} 
\newcommand{\ndist}{$112$} 
\newcommand{\nassocprobclassified}{$36$} 
\newcommand{\nclassifiedsnrs}{$30$} 
\newcommand{\nnotsnrs}{four} 
\newcommand{\nnotsnrsCap}{Four} 
\newcommand{\nmarginal}{$14$} 
\newcommand{\nclassifiedandmarginal}{$48$} 
\newcommand{\nother}{$54$} 
\newcommand{\ndetected}{$102$} 
\newcommand{\nULs}{$245$} 
\newcommand{\nextended}{$17$} 
\newcommand{\nnewpointlike}{$10$} 
\newcommand{\nnewextended}{four} 
\newcommand{\notherLogPpct}{$7\%$} 
\newcommand{\nclassLogPpct}{$6\%$} 

\newcommand{\nmock}{two} 
\newcommand{\nmockN}{2}  
\newcommand{\nmockpercent}{$\sim6\%$} 
\newcommand{\nmockclassifiedandmarginal}{nine} 
\newcommand{\nmockclassifiedandmarginalpercent}{$19\%$} %
\newcommand{\nninetyfivemock}{eight} 
\newcommand{\nninetyfivemockpercent}{$22\%$} 
\newcommand{\nninetyfivemockclassifiedandmarginal}{$18$} 
\newcommand{\nninetyfivemockclassifiedandmarginalpercent}{$38\%$} 

\newcommand{\nmarginalAIEM}{two} 
\newcommand{\notherAIEM}{nine} 
\newcommand{\notherAIEMcap}{Nine} 
\usepackage[
breaklinks=true,%
colorlinks=true,%
linktocpage=true,
pdfauthor={T. J. Brandt, J. W. Hewitt, F. de Palma, et al.},%
pdftitle={\FermiLat{} Supernova Remnant Catalog}%
]{hyperref}

\begin{document}

\title{The $1$\st{} \FermiLat{} Supernova Remnant Catalog} %

\author{
F.~Acero\altaffilmark{2}, 
M.~Ackermann\altaffilmark{3}, 
M.~Ajello\altaffilmark{4}, 
L.~Baldini\altaffilmark{5,6}, 
J.~Ballet\altaffilmark{2}, 
G.~Barbiellini\altaffilmark{7,8}, 
D.~Bastieri\altaffilmark{9,10}, 
R.~Bellazzini\altaffilmark{11}, 
E.~Bissaldi\altaffilmark{12}, 
R.~D.~Blandford\altaffilmark{6}
% ...
}
% ...
\altaffiltext{2}{Laboratoire AIM, CEA-IRFU/CNRS/Universit\'e Paris Diderot, Service d'Astrophysique, CEA Saclay, F-91191 Gif sur Yvette, France}
\altaffiltext{3}{Deutsches Elektronen Synchrotron DESY, D-15738 Zeuthen, Germany}
\altaffiltext{4}{Department of Physics and Astronomy, Clemson University, Kinard Lab of Physics, Clemson, SC 29634-0978, USA}
\altaffiltext{5}{Universit\`a di Pisa and Istituto Nazionale di Fisica Nucleare, Sezione di Pisa I-56127 Pisa, Italy}
\altaffiltext{6}{W. W. Hansen Experimental Physics Laboratory, Kavli Institute for Particle Astrophysics and Cosmology, Department of Physics and SLAC National Accelerator Laboratory, Stanford University, Stanford, CA 94305, USA}
\altaffiltext{7}{Istituto Nazionale di Fisica Nucleare, Sezione di Trieste, I-34127 Trieste, Italy}
\altaffiltext{8}{Dipartimento di Fisica, Universit\`a di Trieste, I-34127 Trieste, Italy}
\altaffiltext{9}{Istituto Nazionale di Fisica Nucleare, Sezione di Padova, I-35131 Padova, Italy}
\altaffiltext{10}{Dipartimento di Fisica e Astronomia ``G. Galilei'', Universit\`a di Padova, I-35131 Padova, Italy}
\altaffiltext{11}{Istituto Nazionale di Fisica Nucleare, Sezione di Pisa, I-56127 Pisa, Italy}
\altaffiltext{12}{Istituto Nazionale di Fisica Nucleare, Sezione di Bari, I-70126 Bari, Italy}
% ...

\begin{abstract}
To uniformly determine the properties of supernova remnants (SNRs) at high energies, we have developed the first systematic survey at energies from $1$ to $100$\,GeV using data from the \Fermi{} Large Area Telescope. Based on the spatial overlap of sources detected at GeV energies with SNRs known from radio surveys, we classify \nclassifiedsnrs~sources as likely GeV SNRs. We also report \nmarginal~marginal associations and \nULs~flux upper limits. A mock catalog in which the positions of known remnants are scrambled in Galactic longitude, allows us to determine an upper limit of \nninetyfivemockpercent{} on the number of GeV candidates falsely identified as SNRs. We have also developed a method to estimate spectral and spatial systematic errors arising from the diffuse interstellar emission model, a key component of all Galactic \FermiLat{} analyses. By studying remnants uniformly in aggregate, we measure the GeV properties common to these objects and provide a crucial context for the detailed modeling of individual SNRs. Combining our GeV results with multiwavelength (MW) data, including radio, X-ray, and TeV, demonstrates the need for improvements to previously sufficient, simple models describing the GeV and radio emission from these objects. We model the GeV and MW emission from SNRs in aggregate to constrain their maximal contribution to observed Galactic cosmic rays.
\end{abstract}


\parindent=0.5 cm

\section{Introduction}

The highly energetic nature of supernova remnants (SNRs) has been long known from evidence of nonthermal particle acceleration. 
Synchrotron emission from relativistic electrons was first detected at radio wavelengths, where SNRs have been most extensively cataloged \citep{Green91,Green04,Green09-GreensCat}. X-ray telescopes of the last three decades have detected both thermal bremsstrahlung emission, a product of gas heated by expanding blast waves, and nonthermal X-ray synchrotron emission. The nonthermal X-rays suggest a population of $\sim$~TeV electrons accelerated at the shock front \citep{Seward90,Vink12}. These multiwavelength (MW) observations from radio to X-rays have provided significant insights into SNRs as drivers of galactic evolution, as well as sources of relativistic particles. 
Yet it has been problematic to observe on-going particle acceleration in situ and determine the partitioning and flow of energy through many of these systems. 
The complexities of SNRs and their interactions with diverse environments has made it difficult to both predict properties from shock acceleration theory, e.g. specific hadronic and leptonic acceleration efficiencies, and to infer them from observations.

The origin and acceleration process(es) of cosmic rays (CRs), which are highly energetic particles mainly comprised of protons and nuclei with a small fraction ($\lesssim1\%$)  of leptons \citep{Olive14-PDG}, have remained a mystery for over $100$\,years. Energetic arguments indicate that SNRs are probable sources of Galactic hadrons even up to PeV energies due to their strong shocks \citep[e.g.][]{Helder12}. However it remains difficult to conclusively demonstrate that individual Galactic accelerators supply the Galactic CR population.

Of all the wavelengths, \g-rays offer the most readily accessible window into
energetic particles available to date due to the variety of processes producing high energy photons \citep{Stecker71,Gaisser98}. Relativistic leptons can produce \g-rays by inverse Compton (IC) scattering low energy photons or by interacting with atomic nuclei, producing bremsstrahlung radiation. Relativistic hadrons may interact with subrelativistic nuclei, creating both neutral pions which decay to two \g-rays and charged pions which decay to energetic leptons and neutrinos. 

Only recently have \g-ray telescopes obtained sufficient spatial and spectral resolution to distinguish SNR-produced high energy photons from the backgrounds. The EGRET instrument detected several Galactic sources, but was unable to unambiguously identify SNRs \citep{Sturner95,Esposito96}. Imaging air Cherenkov telescopes successfully identified extended emission from several bright SNRs at TeV energies \citep{Carrigan13-HESSgpsurvey}. However, these telescopes do not provide data across the large energy range needed to discriminate between possible emission mechanisms, nor do they provide full sky coverage. The launches of  AGILE in 2007 and \Fermi{} in 2008 finally provided the capability to unambiguously identify SNRs in \g-rays and to detect the spectral signature of accelerated protons from the brightest of them \citep{Giuliani11-W44agile,Ackermann13-pionBump}. 

Surveys in the GeV energy range have now identified hundreds of sources in the Galactic plane~\citep[e.g.][]{Nolan12-2FGL,Acero15-3FGL}, with SNRs being one of many observed source classes. Pulsars, pulsar wind nebulae (PWNe), and binaries have all been identified as \g-ray sources spatially coincident with known Galactic SNRs. Many studies with the \Fermi{} Large Area Telescope (LAT) have been able to spatially resolve extended emission from SNRs, making definite identification possible despite the plethora of potentially plausible counterparts in the Galactic plane~\citep[e.g.][]{Katagiri11-CygLoop}. Individual studies have found SNRs spanning a range of ages interacting with the ambient interstellar medium (ISM) or dense molecular clouds (MCs) \citep[e.g. as noted in][]{Thompson12-FermiCRsReview}. While these GeV SNRs display many similar characteristics, no systematic analysis has yet been undertaken. Understanding the properties of SNRs as a class of \g-ray emitters and as potential CR sources motivates this uniform study of all known SNRs in our galaxy. 

To improve our understanding of \g-ray SNR properties and SNRs' potential contribution to the Galactic CR population, we have created the first \FermiLat{} catalog of SNRs. The systematic characterization of GeV emission in regions containing known SNRs is described in Section~\ref {Sec:AnalysisMethod}, with details on the input source model in Section~\ref{Sec:AddSrcs} and a description of the general analysis method in Section~\ref{Sec:DetectMethod}. We discuss sources of systematic error in Section~\ref{Sec:SysErr} and describe our findings in Section~\ref{Sec:CatDescription}, with details of the method used for association in Section~\ref{Sec:Class}. 
We created a number of methods to allow us to uniformly address complications usually treated within the context of an individual region. Further details on these methods for iteratively adding sources to a region's model, estimating the error due to the interstellar emission modeling, and estimating the chance spatial coincidence of a GeV source, can be found in Appendices~\ref{appen:addSrcs2FGL}, \ref{appen:aIEM}, and \ref{Appen:ChanceCoinc}, respectively. 
To better understand the \g-ray properties of SNRs, we compare the \g-ray results to MW data assembled for all Galactic SNRs, including a detailed comparison with radio and TeV counterparts, in Section~\ref{Sec:GeVSNRDiscussion}. Finally, in Section~\ref{Sec:CRs} we explore whether the SNR paradigm for CR origins is consistent with our catalog results. To facilitate further study, we have provided a number of online data products, described in Appendix~\ref{Sec:CatProds}.

\subsection{The \FermiLat{} Instrument}\label{Sec:FermiLAT}

The \FermiLat{} is a pair-conversion \g-ray telescope that observes photons from $20$\,MeV to $>300$\,GeV. Launched on 2008 June 11, the default observing mode is an all-sky survey optimized to provide relatively uniform coverage of the entire sky every three hours, including the Galactic plane where most known SNRs are located. Further details of the instrument can be found in \citep{Atwood09-FermiLAT}.

\subsection{Galactic Supernova Remnants}\label{Sec:SNRSelection}

In this work we focus on the \nGalSNRs~currently known Galactic SNRs. They are derived from the $274$~SNRs noted in the catalog of \citet[hereafter Green's catalog]{Green09-GreensCat}, plus five additional SNRs identified following its publication. 
All but $16$ of these SNRs have been identified by their radio synchrotron emission, so their centroids and extensions are primarily determined from the radio. When the radio detection is not securely identified through the synchrotron emission, positional information is obtained from the optical, X-ray, or TeV observations that identified the SNR, as noted in Green's catalog. The catalog is thought to be complete down to a $1$\,GHz radio surface brightness limit of $\approx 10^{-20}$\,W\,m$^{-2}$\,Hz$^{-1}$\,sr$^{-1}$ (i.e.\,$1$\,MJy\,sr$^{-1}$). However, selection effects are known to bias radio surveys against the identification of radio faint and small angular size remnants \citep{Green04,Brogan06}. We note that as this work neared completion, a revised catalog of $294$~SNRs was published \citep{Green14-GreensCat}, representing only a small increase ($<10\%$) over the previous catalog.

We briefly describe the five SNRs added to our catalog since the publication of Green's catalog. For the purposes of this work, these are implicitly included when we refer to Green's catalog and are also in the 2014 catalog unless otherwise noted.
\begin{description}
\item[SNR G5.7$-$0.0:] 
Identified in the radio by \cite{Brogan06}, this remnant is known to be interacting with a nearby dense cloud due to the presence of OH\,($1720$\,MHz) masers \citep{2009ApJ...694L..16H}. The TeV source HESS J1800$-$240C is coincident with the SNR, though it is unclear whether the \g-ray emission is attributable to SNR G5.7$-$0.0 or escaping CRs from SNR W28 \citep{Aharonian08-w28,Hanabata14}. This SNR was included in \cite{Green14-GreensCat} as a probable SNR, but was not included in the final list of $294$~firmly identified SNRs.

\item[SNR G35.6$-$0.4:] 
Re-identified as an SNR by \cite{Green09-g35.6} but not included in Green's 2009 catalog, this is a middle-aged remnant with nearby MCs thought to lie at a distance of $3.6\pm0.4$\,kpc \citep{2013ApJ...775...95Z}. The nearby TeV source HESS J1858+020 \citep{Aharonian08-HESSJ1858+020}, has been proposed to originate from CRs escaping from the SNR and illuminating nearby clouds \citep{Paron10-G35.6}.

\item[SNR G213.3$-$0.4:] 
A very low radio surface brightness SNR initially designated as G213.0$-$0.6 by \cite{Reich03}, the SNR identification was later confirmed by optical line observations \citep{Stupar12}. The SNR lies near the \Hii\ region S284, which is coincident with the \g-ray source 2FGL J0647.7+0032. No conclusive evidence for interaction between the SNR and S284 has been presented. The X-ray source 1RXS J065049.7$-$003220 lies near the center of the SNR.

\item[SNR G306.3$-$0.9:] 
This X-ray source was first reported by \cite{Miller11} with the designation Swift J132150.9$-$633350. This is a small-diameter SNR with a radius of $110$\arcsec . X-ray observations indicate a young SNR of age $1300-4600$\,years in the Sedov phase and at a distance of $8$\,kpc \citep{Reynolds13}. The SNR also shows $24\,\mu$m emission, indicating shocked or irradiated warm dust.

\item[SNR G308.4$-$1.4:] 
This shell-type SNR was initially identified in radio surveys due to its steep radio spectral index $\alpha$=$-$0.7$\pm$0.2, and confirmed by its detection as an extended X-ray source \citep{Reynolds12}. The eastern part of the remnant shows enhanced radio, infrared and X-ray emission, which may signal the shock-wave is expanding into a denser region to the east \citep{2012AA...544A...7P,DeHorta13}. Chandra observations also revealed a bright X-ray point source near the geometrical center with a soft spectrum and putative periodicity that make it a candidate compact binary \citep{Hui12}. Given a distance estimate of $6$ to $12$\,kpc and an age of $5,000$ to $7,500$\,years for the SNR \citep{2012AA...544A...7P}, the point source and remnant may have originated from the same progenitor system.

\end{description}

\section{Analysis Methods}\label{Sec:AnalysisMethod}

To systematically analyze the \FermiLat{} \g-ray data, we apply a maximum likelihood \citep{Mattox96-Likelihood} framework to Regions of Interest (RoIs) centered on known SNRs \citep{Green09-GreensCat}. For each SNR, we begin by constructing a model for the spectral and spatial dependence of the \g-ray emission which includes significant point sources in the RoI. We then test for the existence of a \g-ray source near the center. This includes determining the most likely position and extension of the candidate source and testing for spectral curvature, rather than assuming it follows a power law across the energy range studied. In cases where we find no significant source associated with the SNR, we calculate upper limits on the flux. We calculate both statistical and systematic errors, where the latter are estimated from both the uncertainty in the effective area and the effects of changing the interstellar emission model (IEM), which accounts for \g-rays produced by CR interactions with interstellar gas and radiation fields in the Milky Way. 

This analysis uses both the standard Science Tools (version 09-32-05), including \gtlike\footnote{Available at the \Fermi{} Science Support Center: \url{http://fermi.gsfc.nasa.gov/ssc} and described in context at: \url{http://fermi.gsfc.nasa.gov/ssc/data/analysis/documentation/Cicerone/}.}, and the \ptlike{} analysis package~\citep{Kerr10-pointlike} which has been developed and verified for characterizing source extension for \FermiLat{} data \citep{Lande12-extSrcSearch}. Section~\ref{Sec:Data} describes our data selection; Section~\ref{Sec:AddSrcs} details our new method for automatically finding point sources in the \FermiLat{} \g-ray emission; and Section~\ref{Sec:DetectMethod} discusses the detection method. We examine the main sources of systematic error in Section~\ref{Sec:SysErr}.

\subsection{Data Selection}\label{Sec:Data}

This catalog was constructed using 3 years of LAT survey data from the Pass~$7$~(P7) ``Source'' class and the associated P7V6 instrument response functions (IRFs). This interval spans $36$\,months, from $2008$\,August\,$4$ to $2011$\,August\,$4$ (mission elapsed time $239557417-334108806$). The Source event class is optimized for the analysis of persistent LAT sources, and balances effective area against suppression of background from residual misclassified charged particles. We selected only events within a maximum zenith angle of $100$\degr{} and use the recommended filter string ``DATA\_QUAL==1 \&\& LAT\_CONFIG==1'' in {\tt gtmktime}\footnote{See the LAT data selection recommendations at the \Fermi{} Science Support Center: \url{http://fermi.gsfc.nasa.gov/ssc/data/analysis/documentation/Cicerone/Cicerone_Data_Exploration/Data_preparation.html}.}. 
The P7 data and associated products are comparable to  those used in the other \g-ray catalogs employed in this work. We used the first three years of science data for which the associated IEM is suitable for measuring sources with extensions $>2$\degr\footnote{See the LAT caveats, \url{http://fermi.gsfc.nasa.gov/ssc/data/analysis/LAT_caveats.html}, particularly those for the IEM developed for Pass~$7$ reprocessed data described in \url{http://fermi.gsfc.nasa.gov/ssc/data/access/lat/Model_details/FSSC_model_diffus_reprocessed_v12.pdf}.}. A detailed discussion of the instrument and event classes can be found in \cite{Atwood09-FermiLAT} and at the \Fermi{} Science Support Center\footnotemark[1].

For each of the \nGalSNRs~SNRs we modeled emission within a $10$\degr{}\,radius of the SNR's center. As a compromise between number of photons collected, spatial resolution, and the impact of the IEM, we chose $1$\,GeV as our minimum energy threshold. The limited statistics in source class above $100$\,GeV motivated using this as our upper energy limit. 

To avoid times during which transient sources near SNRs were flaring, we removed periods with significant weekly variability detected by the \Fermi{} All-sky Variability Analysis (FAVA) \citep{Ackermann13-FAVA}. We conservatively defined a radius within which a flaring source may significantly affect the flux of a source at the center. We take this distance to be the radio radius of an SNR plus $2.8$\degr, corresponding to the overall $95\%$~containment radius for the \FermiLat{} point spread function (PSF) for a $1$\,GeV photon at normal incidence \citep{LAT-instrument-paper}. The time ranges of FAVA flares within this distance were removed in $23$~RoIs, leaving $\geq 98.9\%$ of the total data in each RoI.

\subsection{Input Source Model Construction}\label{Sec:AddSrcs}

To characterize each candidate SNR we constructed a model of \g-ray emission in the RoI which includes all significant sources of emission as well as the residual background from CRs misclassified as \g-rays. We implemented an analysis method to create and optimize the \nGalSNRs~models for each of the $279$~RoIs. For each RoI, we initially included all sources within the $10$\degr{} RoI listed in the Second \FermiLat{} catalog (2FGL) \citep{Nolan12-2FGL}, based on $2$\,years of Source class data. To this we added pulsars from the LAT Second Pulsar Catalog (2PC) \citep{Abdo13-2PC}, based on $3$\,years of source class data, with 2PC taking precedence for sources that exist in both. 
For the diffuse emission we combined the standard IEM corresponding to our P7 data set, gal\_2yearp7v6\_v0.fits, with the standard model for isotropic emission, which accounts for extragalactic diffuse \g-ray emission and residual charged particles misclassified as \g-rays. Both the corresponding isotropic model, iso\_p7v6source.txt, and the IEM are the same as used for the 2FGL catalog analysis\footnote{Further details on the diffuse emission models are available at \url{http://fermi.gsfc.nasa.gov/ssc/data/access/lat/BackgroundModels.html}}.

Compared to 2FGL, we used an additional year of data and limited the energy range to $1-100$\,GeV. This can result in different detection significances and localizations than previously reported in 2FGL. To account for these effects, we recreated the RoIs' inner $3$\degr{}\,radius regions, which encompass the radio extents of all known SNRs, observed to be $\leq 2.6$\degr{} and allows a margin for the LAT PSF. The weighted average $68\%$~containment radius of the LAT PSF for events at $1$\,GeV is $\sim0.7$\degr{} \citep{LAT-instrument-paper}. We note that this implicitly assumes that an SNR's GeV extent should not be more than about an order of magnitude larger than its radio extension and also note that the selection biases stated in Green's catalog limit the range of known SNRs' radio extensions. 

To build the inner $3$\degr{}\,radius model of each RoI, we first removed all sources except identified Active Galactic Nuclei (AGN) and pulsars, whose positions on the sky are independently confirmed by precise timing measurements \citep{Abdo13-2PC}. Retained AGN were assigned their 2FGL positions and spectral model forms. Pulsars' positions and spectral forms were taken from 2PC. 2FGL sources identified or associated with SNRs are removed when they lie within the inner $3$\degr{}. 

We generated a map of source test statistic (TS) defined in \citet{Mattox96-Likelihood} via \ptlike{} on a square grid with $0.1$\degr{}\,$\times$\,$0.1$\degr{} spacing that covers the entire RoI. \ptlike{} employs a binned maximum likelihood method. The source TS is defined as twice the logarithm of the ratio between the likelihood~$\mathcal{L}_1$, here obtained by fitting the model to the data including a test source, and the likelihood~$\mathcal{L}_0$, obtained here by fitting without the source, i.e., $\textrm{TS = 2 log}(\mathcal{L}_1/\mathcal{L}_0$). At the position of the maximum TS value, we added a new point source with a Power Law (PL) spectral model:
\begin{equation}
\label{eqn:PL}
\frac{dN}{dE} = N\frac{(-\Gamma+1)E^{-\Gamma}} {E_{\rm max}^{-\Gamma+1} - E_{\rm min}^{-\Gamma+1}}
\end{equation}
where $N$ is the integrated photon flux, $\Gamma$ is the photon index, and $E_{\rm min}$ and $E_{\rm max}$ are the lower and upper limit of the energy range in the fit, set to $1$\,GeV and $100$\,GeV, respectively.
We then performed a maximum likelihood fit of the RoI to determine $N$ and $\Gamma$ and localized the newly added source. 
The significance of a point source with a PL spectral model is determined by the $\chi^2_n$ distribution for $n$ additional degrees of freedom for the additional point source, which is typically slightly less than~$\sqrt{\textrm{TS}}$\footnote{See \url{http://fermi.gsfc.nasa.gov/ssc/data/analysis/documentation/Cicerone/Cicerone_Likelihood/TS_Maps.html} for further details.}.

To promote consistent convergence of the likelihood fit, we limited the number of free parameters in the model. For sources remaining after the removal step, described above, we freed the normalization parameters for the sources within $5$\degr{} of the RoI center, including identified AGN and pulsars. For 2FGL sources between $5$\degr{} and $10$\degr{}, we fixed all parameters. The spectrum of the IEM was scaled with a PL whose normalization and index were free, as done in 2FGL. For the isotropic emission model, we left the normalization fixed to the global fit value since the RoIs are too small to allow fitting the isotropic and Galactic IEM components independently. The isotropic component's contribution to the total flux is small compared to the IEM's at low Galactic latitudes.

After localizing them, the new sources were tested for spectral curvature. In each of the four~energy bands between $1$ and $100$ GeV, centered at $1.8$, $5.6$, $17.8$ and $56.2$\,GeV, we calculated the TS value for a PL with spectral index fixed to $2$ and then summed the TS values. We refer to this as $\mathrm{TS_{band fits}}$. A value for $\mathrm{TS_{band fits}}$ much greater than the TS calculated with a PL ($\mathrm{TS_{PL}}$) suggests with a more rapid calculation that the PL model may not accurately describe the source. Analogously to 2FGL \citep{Nolan12-2FGL}, we allow for deviations of source spectra from a PL form by modeling sources with a log-normal model known colloquially as LogParabola or logP:
 \begin{equation}
\newcommand{\pfrac}[2]{\left(\frac{#1}{#2}\right)} \frac{dN}{dE} = N_0\pfrac{E}{E_b}^{-(\alpha + \beta\log(E/E_b))}
\label{eqn:logP}
\end{equation}
where $N_0$ is the normalization in units of photons/MeV, $\alpha$ and $\beta$ define the curved spectrum, and $E_b$ is fixed to $2$\,GeV\footnote{Note: $E_b$ is a scale parameter which should be set near the lower energy range of the spectrum being fit and is usually fixed, see \citet{Massaro04-logp}}. If $\mathrm{TS_{band fits} - TS_{PL}} \geq 25$, we replaced the PL spectral model with a logP model and refit the RoI, including a new localization step for the source. We retained the logP model for the source if the global \logL{} across the full band improved sufficiently: 
$\mathrm{TS_{curve}} \equiv 2 ($\logL{}$_{\mathrm{logP}}-$\logL{}$_{\mathrm{PL}}) \geq 16$. 
Otherwise we returned the source to the PL model which provided the better global \logL. Across all RoIs, less than $2\%$ of the newly added sources retained the logP model. 
 
We continued iteratively generating TS maps and adding sources within the entire RoI until additional new sources did not significantly change the global likelihood of the fit. The threshold criterion was defined as obtaining $\mathrm{TS < 16}$ for three consecutively added new sources, denoted as $\mathrm{N_{TS < 16} = 3}$. Despite iteratively adding a source at the location of the peak position in the TS map, the TS values of new sources may not decrease monotonically with iteration for several reasons. First, source positions were localized after fitting the RoI and generating the TS map. Second, some added sources were fit with a more complex spectral model than a simple PL. Finally, when creating the TS map, we fixed the source's spectral index to $2$, whereas when adding the actual source to the model, we allowed its index to vary. 

The specific value of $\mathrm{N_{TS < 16} = 3}$ was chosen to avoid missing sources with $\mathrm{TS \geq 25}$, the threshold commonly used for source detection in LAT data, and to optimize computation time. We tested the threshold by selecting eight representative SNRs from both complex and relatively simple regions of the sky, with both hard and soft spectral indices. We applied the above procedure to the test RoIs using a criterion of $\mathrm{N_{TS < 16} = 6}$ and counted how many $\mathrm{TS \geq 25}$ sources would be excluded if a smaller $\mathrm{N_{TS < 16}}$ criterion was used. Reducing the threshold to $\mathrm{N_{TS < 16} = 3}$ cut only one significant source in any of the regions. Since the maximum number of sources added in any test RoI was $38$, the minimum $14$, and the total number of sources added across all test regions was $221$, we chose to use $\mathrm{N_{TS < 16} = 3}$ for the full sample. To allow for proper convergence of the likelihood fit, we reduced the number of free parameters prior to each new source addition. If the previously added source was between $3$\degr{} and $5$\degr{} of the center of the RoI, just its normalization was freed, and if greater than $5$\degr{} all its source parameters were fixed.

To avoid having newly added sources overlap with pulsars, we deleted new sources from the RoI if they were within~$0.2$\degr{} of a \g-ray pulsar and refit the pulsar in the $1-100$\,GeV range following the 2PC conventions. 
2PC modeled pulsar spectra as PL with an exponential cutoff (PLEC),
\begin{equation}
\newcommand{\pfrac}[2]{\left(\frac{#1}{#2}\right)} \frac{dN}{dE} = N_0 \pfrac{E}{E_0}^{-\Gamma} \exp\left(-\frac{E}{E_c}\right)^{b},
\label{eqn:PLEC}
\end{equation}
where \textit{$N_0$} is the normalization factor, \textit{$\Gamma$} is the photon spectral index, \textit{$E_c$} the cutoff energy, and $b$ determines to the sharpness of the cutoff. 2PC assessed the validity of fixing $b$ to $1$ in Equation~\ref{eqn:PLEC} (PLEC1) by repeating the analysis using a PL model, as well as the more general exponentially cut off PL form, allowing the parameter $b$ in Equation~\ref{eqn:PLEC} to vary. For the pulsar spectra in this analysis, we compared the maximum likelihood values for spectral models with and without a cutoff and with and without the value of $b$ being free, via $\mathrm{TS_{cut}} \equiv 2 ($\logL{}$_{\mathrm{PLEC1}}-$\logL{}$_{\mathrm{PL}})$ and $\mathrm{TS}_{b} \equiv 2 ($\logL{}$_{\mathrm{PLEC}}-$\logL{}$_{\mathrm{PLEC1}})$ to determine which to use. If $\mathrm{TS_{cut}} < 9$ is reported for the pulsar in 2PC then a PL model is used. If TS$_{\mathrm cut} \geq 9$, we then check to see if the cutoff energy fit in 2PC lies within the restricted energy range of $1-100$\,GeV used in this work. For pulsars with cutoffs $\geq 1$\,GeV, we then use the PLEC model if TS$_{\mathrm b} \geq 9$, and the PLEC model with cutoff freed otherwise. For those pulsars with cutoffs less than 1 GeV the spectral parameters are fixed to the 2PC values.

To complete the construction of our point source RoI model, we took the output of the previous steps and removed all sources with TS~$< 16$. This final model was then used as the starting model for analyzing candidate SNR emission. We conservatively allow sources with TS down to $16$ $(\sim4\,\sigma)$ in order to account for the effects of at least the brightest sub-threshold sources on the parameter fits for the other sources in the model. Furthermore, while the SNR analysis method described in the next subsection (\ref{Sec:DetectMethod}) is allowed to remove sources, it cannot add them. Thus we start from a set of sources designed to allow the final model to capture all significant emission within the central region. To corroborate our method of systematically adding sources to a region, we compare our RoI source models with those found by the 2FGL approach in Appendix~\ref{appen:addSrcs2FGL}. 
 

\subsection{Detection Method}\label{Sec:DetectMethod}

For each SNR, we characterize the morphology and spectrum of any \g-ray emission that may be coincident with the radio position reported in Green's catalog. This was achieved by testing multiple hypotheses for the spatial distribution of \g-ray emission: a point source and two different algorithms for an extended disk. The best fit was selected based on the global likelihoods of the fitted hypotheses and their numbers of degrees of freedom. The hypothesis with the best global likelihood was then evaluated using a classification algorithm described in Section~\ref{Sec:Class} to determine whether the radio SNR could be associated with the detected \g-ray emission. 

\begin{figure}[h!]
\centering
\includegraphics[width=0.8\columnwidth]{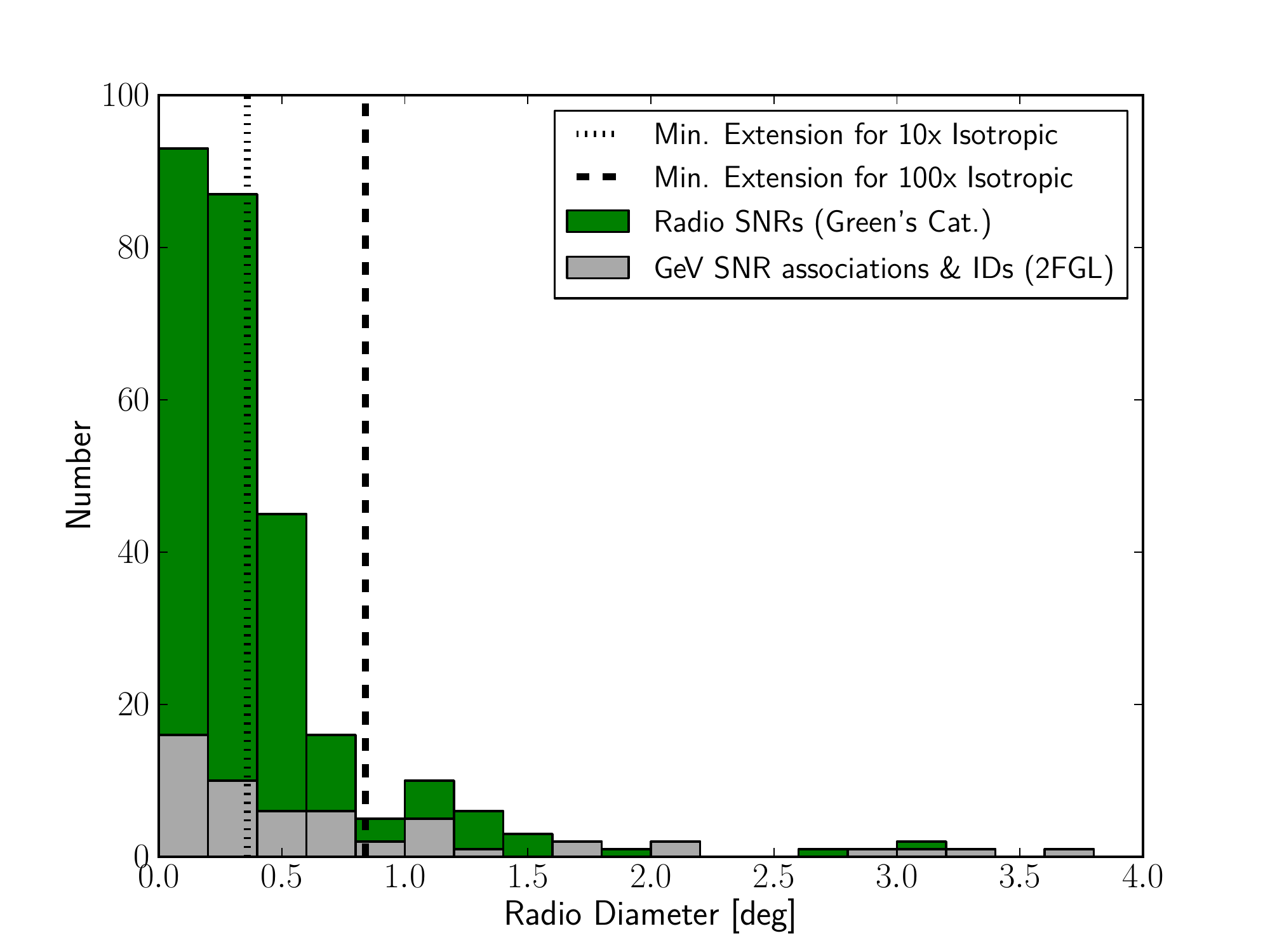} 
\caption{Distribution of SNR radio diameters from Green's catalog. The vertical dashed lines indicate the minimum detectable extension for a source with a photon flux of $10^{-8}$\,ph\,cm$^{-2}$\,s$^{-1}$ in the $1-100$\,GeV energy range and a PL index of $-2.5$, from simulations of $2$\,years of data and the P7V6 IRFs \citep{Lande12-extSrcSearch}. In that work, simulations using $10$x and $100$x the isotropic background level (thin-dotted and thick-dashed lines) are used to estimate a reasonable background range for sources in the Galactic plane.}
\label{fig:size_hist_p}
\end{figure}

Spatial coincidence is a necessary but not sufficient criterion to identify a \g-ray source with a known SNR. The detection of spatially extended \g-ray emission increases confidence in an identification, especially if GeV and radio sizes are similar, as has been observed on an individual basis for several extended SNRs \citep[e.g.][]{Lande12-extSrcSearch}. The LAT has sufficient spatial resolution to detect many Galactic SNRs as extended. Figure~\ref{fig:size_hist_p} shows the distribution of radio diameters from Green's catalog. Vertical dashed lines show the minimum detectable extension for sources with flux and index typical of those observed in this catalog, based on simulations using the P7V6 IRFs~\citep{Lande12-extSrcSearch}. The minimum detectable extension depends not only on the source's flux and spectrum, but also the flux of the background, which was estimated by scaling the average isotropic background level by factors of 10 and 100 to be comparable to the Galactic plane. As figure~\ref{fig:size_hist_p} illustrates, roughly one third of the known Galactic SNRs may be resolved by the LAT if they are sufficiently bright GeV sources.

In order to determine the best representation for each SNR, we analyzed each SNR-centered RoI using multiple hypotheses for the spatial and spectral form. We used \ptlike{}~\citep{Kerr10-pointlike} to compare PL and logP spectral forms, to compare point source versus extended source hypotheses, and to analyze the robustness of sources near the extended source.
 
For each hypothesis, we started with the input model described in Sections~\ref{Sec:Data} and~\ref{Sec:AddSrcs}. We removed sources falling within the SNR's radio disk unless they had been identified as an AGN or pulsar, as described in Section~\ref{Sec:AddSrcs}. We then proceeded to evaluate the following point and extended source hypotheses. For the point source hypothesis, a point source with a PL index initialized to $2.5$ was placed at the radio centroid of the SNR. The positions, spectral index, and spectral normalization of the point source were then fit. As for the initial input model described in Section~\ref{Sec:AddSrcs}, we tested the source for spectral curvature. To test the extended source hypothesis, we employed two separate procedures. Both employed a uniform disk model initially placed at the center of the RoI with a radius equal to that observed in the radio. In the first procedure, called the ``disk" hypothesis, we fit both the position and extension of the disk, as well as tested for spectral curvature. A second procedure, which results in a model we call the ``neardisk" hypothesis, additionally examines the significance of sources nearby the disk, removing those which are not considered independently significant and refitting the disk position and radius. This procedure is described in Section~\ref{Sec:LocExtSpec}.

Having evaluated these hypotheses, we compared the global likelihood values of the final extended hypothesis and of the point source hypothesis to determine which model had the largest maximum likelihood. If the source is significant in the best hypothesis, the model parameters are reported in Tables~\ref{Tab:ResultsSpat} and \ref{Tab:ResultsSpec}. If no hypothesis had a significant \g-ray source coincident with the radio SNR, we calculated the upper limit on the flux from a region consistent with the radio SNR, described in Section~\ref{Sec:FluxULs}, and report  the results in Table~\ref{Tab:ResultsULs}. 

\subsubsection{Localization, Extension, and Spectral Curvature}\label{Sec:LocExtSpec}

To test our hypotheses, we combined the initial model of point sources (Section~\ref{Sec:AddSrcs}) and the Galactic and isotropic diffuse contributions (Section~\ref{Sec:Data} and \ref{Sec:AddSrcs}) with a test source at the center of each RoI. All sources that fell within the radio SNR radius other than previously identified AGN or pulsars were removed, as was done for the input source model (Section~\ref{Sec:AddSrcs}). We note that multiple point sources removed within a single radio SNR radius may represent substructure within the source itself. This process conservatively assigns the majority of the flux to a single source, rather than decomposing it. We optimized the position of the test source with \ptlike{}, iteratively allowing other model parameters to vary. For all hypotheses, the normalizations of all sources within $5$\degr{} of the radio SNR center were fit while all other spectral parameters were fixed. The parameters for sources outside $5$\degr{} were also fixed.

For the point source hypothesis, a point source was placed at the radio centroid of the SNR. For the disk hypothesis, a uniform disk with radius equal to the radio radius was placed at the center. In both hypotheses, the normalization, index, and position of the candidate source were fit. For the disk hypothesis, the extension was also fit. Previous analyses of a range of possible Galactic SNR sources with similar data sets \citep[e.g.][]{Lande12-extSrcSearch} typically showed no differences in global likelihood significant enough to justify choosing a Gaussian over a uniform disk template or vice versa. In addition, there was typically little difference in spectral parameters for the two spatial forms. For simplicity and clarity, we thus test only the uniform disk hypothesis. We allowed the localization to wander up to  $5$\degr{} in the fits as a reasonable upper limit on what might later be associated with the SNR. This is roughly twice the radius of largest radio SNR.

We included an additional disk hypothesis in which we recalculated the significance of each nearby point source. Because neighboring sources can influence the best fit disk parameters, we iteratively evaluated the significance of the neighboring source by calculating TS$_{\rm nearby}$, defined as twice the difference between the model's log-likelihood (\logL{}) with the nearby point source and the model without the source, as determined by \ptlike. Starting from the fitted disk model, for each neighboring point source we refit the position, extension, normalization, and spectrum of the uniform disk after removing the source. A nearby source was considered to be significant and thus kept if TS$_{\rm nearby} \geq 9$. Each point source was evaluated individually, starting with the closest point source and extending radially outward to all sources within $1$\degr{} of the furthest edge of the SNR's radio disk. The final result of this iterative process is called the ``neardisk" hypothesis which, for cases where neighboring source(s) were removed, can have different best fit disk parameters. As a final step we refit the region with \gtlike, using the neardisk model.

We chose the best extended source hypothesis by comparing the final disk and neardisk \gtlike{} \logL{} values. Since the neardisk hypothesis can have fewer degrees of freedom, we chose the final disk hypothesis only if $2\times$(\logL{}$_{\rm disk}$-\logL{}$_{\rm neardisk}$) $\geq 9$. Otherwise, we used the neardisk model as the final extended source hypothesis, hereafter referred to as the ``disk hypothesis''.

In some cases a point source could not be localized starting at the SNR center. If the \ptlike{} localization failed to converge when starting at the SNR center, we placed the candidate at the position of the most significant source removed from within the radio SNR radius and followed the procedure outlined above. For $69$~RoIs there was either no source removed within the radio SNR or localization failed. For $31$~RoIs, the candidate found had a TS~$<1$ and was removed from the model so as not to cause instabilities in the minimization. If the disk hypotheses converged and the final candidate was significant (TS~$\geq 25$) in both the localization and spectral fits, the best extended hypothesis was selected. 

Prior to the final fit of the region, sources were tested for spectral curvature using $\mathrm{TS_{band fits} - TS_{PL}}\geq~25$. If this criterion was satisfied then we replaced the PL spectral model with a logP model and refit the RoI. The final spectral model was selected, as for the input model, by comparing the \logL{} values, in this case $\mathrm{TS_{curve}} \geq 16$, as defined in Section~\ref{Sec:AddSrcs}. Seven sources were found to be significantly better fit by a logP spectrum. To obtain final spectral parameters, we performed a final fit using the standard likelihood analysis tool \gtlike. The normalization and index parameters were constrained to lie within a physically reasonable range. 


We determined the final RoI model by selecting the most likely hypothesis based on a comparison of the \gtlike{} global \logL{} of the point source hypothesis with the most likely extended source hypothesis. An extended hypothesis was considered significantly more likely if $\mathrm{TS_{ext}}$ was $\geq 16$, where $\mathrm{TS_{ext}}$ is defined as twice the difference between the \logL{} of the final model from the disk hypothesis and that of the point source hypothesis, $\mathrm{TS_{ext}} =  2 ($\logL{}$_{\mathrm{disk}}-$\logL{}$_{\mathrm{point}})$, as in~\citet{Lande12-extSrcSearch}. Otherwise, if the point source itself had TS\,$>25$, we chose the point source hypothesis. In cases in which the optimization for the position of the point source did not converge but an extended disk was detected, we calculated the global \logL{} of the region without any source and with a point source at the center of the extended source. We then use the latter value to calculate $\mathrm{TS_{ext}}$ reported in Table \ref{Tab:ResultsSpat}. For these candidates, if the source was significantly extended in both cases, we select the extended hypothesis. If none of the criteria were met, the candidate was considered undetected and we calculated an upper limit on the flux. Both the upper limits and flux calculation are described in the following subsection.

\subsubsection{Fluxes and Upper Limits}\label{Sec:FluxULs}

Fluxes in the $1-100$\,GeV band are determined using the standard analysis tool \gtlike{} by a final fit of the model chosen to have the overall maximum likelihood characterization of the morphology and spectrum of the candidate source from the analysis detailed in Sections~\ref{Sec:DetectMethod} and \ref{Sec:LocExtSpec}. For those RoIs where no significant source was detected, we computed Bayesian upper limits on the flux using the method in described in \citet{Helene83} excluding any overlapping sources in the model that have not been identified as AGN or pulsars, as described in Section~\ref{Sec:AddSrcs}. As a spatial model we used a uniform disk equal in position and radius to that reported in Green's catalog. We assumed the spectral model to be a PL and report upper limits for indices of $2.0$ and $2.5$ at $95\%$ and $99\%$ confidence levels. The choice of indices was motivated by the distribution of PL indices for classified sources, those passing the most stringent $0.4$ thresholds (see Section~\ref{Sec:Class} and index distribution in Figure~\ref{fig:GeVFluxGeVIndex}), which have an approximately Gaussian distribution with a mean of~$2.5$ and a standard deviation of~$0.5$. The results are reported in Section~\ref{Sec:TheCatalog}.

\subsection{Sources of Systematic Error}\label{Sec:SysErr}
For sources with significant emission, we estimated the systematic error propagating from the systematic uncertainty of the effective area and from the choice of IEM. For the former, we propagated the error using the standard bracketing IRF procedure, described in Section~\ref{Sec:EffAreaSysErr}. For the latter, we developed a new method in which we vary the underlying IEM, described in Section~\ref{Sec:IEMSysErr} with further details in Appendix~\ref{appen:aIEM}. As we take the effective area and underlying IEM systematic errors to be independent, when we can evaluate both components of the systematic uncertainty, we added them in quadrature and report them in Section~\ref{Sec:CatDescription}. We briefly compare the total systematic and statistical errors in Section~\ref{Sec:SysvStatErr}.

\subsubsection{Effective Area Systematic Error}\label{Sec:EffAreaSysErr} 

Following the standard method \citep{LAT-instrument-paper}, we estimated the systematic error associated with the effective area by calculating uncertainties in the IRFs which symmetrically bracket the standard effective area. The final spectral fit of each candidate's region was performed with each of the bracketing IRFs, as these changes primarily affect the spectral fit and have a minimal effect on the source localization and extension. 

To estimate the systematic error on the spectral normalization, we created bracketing IRFs which uniformly scaled the effective area to its maximal systematic values. Using a $(E^2 - E_0^2)/(E^2 + E_0^2)$ function smoothly switches the effective area between its maximal systematic values at the pivot energy, providing a better estimate of the spectral index's systematic error. The pivot energy $E_0$ is defined as the energy at which the error on the differential flux is minimal and the errors on the spectral index and flux normalizations are decorrelated. For each detected candidate, the pivot energy was calculated from the \gtlike{} fit's covariance matrix. 
For PL spectra, the pivot energy can be calculated as:
\begin{equation}
\log(E_0) = \frac{\log(E_1)+\log(E_2)}{2} + \frac{C_{K\Gamma}}{K C_{\Gamma\Gamma}}
\end{equation}
where $E_1$ and $E_2$ are the end points of the energy range, $\Gamma$ the PL index defined as in Equation \ref{eqn:PL}, and $K$ the integral of the source spectrum. $C_{\Gamma\Gamma}$ and $C_{K\Gamma}$ are the covariance matrix terms associated with the index and normalization, respectively.  For the logP sources, as the covariance matrix for spectral models with $\geq 3$ parameters requires a more complex transformation than a simple shift in reference energy and our energy range is relatively small, we estimate these sources' pivot energies from the covariance matrices for the best fit PL models. Since the standard IEM was constructed from the data using the standard IRF, we only used the standard IRF with the IEM, rather than the bracketing IRFs, while the remaining components were fit with the bracketing IRFs. 

Estimates of the systematic error on the candidates' flux and index due to the effective area's systematic uncertainty are reported in Table~\ref{Tab:ResultsSpec} of Section~\ref{Sec:TheCatalog}. Effective area systematic errors were not calculated for the candidates detailed in Section~\ref{Sec:NotSNRs}, identified as not SNRs, and for the candidate for SNR G5.2$-$2.6 as the index remained at the extreme value allowed in the fit and thus poorly determined, as noted in Section~\ref{Sec:CatCaveats}.

\subsubsection{Systematic Error from the Choice of IEM}\label{Sec:IEMSysErr}

Interstellar emission contributes substantially to LAT observations in the Galactic plane, where the majority of SNRs are located. Moreover, interstellar \g-ray emission is highly structured on scales smaller than the RoIs typically used for this analysis. To explore the systematic effects on SNRs' fitted properties caused by interstellar emission modeling, we have developed a method employing alternative IEMs. By comparing the source analysis results using these alternative models to the results obtained with the standard IEM, we can approximate the systematic uncertainty. An earlier version of this method was described in \cite{dePalma13-AltIEMSystematics_FSymp}.

The alternative IEMs were built using a different approach than the standard IEM. The work in \citet{Ackermann12-aIEMs}, using the GALPROP\footnote{\url{http://galprop.stanford.edu/}} CR propagation and interaction code, was the starting point for our alternative IEM building strategy. We varied the values of three input parameters that were found to be the most relevant in modeling the Galactic plane: CR source distribution, height of the CR propagation halo, and \hi{} spin temperature \citep{Ackermann12-aIEMs}. In this way, we obtained eight alternative IEMs. The models were constructed to have separate templates for emission associated with gas traced by \hi{} and CO in four Galactocentric rings and an IC template covering the full sky. By allowing separate scaling factors for these different components of the model, we allowed many more degrees of freedom in fitting the diffuse emission to each RoI.

For each candidate SNR we considered two hypotheses: the point source and that preferred by the previous fit between the disk and neardisk hypotheses, as described in Section~\ref{Sec:LocExtSpec}. In both cases we started from the output model of the previous analysis so the fitted parameters are as close as possible to their best values, replacing the standard IEM with the alternative ones.
For each candidate SNR we performed independent fits for each hypothesis for each of the eight alternative IEMs as well as the standard IEM, for a total of $18$ fits of the region. For each of the IEMs we chose the best extension model using the method described in Section \ref{Sec:LocExtSpec}. Appendix~\ref{appen:aIEM} contains further details on the alternative IEMs and their use in the likelihood analysis.

For each fitted parameter $P$ we obtain a set of $M$\,$=8$ values $P_i$ that we compare to the value obtained with the standard model $P_\mathrm{STD}$. Our estimate of the systematic uncertainty on $P$ due to the modeling of interstellar emission is:
\begin{equation}
\label{eq:sys_err_weighted}
 \sigma_\mathrm{sys,w}=\sqrt{\frac{1}{\sum^M_i w_i} \sum^M_i w_i(P_i-P_\mathrm{STD})^2},
\end{equation}
where the weights are:
\begin{equation}
\label{eq:sys_weight}
 w_i=1/\sigma_i^2.
\end{equation}
In Equation~\ref{eq:sys_weight}, $\sigma_i$ is the statistical error of a parameter with a particular alternative IEM, used as a weight in Equation~\ref{eq:sys_err_weighted}. We use the parameter value $P_\mathrm{STD}$  for the standard model calculated identically to that of the alternative IEMs, to ensure congruity given the necessary differences in degrees of freedom between this error analysis and the standard analysis.
To estimate the systematic errors on the extension, we substitute $0.2$\degr{} for $\sigma_i$ in calculating the weight when the point hypothesis is preferred, as a proxy for the smallest extension resolvable by the LAT for this analysis \citep[][and further discussed in Section~\ref{Sec:SpatCoinc}]{Lande12-extSrcSearch}. In cases where the best hypothesis with all the alternative IEMs is the point hypothesis, we report this rather than an extension error estimate. 

We weighted the parameters by their statistical error to prevent those with values statistically compatible with the other alternative IEMs' parameters from causing overly large systematic errors. This was particularly important for the spectral index of a candidate. We exclude from our error calculation cases when during the likelihood fit, the index was at or close to the upper or lower limit of its allowed range of variation, indicating a fit convergence problem, further described in Section~\ref{Sec:CatCaveats}. In the cases where the index was at the limit, we fixed it to $2.5$ and fit a flux which is used in the flux limit calculation. We tabulate these and other cases where the fit did not converge in the third column of the ``Alt IEMs Effect'' section in Tables~\ref{Tab:ResultsSpat} and~\ref{Tab:ResultsSpec}. We discuss the implications of the convergence problems further in Section~\ref{Sec:CatCaveats}.

The mathematical minimum required number of alternative IEMs with solutions for a given parameter to calculate the average and the standard deviation and thus the systematic error in Equation~\ref{eq:sys_err_weighted} from the choice of IEMs is two. We include in the quoted error for all parameters listed in Tables~\ref{Tab:ResultsSpat} and~\ref{Tab:ResultsSpec}, those which satisfy the mathematical minimum (two). Care should be used with all candidates for which the alternative IEMs had convergence problems. As IEM substructure can affect the significance of the final extension measured as well as the extension itself, we expect changes in the extension hypothesis, such as seen for SNR G296.5$+$10.0, which is only just above the extension threshold when using the standard IEM. Such changes are also reflected in the size of the systematic error on the flux.

We note that this strategy for estimating systematic uncertainty from interstellar emission modeling does not represent the complete range of systematics involved. In particular, we have tested only one alternative method for building the IEM and varied only three of the input parameters. This ensemble of models therefore cannot be expected to encompass the full uncertainty associated with the IEM. Further, as the alternative method differs from that used to create the standard IEM, the parameters estimated with the alternative IEM may not bracket the value determined using the standard IEM. Our estimate of the systematic error in Equation~\ref{eq:sys_err_weighted} accounts for this. Moreover, the estimated uncertainty does not contain other possibly important sources of systematic error in the definition of the IEMs (see Appendix~\ref{appen:aIEM} for details). While the resulting uncertainty should be considered a limited estimate of the systematic uncertainty due to interstellar emission modeling, rather than a full determination, it is critical for interpreting the data. This work represents our most complete and systematic effort to date.

\section{The \FermiLat{} SNR Catalog}\label{Sec:CatDescription}

We determined the \g-ray characteristics of the candidate sources following the analysis method described in Section~\ref{Sec:AnalysisMethod}. To estimate the probability that the \g-ray candidate is associated with the SNR, we quantified their spatial overlap, discussed in Section~\ref{Sec:Class}. As spatial coincidence can lead to associations with non-SNR sources, particularly in the rich Galactic plane, we estimated the rate of false discovery (Section~\ref{Sec:ChanceCoinc}). Section~\ref{Sec:TheCatalog} contains all candidates' GeV properties as well as upper limits on the detected flux for all remaining known SNRs. It also includes a discussion of newly detected sources that are likely to be associated with an SNR. Section~\ref{Sec:Validation} details verifications that our automated analysis completed successfully for all RoIs and compares the results to other existing LAT analyses, finding the expected agreement.


\subsection{Source Classification}\label{Sec:Class}

In the past, \g-ray sources have been associated with radio SNRs based on characteristics including spatial coincidence, the lack of variability or pulsation, and spectral form. The degree of spatial coincidence is generally defined in terms of positional coincidence and size or morphology. In order to test for GeV emission associated with SNRs from Green's catalog, we searched for GeV emission in the region of each SNR (Section~\ref{Sec:AnalysisMethod}) and used the spatial overlap between the radio SNR and \g-ray candidate to evaluate the probability that the same source gives rise to the emission in both bands. Other SNR MW properties, such as evidence for interaction with MCs or the presence of non-thermal X-ray sources, can also help to identify counterparts. However, these tracers are incomplete and exhibit strong selection effects. Consequently, we do not use them in this general study. 

By including all identified AGN and pulsars in our models of the GeV emission using the 2FGL catalog and 2PC, we exclude these known sources from being identified as SNRs (Section~\ref{Sec:AddSrcs}). 
We also verified that candidates were not associated with already identified \g-ray sources, including binaries, pulsars, or PWNe (see Section~\ref{Sec:NotSNRs}). We further removed all time periods when a flaring source may affect our GeV SNR candidates (Section~\ref{Sec:Data}).

\subsubsection{Spatial Coincidence}\label{Sec:SpatCoinc}

To define association probabilities for the candidates, we compared spatial information available from Green's catalog, namely the SNRs' radio positions and radii, with that of the GeV candidates' localizations, localization errors, and extensions. When a GeV candidate is significantly extended, we use the maximum likelihood disk radius as the measure of the extension (see Section~\ref{Sec:LocExtSpec}).

\begin{figure}[ht]
  \centering
  \includegraphics[width=0.495\columnwidth]{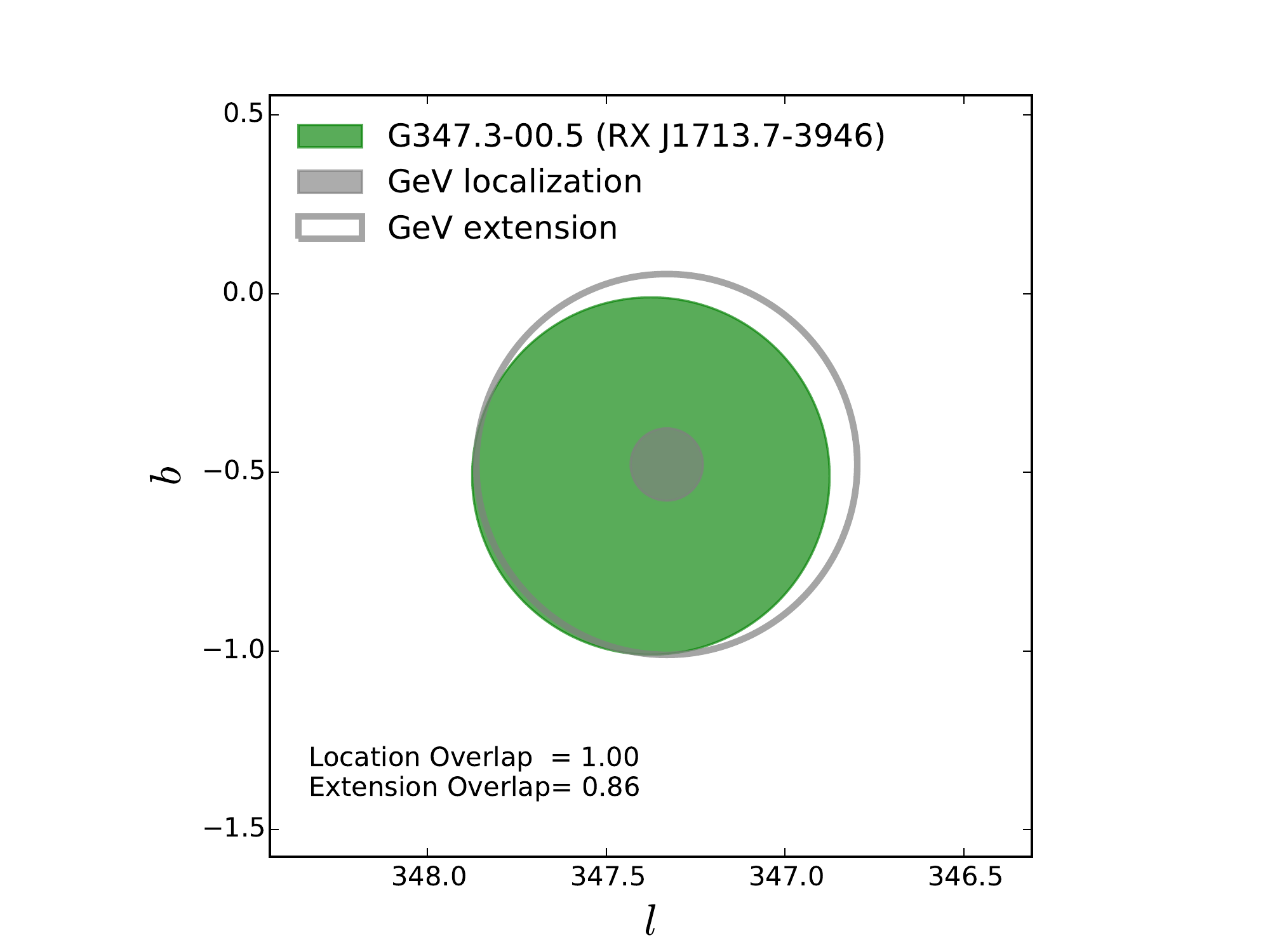} 
  \includegraphics[width=0.495\columnwidth]{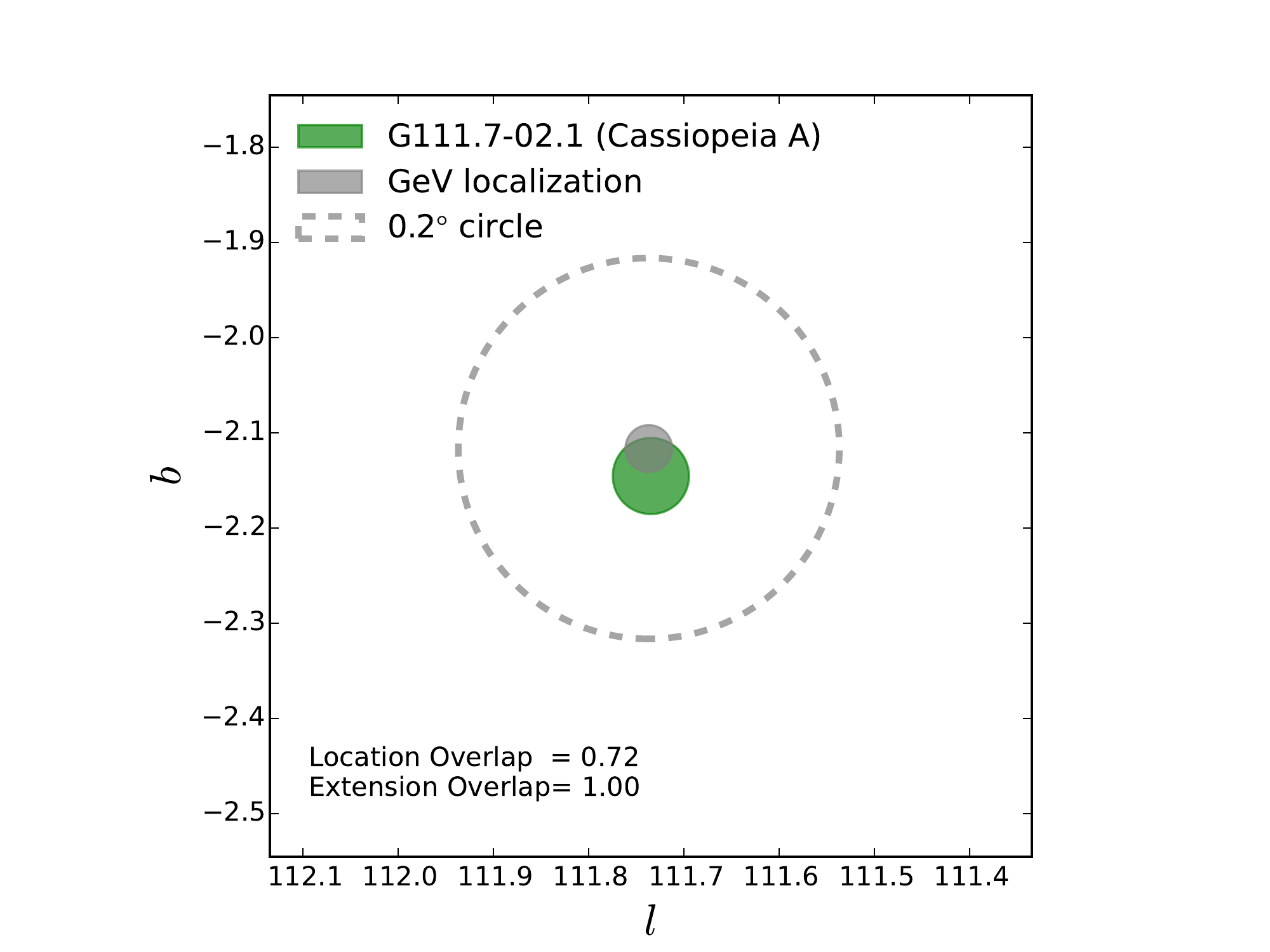} 
   \caption{Illustration of the overlap method for an extended GeV candidate (G347.3$-$0.5, left) and a point GeV candidate (G111.7$-$2.1, right). In both cases, the center of the GeV emission is located within the SNR's radio boundaries (shown in green), i.e. the parameter $\overlaploc$ is close to unity. In addition, the GeV extension is compatible with the SNR's radio disk, with $\overlapext$ close to unity. For a point GeV candidate (right panel), the radio extent is compared to the GeV minimum resolvable radius (here taken as $0.2$\degr; see text for further details).
}
  \label{fig:Overlap-example}
\end{figure}

Using this position and extension information, we derived two parameters. The first, called $\overlaploc$, provides a quantitative measure of whether the GeV localization is within the SNR's angular extent in radio. This parameter, ranging from $0$ to $1$, is calculated as:
\begin{equation}\label{eq:overlaploc}
 \overlaploc=\frac{\overline{Radio \cap
GeV_\mathrm{loc}}}{\min(\overline{Radio},\overline{GeV_\mathrm{loc}})}
\end{equation}
where $Radio$ represents the SNR's radio disk and $GeV_{\rm loc}$ is the GeV $95\%$ error circle.  The notation $\overline{X}$ represents the area of $X$. The second parameter, $\overlapext$, quantifies whether the GeV candidate's localization and extension is consistent with the location and extension of the radio SNR:
\begin{equation}\label{eq:overlapext_ext}
 \overlapext=\frac{\overline{Radio \cap
GeV_\mathrm{ext}}}{\max(\overline{Radio},\overline{GeV_\mathrm{ext}})}.
\end{equation}
where $Radio$ is again the SNR's radio disk and  $GeV_{\rm ext}$ is the best fit GeV disk if the GeV candidate is significantly extended. If the GeV detection is consistent with a point source, we determined if the corresponding SNR's radio size was consistent with it by redefining the $\overlapext$ parameter as: 
\begin{equation}\label{eq:overlapext_pt}
 \overlapext=\frac{\overline{Radio \cap GeV_\mathrm{min}}}{\overline{Radio}}.
\end{equation}
where $GeV_\mathrm{min}$ is the minimum resolvable radius, $GeV_\mathrm{min} \equiv 0.2^\circ$ for this analysis.
Illustrations of the $\overlaploc$ and $\overlapext$ parameters in the cases of an extended and a point GeV detection are shown in Figure~\ref{fig:Overlap-example}.

\begin{figure}[h!]
  \centering
  \includegraphics[width=0.8\columnwidth]{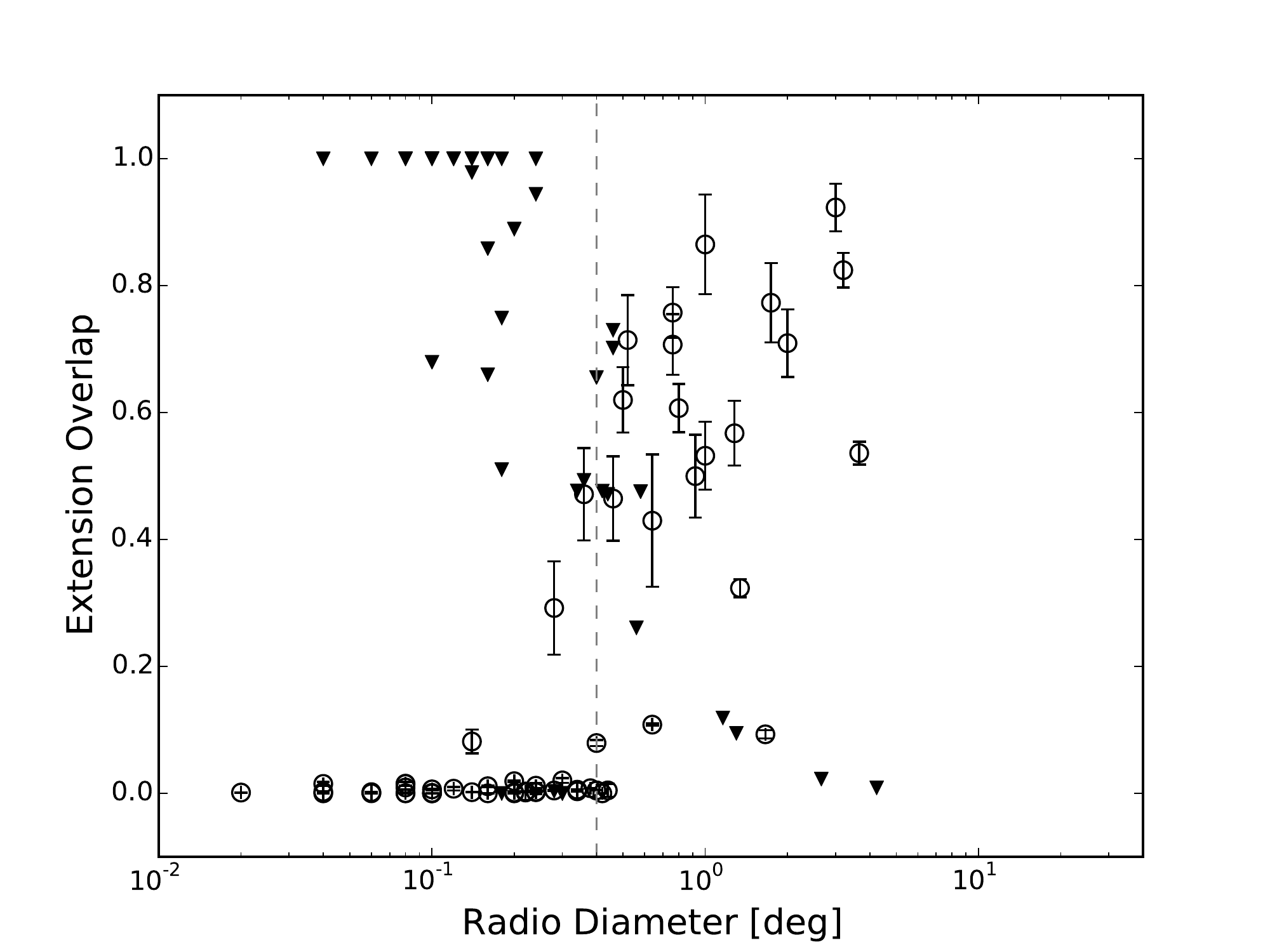} 
  \caption{The extension overlap as defined in Equations~\ref{eq:overlapext_ext} and \ref{eq:overlapext_pt} for all sources with significant GeV emission, plotted as a function of the radio diameter. The GeV candidates with significant extension are shown as open symbols; those consistent with the point hypothesis are filled. The vertical dashed line represents the minimum resolvable GeV diameter ($\equiv0.4$\degr) used in Equation~\ref{eq:overlapext_pt}.} 
  \label{fig:radiusOverlapExt}
\end{figure}

The value adopted for $GeV_\mathrm{min}$ is close to the smallest angular extension measured in our sample (Table~\ref{Tab:ResultsSpat}, candidate for SNR~G$0.0+0.0$). 
The minimum resolvable radius ($GeV_\mathrm{min}$) in our analysis is also visible in Figure~\ref{fig:radiusOverlapExt}, and also effects the separation in extension overlap below the minimum radius. In addition, Monte Carlo simulations presented in \citet[][Section~4]{Lande12-extSrcSearch} showed that, for a similar data set with \,$E>1$\,GeV, the extension detection threshold is $0.2-0.3$\degr{}, depending on the source's spectral index and the diffuse background level. 

\begin{figure}[ht]
  \centering
  \includegraphics[width=0.8\columnwidth]{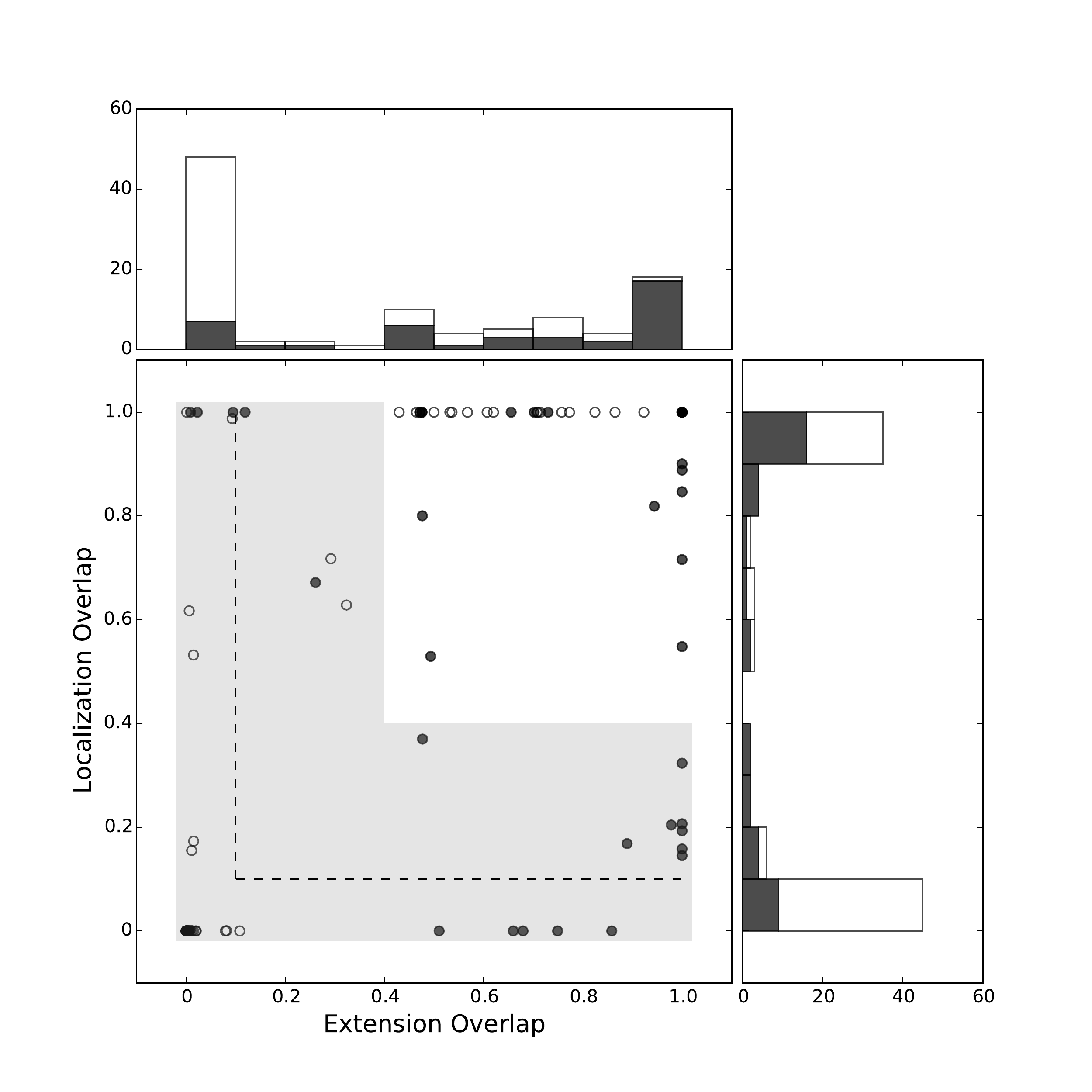} 
  \caption{Distribution of $\overlaploc$ and $\overlapext$ as defined in Equations~\ref{eq:overlaploc} and \ref{eq:overlapext_ext} for all significant GeV sources. GeV candidates with significant extension are shown as open circles and GeV candidates consistent with the point hypothesis are filled. Stacked bar histograms of both parameters are also shown. The classified candidate region discussed in the main text is shown with a white background. Those points in the grey region to the upper right of the dashed line are marginally classified candidates.}
  \label{fig:OverlapLocExt}
\end{figure}

\begin{figure}[h!]
  \centering
  \includegraphics[width=0.6\columnwidth]{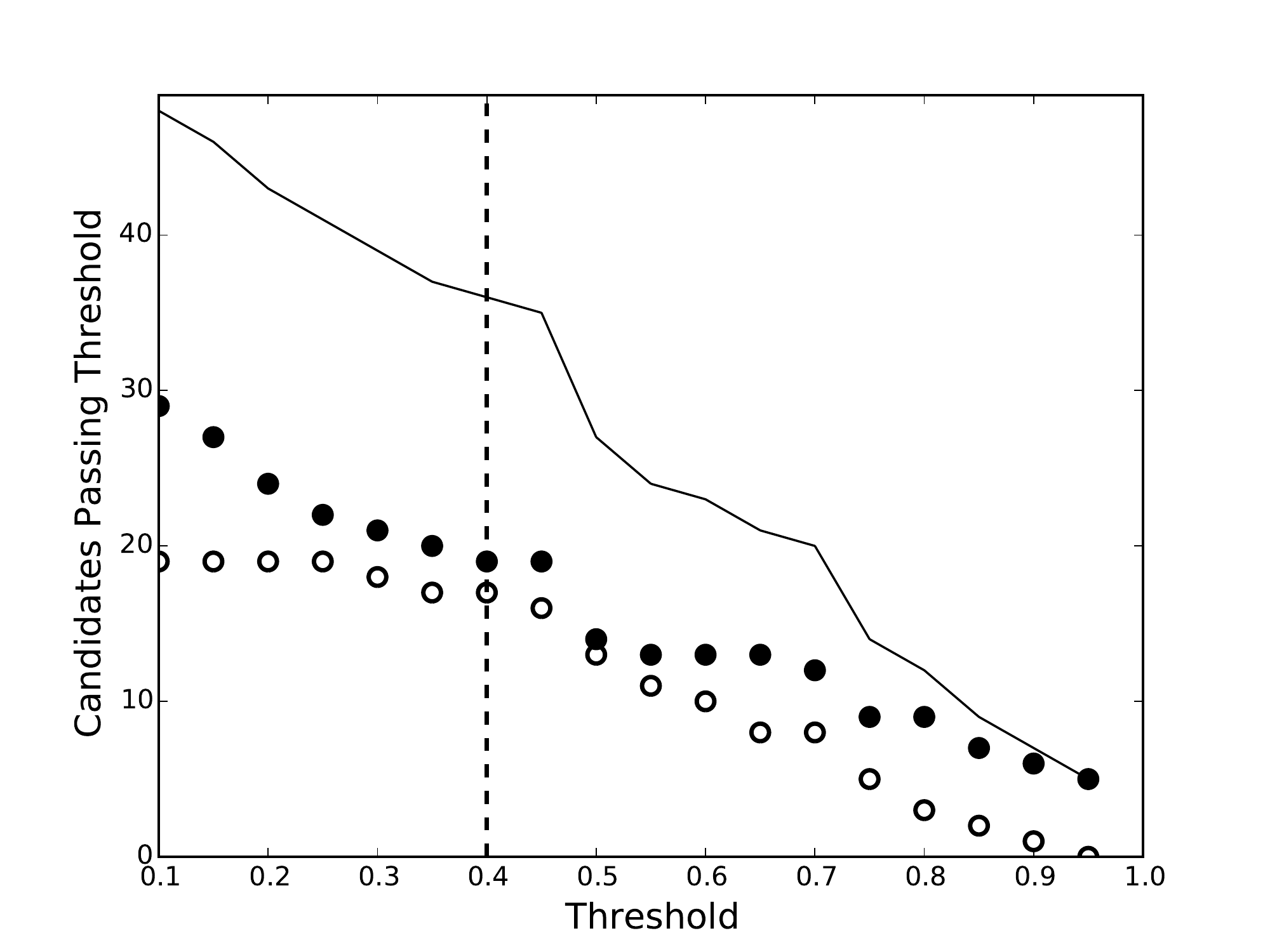} 
  \caption{Numbers of GeV candidates passing the selection criteria as a function of the threshold, defined as $\overlapext$ and  $\overlaploc >$\,threshold. The GeV candidates with significant extension are shown as open symbols; those consistent with the point hypothesis are filled. The solid line indicates the sum of the two populations. 
The vertical dashed line indicates the threshold value above which sources are classified as SNRs, discussed in Section~\ref{Sec:ChanceCoinc}.
  }
  \label{fig:Ngreen_threshold}
\end{figure}

When comparing radio SNRs with GeV candidates, these overlap parameters are used to require that both the GeV centroid is within the SNR's radio area and that their extensions are comparable. The distributions of the $\overlapext$ and $\overlaploc$ parameters for all GeV detections are shown in Figure \ref{fig:OverlapLocExt}. 
While all GeV candidates are listed in Table~\ref{Tab:ResultsSpat}, we label the GeV detections with the most likely chance of true association as ``classified candidates,'' defined as those sources with $ \overlapext > 0.4$ and $\overlaploc > 0.4 $. ``Marginally classified candidates'' are those GeV sources with a moderate chance of true association, defined as $\overlapext > 0.1$ and $\overlaploc > 0.1$ and at least one overlap estimator $<0.4$. Candidates which have overlap parameters outside of these categories are referred to as ``other'' sources. The number of classified GeV candidates as a function of overlap threshold value is shown in Figure~\ref{fig:Ngreen_threshold} while the choice of threshold value is discussed in the context of chance spatial coincidence in Section~\ref{Sec:ChanceCoinc}.

\FloatBarrier

\subsubsection{Chance Coincidence}\label{Sec:ChanceCoinc}
As we rely on spatial overlap for our GeV classification (see Section~\ref{Sec:Class}), we estimate the probability that any particular coincidence occurs by chance. In order to do so, we created a mock SNR catalog derived from Green's catalog with SNR positions randomized in longitude, while retaining latitude and extension information as well as a sufficiently similar distribution of diffuse fluxes under the mock and Green's SNR positions. Performing the standard analysis on the mock catalog should produce no classified or marginally classified candidates; any found would therefore be by chance. While this technique gives no information about whether any particular overlap is real, it does allow us to estimate the global rate of false discovery. Details of the mock catalog construction and false discovery rate derivation are in Appendix~\ref{Appen:ChanceCoinc}.

\begin{figure}[h!]
  \centering
  \includegraphics[width=0.6\columnwidth]{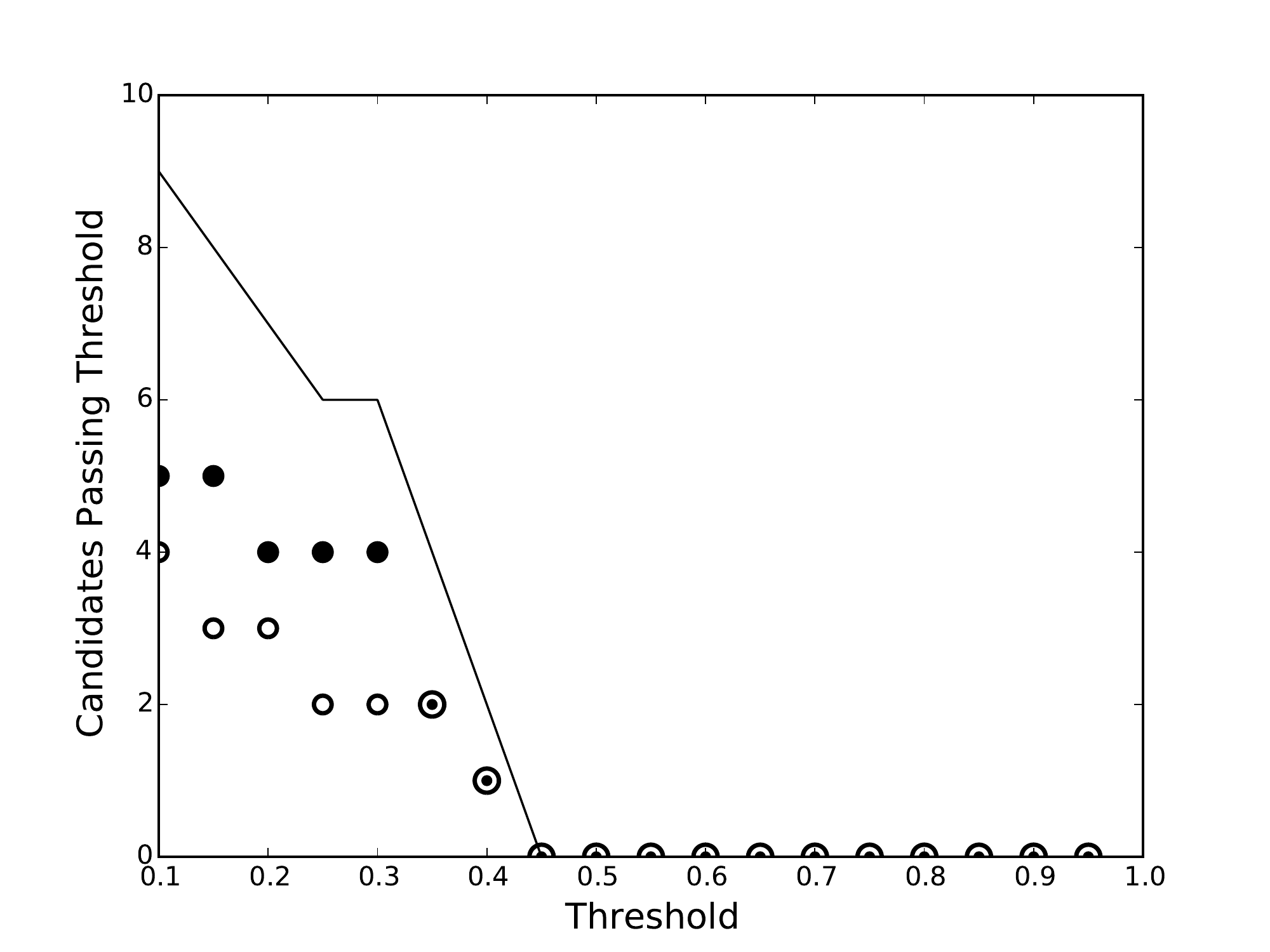} 
  \caption{Number of GeV coincidences from the mock SNR catalog ($N_{\rm mock}$) as a function of threshold value. The mock candidates with significant extension are shown as open symbols; those consistent with the point hypothesis are filled. The solid line indicates the sum of the two populations. 
}
  \label{fig:Nmock_threshold}
\end{figure}

After running the standard analysis on a mock catalog and applying the overlap classification, we find that only \nmock~out of \nGalSNRs~mock SNRs are spatially coincident with a GeV excess, as shown in Figure~\ref{fig:Nmock_threshold}, given the classification criteria in Section~\ref{Sec:SpatCoinc}. Comparing the $N_{\rm mock}=$\,\nmockN~mock coincidences to the $N_{\rm Green}=$\,\nassocprobclassified~candidates passing the association probability threshold prior to removing known sources and accounting for IEM systematics (described in Section \ref{Sec:TheCatalog}), we estimate a false discovery rate of \nmockpercent{} for this specific realization of the mock catalog. Including the sources which meet the more lenient marginal classification criteria gives \nmockclassifiedandmarginal~mock coincidences compared to \nclassifiedandmarginal~real, for a false discovery rate of~$\sim$\nmockclassifiedandmarginalpercent{} with this realization. 

Analyzing many more realizations of the mock catalog would improve our understanding of the result, but is prohibitively expensive in CPU time. However, using the result of this specific realization, we can estimate an upper limit on the number of false discoveries for any trial (see Appendix~\ref{Sec:ULmock}). We determined that, at $95\%$~confidence, the number of false discoveries will be less than~\nninetyfivemock{} for any mock catalog prepared as described above, corresponding to an upper limit of~\nninetyfivemockpercent{} for the false discovery rate. With the marginally classified mock candidates, the $95\%$ confidence upper limit is \nninetyfivemockclassifiedandmarginal~mock coincidences, or \nninetyfivemockclassifiedandmarginalpercent~false discovery rate for marginally classified candidates.

\begin{figure}[h!]
  \centering
  \includegraphics[width=0.6\columnwidth]{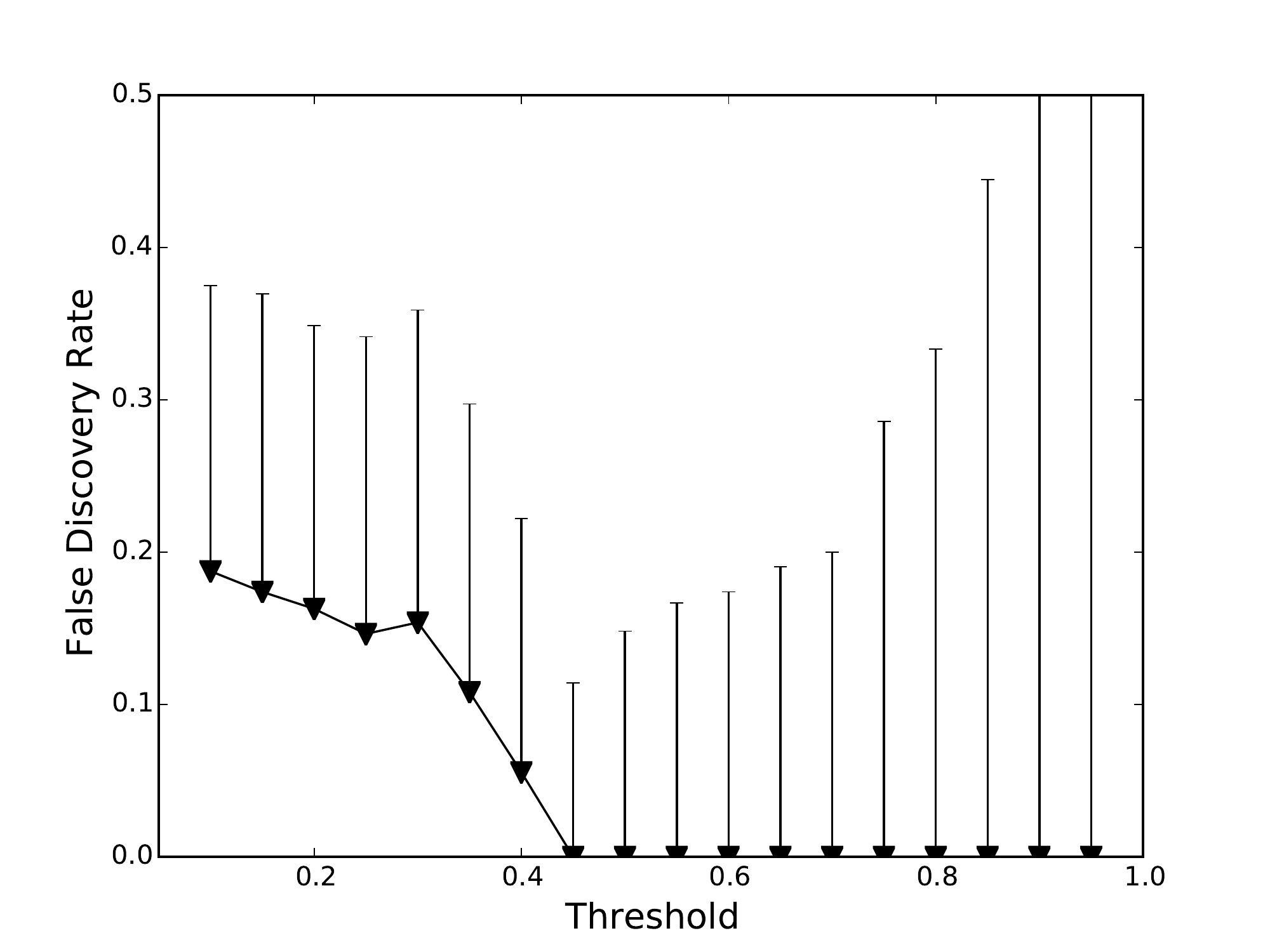} 
  \caption{ The numbers of coincidences in the mock catalog $N_{\rm mock}$ can be used to estimate the false discovery rate ($\equiv N_{\rm mock} / N_{\rm Green}$, triangles) for a given threshold value. The solid error bars show the $95\%$~upper limits on the false discovery rate (see Section~\ref{Sec:ChanceCoinc}). The apparent increase in upper limit for thresholds larger than $\sim0.45$ is because no mock sources pass the threshold (Figure~\ref{fig:Nmock_threshold}) while the number of GeV candidates $N_{\rm Green}$ decreases and remains greater than zero (Figure~\ref{fig:Ngreen_threshold}). 
}
  \label{fig:rate_threshold}
\end{figure}

The evolution of the false discovery rate, $N_{\rm mock} / N_{\rm Green}$, for classified candidates as a function of threshold is represented in Figure~\ref{fig:rate_threshold}. The estimated rate and the $95\%$ confidence upper limit decrease through a threshold of $\sim0.45$. The increase in rate and upper limit for thresholds $\gtrsim0.45$ is because no mock sources pass the threshold (Figure~\ref{fig:Nmock_threshold}) while the number of GeV candidates $N_{\rm Green}$ decreases and remains greater than zero (Figure~\ref{fig:Ngreen_threshold}). Thus, we conservatively take $0.4$ for our overlap threshold, noting its relatively small impact on our final conclusions; see Section~\ref{Sec:Class} for further discussion.

\FloatBarrier

\subsection{Catalog Results}\label{Sec:TheCatalog}

We detected \ndetected~candidates with a final source TS\,$>25$ in the \nGalSNRs~SNR RoIs (see Section~\ref{Sec:DetectMethod}). Of the \ndetected~detected candidates, \nassocprobclassified{} passed the association probability threshold (Section~\ref{Sec:SpatCoinc}). Of these, \nclassifiedsnrs~SNRs ($\sim11\%$ of the total) show significant emission for all alternative IEMs and are classified as likely GeV SNRs. An additional \nnotsnrs{} were identified as sources which are not SNRs, as detailed in Section~\ref{Sec:NotSNRs}; \nmarginalAIEM~other candidates were demoted to marginal due to their dependence on the IEM, as described in the next paragraph. Of the sources likely to be GeV SNRs, \nextended~show evidence for extension (TS$_{ext} > 16$). Only sources associated with SNRs G34.7$-$0.4 and G189.1+3.0 show evidence of significant spectral curvature in the $1-100$\,GeV range and are fit with logP spectra. Of the classified candidates, \nnewextended~extended and \nnewpointlike~point SNRs are new and published here for the first time. We describe the \nnewextended~new extended SNRs, G24.7+0.6, G205.5+0.5, G296.5+10.0, and G326.3$-$1.8, in Section~\ref{Sec:NewSNRs}. 

\begin{figure}[h!]
\centering
\includegraphics[width=0.8\columnwidth]{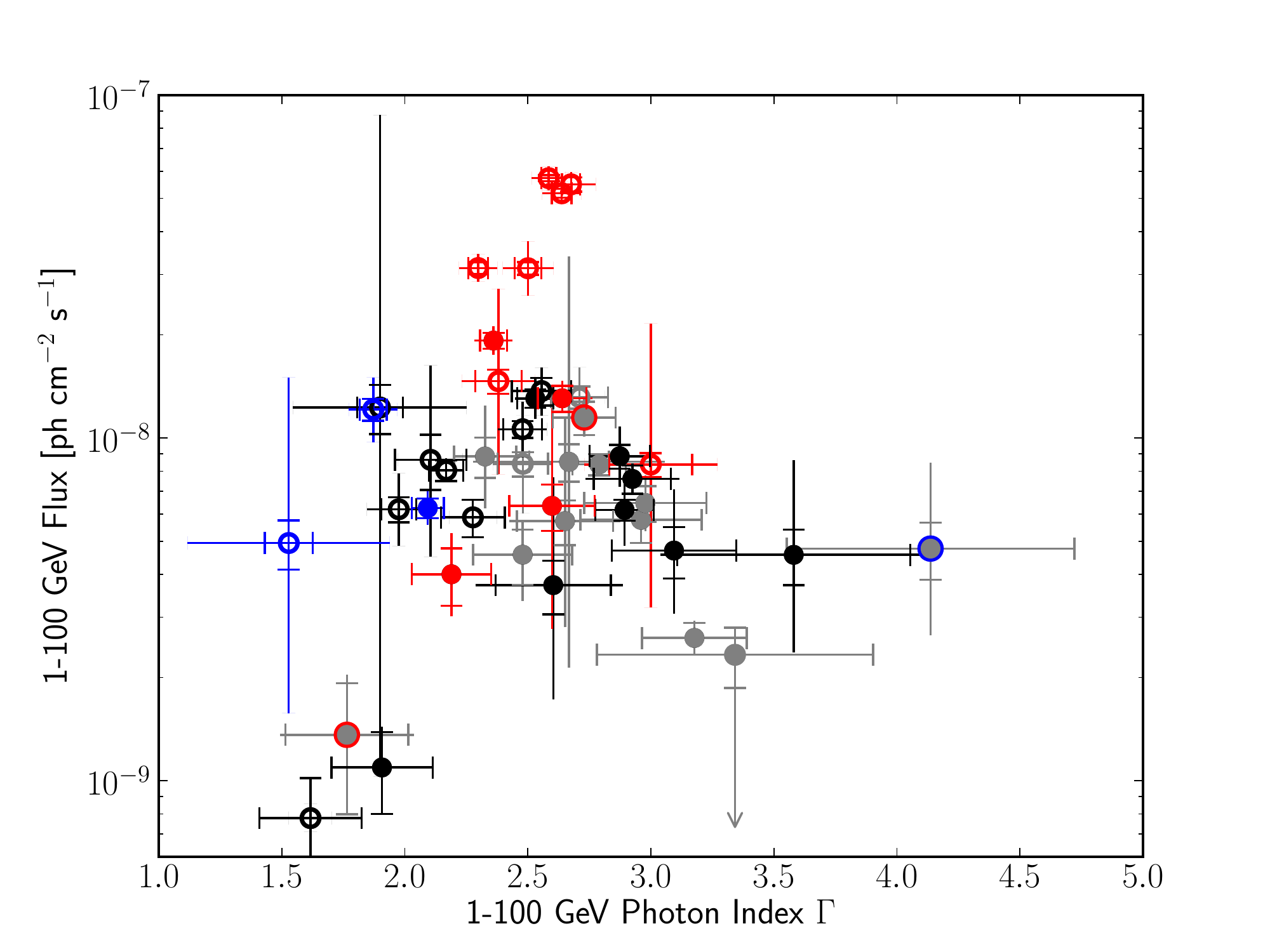} 
\caption{
The distribution of fitted photon index and flux in the energy range $1-100$\,GeV. The index shown for sources for which the logP form is more significant is determined from re-fitting the sources with a PL spectral form rather than their parabolic index $\alpha$, for consistency. Open circles indicate extended SNRs while filled circles indicate point-like sources. All SNRs that passed classification are shown as black unless also classified as young non-thermal X-ray SNRs (blue) or as interacting with MCs (red). Candidates which did not pass classification but still had both fractional overlaps $>0.1$ are grey. If they are also young or interacting, they are outlined in blue or red, respectively\protect\footnotemark. These classes are further defined in Section~\ref{Sec:GeVSNRDiscussion}. Statistical error bars have caps; error bars without caps represent the systematic error, described in Section~\ref{Sec:SysErr}. 
\label{fig:GeVFluxGeVIndex}}
\end{figure}

The results of our spatial and spectral analyses for all significant sources are reported in Tables~\ref{Tab:ResultsSpat} and~\ref{Tab:ResultsSpec}. These include the \nmarginal~candidates whose classifications as SNRs were marginal (Section~\ref{Sec:SpatCoinc}) and those demoted to marginal based on their lower significance with at least one alternative IEM (Section~\ref{Sec:IEMSysErr}). The candidates associated with SNRs G73.9+0.9 and G32.4+0.1 were demoted this way. Two of the marginal candidates showed evidence of extension; none showed evidence of spectral curvature over the energy range studied. The remaining \nother~candidates had little spatial overlap with the radio SNR: $\overlaploc$ and $\overlapext < 0.1$. Being significant sources, notably within $5$\degr{} of known SNRs, their parameters are also reported in Tables~\ref{Tab:ResultsSpat} and \ref{Tab:ResultsSpec}, but they are not considered in the discussion of GeV SNR candidates. Four of these, $\sim$\notherLogPpct, showed evidence of spectral curvature, preferring the logP form over the PL, compared to $\sim$\nclassLogPpct~of classified candidates. The best hypotheses for candidates give a distribution of parameters in the $1-100$\,GeV range that span more than two orders of magnitude in photon flux and range from $1.5$ to as great as $4.0$ in index (Figure~\ref{fig:GeVFluxGeVIndex}). 

For those \nULs~SNRs that are either not detected by this analysis or which fail to meet the most stringent threshold for classification as a detected SNR (i.e. $\overlapext$ and $\overlaploc \geq~0.4$), upper limits assuming the radio disk morphology of Green's catalog with PL indices of $2.0$ and $2.5$ are reported in Table~\ref{Tab:ResultsULs}. For those candidates which fail to meet the most stringent threshold, we replaced the source with the radio disk. We do not calculate upper limits for the \nnotsnrs~sources which are identified as not SNRs (Section~\ref{Sec:NotSNRs}). A FITS version of the catalog is available through the \Fermi{} Science Support Center, as described in Appendix~\ref{apen:SupMat}.

\oddsidemargin -0.5in

\begin{landscape}
\pagestyle{empty}
\begin{deluxetable}{lrrrrrccccc}
\tablewidth{0pt}
\setlength{\tabcolsep}{0.05in}
\singlespace
\tabletypesize{\scriptsize}
\tablecaption{LAT SNR Catalog: Results of Spatial Analysis\label{Tab:ResultsSpat}}
\tablewidth{7.7 in}
\tablehead{
\colhead{SNR Name} & 
\colhead{TS} & 
\colhead{RA (J2000)} & 
\colhead{Dec (J2000)} & 
\colhead{$l$} & 
\colhead{$b$} & 
\colhead{$\mathrm{TS_{ext}}$} & 
\colhead{Radius} & 
\colhead{Alt IEMs} & 
\colhead{Location} & 
\colhead{Extension} \\
\colhead{ } & 
\colhead{ } & 
\colhead{[deg] $\pm$ stat $\pm$ sys} & 
\colhead{[deg] $\pm$ stat $\pm$ sys} & 
\colhead{[deg]} & 
\colhead{[deg]} & 
\colhead{ } & 
\colhead{[deg] $\pm$ stat $\pm$ sys} & 
\colhead{Effect} & 
\colhead{ Overlap } & 
\colhead{ Overlap }
}
\startdata
\multicolumn{9}{c}{{{\bf Classified Candidates:}}}\\
G006.4$-$00.1  & 1622 & 270.36 $\pm$ 0.01 $\pm$ 0.02 & $-$23.44 $\pm$ 0.01 $\pm$ 0.01 & 6.54 & $-$0.26 & 168 & 0.37 $\pm$ 0.02\phn$_{-0.00}^{+0.00}$ & 0, 0, 0 & 1.00 & 0.61 \\
G008.7$-$00.1  & 788 & 271.36 $\pm$ 0.02 $\pm$ 0.03 & $-$21.59 $\pm$ 0.02 $\pm$ 0.01 & 8.60 & $-$0.16 & 123 & 0.37 $\pm$ 0.02\phn$_{-0.03}^{+0.03}$ & 0, 0, 0 & 1.00 & 0.71 \\
G020.0$-$00.2  & 32 & 277.11 $\pm$ 0.09 $\pm$ 0.22 & $-$11.50 $\pm$ 0.05 $\pm$ 0.07 & 20.09 & $-$0.20 &  \nodata  &  \nodata  & 0, 4, 0 & 0.93 & 1.00 \\
G023.3$-$00.3  & 208 & 278.57 $\pm$ 0.03 $\pm$ 0.07 & $-$8.75 $\pm$ 0.03 $\pm$ 0.04 & 23.20 & $-$0.20 & 25 & 0.33 $\pm$ 0.04\phn$_{-0.00}^{+0.00}$ & 0, 7, 0 & 1.00 & 0.46 \\
G024.7$+$00.6  & 89 & 278.60 $\pm$ 0.03 $\pm$ 0.11 & $-$7.17 $\pm$ 0.03 $\pm$ 0.03 & 24.61 & 0.50 & 25 & 0.25 $\pm$ 0.04\phn$_{-0.12}^{+0.21}$ & 0, 4, 0 & 1.00 & 0.47 \\
G034.7$-$00.4  & 832 & 284.05 $\pm$ 0.01 $\pm$ 0.01 & 1.34 $\pm$ 0.02 $\pm$ 0.00 & 34.66 & $-$0.45 & 61 & 0.31 $\pm$ 0.02\phn$_{-0.01}^{+0.01}$ & 0, 0, 0 & 1.00 & 0.71 \\
G043.3$-$00.2  & 787 & 287.75 $\pm$ 0.01 $\pm$ 0.00 & 9.09 $\pm$ 0.01 $\pm$ 0.00 & 43.25 & $-$0.17 &  \nodata  &  \nodata  & 0, 1, 0 & 0.55 & 1.00 \\
\enddata
\tablecomments{Results of the spatial analysis for the \ndetected~detected candidates, ordered by increasing Galactic longitude within their classification category. The detected candidate's test statistic is listed in the TS column; the best fit position in the RA and Dec columns, reproduced as Galactic longitude and latitude $(l, b)$ for ease of use. Those candidates with errors listed as $0.00$ had shifts in position $<0.005\degr$. The test statistic for extension is listed under $\mathrm{TS_{ext}}$. 
The column entitled Alt IEMs Effect lists the number of alternate IEMs for which a candidate's significance became less than $9$; for which the best extension hypothesis changed between extended and point; and for which the likelihood maximization had convergence problems (see Section~\ref{Sec:CatCaveats}). Location and Extension Overlap columns list the fractional overlap in the GeV candidate's position and extension with respect to Green's catalog's radio SNR, Equations~\ref{eq:overlaploc} and \ref{eq:overlapext_ext}~or~\ref{eq:overlapext_pt}, respectively, described in detail in Section~\ref{Sec:SpatCoinc}.}
\tablenotetext{a}{Candidates with emission consistent with a known, non-SNR source, detailed in Section \ref{Sec:NotSNRs}.}
\tablenotetext{b}{Emission from regions that should be considered especially uncertain; see Section~\ref{Sec:CatCaveats} for details. }
\end{deluxetable}
\end{landscape}

\clearpage

\oddsidemargin -0.3in
\pagestyle{empty}
\begin{deluxetable}{lrcrcccrr}
\tablecolumns{9}
\tablewidth{0pt}
\setlength{\tabcolsep}{0.05in}
\singlespace
\tabletypesize{\scriptsize}
\tablecaption{LAT SNR Catalog: Results of Spectral Analysis\label{Tab:ResultsSpec}}
\tablewidth{7.0in}
\tablehead{
\colhead{SNR Name} & 
\colhead{TS} & 
\colhead{Spectral} & 
\colhead{Flux $\pm$ stat $\pm$ sys} & 
\colhead{Index} & 
\colhead{$\beta$} &
\colhead{Alt IEM} & 
\colhead{Location} & 
\colhead{Extension} \\
\colhead{ } & 
\colhead{ } & 
\colhead{Form} & 
\colhead{[10$^{-9}$ ph cm$^{-2}$ s$^{-1}$]} & 
\colhead{$\pm$ stat $\pm$ sys} & 
\colhead{$\pm$ stat $\pm$ sys} & 
\colhead{Effect} & 
\colhead{ Overlap } & 
\colhead{ Overlap }
}
\startdata
\multicolumn{9}{c}{{{\bf Classified SNRs:}}}\\
G006.4$-$00.1  & 1622 &  PL  & 51.77 $\pm$ 1.54\phn\phn$_{-3.72}^{+4.25}$ & 2.64 $\pm$ 0.04\phn$_{-0.08}^{+0.08}$ &  \nodata & 0, 0, 0 & 1.00 & 0.61 \\
G008.7$-$00.1  & 788 &  PL  & 31.27 $\pm$ 1.32\phn\phn$_{-5.27}^{+6.22}$ & 2.50 $\pm$ 0.05\phn$_{-0.10}^{+0.10}$ &  \nodata & 0, 0, 0 & 1.00 & 0.71 \\
G020.0$-$00.2  & 32 &  PL  & 4.56 $\pm$ 0.84\phn\phn$_{-2.20}^{+4.06}$ & 3.58 $\pm$ 0.47\phn$_{-0.54}^{+0.54}$ &  \nodata & 0, 4, 0 & 0.93 & 1.00 \\
G023.3$-$00.3  & 208 &  PL  & 14.64 $\pm$ 1.18\phn$_{\phn-6.81}^{+12.56}$ & 2.38 $\pm$ 0.09\phn$_{-0.15}^{+0.15}$ &  \nodata & 0, 7, 0 & 1.00 & 0.46 \\
G024.7$+$00.6  & 89 &  PL  & 8.63 $\pm$ 1.58\phn\phn$_{-4.13}^{+7.65}$ & 2.10 $\pm$ 0.15\phn$_{-0.13}^{+0.13}$ &  \nodata & 0, 4, 0 & 1.00 & 0.47 \\
G034.7$-$00.4  & 832 &  logP  & 54.95 $\pm$ 2.68\phn\phn$_{-4.05}^{+4.71}$ & 2.27 $\pm$ 0.09\phn$_{-0.10}^{+0.10}$ & 0.41 $\pm$ 0.09\phn$_{-0.05}^{+0.05}$ & 0, 0, 0 & 1.00 & 0.71 \\
G034.7$-$00.4  &  \nodata  &  PL  &  \nodata \phantom{1$_{-9.99}^{+9.99}$}  & 2.34 $\pm$ 0.09\phn$_{-0.07}^{+0.07}$ &  \nodata & 0, 0, 0 &  \nodata  &  \nodata \\
\enddata
\tablecomments{Results of the spectral analysis for the \ndetected~detected candidates, ordered by increasing Galactic longitude within their classification category. Spectral Form denotes if the candidate was best fit with a PL or a logP. Index reflects the PL index or logP~$\alpha$ (Equation~\ref{eqn:logP}) for the specified spectral form. 
For those classified candidates with logP spectral forms and which appear in plots showing other candidates' (PL) indexes, we also report here their best fit PL index and errors for completeness. The $\beta$~column lists the logP~$\beta$ specified in Equation~\ref{eqn:logP}, as applicable. 
The column entitled Alt IEM Effect lists the number of alternate IEMs for which a candidate's significance became less than $9$; for which the best extension hypothesis changed between extended and pointlike; and for which the likelihood maximization had convergence problems (see Section~\ref{Sec:CatCaveats}). Location and Extension Overlap columns list the fractional overlap in the GeV candidate's position and extension with respect to Green's catalog's radio SNR, Equations~\ref{eq:overlaploc} and \ref{eq:overlapext_ext}~or~\ref{eq:overlapext_pt}, respectively, described in detail in Section~\ref{Sec:SpatCoinc}.
}
\tablenotetext{a}{Candidates with emission consistent with a known, non-SNR source, detailed in Section~\ref{Sec:NotSNRs}.}
\tablenotetext{b}{Emission from regions that should be treated with care; see Section~\ref{Sec:CatCaveats} for details.}
\end{deluxetable}

\clearpage

\pagestyle{empty}
\begin{deluxetable}{lccrrcrrrr}
\tablewidth{0pt}
\setlength{\tabcolsep}{0.05in}
\singlespace
\tabletypesize{\scriptsize}
\tablecaption{LAT SNR Catalog: SNRs Not Detected by the LAT\label{Tab:ResultsULs}}
\tablehead{
\colhead{SNR} &		
\colhead{RA$_{radio}$} &		
\colhead{Dec$_{radio}$} &		
\colhead{$l$} &		
\colhead{$b$} &		
\colhead{Radius$_{radio}$} &		
\multicolumn{2}{c}{$\Gamma = 2.0$}&
\multicolumn{2}{c}{$\Gamma = 2.5$} \\
\colhead{ } &		
\colhead{[deg] (J2000)} &		
\colhead{[deg] (J2000)} &		
\colhead{[deg]} &		
\colhead{[deg]} &		
\colhead{[deg]} &		
\colhead{95\% UL} &		
\colhead{99\% UL} &		
\colhead{95\% UL} &		
\colhead{99\% UL} 
}
\startdata
G000.0$+$00.0  & 266.43 & $-$29.00 & 359.96 & $-$0.05 & 0.02 & 3.0E$-$08 & 3.1E$-$08 & 4.4E$-$08 & 4.5E$-$08 \\
G000.3$+$00.0  & 266.56 & $-$28.63 & 0.33 & 0.04 & 0.09 & 7.0E$-$09 & 7.6E$-$09 & 1.2E$-$08 & 1.3E$-$08 \\
G000.9$+$00.1  & 266.84 & $-$28.15 & 0.87 & 0.08 & 0.07 & 7.3E$-$10 & 1.0E$-$09 & 1.2E$-$10 & 1.9E$-$10 \\
G001.0$-$00.1  & 267.12 & $-$28.15 & 1.00 & $-$0.13 & 0.07 & 7.6E$-$10 & 1.0E$-$09 & 9.7E$-$10 & 1.4E$-$09 \\
G001.4$-$00.1  & 267.41 & $-$27.77 & 1.46 & $-$0.15 & 0.08 & 2.6E$-$10 & 4.0E$-$10 & 3.0E$-$10 & 4.6E$-$10 \\
G001.9$+$00.3  & 267.19 & $-$27.17 & 1.87 & 0.32 & 0.01 & 1.4E$-$10 & 2.2E$-$10 & 1.8E$-$10 & 2.7E$-$10 \\
G003.7$-$00.2  & 268.86 & $-$25.83 & 3.78 & $-$0.28 & 0.10 & 6.7E$-$10 & 9.4E$-$10 & 1.1E$-$09 & 1.4E$-$09 \\
G003.8$+$00.3  & 268.23 & $-$25.47 & 3.81 & 0.39 & 0.15 & 4.4E$-$10 & 6.4E$-$10 & 5.3E$-$10 & 7.9E$-$10 \\
\enddata
\tablecomments{Upper limits for the \nULs~undetected SNRs in flux units of ph cm$^{-2}$ s$^{-1}$ in the $1-100$\,GeV energy range. Both 95\% and 99\% confidence upper limits are provided for two different GeV PL indices, $\Gamma=\{2.0, 2.5\}$.}
\end{deluxetable}

\clearpage

\footnotetext{No extended marginally classified candidates were also identified as young or interacting.} 

\FloatBarrier

\subsubsection{New Extended SNRs}\label{Sec:NewSNRs}

We identified \nnewextended~new extended SNRs in this work which pass the classification threshold to be associated with the radio SNR and have sufficient stability with the alternative IEMs (Section~\ref{Sec:IEMSysErr}). We caution that complex, diffuse emission may cause sources with nominally detectable extensions to be either detected as point-like objects or not detected at all. 

\begin{description}
\item[SNR G24.7+0.6:] 
First identified by radio observations \citep{1984A&A...133L...4R}, SNR G24.7+0.6 has been called ``Crab-like'' due to its composite radio morphology: a bright central core with a flat radio spectrum ($\alpha=-0.17$) surrounded by a $\sim0.5$\degr{} diameter shell \citep{1987ApJ...316..660B}. A luminous blue variable star G24.73+0.69 is located to the south, just outside the remnant shell \citep{2012A&A...538A..14P}. The SNR has not been well studied in wavelengths other than radio, and no compact object has been identified to power the putative central nebula.

The candidate associated with SNR G24.7+0.6 is extended with a radius of $0.25$\degr{} and a relatively hard index of $2.1$. The source had possible counterparts in previous \FermiLat{} catalogs (1FGL J1834.7$-$0709c, 2FGL J1834.7$-$0705c, 1FHL J1834.6$-$0703). The extension we find above $1$\,GeV is consistent with the radio size, but offset toward the southern massive star-forming region. There is no clear evidence of interaction between the SNR and this star-forming region. We note that the SNR is embedded in a region of the Galactic plane with bright emission, so a more detailed study would be needed to determine how much emission originates from the SNR shell as opposed to the putative central nebula or other sources of plane emission.

\item[SNR G205.5+0.5 (Monoceros Loop):] 
The Monoceros Loop is a large radio SNR identified toward the Galactic anticenter.
The $2.3$\degr{} extension of the SNR above $1$\,GeV is larger than the $1.8$\degr{} radius observed in the radio. The candidate is also offset slightly in the direction of the Rosette nebula, a massive MC with star formation. A recent study of neutral hydrogen gas in the vicinity of the Monoceros Loop suggests it may be interacting with the Rosette nebula \citep{2012AA...545A..86X}. A detailed analysis to study the GeV morphology of this remnant is currently underway (H. Katagiri, in prep.).

\item[SNR G296.5+10.0 (PKS 1209$-$51/52):]  
This bilateral shell SNR has a large angular extension, detected in the radio, optical, and X-rays. A pulsar with a period of $424$\,ms lies near the center of the SNR \citep{Zavlin00-PSR}, though no radio or \g-ray emission is detected. The SNR was previously studied in \g-rays by \cite{Araya13-G295.6+10.0} using Pass~$7$ data, who found that an extended disk improved the model, but not sufficiently to claim a clear detection of extension for the $0.2-100$\,GeV energy range. The best radius found here is $0.7\pm0.1$\degr, consistent with the radio extension. However this may be spurious as the significance of extension falls below threshold for all alternative IEMs. 

\item[SNR G326.3$-$1.8 (MSH 15$-$56):] 

MSH 15$-$56 is a composite SNR with a PWN at the southwestern rim of the radio shell \citep{1998MmSAI..69..939P}. Recent X-ray studies of the SNR and its PWN indicate an age of $\sim10$\,kyr for the remnant, which is expanding at~$860$\,\kms{} \citep{2013ApJ...773...25Y}. \FermiLat{} observations of MSH 15$-$56 have been previously analyzed by \cite{2013ApJ...768...61T}, finding a hard PL source consistent with the spectral parameters of this work. 
The extension measured in the present work is 0.42\degr, somewhat smaller than the 0.63\degr{} diameter of the radio shell, but slightly larger than the X-ray PWN. While \citet{2013ApJ...768...61T} 's evolutionary modeling of X-ray and \g-ray emission from the PWN can explain the LAT source, an SNR origin could not be ruled out. Further detailed studies of the \g-ray morphology may clarify this.

\end{description}

\subsubsection{Sources Determined To Be Not SNRs}\label{Sec:NotSNRs}

\nnotsnrsCap{} detected sources pass the classification criteria, but have been identified, primarily through temporal changes in their \g-ray flux, as GeV sources other than SNRs. Here we briefly detail these detections and the reasons for excluding them from further discussion in this catalog. 

\begin{description}
\item[SNR G184.6$-$5.8 (Crab Nebula):] 
The Crab is among the brightest persistent sources of \g-rays, with the radiation and dynamics of the nebula dominated by the pulsar wind, not by the historical supernova. The automated analysis found a source of emission above $1$\,GeV consistent with previously reported emission from the PWN \citep{2010ApJ...708.1254A}.

\item[SNR G284.3$-$1.8 and \g-ray binary 1FGL J1018.6$-$5856:] 
The discovery of periodic emission from 1FGL J1018.6$-$5856 led to its identification as a \g-ray binary \citep{Abdo12-1FGLJ1018}. The TeV source HESS J1018$-$589A is also coincident with the binary, though no periodicity is observed and several associations appear plausible \citep{HESS-J1018}. The source detected in this analysis for SNR G284.3$-$1.8 (MSH 10$-$53) matches both spatially and spectrally with the analysis from $1-10$\,GeV presented in \cite{Abdo12-1FGLJ1018} for the binary. We therefore conclude that the detected source is the previously identified binary, and not a new source potentially associated with the SNR.

\item[SNR G320.4$-$1.2 (MSH 15$-$52):] 
Identified as a PWN in \cite{Abdo10-msh15--52}, the emission from GeV to TeV energies arises from IC emission from the wind nebula of PSR B1509$-$58. As with the Crab PWN, the source detected in our automated analysis is consistent with the previously identified bright pulsar and its nebula rather than being related to the SNR.

\item[SNR G5.4$-$1.2 / PSR J1801$-$2451:] 
The spectral parameters reported in 2PC are fit only for the peak flux during the pulsation period and are therefore not appropriate for our automated analysis performed using all rotation phase intervals. The candidate was spatially coincident with PSR J1801$-$2451 (B1757$-$24) and had a steep photon index of $2.9\pm0.3$, consistent with the $3\pm2$\,GeV cutoff measured for the pulsar \citep{Abdo13-2PC}. We therefore considered this emission to be due to the pulsar and not SNR G5.4$-$1.2. A detailed study of the phase-selected emission from the pulsar could determine whether any emission may be attributed to the SNR, and is beyond the scope of this work.

\end{description}

\subsubsection{Caveats}\label{Sec:CatCaveats}

Due to the inherent complexity of the \g-ray data and the necessity of treating regions and candidates uniformly, we note here important cautions regarding the results for some candidates. Two classified candidates have peculiar GeV features while \nmarginalAIEM~marginal and \notherAIEM~other candidates show significant changes in a candidate's significance when using the alternative IEMs as described in Section~\ref{Sec:IEMSysErr}. 

\begin{description}

\item[SNR G5.2$-$2.6:] 
The automated analysis detected a candidate in the region studied, but the source's index always reached the extreme value ($5$) allowed in the fit. We attempted to achieve convergence by limiting the number of free parameters in the fit and by extending the parameter boundaries; in all cases, the index tended to an extremely soft value. The candidate itself has small location and extension overlap fractions, making it unlikely to be the SNR.

\item[SNR G6.5$-$0.4:] 
This SNR has been identified in the radio as adjacent to and overlapping the much brighter SNR G6.4$-$0.1 (W28). Our automated analysis detected a point source at the position of SNR G6.5$-$0.4 only when an extended disk template for W28 was not included; when W28 was included in the region, no additional source was found at the position of SNR G6.5$-$0.4. The GeV emission from this region has previously been studied in detail by \citet{Abdo10-W28} who concluded that the extended emission  in this region was consistent with the SNR W28. Nearby GeV and TeV sources outside this immediate vicinity may be attributed to MCs illuminated by escaping CRs; these sources are not coincident with SNR G6.5$-$0.4 \citep{Hanabata14}. We therefore exclude the point candidate overlapping SNR G6.5$-$0.4 found without the extended template for W28 from the catalog. Instead, including the extended disk template for W28 in the region's model, we report upper limits on the flux from SNR G6.5$-$0.4 in Table~\ref{Tab:ResultsULs}. 

\item[SNR G32.4+0.1:] This candidate has a TS $<$ 9 with two alternative IEMs and is demoted from classified to marginally classified. Additionally, it changed from a point source to an extended source when using one of the alternative IEMs. 

\item[SNR G73.9+0.9:] 
This candidate showed no significant detection for any of the alternative IEMs and is therefore downgraded from a classified to a marginally classified source. For three of the alternative IEMs the best fit was for an extended source. We represent the lack of significant source detection with any of the alternative IEMs as a downward pointing arrow in place of its systematic flux uncertainty in all plots where a flux point would appear. We also exclude the alternative IEM systematic error from the error on the index and other quantities since such parameters are not relevant for an undetected source. The systematic error for this source still includes the propagated uncertainty from the systematic uncertainty on the effective area, derived using the bracketing IRFs (Section~\ref{Sec:EffAreaSysErr}).

\item[SNR G263.9$-$3.3:] 
The automated analysis found a point source within the Vela SNR near the position of the Vela pulsar with a very soft photon index ($3.9\pm0.5$) and a TS of only $28$. The flux of this source is two orders of magnitude lower than the pulsar and, given its location and soft spectral index, the source may be attributable to residuals in the fit of the bright pulsar's all-phase spectrum or to the Vela-X nebula. A detailed study of $4$~years' data using just the off-peak phase interval found only emission associated with the Vela-X PWN, which has a harder photon index than the source detected in our catalog analysis \citep{Grondin2013}. In addition, analyzing the candidate with the bracketing IRFs indicated that the candidate was insignificant in light of these systematic uncertainties. Coupled with having only one case with a reasonable PL index value for the alternative IEMs, as described in Section~\ref{Sec:IEMSysErr}, we are unable to estimate an upper limit on the index.

\end{description}

As noted in Section~\ref{Sec:IEMSysErr}, we excluded from our systematic error estimate those solutions with alternative IEMs for which the resulting candidate had an index at or near a limit or otherwise had convergence problems. While this may in some cases cause the systematic error to be underestimated, it often leads to a larger error due to the inability of the fitting algorithm to estimate the statistical error giving the weight in Equation~\ref{eq:sys_weight}. The number of alternative IEMs for which the fit had convergence problems for a given candidate is listed in the first column of the Alt IEM Effect column in Tables~\ref{Tab:ResultsSpat} and \ref{Tab:ResultsSpec}. In addition to the two candidates explicitly described above and treated as marginal rather than classified candidates, this happens only for \notherAIEM~candidates classified as ``other". Care should be taken with these other candidates, as with the candidates associated with SNRs G32.4+0.1 and G73.9+0.9. 

\subsection{Verifications}\label{Sec:Validation}
To verify our analysis, we both ensured that our analysis was internally consistent (Section~\ref{Sec:InternalValid}) and sensible in comparison to previously published analyses (Section~\ref{Sec:ExternalValid}). 

\subsubsection{Internal Verification}\label{Sec:InternalValid}

We internally verified our results in several ways. 
Within the analysis itself, we ensured that the values of fitted parameters were not at extrema and that the \logL{} had improved following the fit. Only SNR G5.2$-$2.6 had a parameter at the extreme limit (in this case the index), as discussed in Section~\ref{Sec:CatCaveats}.

The statistical errors for all parameters are within a reasonable range relative to their values. Errors on the source position are all positive values $<0.25$\degr{}. As expected, positional errors tended to decrease with increasing source TS and tended to be larger when the source was closer to the Galactic plane and to the Galactic center. Errors on the disk radius ranged between $0.01$\degr{} and $0.2$\degr{}, $\lesssim15\%$ of the radius measured. For the flux, fractional errors ranged from $\sim2\%$ for the brightest candidate to not more than $50\%$ for the dimmest candidate, showing the expected decrease in statistical error for sources with higher fluxes and improved TS. Index errors never exceeded $0.8$ and were $<0.3$ for nearly all candidates.

For the logP sources, statistical errors on $\alpha$ tended to increase with decreasing values of~$\alpha$ while $\beta$ errors increased with increasing values of $\beta$. This arises from the underlying trend that the statistical error tends to increase as the parameter becomes less significant. All candidates except the one near SNR G358.1+0.1 had statistical errors on $\alpha$ less than $\alpha$ itself. The candidate in the region of G351.2+0.1 also had a higher error relative to its $\alpha$ and $\beta$ parameters than the others, though never exceeding the values themselves. Both candidates lie in complex regions and are subsequently excluded by the classification process. In all cases the $\beta$ statistical errors were less than the $\beta$ values themselves. 

We take these verifications as positive indication that the analysis process completed successfully for all candidates. For all non-detections, we also verified that convergence was achieved when calculating the Bayesian upper limits.

\subsubsection{Comparisons with Published LAT Analyses}\label{Sec:ExternalValid}

Previous LAT source catalogs, including the 2FGL catalog \citep{Nolan12-2FGL} and the \FermiLat{} TeV PWN study \citep{Acero13-TevPwnCat}, provide a basis for comparison with the results of this work. First we consider SNR associations reported in 2FGL. All point source, classified SNRs in our work have $1-100$\,GeV fluxes consistent within the errors of the source fluxes reported in 2FGL. Additionally, all but five extended sources classified as SNRs ($\sim16\%$) have fluxes consistent with their 2FGL counterparts. Of these, two were only marginally lower than their 2FGL counterparts while three were not identified as extended sources at the time of 2FGL publication. Sources with PL spectra in both this work and the 2FGL catalogs have consistent spectral indices, despite the larger energy range examined in 2FGL. Our results are also consistent with the previous examination of Galactic TeV PWNe in~\citet{Acero13-TevPwnCat}. In particular, all spectra were consistent except for HESS J1804$-$216, which may be associated with SNR~G8.7$-$0.1 but has a smaller extension in the TeV than reported here in the GeV. The difference also may be due to the narrower energy range ($10-316$\,GeV) examined by~\cite{Acero13-TevPwnCat}.

Several remnants have also been studied individually, so we compared our results to the fluxes, indices, localizations, and extensions for extended sources that were reported in these publications \citep{Abdo10-CasA, Lande12-extSrcSearch, Castro10-interactMCs, Abdo10-W28, Abdo10-W49B, Abdo10-W44, Giordano12-Tycho, Katsuta12-S147, Brandt13-CTB37A, HESS2015-J1834-087}. The majority of results, even including for candidates determined not to be SNRs such as the Crab Nebula (Section~\ref{Sec:NotSNRs}), were consistent within errors. This increases confidence in the methods developed and employed. 

The majority of differences are ascribable to differences in data sets, particularly their associated IRFs and IEMs, in time periods, and in different energy ranges studied with different localization and extension methods. 
For instance, while all candidates with the extended hypothesis preferred had extensions similar to those found in published studies of individual SNRs after accounting for the difference between Gaussian and uniform disks, only one candidate, SNR G348.5+0.1, with a published extension \citep{Brandt13-CTB37A} was detected as a nominal point source. The extension measured for this point source was within errors of our extension detection threshold, $0.2$\degr, as described in Section~\ref{Sec:SpatCoinc}. Further, the published extension was smaller than the smallest extension reported for any candidate. Even in the case of similar data sets, fitting over a different energy range can lead to a different spectral index, which, when extrapolated to the $1-100$\,GeV energy reported here, sometimes resulted in differing fluxes.

We finally note that some SNR candidates were not detected in this work but appear in other publications that study differing data sets and energy ranges. In particular, the historical remnant Tycho (SNR G120.1+1.4), which is detected with \FermiLat{} using Pass 6 Diffuse class data at energies of $0.4-100$\,GeV \citep{Giordano12-Tycho}, falls just below the detection threshold ($\mathrm{TS}=19<25$) in this work, with a photon index slightly harder than but consistent within statistical errors of that found for the lower energy range in \citet{Giordano12-Tycho}. 

Given the consistency of the results herein with the results for the corresponding sources in 2FGL and \cite{Acero13-TevPwnCat}, as well as the majority of individually studied sources, a further discussion of specific discrepancies is beyond the scope of this catalog. We also preserve the uniformity of our sample by not modifying the list of candidates determined by the automated procedure.

\subsection{Comparison of Systematic and Statistical Errors}\label{Sec:SysvStatErr}

The systematic nature of this work allows us a unique opportunity to study trends in errors for these candidates in the Galactic plane and in the $1-100$\,GeV energy range. We exclude from this discussion all candidates that were associated with some other object and, for consistency, show only PL indices on the relevant plots. Further details on the derivation of systematic errors can be found in Section~\ref{Sec:SysErr}.

\begin{figure}[h!]
  \centering 
  \includegraphics[width=0.95\columnwidth]{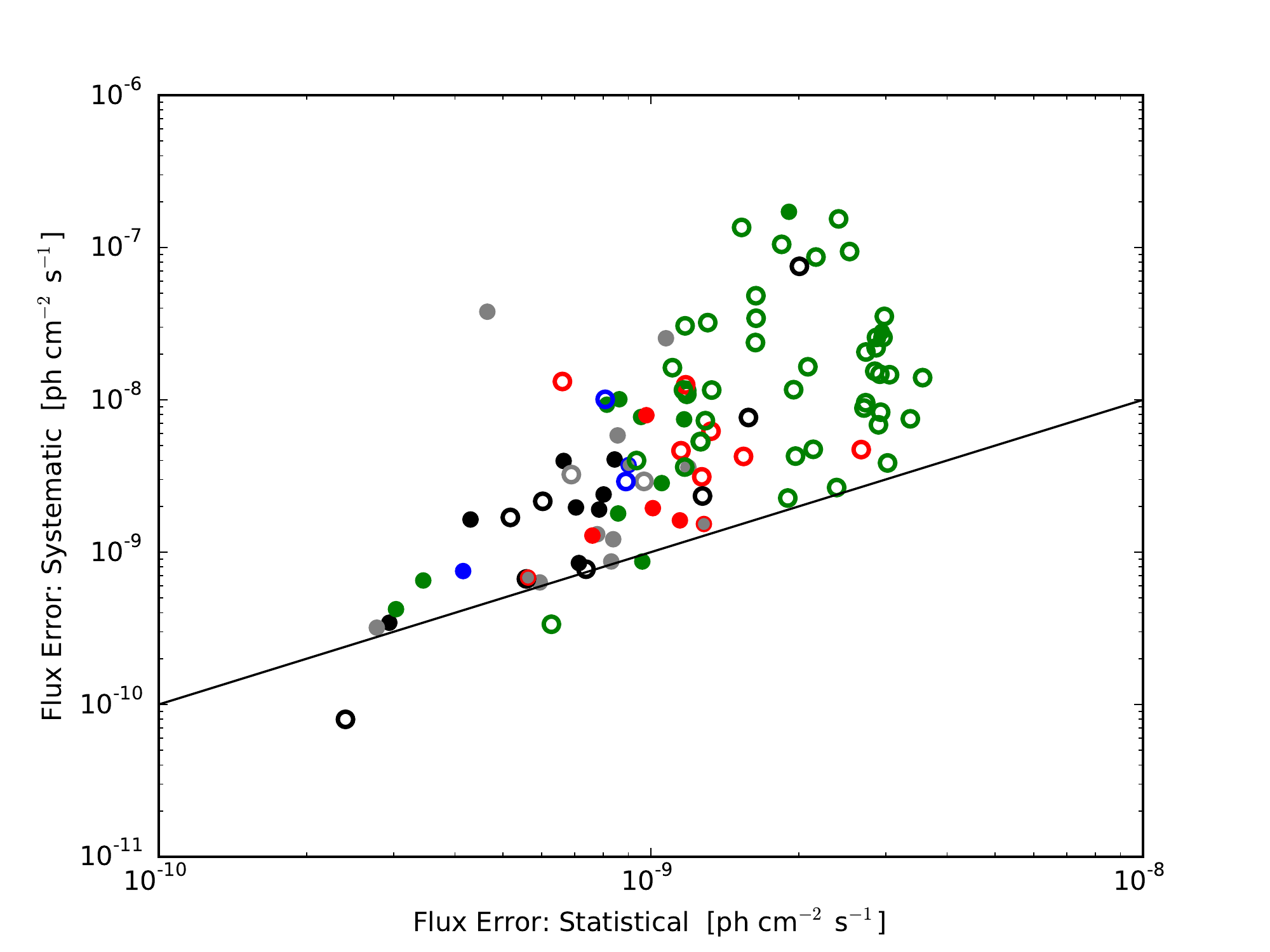} 
  \caption{Comparison of the statistical and systematic errors, the latter derived from the alternative IEMs and the bracketing IRFs, for the flux. The line indicates $1:1$ correspondence. Symbols and colors are as in Figure \ref{fig:GeVFluxGeVIndex} with the addition that all candidates classified as ``other'' are shown in green.  The systematic error typically dominates the statistical error on the flux for all classes of candidates. } 
  \label{fig:sysvStatFluxErrs}
 \end{figure}

We find that the total systematic error, derived using the bracketing IRFs and alternative IEMs and added in quadrature, dominates the statistical error on a candidate's flux, as seen in Figure~\ref{fig:sysvStatFluxErrs}. The systematic error ranges over about three decades in flux, while the statistical error covers only two. Both the systematic and statistical errors tend to be smaller for point-like candidates. The extended candidates with larger errors tend to not be classified, even marginally, as SNR candidates. This is in part driven by the alternative IEM error estimate, described further in Section~\ref{aIEMsContext}. From this we also infer that simply having larger error bars does not make a candidate more likely to be classified as an SNR.

\begin{figure}[h!]
  \centering
  \includegraphics[width=0.95\columnwidth]{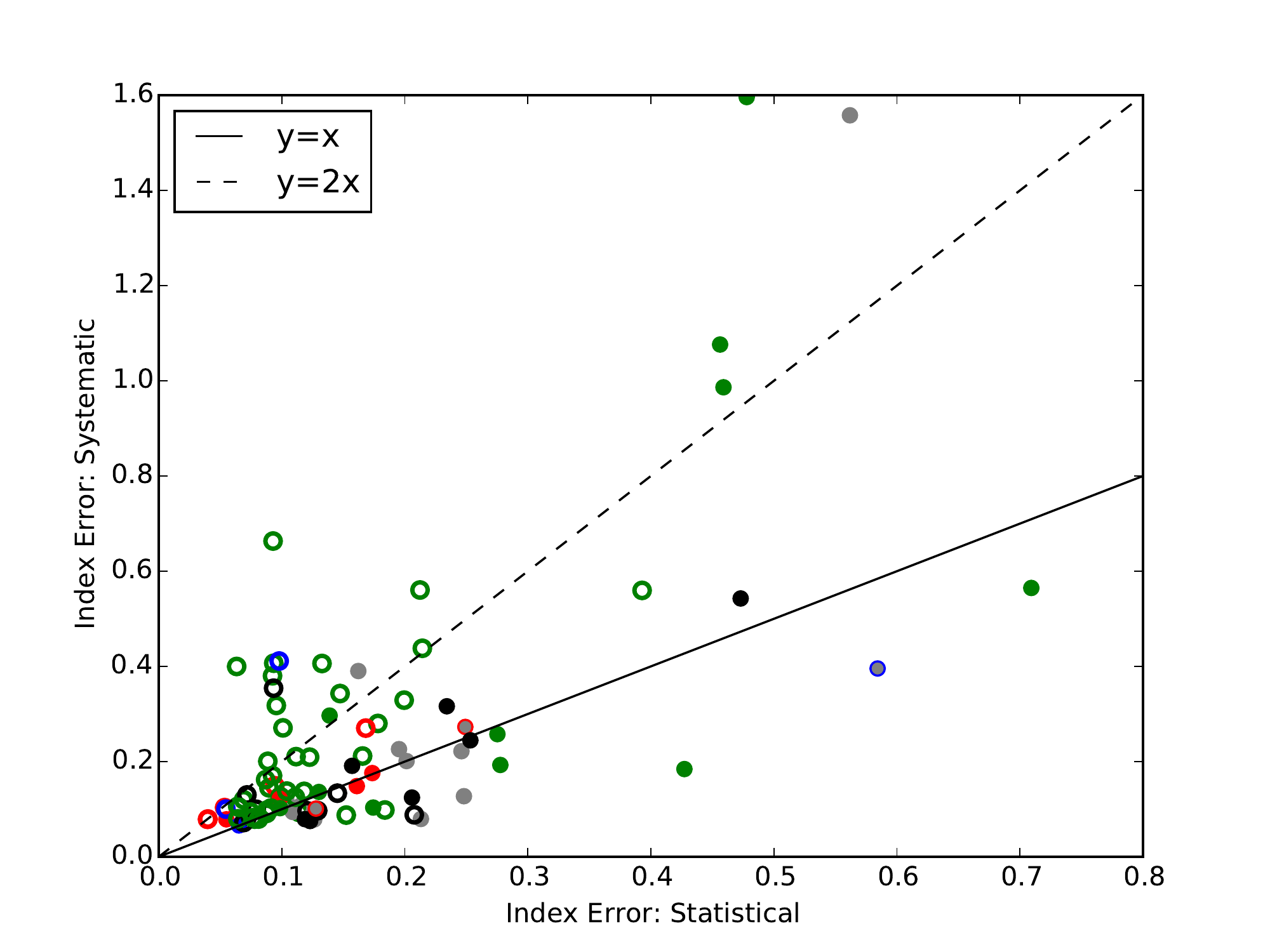} 
  \caption{Comparison of the systematic and statistical errors on the PL index. The symbols and colors are as in Figure~\ref{fig:sysvStatFluxErrs}. The low number of candidates above the dashed line shows that the systematic error is usually less than twice the statistical error. The solid line indicates equal systematic and statistical errors. Unlike for the flux, in a number of cases the statistical error dominates the systematic.}

  \label{fig:sysvStatIndexErrs}
 \end{figure}

Unlike for flux, Figure~\ref{fig:sysvStatIndexErrs} shows that statistical errors for PL index dominate over systematics for a number of candidates. In the cases where it does not, the index's systematic error is typically not more than twice the statistical error and always within an order of magnitude. Also unlike the flux errors, there is no obvious trend with extension or classification. 

\FloatBarrier


\section{The GeV SNR Population in a MW Context}\label{Sec:GeVSNRDiscussion}

In order to better understand both the GeV characteristics of SNRs and their potential for accelerating CRs, we examine the population of classified sources within the context of MW observations. As we began our analysis using Green's catalog, derived mainly from radio observations, in Section~\ref{Sec:GeVRadio} we examine the data for correlations between the radio and GeV measurements. We compare fluxes, luminosities, and indices and note that the radio and GeV sizes are similar for all candidates, including marginal candidates with the relaxed classification thresholds of $0.1$ (see Section~\ref{Sec:SpatCoinc}). 

The GeV-radio comparisons demonstrate that, with the (relative) wealth of data now available, the simplest models are no longer sufficient. Underlying particle populations may have spectral curvature, reflected in a comparison between the GeV and extant TeV data, described in Section~\ref{Sec:GeVTeV}. As suggested by the clustering of sources by class in Figures~\ref{fig:GeVFluxGeVIndex}, \ref{fig:GeVradioIndex}, \ref{fig:GeVTeVIndex} and \ref{fig:AgeGeVIndex}, earlier works, e.g. \cite{Thompson12-FermiCRsReview, Dermer13-CRsSNRs, Hewitt13-snrCat_ICRC, Brandt13-snrCatCrConst_ICRC}, and \cite{Slane14-GeVSNRs}, have also noted possible trends for young SNRs and those interacting with dense MCs. We examine the possibilities for disentangling the effects of evolution and environment for our statistically significant, uniformly measured set of candidates in Section~\ref{Sec:EvolEnviron}. This required an updated list of SNR properties, which utilized both the Galactic SNR high energy observations\footnote{www.physics.umanitoba.ca/snr/SNRcat/} published by~\cite{Ferrand12-XraySNRs} and the online TeVCat\footnote{http://tevcat.uchicago.edu/}.

The commonly used ``young'' and ``interacting'' SNR subclasses are defined here for clarity and used in the figures. A ``young'' SNR is typically defined as being in the Sedov phase or younger. We use X-ray synchrotron emission associated with high-velocity shocks ($\gtrsim$2,000\,\kms) to indicate this observationally as the age at which an SNR leaves the Sedov phase depends on its surrounding environment. Table~\ref{Tab:YoungSNRs} lists all remnants that have clearly identified X-ray synchrotron emission associated with the SNR shock front. Three classified sources, G111.7$-$2.1 (Cas A), G347.3$-$0.5, and G266.2$-$1.2, and the marginally classified G32.4+0.1, are young, non-thermal X-ray SNRs and colored blue in the figures (in outline for the marginal candidate). 

Sources associated with SNRs interacting with dense ($\gtrsim100$\,cm$^{-3}$) gas in large MCs are another important subclass of \g-ray SNRs, colored red in the figures in this paper. Eleven sources classified as SNRs show clear evidence of interaction based on the detection of at least one molecular species: OH ($1720$\,MHz) masers \citep{1996AJ....111.1651F,Green97,2009ApJ...694L..16H}; H$_2$ atomic and molecular vibrational and rotational lines in the infrared \citep{Reach00-IR-MC-SNRs}; and molecular line broadening of $\geq10$\kms{} \citep{White87} for molecules such as CO or HCO$^{+}$ which are easily detected by radio telescopes. We have expanded on the early list in the appendix of \cite{2010ApJ...712.1147J} by including more recent publications and excluding identifications based solely on morphology, which are difficult to establish with certainty. Table~\ref{Tab:InteractingSNRs} lists the GeV candidates associated with these interacting SNRs along with the identifying MW tracers. The majority of the evidence for these interactions arises from OH masers, which are particularly robust tracers that require a narrow range of physical conditions that only arise in slow SNR shocks into dense MCs.

It is important to account for the distances of the SNRs when comparing physical quantities such as luminosity. Table~\ref{Tab:SNRinfo} records distance from the literature, including the most recent and/or most certain distance estimates adopted in this work. Of the \nGalSNRs~SNRs studied, only \ndist{} have published distance estimates. Most often these distances are determined from observed line-of-sight velocities using an assumed Galactic rotation curve. Furthermore, kinematic distance estimates have largely been done on an individual basis, and are not uniformly determined for all SNRs. We do not consider distances derived using the ``$\Sigma$-D relation" because SNRs show a wide range of physical diameters (D) for a given surface brightness ($\Sigma$), limiting the utility of such a relationship for determining the distances to individual SNRs \citep{Green12-distances}.

\pagestyle{empty}
\begin{deluxetable}{ll}
\tablewidth{350pt}
\singlespace
\tabletypesize{\scriptsize}
\tablecaption{X-ray Synchrotron SNRs \label{Tab:YoungSNRs}}
\tablehead{
\colhead{Name} & 
\colhead{Reference(s)} 
}
\startdata
G001.9$+$00.3 & \cite{2008ApJ...680L..41R} \\
G004.5$+$06.8 & \cite{2005ApJ...621..793B} \\
G021.5$-$00.9 & \cite{2010ApJ...724..572M}, \cite{2014ApJ...789...72N} \\
G028.6$-$00.1 & \cite{2003ApJ...588..338U} \\
G032.4$+$00.1 & \cite{2012PASJ...64...61U} \\
G111.7$-$02.1 & \cite{2006ApJ...647L..41R}, \cite{2008ApJ...686.1094H}, \cite{2009PASJ...61.1217M} \\
G120.1$+$01.4 & \cite{2011ApJ...728L..28E} \\
G266.2$-$01.2 & \cite{2005ApJ...632..294B}, \cite{2007ApJ...661..236A}, \cite{2010ApJ...721.1492P} \\
G315.4$-$02.3 & \cite{2012AA...545A..28L} \\
G327.6$+$14.6 & \cite{2008PASJ...60S.153B} \\
G330.2$+$01.0 & \cite{2006PASJ...58L..11T} \\
G347.3$-$00.5 & \cite{2009AA...505..157A}, \cite{2003ApJ...593..377P} \\
G348.7$+$00.3 & \cite{2012PASJ...64...61U} \\
G353.6$-$00.7 & \cite{2012ApJ...756..149B} \\
\enddata
\tablecomments{This list contains SNRs that have clearly detected X-ray synchrotron emission. We have excluded the interacting SNR G6.4-0.1 \citep[W28]{2014ApJ...791...87Z} from this list, which has been cited as non-thermal X-ray emitter, but for which the evidence is not yet sufficiently clear. }
\end{deluxetable}


\pagestyle{empty}
\begin{deluxetable}{lcl}
\tablewidth{350pt}
\singlespace
\tabletypesize{\scriptsize}
\tablecaption{Interacting SNRs \label{Tab:InteractingSNRs}}
\tablehead{
\colhead{Name} & 
\colhead{Evidence of Interaction} & 
\colhead{Reference(s)} 
}
\startdata
G006.4$-$00.1 & OH, LB, H2 & \cite{2002AJ....124.2145V}, \cite{2011JCAP...05..026C} \\
G008.7$-$00.1 & OH & \cite{2011JCAP...05..026C} \\
G023.3$-$00.3 & OH & \cite{2013ApJ...773L..19F} \\
G034.7$-$00.4 & OH, LB, H2 & \cite{1999ApJ...522..349C}, \cite{2005ApJ...618..297R} \\
G043.3$-$00.2 & H2 & \cite{2011JCAP...05..026C}, \cite{2011ApJ...732..114L} \\
G049.2$-$00.7 & OH, LB & \cite{2011JCAP...05..026C} \\
G089.0$+$04.7 & LB, H2 & \cite{2006ApJ...637..283B} \\
G189.1$+$03.0 & OH, LB, H2 & \cite{2011JCAP...05..026C} \\
G348.5$+$00.1 & OH, H2 & \cite{2011JCAP...05..026C} \\
G349.7$+$00.2 & OH, LB, H2 & \cite{2011JCAP...05..026C} \\
G357.7$-$00.1 & OH, H2 & \nodata \\
\multicolumn{3}{c}{{{\bf SNRs with No GeV Candidate:}}}\\ 
G000.0$+$00.0 & OH, LB, H2 & \cite{2011JCAP...05..026C} \\
G001.0$-$00.1 & OH & \nodata \\
G001.4$-$00.1 & OH & \nodata \\
G005.4$-$01.2 & OH & \nodata \\
G005.7$-$00.0 & OH & \nodata \\
G009.7$-$00.0 & OH & \nodata \\
G016.7$+$00.1 & OH & \nodata \\
G018.8$+$00.3 & LB & \cite{2004AA...426..201D}, \cite{1999AJ....118..930D} \\
G021.8$-$00.6 & OH, LB, H2 & \nodata \\
G029.7$-$00.3 & LB & \cite{2006ApJ...647.1286L} \\
G031.9$+$00.0 & OH, LB, H2 & \cite{2011JCAP...05..026C} \\
G032.8$-$00.1 & OH & \cite{2011ApJ...743....4Z} \\
G039.2$-$00.3 & LB, H2 & \nodata \\
G041.1$-$00.3 & LB & \nodata \\
G054.4$-$00.3 & LB, H2 & \nodata \\
G304.6$+$00.1 & H2 & \cite{2010AA...523A..76C} \\
G332.4$-$00.4 & H2 & \cite{2011ApJ...732..114L} \\
G337.0$-$00.1 & OH & \nodata \\
G337.8$-$00.1 & OH & \nodata \\
G346.6$-$00.2 & OH, H2 & \nodata \\
G348.5$-$00.0 & OH, H2 & \nodata \\
G357.7$+$00.3 & OH & \nodata \\
G359.1$+$00.9 & OH, LB, H2 & \nodata \\
G359.1$-$00.5 & OH & \cite{2011JCAP...05..026C}, \cite{1995Sci...270.1801Y}
\enddata
\tablecomments{Table of SNRs which show evidence of interaction. ``OH'' refers to observations of OH ($1720$\,MHz) masers associated with the remnant. ``H2'' refers to observations of vibrational and/or rotational lines of shocked H$_2$. ``LB'' indicates evidence of molecular line broadening (e.g. CO, HCO$^+$) with velocities $>$10 \kms\ associated with the remnant. See Section \ref{Sec:GeVSNRDiscussion} for details. Those SNRs without references listed are from the initial interacting SNRs list compiled by \cite{2010ApJ...712.1147J}. See that work for those additional references.}
\end{deluxetable}

\clearpage


\pagestyle{empty}
\begin{deluxetable}{lclll}
\tablewidth{450pt}
\setlength{\tabcolsep}{0.02in} 
\singlespace
\tabletypesize{\scriptsize}
\tablecaption{Distances to SNRs \label{Tab:SNRinfo}}
\tablehead{
\colhead{Name} & 
\colhead{d [kpc]} & 
\colhead{Method} & 
\colhead{Reference(s)}   
}
\startdata
G000.0$+$00.0 & 8.5                           & IAU value & \cite{1986MNRAS.221.1023K} \\
G000.3$+$00.0 & 8.5$^{+3.0}_{-3.0}$          & H\,{\sc i} & \cite{2010ApJS..191..275L} \\
G000.9$+$00.1 & 8.5$^{+7.5}_{-1.5}$          & PSR & \cite{2009ApJ...700L..34C} \\
G001.0$-$00.1 & 8.5                           & Maser & \cite{1999ApJ...527..172Y} \\
G001.4$-$00.1 & 8.5$^{+5.6}_{-0.0}$          & Maser & \cite{1999ApJ...527..172Y} \\
G004.5$+$06.8 & 7.0$^{+2.0}_{-0.6}$          & H\,{\sc i} & \cite{1999AJ....118..926R}, \cite{2005AdSpR..35.1027S}, \cite{2008AA...488..219A} \\
G005.4$-$01.2 & 4.75$^{+0.45}_{-0.45}$       & Maser & \cite{2009ApJ...694L..16H} \\
G005.7$-$00.0 & 8.4$^{+5.3}_{-5.3}$          & Maser & \cite{2009ApJ...694L..16H} \\
G006.4$-$00.1 & 1.9$^{+0.4}_{-0.4}$          & Maser, CO & \cite{2002AJ....124.2145V} \\
G008.7$-$00.1 & 4.5                           & Maser & \cite{1990Natur.343..146K} \\
G009.7$-$00.0 & 4.7                           & Maser & \cite{2009ApJ...694L..16H} \\
G011.2$-$00.3 & 5$^{+21}_{-0.5}$             & H\,{\sc i} & \cite{1972ApJS...24...49R}, \cite{1985ApJ...296..461B}, \cite{1988MNRAS.231..735G} \\
G012.8$-$00.0 & 4.7$^{+1.3}_{-1.1}$          & PSR & \cite{2012ApJ...753L..14H} \\
G013.3$-$01.3 & 3.3$^{+1.8}_{-1.7}$          & CO & \cite{1995ApJ...449..681S}, \cite{1998AJ....116.1323K} \\
G015.1$-$01.6 & 5.7$^{+1.3}_{-3.5}$          & NH & \cite{2008AA...481..705B} \\
G015.4$+$00.1 & 4.8$^{+1.0}_{-1.0}$          & CO & \cite{2013AA...557L..15C} \\
G016.7$+$00.1 & 10.0$^{+3.7}_{-7.4}$         & Maser, CO & \cite{2008ApJ...683..189H}, \cite{2000ApJ...545..874R} \\
G016.8$-$01.1 & 5.1$^{+4.6}_{-1.8}$          & H\,{\sc i} & \cite{2011AA...536A..83S} \\
G018.1$-$00.1 & 5.58$^{+0.24}_{-0.27}$       & H\,{\sc i} & \cite{2014MNRAS.438.1813L} \\
G018.6$-$00.2 & 4.6$^{+0.6}_{-0.6}$          & H\,{\sc i} & \cite{2009AJ....138.1615J} \\
G018.8$+$00.3 & 12.0$^{+3.0}_{-5.1}$         & H\,{\sc i} & \cite{2007AA...474..541T} \\
G021.5$-$00.9 & 4.7$^{+0.4}_{-0.4}$          & PSR & \cite{2006ApJ...637..456C}, \cite{2008MNRAS.391L..54T} \\
G021.8$-$00.6 & 5.35$^{+0.15}_{-0.15}$       & CO, PSR & \cite{2008MNRAS.391L..54T}, \cite{2009ApJ...691..516Z} \\
G023.3$-$00.3 & 4.2$^{+0.3}_{-0.3}$          & H\,{\sc i}, CO & \cite{2008AJ....135..167L}, \cite{2007ApJ...657L..25T} \\
G027.4$+$00.0 & 8.5$^{+0.6}_{-1.0}$          & H\,{\sc i} & \cite{2008ApJ...677..292T} \\
G028.6$-$00.1 & 7.0$^{+1.5}_{-1.0}$          & H\,{\sc i}, NH & \cite{2001PASJ...53L..21B} \\
G028.8$+$01.5 & 4.0                           & NH & \cite{1994AA...286L..47S}, \cite{2010ApJ...725..931M} \\
G029.7$-$00.3 & 7.8$^{+2.8}_{-2.7}$          & H\,{\sc i} & \cite{2008AA...480L..25L} \\
G031.9$+$00.0 & 7.2                           & Maser & \cite{1996AJ....111.1651F} \\
G032.4$+$00.1 & 17                            & NH & \cite{2004PASJ...56.1059Y} \\
G032.8$-$00.1 & 5.2$^{+1.5}_{-0.4}$          & Maser & \cite{2011ApJ...743....4Z} \\
G033.6$+$00.1 & 7.0$^{+1.0}_{-0.5}$          & H\,{\sc i} & \cite{2009AA...507..841G}, \cite{1989ApJ...336..854F} \\
G034.7$-$00.4 & 3.0                           & Maser & \cite{2009AA...498..445P} \\
G035.6$-$00.4 & 3.6$^{+0.4}_{-0.4}$          & H\,{\sc i} & \cite{2013ApJ...775...95Z} \\
G039.2$-$00.3 & 6.5$^{+6.0}_{-0.3}$          & CO & \cite{2009ApJ...694.1266H}, \cite{2011ApJ...727...43S} \\
G041.1$-$00.3 & 10.3$^{+2.5}_{-3.9}$         & CO & \cite{2010ApJ...712.1147J} \\
G043.3$-$00.2 & 10$^{+2}_{-2}$               & H\,{\sc i} & \cite{2001ApJ...550..799B} \\
G049.2$-$00.7 & 4.3$^{+1.7}_{-0.0}$          & Maser, H\,{\sc i} & \cite{1997ApJ...475..194K}, \cite{2009ApJ...706L.270H}, \cite{2013ApJ...769L..17T} \\
G054.1$+$00.3 & 7$^{+2.0}_{-2.5}$            & H\,{\sc i} & \cite{2008AJ....136.1477L} \\
G054.4$-$00.3 & 3.0$^{+0.8}_{-0.8}$          & CO & \cite{1992AAS...96....1J}, \cite{1985AJ.....90.1224C} \\
G069.0$+$02.7 & 1.5$^{+0.6}_{-0.4}$          & H\,{\sc i}, PSR & \cite{2012MNRAS.423..718L} \\
G073.9$+$00.9 & 1.3$^{+0.7}_{-0.8}$          & NH & \cite{1993ARep...37..240L} \\
G074.0$-$08.5 & 0.58$^{+0.06}_{-0.06}$       & PM & \cite{2009ApJ...692..335B} \\
G074.9$+$01.2 & 6.1$^{+0.9}_{-0.9}$          & H\,{\sc i} & \cite{2003ApJ...588..852K} \\
G076.9$+$01.0 & 10.0$^{+5.0}_{-4.0}$         & NH & \cite{2011ApJ...739...39A} \\
G078.2$+$02.1 & 2$^{+2.0}_{-1.5}$            & H\,{\sc i} & \cite{2013MNRAS.436..968L}, \cite{2008AA...490..197L} \\
G089.0$+$04.7 & 1.7$^{+1.3}_{-1.0}$          & CO & \cite{2006ApJ...637..283B} \\
G106.3$+$02.7 & 0.8$^{+1.2}_{-0.1}$          & H\,{\sc i} & \cite{2001ApJ...560..236K} \\
G109.1$-$01.0 & 3.2$^{+0.2}_{-0.2}$          & H\,{\sc i}, CO & \cite{2012ApJ...746L...4K} \\
G111.7$-$02.1 & 3.4$^{+0.3}_{-0.1}$          & PM & \cite{1995ApJ...440..706R} \\
G114.3$+$00.3 & 1.0$^{+1.5}_{-0.3}$          & H\,{\sc i} & \cite{2004ApJ...616..247Y} \\
G116.5$+$01.1 & 1.6                           & H\,{\sc i} & \cite{2004ApJ...616..247Y} \\
G116.9$+$00.2 & 1.6$^{+1.9}_{-0.0}$          & H\,{\sc i} & \cite{2004ApJ...616..247Y}, \cite{1994ApJ...434..635H} \\
G119.5$+$10.2 & 1.4$^{+0.3}_{-0.3}$          & H\,{\sc i} & \cite{1993AJ....105.1060P} \\
G120.1$+$01.4 & 3.0$^{+2.0}_{-0.6}$          & H\,{\sc i} & \cite{2011ApJ...729L..15T}, \cite{2010ApJ...725..894H}, \cite{2008Natur.456..617K} \\
G127.1$+$00.5 & 1.15$^{+0.35}_{-0.25}$       & H\,{\sc i} & \cite{1977AA....59L..13P}, \cite{1993AA...270..393X}, \cite{2006AA...451..251L} \\
G132.7$+$01.3 & 2.2$^{+0.2}_{-0.2}$          & H\,{\sc i} & \cite{1991AA...247..529R} \\
G156.2$+$05.7 & 1.1$^{+1.9}_{-0.8}$          & NH & \cite{1991AA...246L..28P}, \cite{2007MNRAS.376..929G} \\
G160.9$+$02.6 & 0.8$^{+3.2}_{-0.4}$          & H\,{\sc i} & \cite{2007AA...461.1013L}, \cite{1991AJ....101.1033L} \\
G166.0$+$04.3 & 4.5$^{+1.5}_{-1.5}$          & H\,{\sc i} & \cite{1989MNRAS.237..277L} \\
G180.0$-$01.7 & 1.3$^{+0.22}_{-0.16}$        & PSR & \cite{2004AA...426..555S}, \cite{2007ApJ...654..487N}, \cite{2009ApJ...698..250C} \\
G184.6$-$05.8 & 1.93$^{+0.57}_{-0.43}$       & PM & \cite{1973PASP...85..579T} \\
G189.1$+$03.0 & 1.5                           & Maser & \cite{2006ApJ...652.1288H} \\
G205.5$+$00.5 & 1.5$^{+0.1}_{-0.7}$          & H\,{\sc i} & \cite{1986ApJ...301..813O}, \cite{1985ApJ...292...29F}, \cite{2012AA...545A..86X} \\
G260.4$-$03.4 & 2.2$^{+0.3}_{-0.2}$          & H\,{\sc i} & \cite{1988AAS...75..363D}, \cite{2008AA...480..439P} \\
G263.9$-$03.3 & 0.287$^{+0.017}_{-0.021}$    & PSR & \cite{2001PASJ...53.1025M}, \cite{2001ApJ...561..930C}, \cite{2003ApJ...596.1137D} \\
G266.2$-$01.2 & 0.75$^{+0.15}_{-0.25}$       & PM & \cite{2008ApJ...678L..35K} \\
G272.2$-$03.2 & 4.0$^{+1.0}_{-2.2}$          & NH & \cite{2011ApJ...732..114L} \\
G284.3$-$01.8 & 3                             & CO & \cite{1986ApJ...309..667R} \\
G290.1$-$00.8 & 7$^{+4.0}_{-3.5}$            & H\,{\sc i} & \cite{1996AA...315..243R}, \cite{2002ApJ...564..284S}, \cite{2006MNRAS.369..416R} \\
G291.0$-$00.1 & 5$^{+1}_{-1.5}$              & NH & \cite{1998ApJ...499..273H} \\
G292.0$+$01.8 & 6.2$^{+0.9}_{-0.9}$          & H\,{\sc i}, PSR & \cite{2003ApJ...594..326G} \\
G292.2$-$00.5 & 8.4$^{+0.4}_{-0.4}$          & PSR & \cite{2004MNRAS.352.1405C}, \cite{2000ApJ...541..367C} \\
G296.5$+$10.0 & 2.1$^{+1.8}_{-0.9}$          & H\,{\sc i} & \cite{2000AJ....119..281G} \\
G304.6$+$00.1 & 9.7$^{+4.3}_{-1.7}$          & H\,{\sc i} & \cite{1975AA....45..239C} \\
G308.4$-$01.4 & 9.8$^{+0.0}_{-3.9}$          & NH & \cite{2012AA...544A...7P} \\
G309.2$-$00.6 & 4.0$^{+1.4}_{-2.0}$          & NH & \cite{2001ApJ...548..258R} \\
G315.1$+$02.7 & 1.7$^{+3.7}_{-0.3}$          & PM & \cite{2007MNRAS.374.1441S} \\
G315.4$-$02.3 & 2.5$^{+0.3}_{-0.2}$          & PM & \cite{1996AA...315..243R}, \cite{2003AA...407..249S} \\
G315.9$-$00.0 & 8$^{+2}_{-2}$                & PSR & \cite{2009ApJ...703L..55C} \\
G316.3$-$00.0 & 7.2$^{+22.8}_{-2.5}$         & H\,{\sc i} & \cite{1975AA....45..239C} \\
G318.2$+$00.1 & 4.0$^{+5.4}_{-0.7}$          & H\,{\sc i} & \cite{2011arXiv1104.5119H} \\
G320.4$-$01.2 & 5.2$^{+1.4}_{-1.4}$          & H\,{\sc i}, NH & \cite{1999MNRAS.305..724G} \\
G321.9$-$00.3 & 6$^{+4.0}_{-0.5}$            & H\,{\sc i} & \cite{1993MNRAS.261..593S} \\
G326.3$-$01.8 & 4.1$^{+0.7}_{-0.7}$          & NH & \cite{1996AA...315..243R}, \cite{1993ApJ...419..733K} \\
G327.1$-$01.1 & 6.5$^{+6.5}_{-1.5}$          & NH & \cite{1999ApJ...511..274S} \\
G327.4$+$00.4 & 4.3                           & H\,{\sc i} & \cite{2001ApJ...551..394M} \\
G327.6$+$14.6 & 2$^{+0.2}_{-0.4}$            & PM & \cite{2013Sci...340...45N} \\
G328.4$+$00.2 & 17.4$^{+2.6}_{-5.4}$         & H\,{\sc i} & \cite{2001ApJ...551..394M} \\
G330.2$+$01.0 & 4.9                           & H\,{\sc i} & \cite{2001ApJ...551..394M} \\
G332.4$-$00.4 & 3.3                           & H\,{\sc i}, CO & \cite{2006PASA...23...69P}, \cite{2004PASA...21...82R} \\
G332.4$+$00.1 & 7.5$^{+3.5}_{-4.2}$          & NH & \cite{2004ApJ...604..693V} \\
G335.2$+$00.1 & 1.8                           & CO & \cite{2011AA...526A..82E} \\
G337.0$-$00.1 & 11.0                          & Maser & \cite{1996AJ....111.1651F} \\
G337.2$+$00.1 & 14.0$^{+16.0}_{-0.5}$        & H\,{\sc i}, NH & \cite{2005AA...431L...9C}, \cite{2006ApJ...653L..41C} \\
G337.2$-$00.7 & 5.8$^{+3.8}_{-3.8}$          & H\,{\sc i} & \cite{2006ApJ...646..982R}, \cite{2011ApJ...732..114L} \\
G337.8$-$00.1 & 12.3                          & Maser & \cite{1996AJ....111.1651F} \\
G338.3$-$00.0 & 10.0$^{+3.0}_{-2.0}$         & H\,{\sc i} & \cite{2009ApJ...706.1269L} \\
G343.0$-$06.0 & 1.0$^{+0.5}_{-0.5}$          & H\,{\sc i}, NH & \cite{2010ApJ...709..823K}, \cite{2003AA...403..605W}, \cite{2001MNRAS.325..287W} \\
G346.6$-$00.2 & 11.0                          & Maser & \cite{1996AJ....111.1651F} \\
G347.3$-$00.5 & 1.0$^{+0.3}_{-0.2}$          & H\,{\sc i}, CO & \cite{2005ApJ...631..947M} \\
G348.5$+$00.1 & 9$^{+0.5}_{-2.7}$            & H\,{\sc i} & \cite{2012MNRAS.421.2593T} \\
G348.5$-$00.0 & 6.3$^{+7.4}_{-3.3}$          & Maser & \cite{2012MNRAS.421.2593T} \\
G348.7$+$00.3 & 13.2                          & H\,{\sc i} & \cite{2012MNRAS.421.2593T} \\
G349.7$+$00.2 & 11.5$^{+0.7}_{-0.7}$         & Maser & \cite{1996AJ....111.1651F}, \cite{2014ApJ...783L...2T} \\
G350.1$-$00.3 & 4.5$^{+6.2}_{-0.5}$          & H\,{\sc i} & \cite{2008ApJ...680L..37G} \\
G351.7$+$00.8 & 13.2$^{+0.5}_{-11.1}$        & H\,{\sc i} & \cite{2007MNRAS.378.1283T} \\
G352.7$-$00.1 & 7.5$^{+0.9}_{-0.7}$          & H\,{\sc i}, CO & \cite{2009AA...507..841G} \\
G353.6$-$00.7 & 3.2$^{+0.8}_{-0.8}$          & H\,{\sc i}, CO & \cite{2008ApJ...679L..85T} \\
G357.7$+$00.3 & 6.9                           & Maser & \cite{1996AJ....111.1651F} \\
G357.7$-$00.1 & 12                            & Maser & \cite{1996AJ....111.1651F}, \cite{2003ApJ...594L..35G}, \cite{2004MNRAS.354..393L} \\
G359.1$-$00.5 & 4.6                           & Maser & \cite{2007IAUS..242..366Y}, \cite{2008ApJ...683..189H} \\
\enddata
\tablecomments{Table of SNR distances drawn from the literature. Note that ``\nodata'' indicates no data is available. The method for determining the distance is noted as: CO = line-of-sight velocity from molecular CO lines; H\,{\sc i} = kinematic distance from H\,{\sc i} absorption; NH = extinction estimate from optical or X-rays; Maser = kinematic distance from OH maser velocity; PM = Proper motions; PSR = association with pulsar. The d$_{\rm error}$ values indicate the range of uncertainties from the quoted distance values as assessed in the cited publications. The distance uncertainties are often asymmetric.
}
\end{deluxetable}

\FloatBarrier

\subsection{GeV-Radio Comparisons}
\label{Sec:GeVRadio}
\label{Sec:GeVRadioCorr}

We begin our MW correlation study by comparing the GeV and radio emissions from SNRs. The simplest model is that of a single emission zone in which relativistic particles (leptons, or leptons and hadrons with similar particle momentum distributions) are responsible for both radio and GeV emission. Starting from these simple models, we explore to what extent the data motivate more detailed physical models.

\subsubsection{Comparing Extensions}\label{Sec:RadioGeVExtCorrelation}

\begin{figure}[h!]
\centering
\includegraphics[width=0.8\columnwidth]{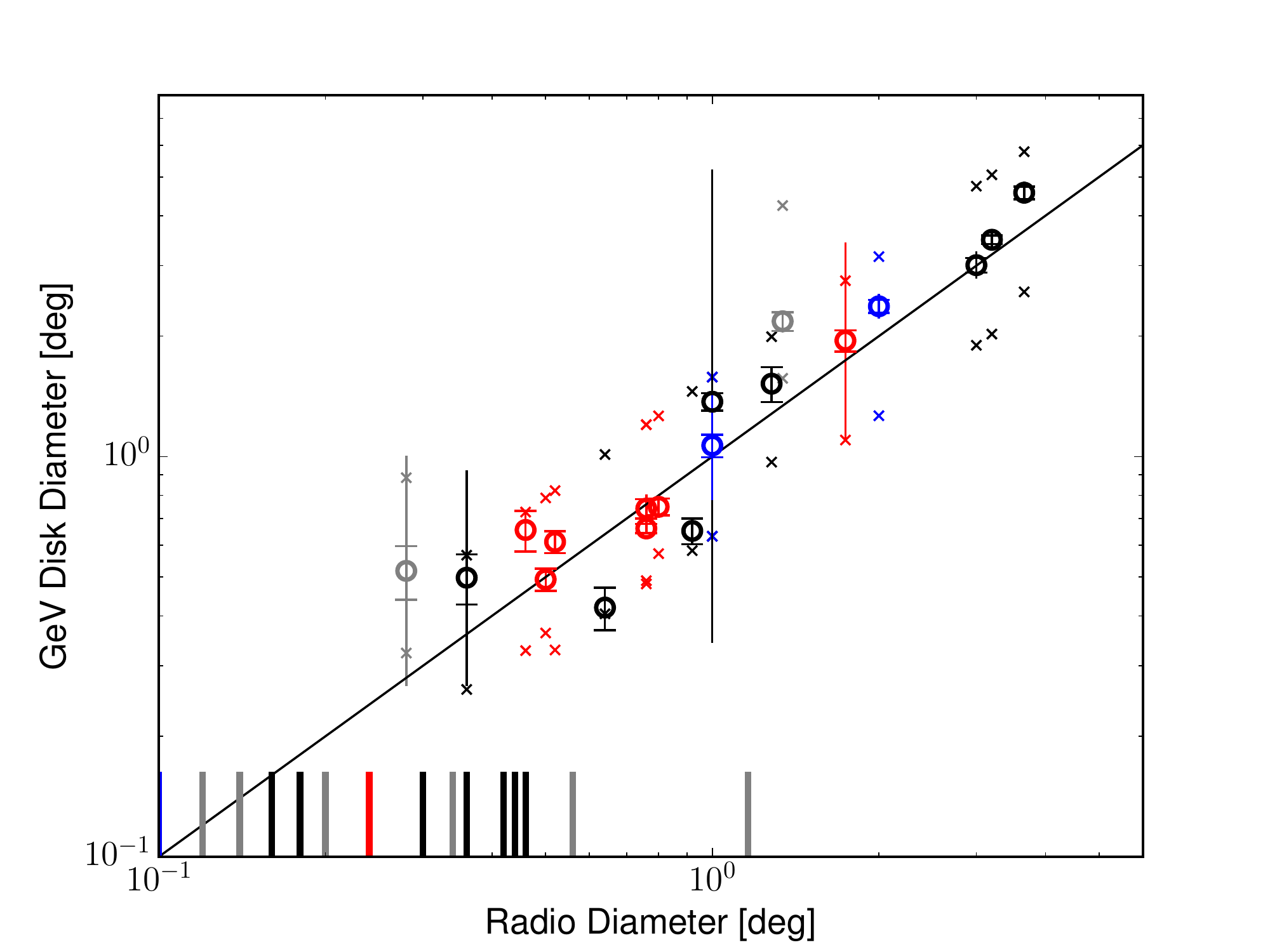} 
\caption{
The radio diameters of the SNRs from Green's catalog are correlated with the fitted GeV diameters for those candidates with significant extension. The solid line represents equal radio and GeV diameters. All cases of detected extension have diameters greater than $0.2$\degr. The ticks denote the radio extension of GeV point-like candidates, colored in order of their characteristics (young or interacting) and by their classifications (well defined or marginal). The small `x's bracketing the points show the minimum and maximum GeV extensions allowed such that the source remains classified or marginally classified (Equation~\ref{eq:overlapext_ext}) given the radio position and extension and best fit GeV position.
Symbols, colors, and error bars are as in Figure~\ref{fig:GeVFluxGeVIndex}.
}
\label{fig:GeVradioSize}
\end{figure}

We find that the best GeV diameter is within errors of the radio diameter for most of the candidates classified as associated with an SNR, as shown in Figure~\ref{fig:GeVradioSize}. All candidates with GeV extension, regardless of classification, have a diameter within $\sim 0.3$\degr{} and $\sim 20$\%\ of the radio diameter. Complex diffuse emission may cause sources with nominally detectable extensions to be either detected as point-like objects or not detected at all. 

The classification requirements defined in Section~\ref{Sec:Class}, namely Equations \ref{eq:overlaploc} and \ref{eq:overlapext_ext}, place constraints on the allowed relationship between the GeV and radio extensions. We calculated the minimum and maximum GeV extension for each radio/GeV candidate combination using the classification thresholds such that the classification remained the same (as classified or as marginally classified). This allowed range for the GeV extension is depicted by the bracketing `x's. The measured GeV extensions are well within the allowed range for the candidates with larger extensions (diameter~$\gtrsim0.5$\degr) and lower systematic errors. The majority of these also have very similar GeV and radio extensions, suggesting that the observed extension correlation is not an artifact of our procedure and in particular the chosen classification thresholds.
As we may over-estimate a candidate's extension rather than decomposing any substructure within it (Section~\ref{Sec:LocExtSpec}), we find conservatively that the GeV extension is not larger than the radio extension for the majority of the candidates.

The similarity observed for radio and GeV extensions is particularly interesting when considering what causes the emission. Our data support hypotheses which have both GeV and radio emission arising from the same location. One such scenario recently discussed is that, for predominately hadronic GeV emission, the \g-ray emission arises from the shock front expanding into and crushing nearby clouds \citep{Uchiyama10-crushedClouds}. With the radio synchrotron emission tracing the shock front, the GeV and radio emission should then be spatially coincident, as presently observed. In contrast, energetic CRs may escape the shock front, traveling ahead of it and illuminating nearby clouds \citep[e.g.][]{2009MNRAS.396.1629G}. In this case, with sufficient resolution, we might expect to see a systematically larger GeV extension than the shock front traced by the radio emission. 

\FloatBarrier

\subsubsection{Searching for Flux and Luminosity Correlations}\label{Sec:RadioGeVCorrelation}

\begin{figure}[h!]
\centering
\includegraphics[width=0.8\columnwidth]{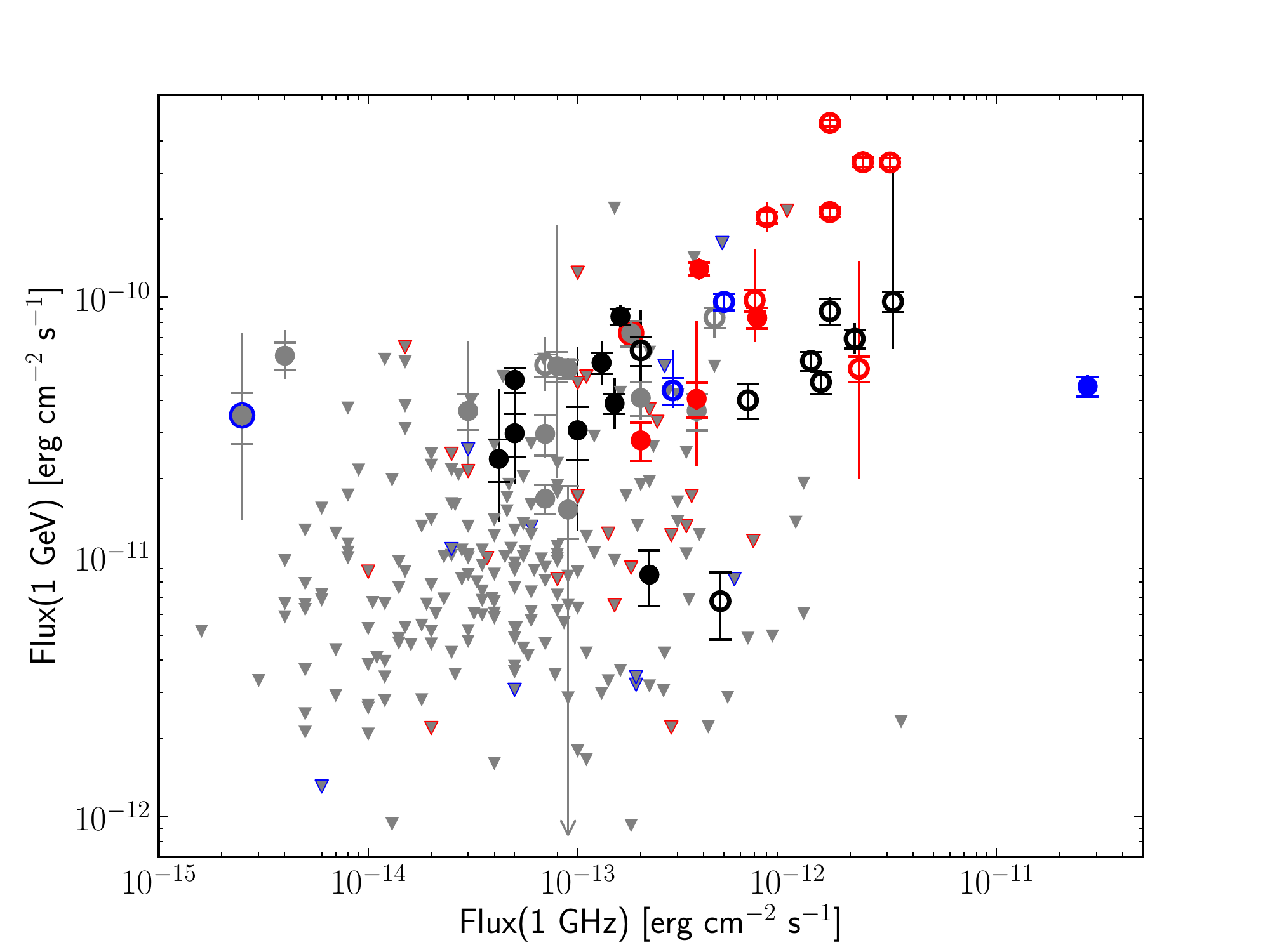} 
\caption{
Comparison of \g-ray and radio spectral flux densities for all SNRs and candidates. For all SNRs that were not detected or which failed classification, grey triangles indicate upper limits at $99\%$ confidence, computed assuming the radio location and extension. Symbols, colors, and error bars are as in Figure \ref{fig:GeVFluxGeVIndex}. 
}
\label{fig:GeVradioFlux}
\end{figure}

It has been suggested, particularly for interacting SNRs, that a correlation may exist between the radio and GeV flux \citep{Uchiyama10-crushedClouds}. Such a correlation could result from the same lepton population directly producing both the radio and GeV emission or because both the radio and GeV emission scale with some underlying physical parameter such as ambient density. Figure~\ref{fig:GeVradioFlux} shows the flux from synchrotron radio emission at $1$\,GHz in comparison to the \g-ray flux at $1$\,GeV. 

To search for evidence of a correlation between the radio and \g-ray flux, we applied Kendall's $\tau$ rank correlation test. We needed a physical \g-ray quantity comparable to the SNR's spectral energy flux density at $1$\,GHz ($\nu$F$_\nu$) in Green's catalog, so we computed the differential \g-ray flux at the reference energy $1$\,GeV from the $1-100$\,GeV band fluxes, indices, and upper limits. We note that radio flux densities are not measured values, but are instead interpolated or extrapolated from the observed radio spectrum of the source, which may or may not include a direct measurement at $1$\,GHz. Green's catalog contains no derived errors on the radio flux density. 

We then applied Kendall's $\tau$ rank correlation test in the same manner as described in \cite{Ackermann12-SFGs} to test for a significant deviation from the null hypothesis that the variables are not correlated. This test can identify even nonparametric correlations, and accounts for identical values and upper limits. The Kendall $\tau$ correlation coefficient is $\tau$ = $0.39$ for the sample of \nclassifiedsnrs~classified SNRs and $\tau$ = $0.17$ when upper limits are included. The significance of these values can be estimated if we assume that both the radio and GeV fluxes are independent variables and the sampling distribution of $\tau$ can be approximated by a normal distribution about a mean of zero with a variance given by $2(2n+5)/9n(n-1)$, where $n$ is the number of SNRs. Under these assumptions we estimate a possible correlation at a significance level of $0.7\sigma$ and $1\sigma$ for the classified SNRs and entire sample, respectively. Both radio and GeV fluxes have an implicit dependence on distance, and our \g-ray catalog is flux limited, so both variables may not be independent. This would only serve to decrease the significance of any positive correlation result. Thus we do not find evidence for a significant correlation between radio and GeV fluxes.

\begin{figure}[h!]
\centering
\includegraphics[width=0.8\columnwidth]{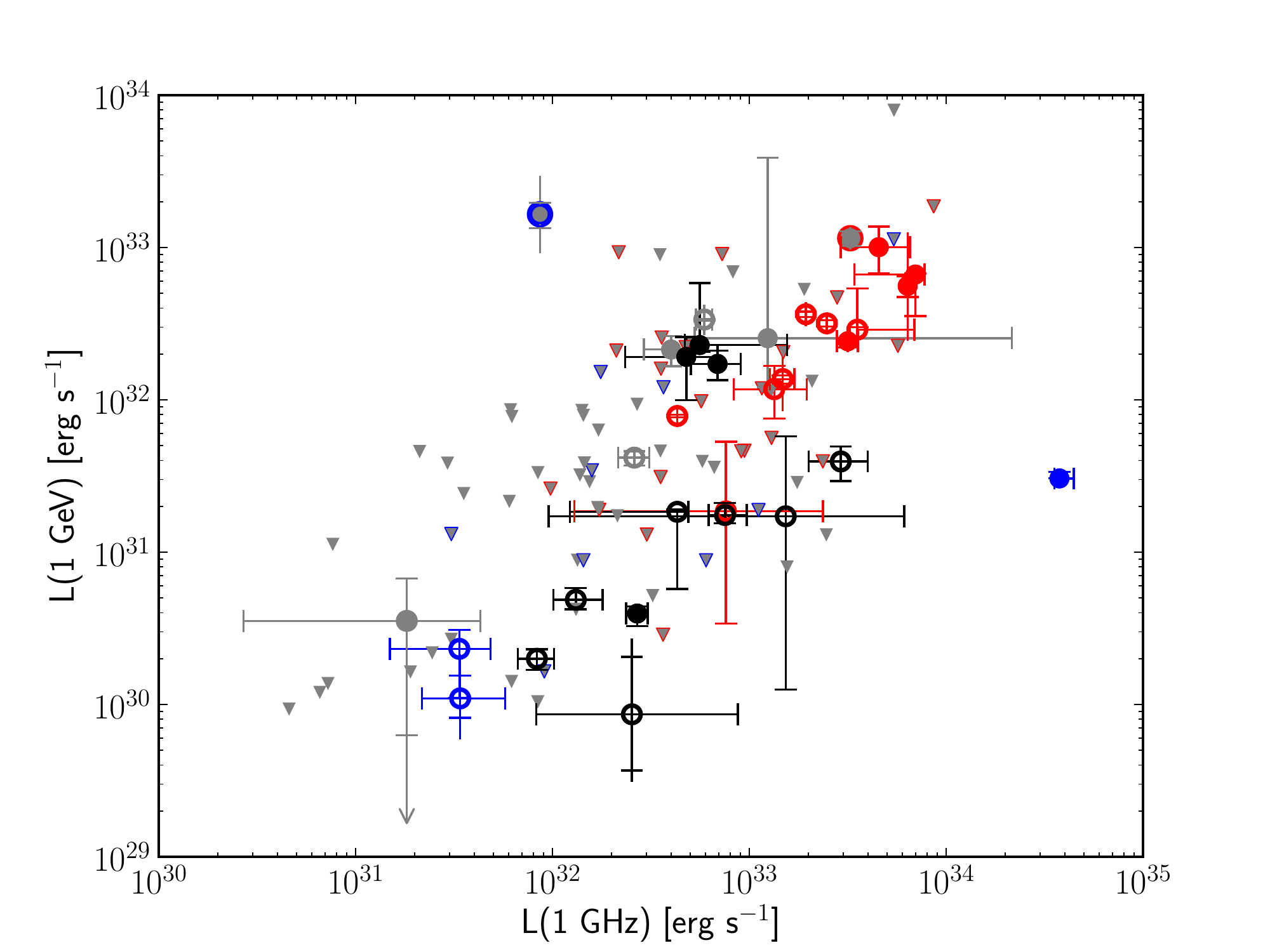} 
\caption{
Comparison of \g-ray and radio luminosities for all candidates and upper limits. The fluxes shown in Figure~\ref{fig:GeVradioFlux} have been converted to luminosities using the distances in Table~\ref{Tab:SNRinfo}. Upper limits do not include potential systematics due to uncertainties in the distance estimates. Symbols, colors, and error bars are as in Figure~\ref{fig:GeVFluxGeVIndex}. 
}
\label{fig:GeVradioLum}
\end{figure}

The lack of evidence for a correlation between the $1$\,GHz and $1$\,GeV fluxes does not mean that no correlation exists however, and a low significance may result from several factors. A physical correlation may exist but be masked by the conversion to flux \citep[for a detailed explanation]{Ackermann12-SFGs}. Figure~\ref{fig:GeVradioLum} shows the $1$\,GHz and $1$\,GeV luminosities. The Kendall $\tau$ rank correlation test indicates $\tau = 0.59$ ($1.0\sigma$) and $\tau$ = $0.22$ ($0.8\sigma$) for the $25$~classified SNRs and $102$~total SNRs with both radio fluxes and distance estimates, respectively. Thus we do not find evidence for a significant correlation between radio and GeV luminosities. We also note that the observed range of fluxes in the radio is approximately two orders of magnitude larger than that currently available with \g-ray observations, adding selection bias. Changes in spectral index at radio and GeV energies, explored in the following section, may also skew any intrinsic correlation. Thus we also cannot strictly rule out an intrinsic correlation. 

Finally, we note that even if a nominal correlation between the radio and GeV fluxes or luminosities was observed, it would not be clear evidence of a physical relationship. Despite its scatter, a correlation between radio surface brightness and diameter (the so called $\Sigma$-D relation) is observed, and the measured GeV diameters tend to correlate with the radio diameters. Therefore, any comparison with radio brightness or luminosity is also expected to show some correlation. Detailed modeling of the observational bias and impact of errors, such as described in \cite{Ackermann12-SFGs} is required to further investigate such a correlation and is beyond the scope of this paper.

\subsubsection{Probing Emission Mechanisms}\label{Sec:EmissionMechs}

\begin{figure}[h!]
\centering
\includegraphics[width=0.8\columnwidth]{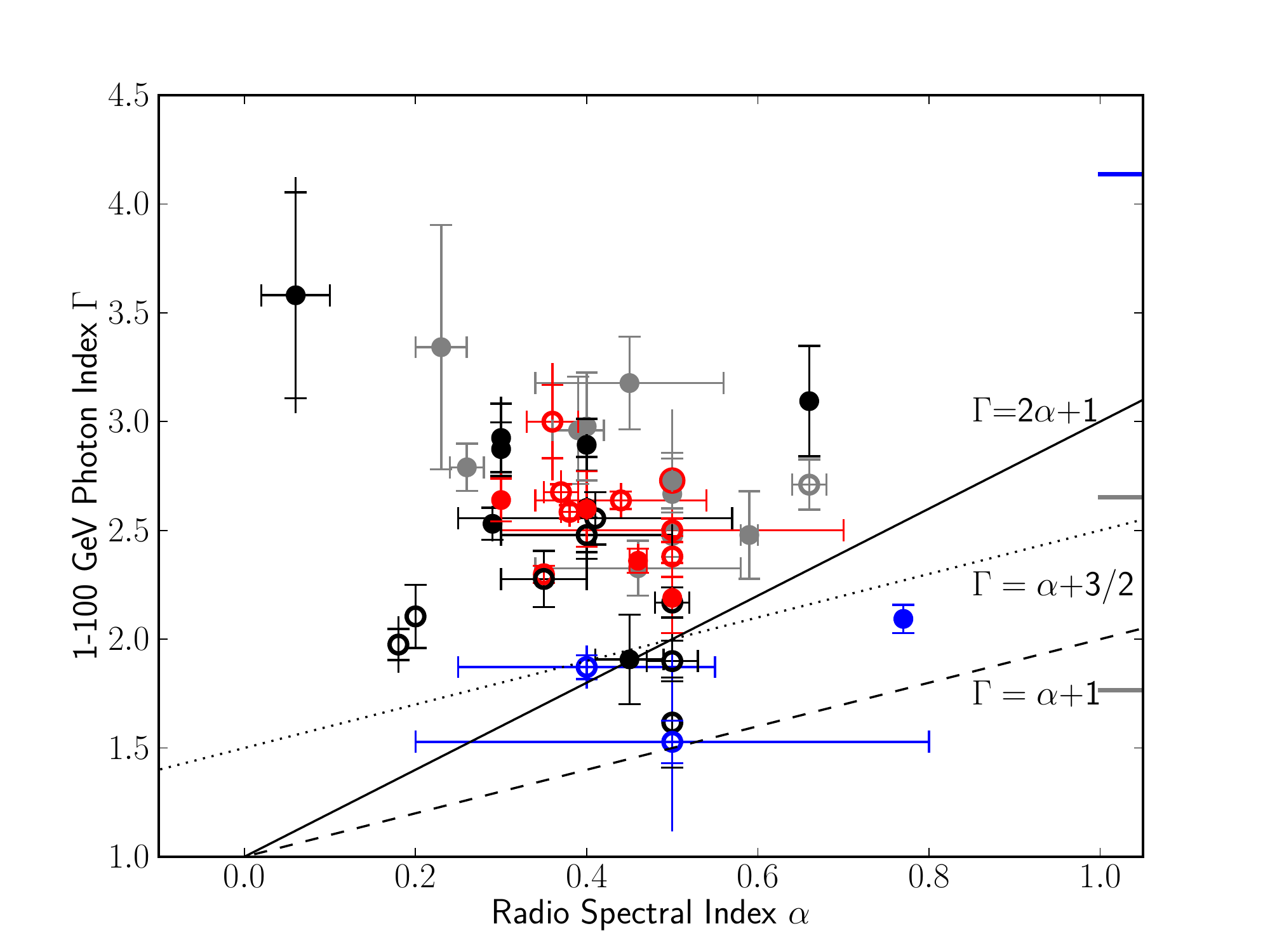} 
\caption{Comparison of radio spectal index, $\alpha$, and GeV photon index, $\Gamma$. The expected correlations are plotted for $\pi^0$ decay or e$^{\pm}$ bremsstrahlung (solid) and IC emission from an electron population that is freshly accelerated (dashed) or cooled by radiative processes (dotted). Emission via a combination of processes would fall between the lines (e.g. between the solid and dashed for a combination of $\pi^0$ decay and IC emission).  Symbols, colors, and error bars are as in Figure \ref{fig:GeVFluxGeVIndex}; ticks along the right hand side show the $1-100$\,GeV photon indices of those SNRs without reported radio spectral indices.
}
\label{fig:GeVradioIndex}
\end{figure}

We test for a relationship between radio and GeV emission and the underlying particle populations through the measured radio and GeV spectral indices. The energy of synchrotron-emitting leptons traced by $1$\,GHz observations depends on the magnetic field. If radio and GeV emission trace the same underlying particle population, then, at energies below the maximum energy reached by the accelerated particles, the photon indices of radio and \g-ray emission should be correlated. For $\pi^0$ decay and e$^\pm$ bremsstrahlung, the GeV and radio photon indices ($\Gamma$ and $\alpha$ respectively) are related as $\Gamma = 2\alpha + 1$. For IC scattering leptons, the GeV and radio photon indices follow $\Gamma = \alpha + 1$, or in the case in which high-energy leptons have been cooled via synchrotron or IC radiation, $\Gamma = \alpha + 3/2$. Figure \ref{fig:GeVradioIndex} compares the deduced radio spectral index $\alpha$ with the $1-100$\,GeV photon index $\Gamma$. 

Nearly all candidates have \g-ray photon indices that are softer than predicted given their radio spectra, regardless of the GeV emission mechanism. The three young SNRs in blue are most consistent with a single underlying particle population, and it has been suggested they emit via IC (dashed line) at GeV energies. The young SNR RX J1713$-$3946 is one of the few examples which bears out this case, being dominated by IC emission \citep{Abdo11-RXJ1713} and falling directly on the $\Gamma = \alpha + 1$ line. 
We also note that one classified extended candidate may be consistent with an IC origin, though no error was reported on the radio spectral index measurement by~\cite{1994MNRAS.270..106M}. This SNR is neither young nor a TeV source. SNRs emitting via a combination of mechanisms under these simple assumptions would have indices falling between the two index relations, that is, they would lie in the region spanned by the $\pi^0$/bremsstrahlung (solid) and IC (dashed) lines. 

The lack of an observed correlation between the indices as expected under these simple assumptions suggests that more detailed physical models are required for the majority of SNR candidates. The observed soft GeV spectra relative to the radio has several potential explanations. 
The underlying leptonic and hadronic populations may have different PL indices. The emitting particle populations may not follow a PL but may instead have breaks or even differing spectral shapes. Finally, there may be different zones with different properties dominating the emission at different wavelengths.

\subsection{GeV-TeV Comparisons}\label{Sec:GeVTeV}

Here we compare the GeV and TeV properties of SNRs to test the second common assumption in SNR models: that momentum distributions of the emitting particle populations do not follow simple PLs but have curvature or breaks. Such changes in spectral slope could also cause breaks in the \g-ray spectra. As TeV emission may originate via the same processes as the \FermiLat-observed GeV emission \citep[e.g.][]{Funk08_GeVTeVGalactic, Tibolla09_HESSUnIds, Tam10_VHEforFermi}, we might expect to see such a change reflected in a spectrum combining \FermiLat{} data with observations from Imaging Air Cherenkov Telescopes (IACTs) such as H.E.S.S., VERITAS, and MAGIC. The converse is also true, where detection predictions in the GeV based on simple PL extrapolation from the TeV have been borne out in GeV studies, e.g. identifications of H.E.S.S. sources from \cite{Tibolla09_HESSUnIds} in 2FGL \citep{Nolan12-2FGL} and \cite{Ackermann12_1FGLUnassoc}. As seen in earlier work on SNRs, particularly on those not clearly interacting with dense gas such as RX~J1713.7$-$3946 \citep{Abdo11-RXJ1713} and Tycho \citep{Giordano12-Tycho}, combining the TeV with the GeV observations significantly constrains the nature of the high energy emission. 

\pagestyle{empty}
\begin{deluxetable}{lcl}
\tablewidth{0pt}
\setlength{\tabcolsep}{0.05in}
\singlespace
\tabletypesize{\scriptsize}
\tablecaption{TeV Spectral Indices for GeV SNR Candidates\label{Tab:TeVSNRs}}
\tablewidth{5in}
\tablehead{
\colhead{Name} & 
\colhead{Index} & 
\colhead{Reference(s)} 
}
\startdata
G006.4$-$00.1 & 2.7$\pm$0.2 & \cite{Aharonian08-w28} \\
G008.7$-$00.1 & 2.7$\pm$0.1 & \cite{2006ApJ...636..777A} \\
G023.3$-$00.3 & 2.5$\pm$0.2 & \cite{2006ApJ...636..777A} \\
G043.3$-$00.2 & 3.1$\pm$0.3 & \cite{2011arXiv1104.5003B} \\
G049.2$-$00.7 & 2.4$\pm$0.1 & \cite{2011ICRC....7..115C} \\
G111.7$-$02.1 & 2.6$\pm$0.2 & \cite{2007AA...474..937A}, \cite{2010ApJ...714..163A} \\
G189.1$+$03.0 & 3.0$\pm$0.4 & \cite{2009ApJ...698L.133A} \\
G266.2$-$01.2 & 2.2$\pm$0.2 & \cite{2007ApJ...661..236A} \\
G347.3$-$00.5 & 2.0$\pm$0.1 & \cite{2007AA...464..235A} \\
G348.5$+$00.1 & 2.3$\pm$0.2 & \cite{2008AA...490..685A} \\
\enddata

\tablecomments{TeV indices reported for GeV SNR candidates, of which all pass the more robust classification threshold ($0.4$). 
}

\end{deluxetable}

\begin{figure}[h]
  \centering
   \includegraphics[width=0.8\columnwidth]{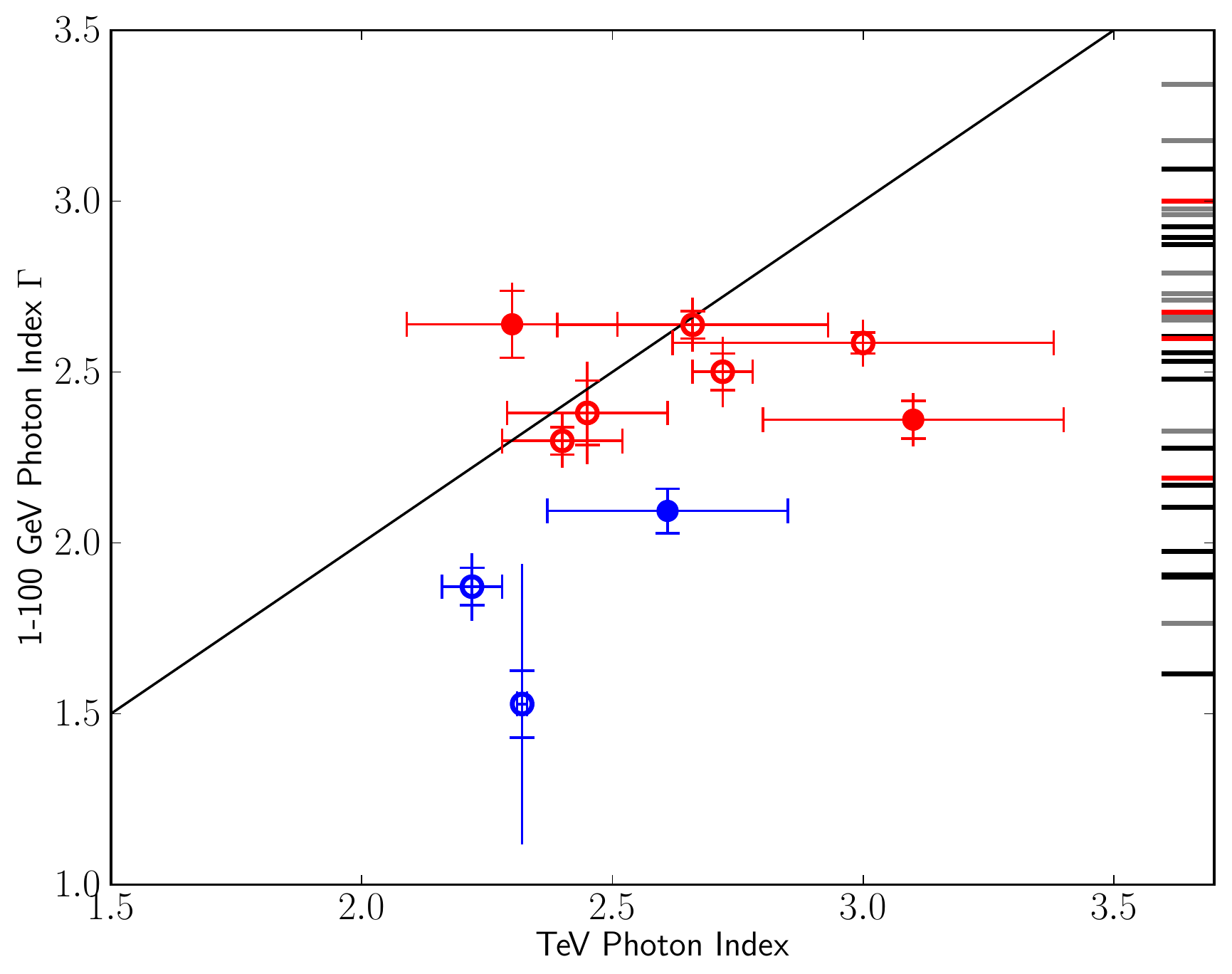} 
  \caption{GeV index compared to published index measurements from IACTs. The line corresponds to equal index values. The predominance of SNRs below the line suggests spectral curvature, potentially reflecting a change in spectral slope of the underlying particle population(s') index or indices. The ticks represent the GeV candidates with indices in the range of those with a TeV counterpart but with no TeV measurements themselves, demonstrating the limitations of the data set. Symbols, colors, and error bars are as in Figure~\ref{fig:GeVFluxGeVIndex}. 
}
  \label{fig:GeVTeVIndex}
 \end{figure}

In Figure~\ref{fig:GeVTeVIndex} we plot the PL index in the GeV versus TeV range for all SNRs observed with both \FermiLat{} and an IACT, tabulated in Table~\ref{Tab:TeVSNRs}. Six of the ten SNR candidates have TeV indices that are softer than their GeV indices, while three have GeV and TeV indices that are consistent with each other, within statistical and systematic errors. The remaining interacting candidate has a somewhat softer index at GeV energies than at TeV. Such a hardening of the index from GeV to TeV suggests that another particle population may dominate at higher energies or that the emission mechanism may change between the GeV and TeV regimes. The majority of the GeV SNR candidates do not have measured TeV indices, as seen by the ticks on the right of Figure~\ref{fig:GeVTeVIndex}, marking their GeV indices. Yet many of these are hard: $12$ candidates and $10$ marginal candidates have indices harder than $2.5$, suggesting they may well be observable by IACTs.

Of the GeV candidates with TeV observations, more than half have possible spectral curvature or breaks at or between GeV and TeV energies. Examples include IC~443 \citep{Abdo10-IC443} with a break at GeV energies and the young SNR RX~J1713$-$3946 \citep{Abdo11-RXJ1713} with a change in spectral slope near TeV energies. Such curvature also may explain the lack of a simple correlation between GeV and radio PL indices, as described in Section~\ref{Sec:EmissionMechs}.   

We note that, as the SNRs are not uniformly surveyed at TeV energies, drawing conclusions about the high energy properties of GeV SNRs requires a careful understanding of the non-TeV observed SNR subsample. Improved TeV studies will clearly provide a more robust comparison and, thereby, better inference of the momenta and any spectral curvature or breaks in the high energy particle population(s) in SNRs. Moreover, TeV studies of SNRs where hadronic emission has been established become crucial for determining the maximum energy to which the hadrons, likely CRs, are accelerated. This will help resolve the question of CR origins (see Section~\ref{Sec:CRs} for further discussion). 

We anticipate growth in this data set for individual, particularly larger SNRs and in the number of constraining upper limits. For example, the large SNR Monoceros is reported here for the first time with a probable GeV counterpart of extension $2.3$\degr. The relatively small fields of view for current IACTs, $\lesssim$\,$5$\degr, and the difficulties in searching for sources larger than the field of view, are both expected to be improved with the advent of Cherenkov Telescope Array (CTA), as well as with better background subtraction techniques for existing telescopes. In addition, we note that water-based Cherenkov observatories such as the High-Altitude Water Cherenkov Observatory (HAWC) have much larger fields of view, instantaneously viewing $\sim2\pi$\,sr, and are thus well suited for the study of large SNRs. The Galactic plane survey anticipated for CTA \citep[e.g.][]{Dubus13_CTA_GalPlane} and the Northern sky survey expected from HAWC \citep[e.g.][]{Westerhoff14_HAWC} could both provide a more complete census of energetic SNR counterparts and yield significantly constraining upper limits. In so doing, they have the potential to measure or constrain underlying particle populations and the maximum energies to which CRs are accelerated, significantly contributing to our knowledge of SNRs' aggregate ability to accelerate CRs.

We also note that the GeV-TeV index plot (Figure~\ref{fig:GeVTeVIndex}) also shows a distinct separation between young and interacting SNRs, which are often older. This suggests an evolution in index with age, from harder when younger to softer when older. 
We explore this further by explicitly investigating the evolution of the GeV index with age and exploring the role of environment in the next section.

\FloatBarrier

\subsection{Evolution or Environment}\label{Sec:EvolEnviron}

\begin{figure}[h!]
\centering
\includegraphics[width=0.8\columnwidth]{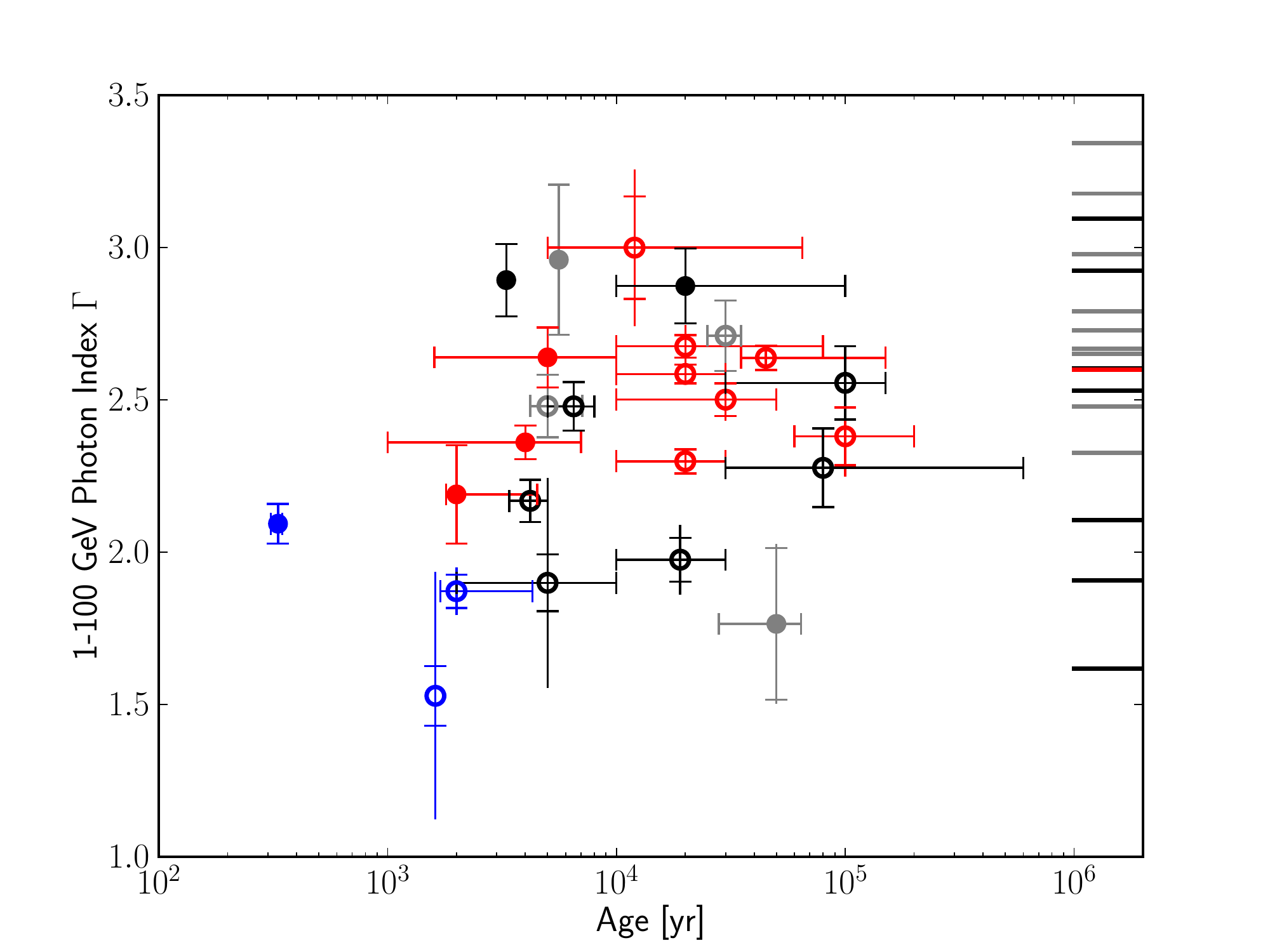} 
\caption{Age versus GeV spectral index. For those with ages in the literature, the young (blue) SNR candidates are separated in this phase space from the identified interacting candidates (red). The ticks on the right show indices for GeV candidates without well-established ages. Symbols, colors, and error bars are as in Figure~\ref{fig:GeVFluxGeVIndex}. 
\label{fig:AgeGeVIndex}}
\end{figure}

Because young SNRs tend to have harder spectral indices than interacting SNRs (Section \ref{Sec:GeVTeV}), in this section we explicitly examine the evolution of GeV index with age of the SNR. We take SNR ages from the literature and plot the $1-100$\,GeV photon index versus age in Figure~\ref{fig:AgeGeVIndex}. For our uniform sample of all GeV SNR candidates, young SNRs tend to have harder GeV photon indices than interacting SNRs, which are likely middle aged, though the scatter in age for the two classes is one to two orders of magnitude. There are two marginal candidates with faint fluxes and no determined ages that provide exceptions to this trend (see Figure~\ref{fig:GeVFluxGeVIndex}). The candidate interacting SNR has a very hard index and the candidate young SNR has a particularly soft index. Due to the lack of MW information, these two marginal candidates (G304.6+0.1 and G32.4+0.1) cannot always be shown on the following plots, but should be borne in mind. 

\begin{figure}[h]
\centering
\includegraphics[width=0.8\columnwidth]{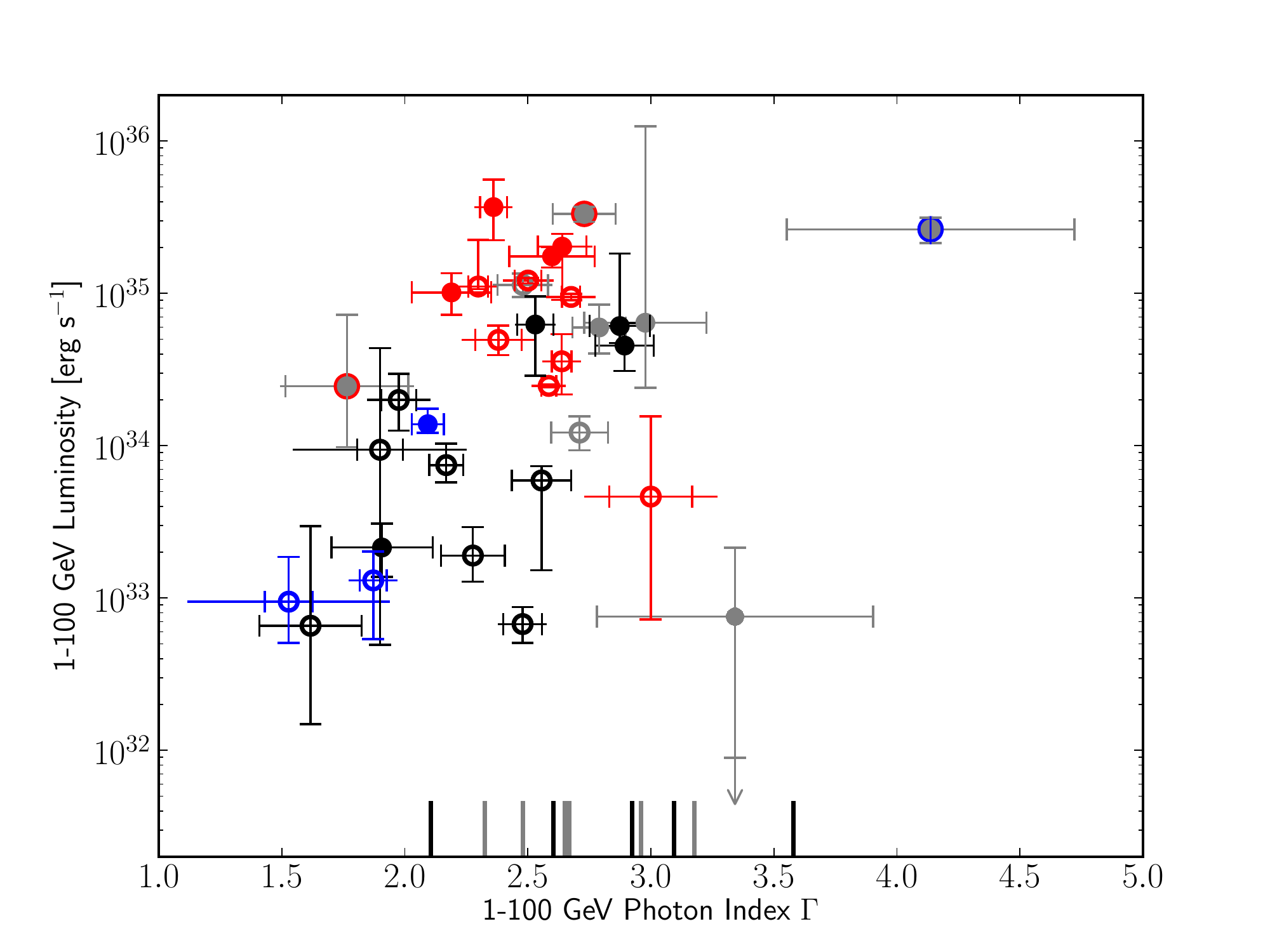} 
\caption{$1-100$\,GeV luminosity versus PL index, with tick marks representing the GeV candidates without reliable distance estimates in the literature. Symbols, colors, and error bars are as in Figure~\ref{fig:GeVFluxGeVIndex}. 
}\label{fig:LumIndex}
\end{figure}

The general trend of younger SNRs having harder indices may be due to the decrease of the maximum acceleration energy as SNRs age and their shock speeds slow down. This would also result in fewer particles being swept up by the shock front, given a constant density, suggesting a corresponding decrease in luminosity with age. For example, \citet{2011JCAP...05..026C} updated a simple evolutionary model for CR acceleration by remnants of a massive progenitor and showed that harder CR spectra result from lower acceleration efficiencies and that the index begins to soften after $\sim10^4$\,yr. This model also predicts a decrease in GeV emission with age until at least $10^4$\,yrs. Figure~\ref{fig:LumIndex} shows the distribution of $1-100$\,GeV luminosity and index for the observed candidates. Harder, fainter young SNRs are clearly separated from brighter and softer older SNRs. Barring the two marginal candidates mentioned before, this is consistent with previous observations, including \cite{Thompson12-FermiCRsReview}, who noted that these older SNRs are also often interacting with large MCs. 

\begin{figure}[h]
\centering
\includegraphics[width=0.8\columnwidth]{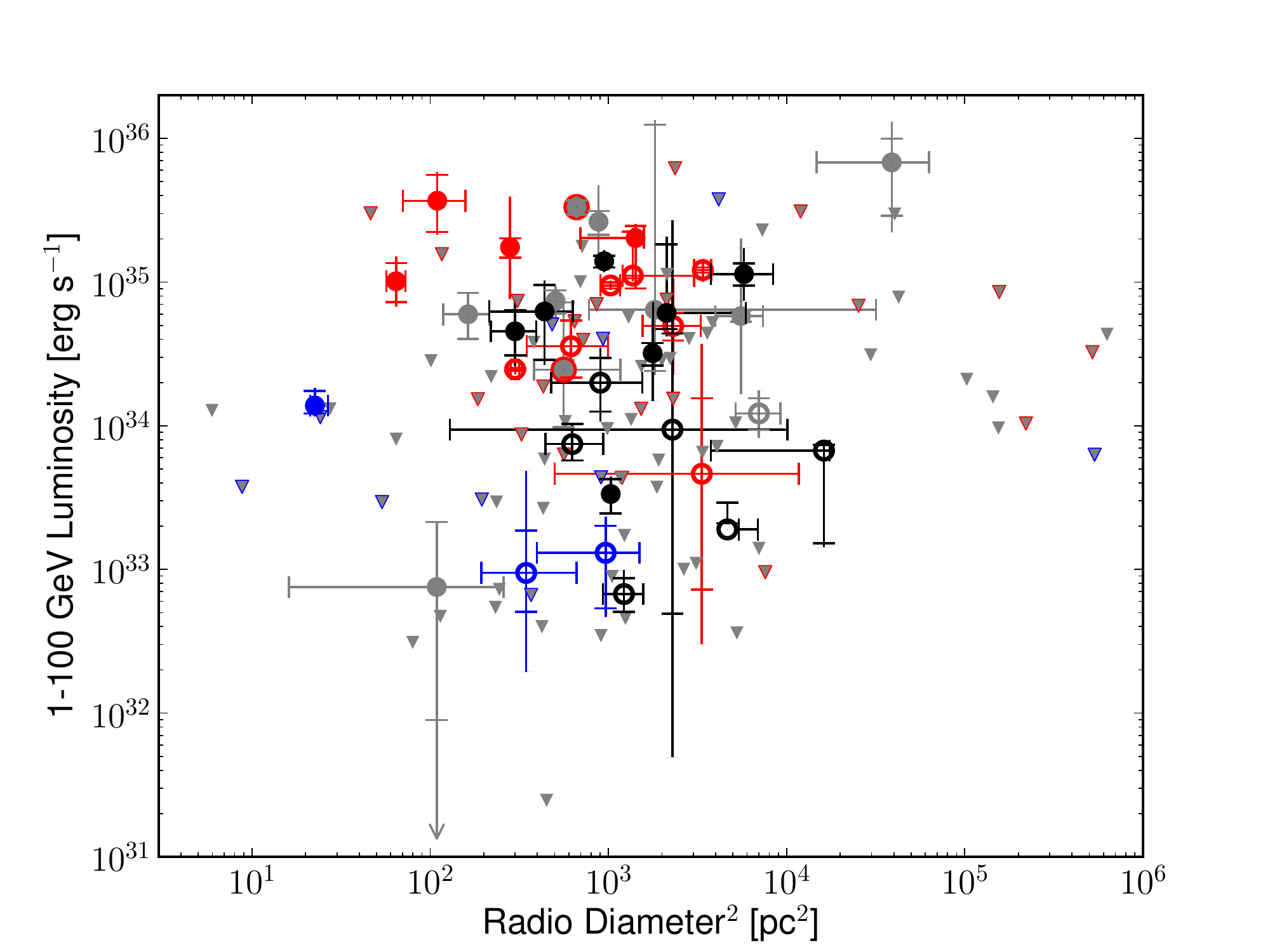} 
\caption{
The $1-100$\,GeV luminosity is plotted against the square of the radio diameters in pc of those SNRs with known distances. Symbols, colors, and error bars are as in Figure~\ref{fig:GeVFluxGeVIndex}. 
 \label{fig:LumDia}}
\end{figure}

To investigate the role of environment in the trends for the young and interacting SNRs, we examined the GeV luminosity versus radio diameter in Figure~\ref{fig:LumDia}. The square of the physical diameter ($D$) can be regarded as a reasonable indicator for SNR age and environment, as its evolution during the Sedov-Taylor phase follows
\begin{equation}
\label{eq:Sedov}
{\it D} \propto n_0^{-1/5} \ {\it E}_{\mathrm{SN}}^{1/5} \ {\it t}^{2/5}
\end{equation}
where $n_0$ is the ambient density of the surrounding medium, $E_{\mathrm{SN}}$ is the supernova energy, and $t$ is the age of the SNR \citep{1950RSPSA.201..159T,1959sdmm.book.....S}. We can thus use the physical diameter as an age proxy: ``effective age". Any apparent correlation between the luminosity and $D^2$ may be due to their inherent dependence on distance (squared). As observed in earlier works, e.g.~\cite{Thompson12-FermiCRsReview}, Figure~\ref{fig:LumDia} shows that, for the detected candidates, interacting SNRs are generally more luminous for a given physical diameter than young SNRs, though there is large scatter. This suggests that SNRs at the same effective age may be more luminous because they have encountered denser gas ($n_0$). With the addition of upper limits, we find some interacting candidates are constrained to lie below the luminosities of most young SNRs. Thus, as we continue to detect SNRs with increasingly fainter \g-ray fluxes, we are likely to find less separation between the luminosities of the two classes. 

\begin{figure}[h]
\centering
\includegraphics[width=0.8\columnwidth]{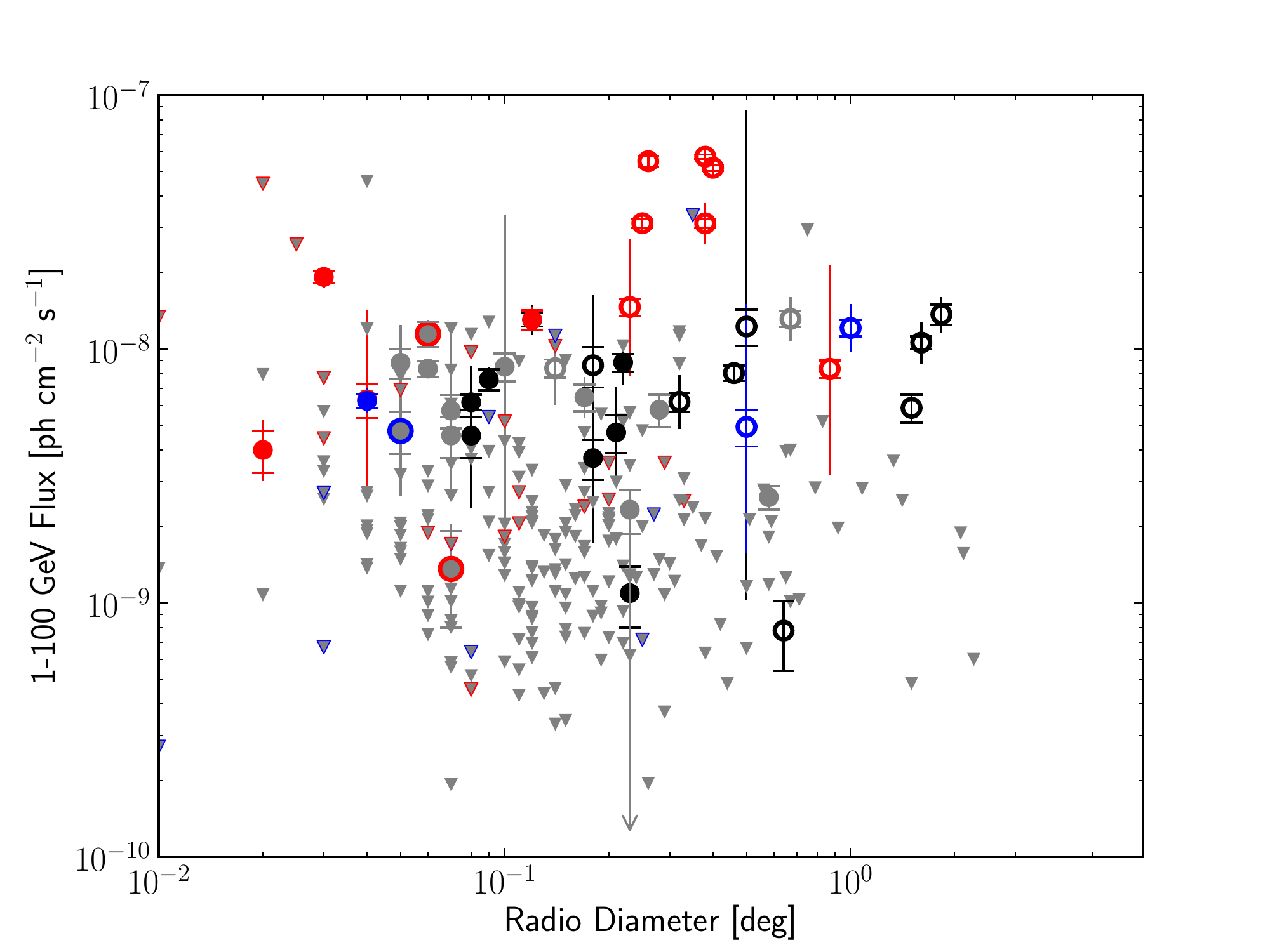} 
\caption{The $1-100$\,GeV flux and upper limits are plotted against the angular diameters of all SNRs observed in the radio. Symbols, colors, and error bars are as in Figure \ref{fig:GeVFluxGeVIndex}. 
}\label{fig:GeVFluxSizeDeg}
\end{figure}

It should also be noted that there is an explicit correlation between the luminosity and physical diameter plotted in Figure~\ref{fig:LumDia} as both are proportional to distance (squared), which is only reliably measured for a subset of our sample. Observational biases, including that young, often smaller and fainter SNRs tend to be more difficult to detect in the radio as well as in \g-rays, may also affect the observed trends. Figure~\ref{fig:GeVFluxSizeDeg} plots the measured GeV flux versus observed radio diameter, and shows no clear separation of classes. No correlation is observed between SNR flux and angular size. 

While there is some separation between young and interacting classes of SNRs in the age, physical radio diameter, GeV luminosity, and GeV index phase space, the scatter is often large. The tendency of younger SNRs to be harder and less luminous in the GeV band than older, often interacting SNRs may be countered as the available MW information increases and as the GeV flux limit decreases as the LAT accumulates additional exposure. This is particularly true in terms of distances and ages. A general trend of softer GeV index and lower GeV luminosity may be caused by the shock front decreasing in speed as the SNR ages. On the other hand, the interacting SNRs may be more luminous due to their interactions with denser surroundings not yet reached by younger SNRs. Ultimately, the large scatter observed in luminosity will likely reflect effects due to both age and environment. A more uniform and complete MW data set will enable significantly greater insight into links between age, environment, and other observed characteristics. 

An SNR's environment may also be explored to some moderate distance beyond its immediate surroundings. In addition to the SNR candidates associated in this catalog, we find a considerable fraction of sources, $\sim55\%$, not spatially coincident but within $5$\degr{} of a radio SNR. Regions containing SNRs are often rich in gas and many of these ``other'' candidates are robust to changes to the IEM and typically have PL spectra at these energies. Nearby regions of high-density gas may be illuminated by CRs escaping the SNR \citep[e.g.][]{2009MNRAS.396.1629G}, which interact to produce detectable GeV emission regardless of whether the SNR itself is sufficiently bright to be detected. Alternatively, studies may show that the SNR shock front is compressing the gas, crushing the cloud, accelerating CRs, and emitting at GeV energies \citep[e.g.][]{Uchiyama10-crushedClouds}. In either case, those ``other'' candidates which are associated with an SNR's environment will help disentangle the various scenarios, their relative rates of occurrence in the Galaxy, and their associated SNR's contribution to the Galactic CR population. In addition, upper limits on SNRs remaining below the resolvable level will constrain the population's contribution to the diffuse Galactic interstellar \g-ray emission.

\FloatBarrier

\section{Constraining SNRs' Cosmic Ray Contribution}\label{Sec:CRs}

SNRs have long been held the most promising candidate sources of Galactic CRs, capable of supplying the flux observed at Earth if they are on average $\sim 5-10\%$ efficient in accelerating CR protons and nuclei \cite[e.g.][]{Strong10-CREnergyBudget}. Recent work examining GeV \g-ray data around half the $\pi^0$ rest mass has led to the detection of the characteristic $\pi^0$ low energy break ($E<100$\,MeV) in two SNRs, IC~443 and W44 \citep{Ackermann13-pionBump}, thus adding another piece to the accumulating evidence \cite[e.g.][]{Thompson12-FermiCRsReview, Brandt13-CTB37A, Castro13-3LatSnrs} that at least some SNRs do accelerate hadrons. However, the question remains as to whether the Galactic SNR population is able to accelerate and release CRs with the appropriate composition and flux up to the transition between the Galactic and extragalactic components.

This systematic search for \g-ray emission from the population of known SNRs in the Galaxy provides the first opportunity to address this question from the perspective of high energy photons. In Section~\ref{Sec:CRConstraintMethod}, we describe the method we use to constrain the CR energy content in SNRs through the measured \g-ray spectral parameters from the classified and marginal SNR candidates as well as the derived upper limits from the others (unclassified or not significant). In Section~\ref{Sec:CRConstraintDiscuss}, we discuss the implications of these constraints within the general SNR-CR paradigm.

\subsection{Method for Constraining CRs from GeV Detections and Upper Limits}\label{Sec:CRConstraintMethod}

In the following, we assume that the \g-ray emission from SNRs probed with \FermiLat{} entirely arises from the interaction of CR protons and nuclei with the surrounding ISM or circumstellar medium through the production and subsequent decay of $\pi^0$. Given that two other emission mechanisms involving accelerated leptons, namely non-thermal bremsstrahlung and IC scattering, could also contribute in the \g-ray domain, the constraints derived from the \FermiLat{} measurements should be considered as upper limits on the CR energy content in SNRs.

For an SNR at a distance $d$ whose forward shock propagates through a medium with a density $n$ and accelerates CR particles following a PL spectrum in momentum with spectral index $\Gamma_{\mathrm{CR}}$, the \g-ray flux resulting from proton-proton (p-p) interactions can be related to the CR energy content, $E_{\mathrm{CR}} \equiv \epsilon_{\mathrm{CR}}E_{\mathrm{SN}}$, as shown in \cite{Drury94-GraySNRsCRs}. The parametrization derived by these authors is valid as long as the effect of the high energy cutoff, related to the maximum particle energy $E_{\mathrm{CR, max}}$, can be neglected. We investigated the effect of $E_{\mathrm{CR, max}}$, taken as an exponential cutoff energy, by computing the \g-ray flux in the $1-100$\,GeV range, F($1-100$\,GeV), for different values of $\Gamma_{\mathrm{CR}}$ and $E_{\mathrm{CR, max}}$ following the model of \g-ray production from p-p interactions developed by \cite{Kamae06-PPinteractions}. To account for the contribution from heavier nuclei in both CRs and ISM, we employed a nominal nuclear enhancement factor of $1.85$ \citep{Mori09-EnhancementFactor}, neglecting the energy dependence of these metallicity effects over the two decades studied \citep{Kachelriess14-EnhancementFactor}. 

\begin{figure}[h]
\centering
\includegraphics[width=0.512\columnwidth]{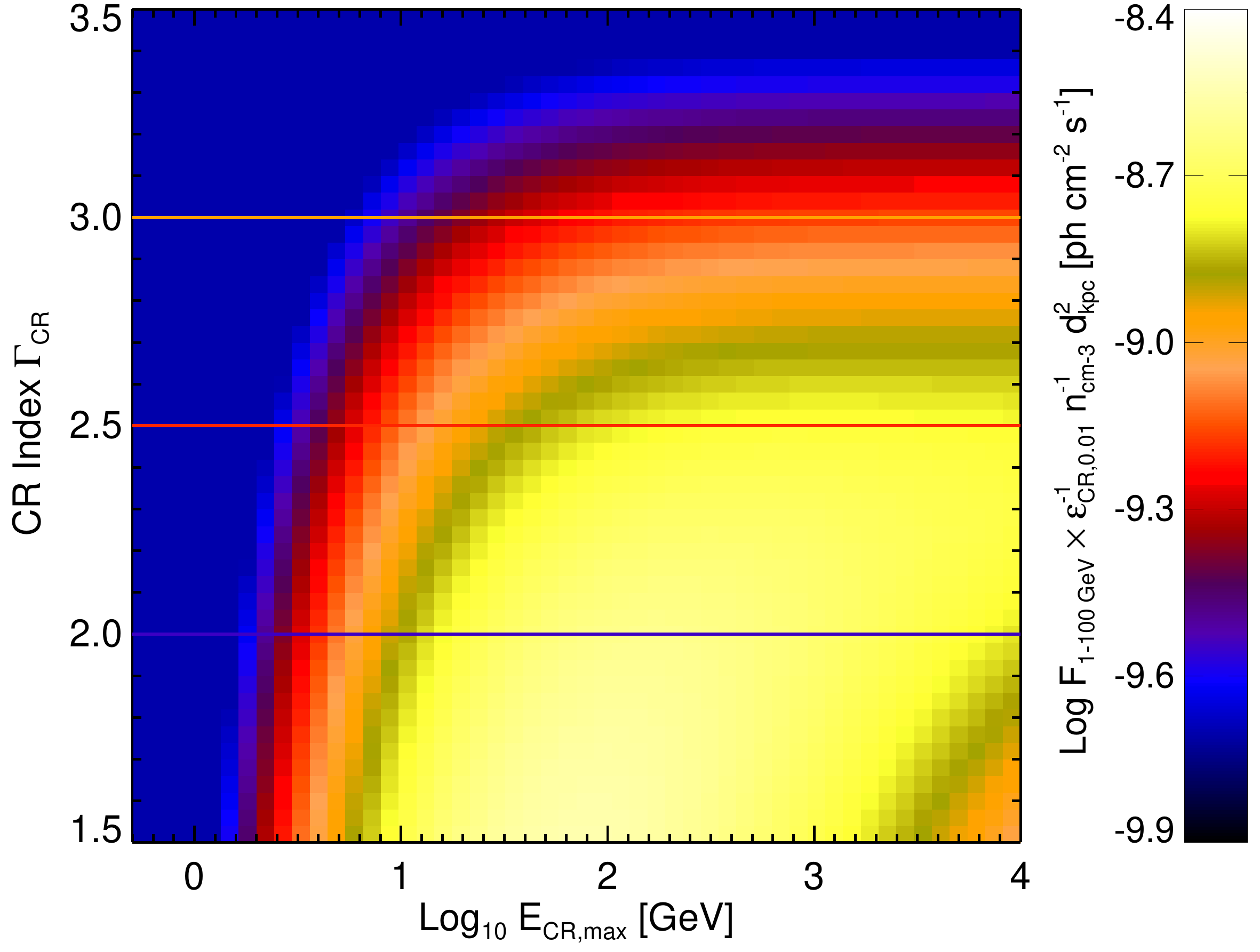} 
\includegraphics[width=0.476\columnwidth]{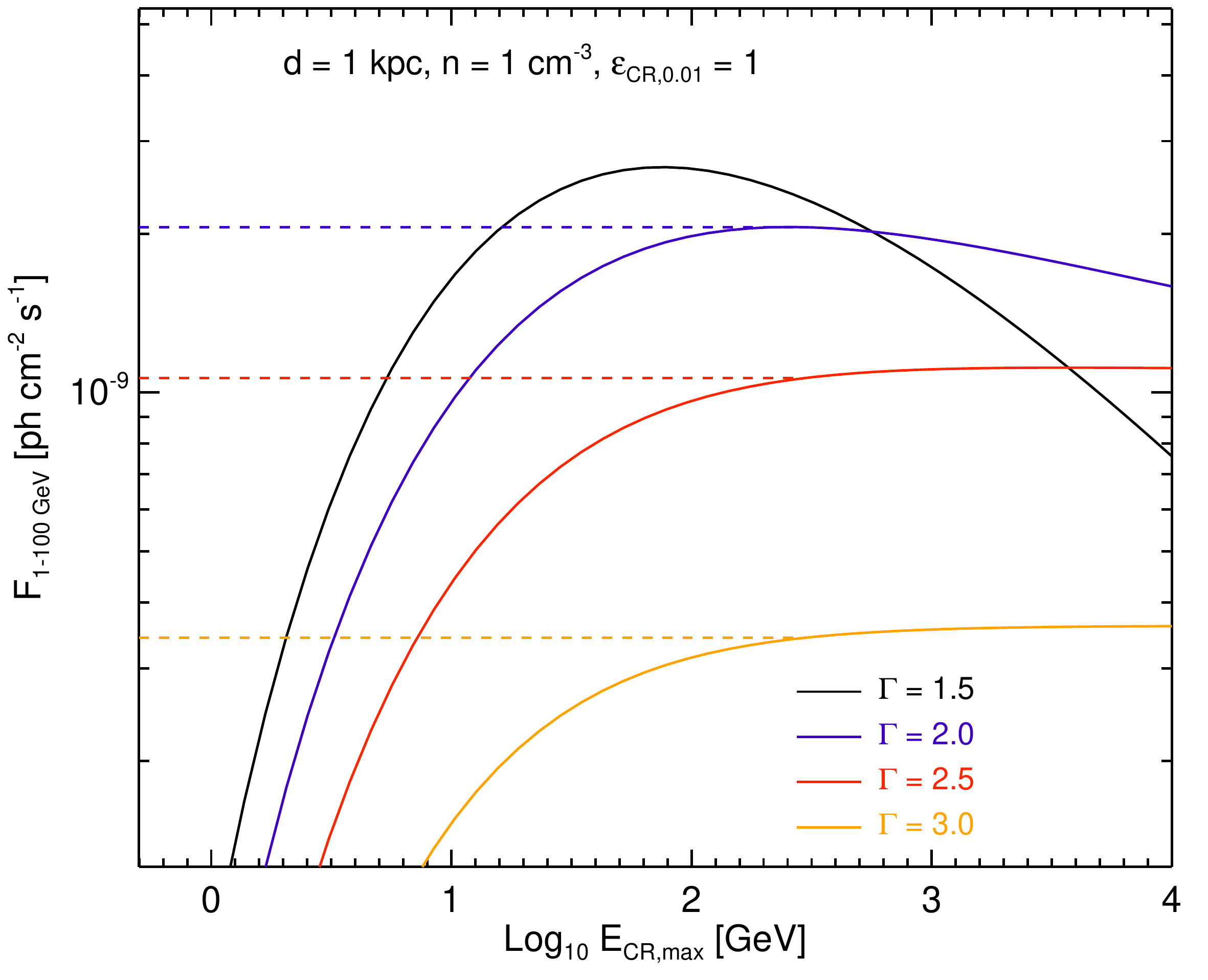} 
\caption{{\it Left:} Under standard assumptions (see text), an SNR's \g-ray flux in the $1-100$\,GeV range can be related to the accelerated CRs' maximal energy $E_{\mathrm{CR, max}}$ and spectral index $\Gamma_{\mathrm{CR}}$ for a given CR energy content above a particle momenta of $10$\,MeV\,$c^{-1}$ ($\epsilon_{\mathrm{CR}} = E_{\mathrm{CR}}/E_{\mathrm{SN}} = 0.01$), effective density ($1$\,cm$^{-3}$), and distance to the SNR ($1$\,kpc). {\it Right:} the relationship between the SNR's \g-ray flux in the $1-100$\,GeV band and $E_{\mathrm{CR, max}}$ for different values of $\Gamma_{\mathrm{CR}}$. For $E_{\mathrm{CR, max}}\gtrsim 200$\,GeV and $\Gamma_{\mathrm{CR}}\gtrsim 2$, the flux is weakly dependent on the CR maximal energy.}
\label{fig:CR_FluxvsEmaxAndIndex}
\end{figure}

Figure~\ref{fig:CR_FluxvsEmaxAndIndex} gives the \g-ray flux in the ($E_{\mathrm{CR, max}}$,~$\Gamma_{\mathrm{CR}}$) plane and shows its dependence on $E_{\mathrm{CR, max}}$ for different CR spectral indices, with $d =1$\,kpc, $n =1$\,cm$^{-3}$, $\epsilon_{\mathrm{CR}} =0.01$, and $E_{\mathrm{SN}}=10^{51}$\,erg. The CR energy content is computed for particle momenta above $10$\,MeV\,$c^{-1}$. The \g-ray flux is nearly independent of the CR maximal energy as long as $E_{\mathrm{CR, max}}\gtrsim 200$\,GeV and $\Gamma_{\mathrm{CR}}\gtrsim 2$. In this case, it can conveniently be approximated using the following expression: 
\begin{equation}
F(1-100\,\textrm{GeV})  \approx f(\Gamma_{\mathrm{CR}}) \times \frac{\epsilon_{\mathrm{CR}}}{0.01} \times \frac{E_{\mathrm{SN}}}{10^{51}\,\textrm{erg}}  \times \frac{n}{1\,\textrm{cm$^{-3}$}}  \times \left(\frac{d}{1\,\textrm{kpc}}\right)^{-2} \times 10^{-9}\,\textrm{cm$^{-2}$ s$^{-1}$} 
\label{eq:accelEfficiency}
\end{equation}
where $f(\Gamma_{\mathrm{CR}})$ is such that $f(2.0) = 2.06, f(2.5) = 1.07$, and $f(3.0) = 0.34$, as shown in Figure~\ref{fig:CR_FluxvsEmaxAndIndex} (right). The above equation is consistent with the expression given in \citet[][see their Table~1 and Equation~9]{Drury94-GraySNRsCRs}, and the resulting estimates of $E_{\mathrm{CR}}$ are in agreement with those obtained from dedicated studies of some well-known \g-ray emitting SNRs such as Cas~A \citep{Yuan13-CasA}. However, such a parametrization does not account for deviations from a PL in the particle and photon spectra, while spectral curvature has been measured for several SNRs interacting with MCs, such as IC~443 and W44 \citep{Ackermann13-pionBump} whose spectra are preferentially fitted with a logP in the present study (see Table~\ref{Tab:ResultsSpec}). Equation~\ref{eq:accelEfficiency} assumes that the (time-dependent) adiabatic and radiative losses do not significantly affect the particle spectra throughout an SNR's evolution. The effect of adiabatic and radiative losses is both complicated and presently under discussion, including in works by \citet{LeeEllison} and \citet{Bell14_CRsFromSNe}, and accounting for them is beyond the scope of this work.

\FloatBarrier

\subsection{CRs from GeV Detections and Upper Limits}\label{Sec:CRConstraintDiscuss}

We use the relationship between an SNR's \g-ray flux, density, and distance shown in Equation~\ref{eq:accelEfficiency} to determine the maximal CR energy content $E_{\mathrm{CR}}$ through $\epsilon_{\mathrm{CR}}$ contributed by every SNR for which we have measured the \g-ray flux and photon index (as reported in Table~\ref{Tab:ResultsSpec}) or derived an upper limit at the $95$\% confidence level (as reported in Table~\ref{Tab:ResultsULs}). In the case of a detected SNR, the photon index, known to reproduce the spectral shape of the parent CR proton/nuclei spectrum above a photon energy of $1$\,GeV \citep[see e.g.][]{Kamae06-PPinteractions}, is taken to be equal to $\Gamma_{\mathrm{CR}}$. In the case of the upper limits, we assume an index of~$2.5$ (i.e.~the average value of the detected SNRs). We use the canonical value of $10^{51}$\,erg for~$E_{\mathrm{SN}}$.

Translating SNRs' \g-ray measurements into constraints on their contribution to $E_{\mathrm{CR}}$ also requires knowledge of their distances and effective densities. For the former, we made use of the distances gathered from the literature and reported in Table~\ref{Tab:SNRinfo}. For the latter, we turn to the $\sim175$ SNRs detected in X-rays \citep{Ferrand12-XraySNRs}, specifically those with thermal emission confidently associated with the forward shock. Following the construction of Equation~\ref{eq:accelEfficiency}, we only considered density estimates based on thermal X-ray emission from the shock-heated ISM or circumstellar medium, and discarded those derived from X-ray emitting regions associated with the ejecta, as in the so-called mixed-morphology SNRs, or with clumps and regions known to be sites of MC interaction. Densities were also obtained from measurements such as IR emission from collisionally heated dust and SNR hydrodynamics. All these constraints on the {\it upstream} density of $25$~SNRs have been gathered from the literature and references are given in Table~\ref{Tab:SNRinfo}. 
Note that we consider the {\it downstream} density to be the relevant value when converting the \g-ray fluxes and upper limits into CR energy contents, and hence, we applied a factor of $4$, i.e.~the compression ratio in the simple case of a strong, unmodified shock\footnote{A larger value could be expected in the case of magnetic field amplification and back reaction of CRs, and the measurements and upper limits on $E_{\mathrm{CR}}$ would then decrease.}, to these {\it upstream} densities before using them in Equation~\ref{eq:accelEfficiency}.

\begin{sidewaysfigure}[h]
\centering
\includegraphics[width=0.9\columnwidth]{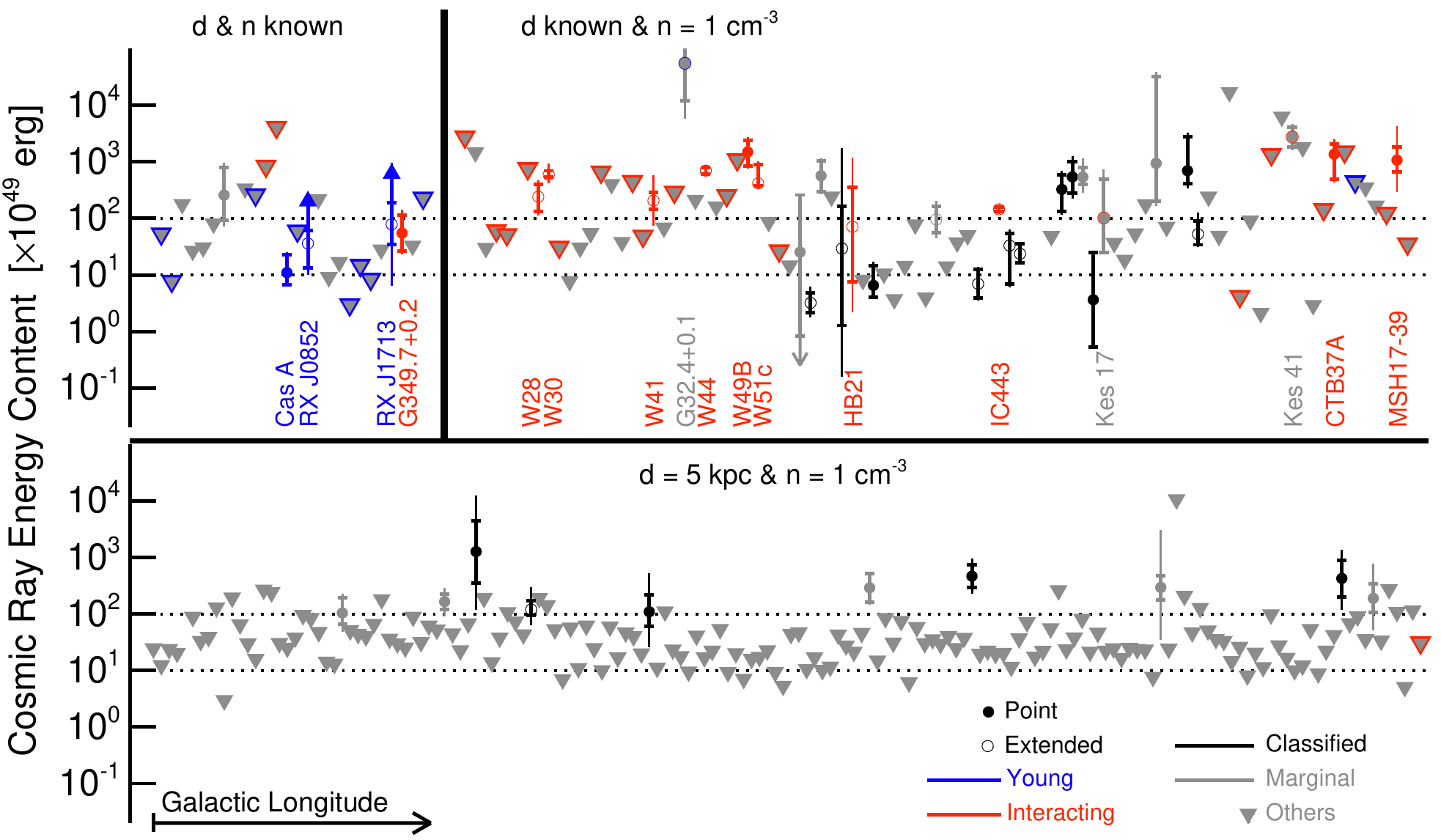} 
\caption{Estimates of the CR energy content (in units of $10^{49}$\,erg) for all Galactic SNRs, divided into three categories according to the level of information on their distances and densities (see text for details), and sorted in Galactic longitude within each subclass. Symbols and color coding are the same as in Figure~\ref{fig:GeVFluxGeVIndex} and are restated in the lower right corner. The names of the young and interacting SNRs ranked as classified or marginal GeV candidates are also given. The two dashed lines indicate a CR energy content of $10$ and $100\%$ of the standard SN explosion energy. Note that we added upward arrows for RX~J1713.7$-$3946 and RX~J0852.0$-$4622 (aka Vela~Jr) given the respective upper limits on the ambient density, based on the absence of thermal X-ray emission in these two SNRs.}
\label{fig:CREfficiencies}
\end{sidewaysfigure}

Figure~\ref{fig:CREfficiencies} shows the constraints on the CR energy content for the population of known Galactic SNRs. The sample has been divided into three subclasses, each sorted by Galactic longitude: SNRs with existing distance and density estimates (upper left panel); SNRs with known distances (upper right panel); and SNRs with unknown distance and density (lower panel). For the latter group, we arbitrarily set the distances and (upstream) densities to representative values of $5$\,kpc and $1$\,cm$^{-3}$, respectively. 

As is clearly visible for the first two subclasses of SNRs, the estimates and upper limits on the CR energy content span more than three orders of magnitude, from a few\,$\times10^{49}$\,erg to several\,$\times10^{52}$\,erg. In particular, a large fraction of the interacting SNRs lie above the $\epsilon_{\mathrm{CR}} = 1$ ($E_{\mathrm{CR}} = E_{\mathrm{SN}} = 10^{51}$\,erg) dashed line. Rather than these SNRs contributing more than their explosion energy to accelerating CRs, the densities experienced by the CR particles in the MC interaction region are likely much larger than those derived from the measurements of X-ray thermal emission or the assumed value of $1$\,cm$^{-3}$. 
\cite{Thompson12-FermiCRsReview} also noted that the luminous interacting SNRs thus far observed exceed the limit of $L_{0.1-100\,\mathrm{GeV}} \sim 10^{34}$\,erg\,s${^{-1}}$ for IC scattering off an interstellar radiation field (ISRF) with intensity similar to the solar neighborhood's for hadrons accelerated with an efficiency of $10\%$ and an electron-to-proton ratio of $1\%$. 
The fact that interacting candidates' lie above this limit, as many in this sample do, thus similarly suggests that they are likely the sites of hadronic interactions in dense environments. This is also consistent with our findings in Section~\ref{Sec:EmissionMechs}. In contrast, most of the young SNRs lie at or below this luminosity limit, suggesting that IC processes may contribute to their measured luminosity, again consistent with our comparison of the radio and GeV indices in Section~\ref{Sec:EmissionMechs}. 

For the group of SNRs with unknown distances and densities, the CR energy content estimates are clustered between the two reference values of $\epsilon_{\mathrm{CR}} = 0.1$ and $1$, their distribution reflecting that of the \FermiLat{} upper limits given in Table~\ref{Tab:ResultsULs}. In particular, the constraints obtained for SNRs lying in the outer Galaxy (in the middle of the panel) whose distances to Earth are likely smaller than $5$\,kpc, potentially fall below the reference value of $\epsilon_{\mathrm{CR}} = 0.1$ provided that the presently unknown effective density is of the order of $1$\,cm$^{-3}$. These and the other limits and detections falling below the nominal value for $\epsilon_{\mathrm{CR}}$ have the potential to significantly constrain the CRs being accelerated by these particular SNRs. By improving these limits and expanding the breadth of MW data used to calculate them, this method will allow us to significantly constrain the ability of known SNRs to provide the observed CRs.

We can estimate the number of GeV emitting SNRs in the Galaxy through the constraint on age from the condition on $E_{\mathrm{CR, max}}\gtrsim 200$\,GeV used to derive Equation~\ref{eq:accelEfficiency}. The maximum energy that CRs can reach throughout an SNR's evolution depends crucially on many factors such as the diffusion regime and, through the development of instabilities, the subsequent level of turbulent amplification of the magnetic field. On a more macroscopic level, $E_{\mathrm{CR, max}}$ can depend on the SNR hydrodynamics, namely the shock velocity and size, as well as on the ambient density \citep{Lagage83-Emax,Bell01-Emax,Bell04-Emax,Ptuskin03-Emax,Ptuskin05-Emax,Blasi07-Emax,Zirakashvili08-Emax,Bell13-Emax}. Nevertheless, according to \cite{Ptuskin03-Emax} and \cite{Ptuskin05-Emax}, the SNR age at which $E_{\mathrm{CR, max}}$ is $\sim200$\,GeV can be conservatively estimated to be $\sim10$\,kyr, after accounting for all the wave damping mechanisms reducing the level of turbulence and hence limiting the maximum CR energy. This value is of the same order as the ages of the oldest GeV-emitting SNRs yet studied, such as Cygnus Loop, W51C, IC\,443, W44 and W28.

Given a rate of $\sim3$~SNe of all types per century in the Milky Way \citep{Li11-SNrate}, there should then be $\sim300$ Galactic SNRs younger than $10$\,kyr, of which $\sim60$ are expected to be catalogued for a fraction of known SNRs of $0.2$, assuming an SNR lifetime of $\sim50$\,kyr. With a mean $95\%$~CL upper limit on the $1-100$\,GeV flux of $\sim10^{-9}$\,cm$^{-2}$\,s$^{-1}$ for index of $2.5$ (see Table~\ref{Tab:ResultsULs}), the horizon of detectability in the \FermiLat~SNR catalog d$_{det}$ amounts to $\sim3.4 \sqrt{n_{cm^{-3}}}$\,kpc, according to Equation~\ref{eq:accelEfficiency} for $\epsilon_{\mathrm{CR}} = 0.1$. Following the standard model of Galactic source distribution presented in \cite{Renaud11-SNRPWNcta}, $\sim5, 50,$ and $100\%$ of the Galactic SNRs are expected to lie at less than $3.4, 10,$ and $20$\,kpc to Earth. These fractions translate into numbers of detectable GeV-emitting SNRs with \FermiLat~through p-p interactions of $\sim3, 30,$ and $60$ for an effective density $n_{cm^{-3}} \sim1, 10,$ and $40$. These estimates are broadly compatible with the number of classified and marginal SNRs in the present study, provided that the average effective density is of the order of tens per cm$^{-3}$. In other words, most of the detected SNRs, apart from the known young and isolated SNRs, should be interacting with dense media, as suggested in Figure~\ref{fig:CREfficiencies}. Thus, the underlying assumption of $\epsilon_{\mathrm{CR}} = 0.1$, required in order for the Galactic SNR population to supply the CR flux observed at Earth, is compatible with the results of this \FermiLat{} SNR catalog. 

MW observations of the GeV-detected SNRs for which we lack information on distances and surrounding densities are encouraged in order to confirm this finding by searching for evidence for SNR-MC interaction and shedding light on the conditions in which the accelerated particles radiate GeV emission. Moreover, as stressed above, the expected large number of middle aged ($\lesssim10$\,kyr) SNRs in the Milky Way yet uncatalogued and potentially emitting in the GeV domain through p-p interactions, suggests that a certain fraction of the unidentified 2FGL sources could actually be unknown shell-type SNRs which could be revealed as such through follow-up radio, IR, and X-ray observations. Finally, this first \FermiLat{} SNR catalog can be used to trigger more detailed systematic studies of the Galactic SNR population, expanding on those performed by \cite{Cristofari13-SNRpop} in the VHE domain and by \cite{Mandelartz13-Neutrino} focusing on the diffuse neutrino flux associated with CR interactions in these sources. In so doing, we will gain significantly greater insight into the possible contribution of Galactic SNRs to the observed CR population.

\FloatBarrier

\section{Conclusions}\label{Sec:Conclusions}

We have systematically characterized the $1-100$\,GeV emission from $36$\,months in \nGalSNRs~regions containing known radio SNRs, identifying sources emitting in the regions and then determining the likelihood that the source nearest the SNR is associated with it. To do so, we developed a new method to systematically characterize emission within $3$\degr{} of each SNR for a data set that is both longer and covering a different energy range in comparison to other source catalogs (2FGL and 2PC). We then localized the candidate \g-ray SNRs, starting from the radio positions, and tested for extension and spectral curvature. 

In this way, we found \ndetected~candidates, \nclassifiedsnrs~of which have sufficient spatial overlap and significance with the alternative IEMs to suggest they are the GeV counterparts to their corresponding radio SNRs and a further \nmarginal~candidates which may also be related to the SNRs. We demonstrate that extension is a powerful discriminator in this regard. Using a mock catalog, we show that $<$\nninetyfivemockpercent{} of the \nassocprobclassified~spatial GeV associations are expected to have a chance coincidence with a radio SNR at the $95\%$ confidence level. The SNR-associated candidates span over two orders of magnitude in flux and a wide range of indices and are split almost equally between those with measurable extension and those seen as unresolved point sources. Of these, \nnewextended~extended and \nnewpointlike~point-like candidates are new associations. 
For the candidates best fit by a PL and passing the most stringent classification threshold, the average flux and index at energies of $1-100$\,GeV is ($8.4 \pm 2.1) \times 10^{-9}$\,ph\,cm$^{-2}$\,s$^{-1}$ and $2.5\pm0.5$. We also report flux upper limits measured at the radio positions and extensions at $95\%$ and $99\%$ confidence levels for indices of $2.0$ and $2.5$ for the \nULs~regions with either no detectable candidate or containing candidates which did not pass the location and extension classification thresholds. 

As Galactic SNRs tend to lie in regions of significant interstellar emission, namely on or near the Galactic plane, in addition to estimating the systematic error from the uncertainty in our knowledge of the instrument's effective area, we also developed a new method to estimate the systematic error arising from the choice of interstellar emission model. This is particularly important as the interstellar \g-ray emission is highly structured on scales smaller than the regions studied. Thus changes to the model-building strategy, i.e., varying the CR source distribution, CR halo height, and \hi{} spin temperature parameters, and separately scaling the \hi{} and CO in Galactocentric rings, play a significant role in interpreting the results for all candidates. In particular, systematic errors estimated from the choice of alternative IEM almost always dominated the flux errors, and while the statistical error on the index was larger for roughly a quarter of the candidates. While this estimate of systematic error from the choice of IEM neither spans the full range of allowed models nor brackets the standard model, it represents our most complete and systematic effort to quantify these errors to date, and is crucial to the interpretation of our results. 

We examined our GeV candidate SNR population in light of MW observations in order to better understand both SNRs' characteristics and potential for accelerating hadrons. While a radio-GeV flux correlation might be expected if the underlying energetic lepton and hadron particle populations had common properties, quantifying such a trend proved elusive. Results from Kendall's $\tau$ rank correlation tests suggested no significant correlation between radio and GeV flux or luminosities. Yet to say there is no correlation requires detailed modeling to account for observational biases in the identified sample of SNRs, errors on distance measurements, and deviations from a simple PL spectrum, all of which may skew an intrinsic correlation. In fact, a correlation can be expected regardless of the physical relationship between the radio- and GeV-emitting particles, because of an existing known $\Sigma$-D correlation between radio surface brightness and physical diameter. We also searched for relations between the GeV and radio properties through the spectral indices of the SNRs, finding that the majority of candidates do not lie in the expected region of phase space under simple assumptions for inverse Compton, bremsstrahlung or $\pi^0$ decay, nor any combination thereof.  Potential extensions to the models include additional spatial zones with different properties or differing spectral indices for the particle populations. The particle populations may also not follow a PL in momentum at all. Indeed, about half of the SNRs measured at TeV energies show indications of a change in spectral slope between their GeV and TeV indices, which may indicate curvature or breaks in the spectral forms of the underlying populations of accelerated particles. We anticipate that data from upcoming TeV instruments will greatly expand the statistical sample of GeV candidates with TeV counterparts.

In examining tracers of age and environment, we found that the classified candidates followed the previously observed trends of young SNRs being harder and fainter at GeV energies than older, often interacting SNRs, though two marginally classified candidates, one young and one interacting, do not follow this trend. Possible explanations for the current observations include the decreasing shock speed and maximum energy causing a softening in the GeV index. Models such as by \cite{2011JCAP...05..026C} suggest that such a scenario would also result in less luminous emission when the SNR is young. Environment likely also contributes as, for instance, older SNRs have a greater chance of interacting with dense ISM. We anticipate that this catalog, combined with more detailed MW studies, will be important to disentangle the effects of evolution and environment.

SNRs have long been considered likely to supply the majority of Galactic CRs, and we estimated the maximal contribution of all remnants assuming their emission (or upper limits) is entirely hadronic. To do so, we combined predictions for CRs with an underlying PL index $\gtrsim2$, commensurate with our measured average GeV index, and a maximum energy $\gtrsim 200$\,GeV, with distance and density estimates from the literature assuming $E_{\mathrm{SN}}=10^{51}$\,erg. We find that the limits on CR energy content span more than three decades, including many interacting candidates for which the densities in the interaction regions are much greater than the nominal density assumed in the calculation, and young candidates with efficiencies below the nominal $\sim10\%$, consistent with possible leptonic emission predictions (e.g. IC). Under the simple assumptions stated in Section~\ref{Sec:CRs}, the contribution from all SNRs, particularly those with flux upper limits, is beginning to constrain the energy content put into CRs from the known SNRs to less than $10\%$, particularly in regions of well characterized IEM background.  
Yet there remains a clear dearth of MW information, particularly regarding ISM densities for the candidates. Improved MW information, coupled with improving GeV flux sensitivity from continued observations with the LAT and the development of the new Pass 8 event selection \citep{Atwood13-pass8} will allow better constraints on SNRs' aggregate ability to accelerate the observed CRs. 

With this first \FermiLat{} SNR Catalog we have systematically characterized GeV emission in regions containing known radio SNRs, creating new methods to systematically address issues associated with these typically complex regions. These include methods for systematically adding sources to a region and better estimating the systematic error due to choice of interstellar emission model. From this, we have determined characteristics of the GeV SNR population, down to our measurement limit, finding \nclassifiedsnrs~classified and \nmarginal~marginal candidates with a false identification limit of $<$\nninetyfivemockpercent. 
This GeV data provide a crucial context for the detailed modeling of individual SNRs. In combination with MW measurements, the GeV data now challenge simple, previously sufficient SNR emission models. Within the limits of existing MW data, our observations generally support previous findings of changes in spectral slope at or near TeV energies and a softening and brightening in the GeV range with age and effective age, yet we see indications that new candidates and new MW data may provide evidence of exceptions to this trend. With uniformly measured data for all known SNRs, we also constrain SNRs' aggregate, maximal contribution to the population of Galactic CRs. With the GeV and other MW data, we find that the candidates and upper limits are generally within expectations if SNRs provide the majority of Galactic CRs and anticipate these limits will improve with both a larger GeV data set with better sensitivity, as will be provided by \FermiLat~Pass 8 data, and with more and better distance and density estimates.

\section{Acknowledgements}
The \FermiLat{} Collaboration acknowledges generous ongoing support
from a number of agencies and institutes that have supported both the
development and the operation of the LAT as well as scientific data analysis.
These include the National Aeronautics and Space Administration and the
Department of Energy in the United States, the Commissariat \`a l'Energie Atomique
and the Centre National de la Recherche Scientifique / Institut National de Physique
Nucl\'eaire et de Physique des Particules in France, the Agenzia Spaziale Italiana
and the Istituto Nazionale di Fisica Nucleare in Italy, the Ministry of Education,
Culture, Sports, Science and Technology (MEXT), High Energy Accelerator Research
Organization (KEK) and Japan Aerospace Exploration Agency (JAXA) in Japan, and
the K.~A.~Wallenberg Foundation, the Swedish Research Council and the
Swedish National Space Board in Sweden.

Additional support for science analysis during the operations phase is gratefully
acknowledged from the Istituto Nazionale di Astrofisica in Italy and the Centre National d'\'Etudes Spatiales in France.

\FloatBarrier
\appendix

\section{Comparison of Source Models with 2FGL}
\label{appen:addSrcs2FGL}

This SNR catalog was constructed using $3$~years of P7 Source class data in the energy range $1-100$\,GeV, whereas 2FGL used $2$\,years of data over the larger energy range $0.1-100$\,GeV. The differences in observing time and energy range resulted in residual, unmodeled emission in some RoIs as well as changes to some 2FGL sources' spectral model, position localization, and detection significance. Here we compare the input source models constructed for this catalog, described in Section~\ref{Sec:AddSrcs}, with 2FGL to better understand the method's ability to describe the regions studied. Since we rederive the input source model only within a $3\degr$\,radius of the center of each RoI, we consider sources only inside that radius.

\begin{figure}[h!]
  \centering
  \makebox[\linewidth]{\includegraphics[width=0.8\columnwidth]{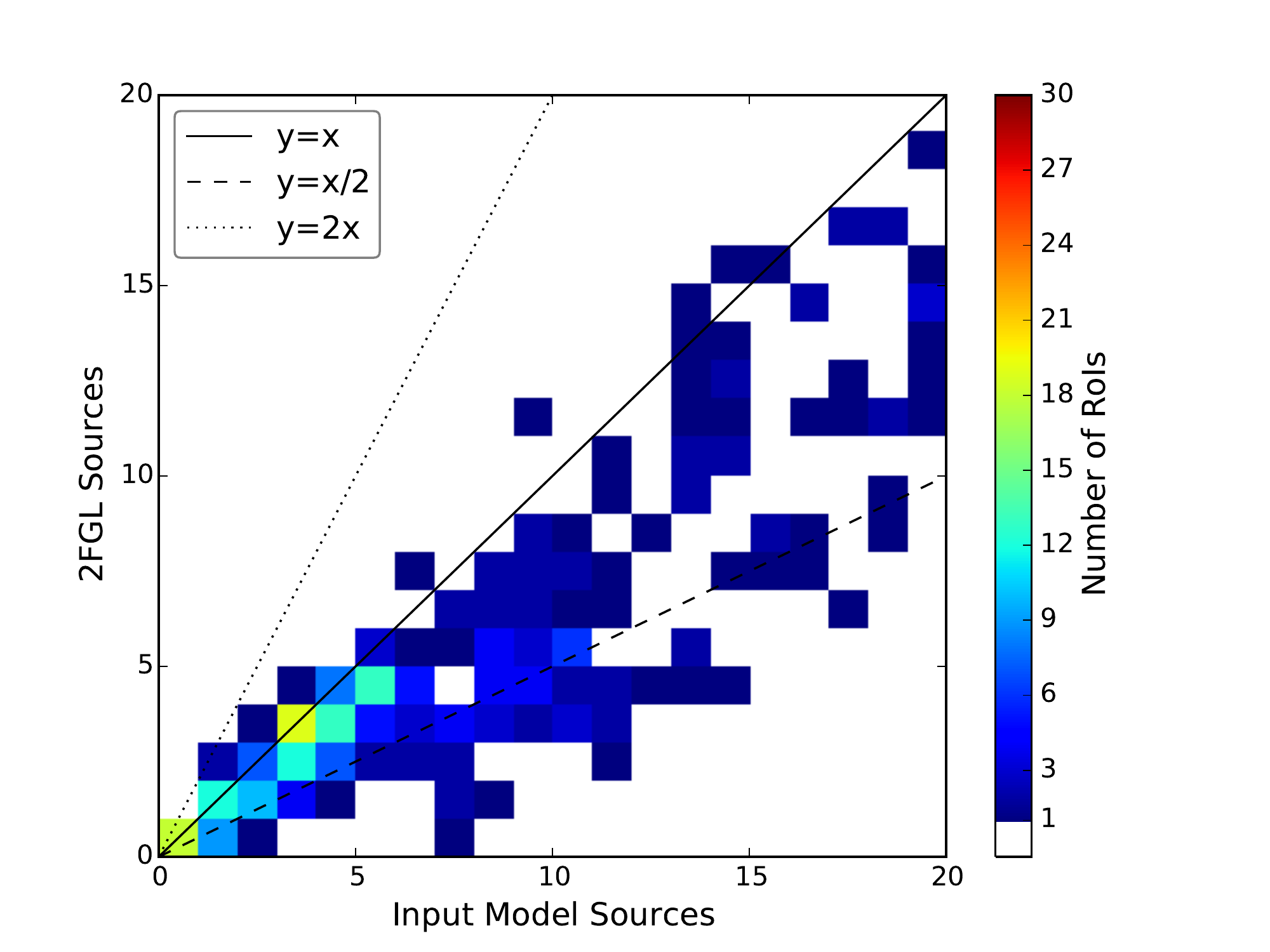}  }  
  \caption{Comparison of the number of 2FGL sources with TS$_{1-100\,\mathrm{GeV}} \geq 25$ (excluding AGN and pulsars) with the number of newly added input model sources in the present analysis, for sources within $3$\degr{} of the center of each RoI. The color scale shows the number of RoIs with a particular combination of numbers of 2FGL sources and new sources. White corresponds to no RoI with that combination of source counts.}
\label{fig:2FGLvAddSrcs} 
\end{figure}

\begin{figure}[h!]
\centering
\makebox[\linewidth]{\includegraphics[width=0.8\columnwidth]{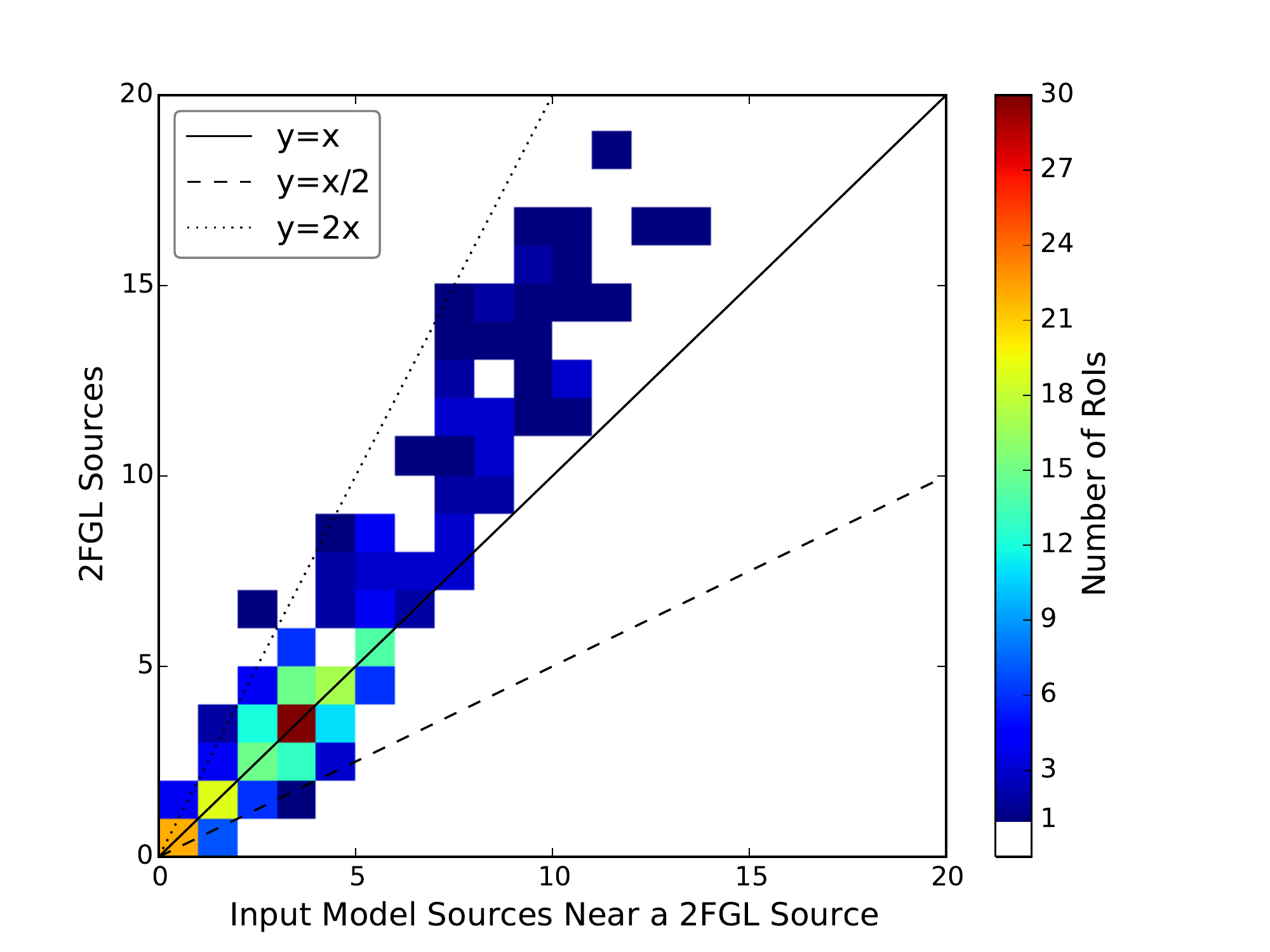}} 
\caption{Same as Figure~\ref{fig:2FGLvAddSrcs}, including only input model sources lying within $0.2$\degr{} of a 2FGL source.}
\label{fig:2FGLvAddSrcsAssoc} 
\end{figure}

Given the data set differences, in each RoI we expect similar but not identical numbers of sources relative to those in 2FGL.
Figures~\ref{fig:2FGLvAddSrcs} and \ref{fig:2FGLvAddSrcsAssoc} show the numbers of significant (TS\,$\geq 25$) 2FGL sources and derived input model sources (excluding 2FGL identified AGN and pulsars kept in the input model) in individual RoIs as 2D histograms. In Figure~\ref{fig:2FGLvAddSrcs}, the number of sources in the derived input model is typically greater than the number of 2FGL sources that are significant at $1-100$\,GeV. 73 of the 279 RoIs studied contain at least one of the the 12 extended 2FGL sources. Since 2FGL extended sources were removed from the inner $3\degr$ of each RoI, and this region was repopulated with point sources, we can detect multiple point sources inside the extent of any removed extended 2FGL sources. This decomposition of extended sources, combined with the longer data set and different energy range compared to 2FGL, contribute to the high ratio of input model to 2FGL sources in some RoI, which demonstrates the need to rederive the source model. 

To more accurately represent the 2FGL sources being reproduced in the central $3\degr$, in Figure~\ref{fig:2FGLvAddSrcsAssoc} we limited the input model sources to those within $0.2\degr$ (approximately the width of the core of the $10$\,GeV PSF) of a 2FGL source, effectively excluding input sources that are not co-spatial with a 2FGL source. Here we see that the majority of 2FGL sources have counterparts in the rederived set. As a region's complexity increases, seen as an increase in numbers of 2FGL sources, up to about half of the 2FGL sources may not have counterparts within $0.2$\degr. Given that in these same regions we have more new sources than 2FGL sources, as seen in Figure~\ref{fig:2FGLvAddSrcs}, we find as expected that the longer data set with improved statistics at higher energies, where the angular resolution of the LAT is the best, allows us to add new sources to account for newly significant excesses in these complex regions. Additionally, sources with low TS in 2FGL are particularly susceptible to having a newly added source which may start at a similar position but then localize further than $0.2\degr$ from the 2FGL source. 

Thus, we find that the method developed and used here produces a model which reproduces the 2FGL sources as expected, including differences that trend as anticipated given the longer data set and modified energy range, yielding better spatial resolution. The new method thus provides reasonable representations of the regions being modeled as input for the final analysis.

\FloatBarrier

\section{Estimating Systematic Uncertainties from Modeling Galactic Diffuse Emission}
\label{appen:aIEM}

As the majority of radio SNRs lie in the Galactic plane, we developed a method to assess the systematic uncertainties due to the modeling of the interstellar emission, which  at low latitudes is the dominant component of the \g-ray flux. To do so, we developed alternative IEMs, described and compared with the standard one in Section~\ref{alIEMconstr}, and available at the FSSC\footnote{\url{http://fermi.gsfc.nasa.gov/ssc/data/access/lat/1st_SNR_catalog/}}. We define the systematic errors of the parameters of a source model using the results of the likelihood fit of a region with the alternative IEMs, detailed in Sections~\ref{alIEMuse} and \ref{alIEM_sys_def}. We illustrate the impact of some input parameters of the alternative models based on the analysis of eight test sources in Section~\ref{aIEMValidation}. In Section~\ref{aIEMsContext} we compare the results with both the statistical errors and the systematic errors due to estimating the effective area, described in Section~\ref{Sec:EffAreaSysErr}. 

\subsection{The Standard IEM and Construction of Alternative IEMs}
\label{alIEMconstr}
Galactic interstellar \g-ray emission is produced through interactions of high energy CR hadrons and leptons with interstellar gas via nucleon-nucleon inelastic collisions and electron Bremsstrahlung, and with low energy radiation fields via IC scattering. The \FermiLat{} collaboration developed the standard IEM for the analysis of P7 data using the simple assumption that energetic CRs uniformly penetrate all gas phases of the ISM. Under this assumption, the Galactic interstellar \g-ray intensities can be modeled as a linear combination of gas column densities and an IC intensity map as a function of energy. The gas column densities are determined from emission lines of atomic hydrogen (\hi{})\footnote{\hi{} column densities for the standard IEM are extracted from the radio data using a uniform spin temperature of $200$\,K.} and CO, the latter a surrogate tracer of molecular hydrogen, and from dust thermal emission maps minus the best fit linear combination of the aforementioned HI and CO maps used to account for gas not traced by the lines (``dark gas''). To account for a possible large scale gradient of CR densities, the gas column density maps were split into six Galactocentric rings using the emission lines' Doppler shifts and a Galactic rotation curve \citep[see Appendix B of][]{Ackermann12-aIEMs}. The IC map is obtained using GALPROP to reproduce the direct CR measurements with a realistic model of the Galactic interstellar radiation field (ISRF), as was done in \cite{Porter08-ISRF}. 
The standard IEM accounts for some extended remaining residuals including, notably, Loop~I \citep{Casandjian09-LoopI} and the \Fermi{} bubbles \citep{Su10-FermiBubbles}, as additional components. Some of these, e.g., the \Fermi{} bubbles, do not have counterparts at other wavelengths and are defined from the LAT data. 

These gas, IC, and additional components, along with sources in the 2FGL Catalog and an isotropic intensity accounting for the extragalactic \g-ray and instrumental backgrounds, were fit to $2$\,years of LAT data. This yielded best fit values of the linear combination coefficients, which can be interpreted as emissivities as a function of energy in the various Galactocentric rings for gas templates. For the IC and isotropic templates, the best fit values renormalize the models as a function of energy. The ratio of the best fit coefficient of the CO map (i.e., the scaling factor for the integrated intensity of the CO line) to twice the coefficient of the \hi{} column density map is commonly referred to as $X_\mathrm{CO}$ or the CO-to-H$_2$ ratio. Similarly, we will refer to the ratio of the dark gas map coefficient to the coefficient of the \hi{} column density map as the dust-to-gas ratio. Formally, these ratios depend on energy and, for CO, on the Galactocentric ring. Only the latter dependence is considered, since no significant variations with energy have been found \citep[e.g.][]{Ackermann12-CygnusCR, 2014arXiv1409.3268P}. The standard IEM is summed over the components and predicts the intensities of Galactic interstellar \g-ray emission in a grid of directions and energies and is used in combination with the isotropic model with which it was fit. Further details are available at the \Fermi{} Science Support Center\footnote{\url{http://fermi.gsfc.nasa.gov/ssc/data/access/lat/Model_details/Pass7_galactic.html}}. 

To explore the uncertainties related to this standard modeling of interstellar emission, we generated eight alternative IEMs to probe key sources of systematics by: 
\begin{itemize}
 \item adopting a different model-building strategy from the standard IEM resulting in different gas emissivities, or equivalently CO-to-H$_2$ and dust-to-gas ratios, and including a different approach for dealing with the remaining extended residuals;
\item varying a few important input parameters in building the alternative IEMs;
\item and allowing more degrees of freedom in the subsequent likelihood analysis of each SNR by separately including and scaling the IC emission and emission traced by \hi{} and CO in four Galactocentric rings. 
\end{itemize}

The alternative model-building strategy starts from the work in \cite{Ackermann12-aIEMs}, using the GALPROP CR propagation and interaction code. The GALPROP output model intensity maps associated with \hi{}, \Hii\footnote{Using the \Hii{} model of \citet{Gaensler08-HIImodel}.}, CO, and IC are then fit simultaneously with an isotropic component and 2FGL sources to $2$\,years of \FermiLat{} data in order to mitigate data-model differences. The intensity maps associated with gas were binned into four Galactocentric rings ($0-4$\,kpc, $4-8$\,kpc, $8-10$\,kpc and $10-30$\,kpc). The intensity from the subdominant \Hii{} component was added to the \hi{} component and scaled together.  We refer to the summed component as \hi{} hereafter. In the all-sky fit, the spectra of all model intensity maps were individually renormalized with a logP function to allow for possible CR spectral variations among the annuli for all  \hi{} and CO maps and to allow for spectral variations in the electron distribution for the IC template.  We also included in the fit an isotropic template, with free normalization determined independently over nine bins from 100~MeV and 300~GeV, and templates for Loop~I and the \Fermi{} bubbles different from those used in the standard IEM. The templates for both Loop~I and the bubbles are based on geometrical models with uniform volumetric luminosity that are integrated along the line of sight to get a sky template.  Loop~I is based on the shell model of \cite{Wolleben07-LoopI} while the bubbles are assumed to be uniform ``balloons'' above and below the plane with edges defined in spherical coordinates centered on the Galactic center by $r = R_0 |\cos\theta|$\footnote{($r$, $\theta$, $\phi$) are the usual spherical coordinates with $\theta=0$ pointing at the north Galactic pole. The model is uniform in $\phi$ and $R_0$ was chosen to be $8$\,kpc to approximately match the shape of the bubbles.}. The spectrum of Loop~I and the bubbles is described with a logP function.

\cite{Ackermann12-aIEMs} explored some systematic uncertainties in modeling interstellar emission by varying a few selected input parameters to GALPROP. Among those, we selected the parameters that, within the range of values allowed by theory and MW/multimessenger observations, were found to cause the largest variations in the predicted \g-ray intensities, and varied them to build our alternative IEMs. These parameters are:
\begin{itemize}
 \item the CR source distribution, for which we adopted the distribution of SNRs according to \cite{Case98-SigmaDSNR} and of pulsars according to \cite{Lorimer06-PSRpop};
 \item the height of the CR propagation halo, for which we adopted the values of $4$\,kpc and $10$\,kpc, the limits of the viable halo sizes from the secondary to primary data used in \cite{Ackermann12-aIEMs};
 \item the uniform spin temperature used to derive the \hi~column density maps from the $21$\,cm line data, for which we adopted the values of $150$\,K and $100,000$\,K (equivalent to an optically thin medium).
\end{itemize}
The values adopted to generate the eight alternative IEMs were chosen to be reasonably extreme; even so they do not reflect the full uncertainty of the IEMs. 
Many other possibly important sources of systematic error exist and include uncertainties in the ISRF model; simplifications of the geometry of the Galaxy, e.g., assuming cylindrical symmetry; small scale non-uniformities in the CO-to-H$_2$ and dust-to-gas ratios; \hi{} spin temperature non-uniformities; and underlying uncertainties in the input gas and dust maps.

Unlike the standard IEM, the alternative IEMs were not summed up into a single sky map.  This allows for more freedom in the IEM when determining the SNR parameters from the likelihood fit in individual RoIs as described in the next section. Separately scaling the IC emission and the \hi{} and CO emission in rings permits the alternative IEMs to better adapt to local structure when analyzing a given source region.

\subsection{Likelihood Analysis Using the Alternative IEMs}\label{alIEMuse}

To estimate the systematics due to the choice of IEM, we characterized the regions containing GeV candidates detected with the standard IEM using the alternative IEMs. We repeat the likelihood analysis for the point and the best extended hypotheses, starting from the best fit model obtained using the procedure described in Section~\ref{Sec:LocExtSpec}, and replacing the standard IEM with the alternative ones and their corresponding isotropic templates. For each candidate's fit with an alternative IEM, we determined the best hypothesis by comparing the point and extended hypotheses' likelihoods as described in Section~\ref{Sec:LocExtSpec}.

The alternative IEMs have nine independent components (four \hi{} rings, four CO rings, and the IC template), contributing more or less significantly to the \g-ray intensity along a given line of sight. To make the fitting procedure more stable, we independently fit the normalization coefficients for components which contributed more than $3\%$ of the initial model's total counts in the RoI, while the remaining components were merged into a single template. If the merged template accounts for more than $3\%$ of the counts in the RoI, we also fit its normalization; otherwise it is kept fixed. Leaving the normalization of templates with a lower percentage of counts free in the fit caused failures in the fitting procedure. The individual IEM component that provided the largest contribution in counts to the RoI was also corrected in spectrum using a PL with free spectral index, which accounted for spectral variations of interstellar emission across the sky analogously to fitting the standard IEM index in the main analysis. The isotropic component was kept fixed, as for the main analysis, because it is more reliably determined in the large-scale fit described in \ref{alIEMconstr} rather than in a relatively small RoI.

Due to the increased number of degrees of freedom when using the alternative IEMs, to make the fitting procedure more stable, we needed to reduce the number of degrees of freedom intrinsic to the background point sources. To do so, we performed an initial spectral fit starting from the same background sources as for the standard IEM analysis, described in Sections~\ref{Sec:AddSrcs} and \ref{Sec:DetectMethod}. For each alternative IEM, we then discarded sources with TS~$< 9$ and allowed only the three sources closest to the RoI center and the two most significant within $5$\degr{} of the RoI center to vary in the following fit procedure. The parameters of all the sources not removed were then reset to the best fit values from the standard analysis, while the IEM components' normalization coefficients were reset to $1$. This approach avoids the possibility that sources with lower significance for a specific IEM, which were obtained using the standard IEM and may now be mismodeled, bias the fit toward inaccurate results from the fitting procedure is unable to recover.

\begin{figure}[h!]
\centering
\includegraphics[width=0.8\columnwidth]{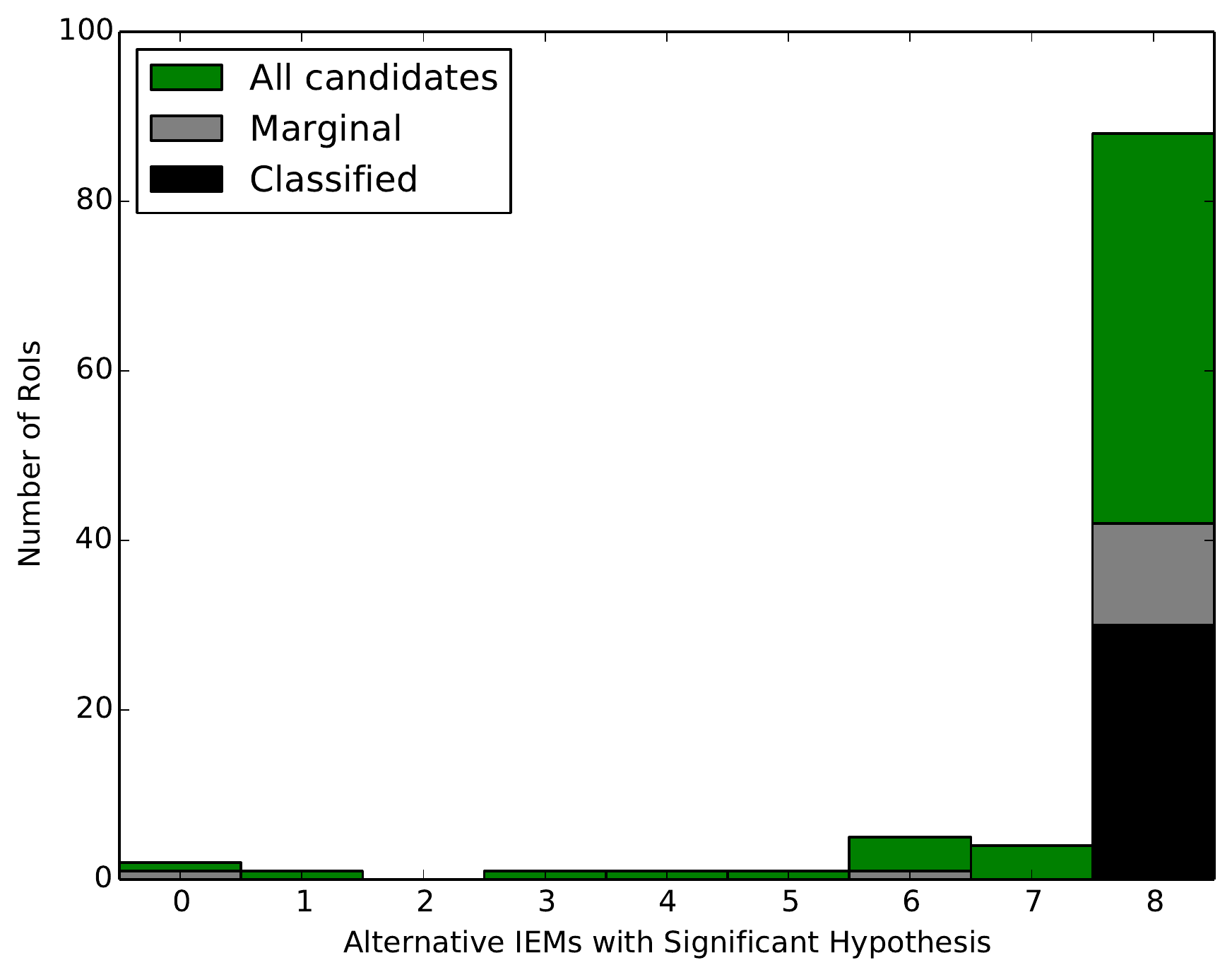} 
\caption{For each candidate's analysis with each of the alternative IEMs, we computed the candidate's significance. 
This histogram shows, for each candidate SNR, the number of analyses with each of the alternative IEMs for which the candidate remained significant (TS$>9$). The histogram reflects the values in Table~\ref{Tab:ResultsSpat}.}
\label{fig:besthyp_sig}
\end{figure}

\begin{figure}[h!]
\centering
\includegraphics[width=0.8\columnwidth]{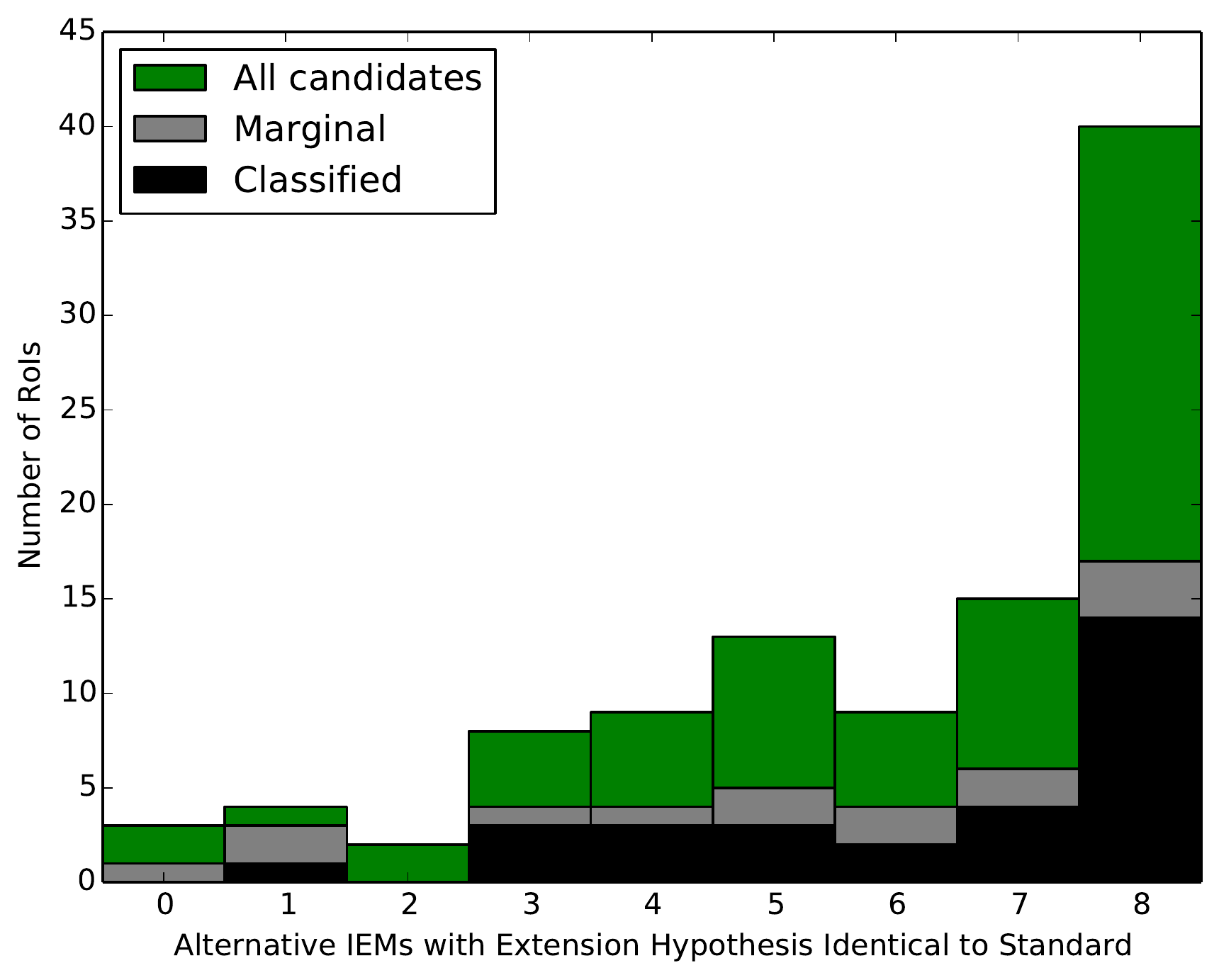} 
\caption{For each candidate's analysis with each of the alternative IEMs, we computed the best extension hypothesis, either point-like or extended. The plot shows the number of analyses of each region for which the best extension hypothesis with the alternative IEM remained the same as that found with the standard one. The values are indicated in Table ~\ref{Tab:ResultsSpat} for each source individually. Since for each alternative IEM we perform a separate localization and spectral fit even if the extension hypothesis is the same, the location and size of the candidate SNR may differ.}
\label{fig:besthyp_sig_ext}
\end{figure}

We then repeated the localization, extension, and spectral analysis as in Section~\ref{Sec:LocExtSpec}, directly adopting the best fit spectral model of the standard analysis rather than evaluating the PL or logP hypotheses. We determined the best extension hypothesis by comparing the maximum likelihoods for the extended and point-like hypotheses using the method described in Section~\ref{Sec:LocExtSpec}. For the best hypothesis, we evaluated the significance of the SNR detection in the overall maximum likelihood. In a few regions when using some of the alternative IEMs, the candidate became less significant, as shown in Figure~\ref{fig:besthyp_sig}. We similarly evaluated the best extension hypothesis for each alternative IEM and compare it to that found using the standard IEM in Figure~\ref{fig:besthyp_sig_ext}. The numbers of alternative IEMs for which the source was found to be insignificant and for which it changed extension hypothesis are listed in Table~\ref{Tab:ResultsSpat}, in the first two columns, respectively, under Alt IEM Effect. As described in Section~\ref{Sec:IEMSysErr}, we demoted classified candidates to marginally classified if their TS was less than nine, to avoid threshold effects, for any of the alternative IEMs. This affected \nmarginalAIEM{} candidates, described in Section~\ref{Sec:CatCaveats}. We also removed from the systematic error calculation, described in Section~\ref{alIEM_sys_def}, any analysis with an alternative IEM for which the fit did not converge sufficiently, also described in Section~\ref{Sec:CatCaveats}. \notherAIEMcap~candidates classified as ``other'' were affected by this. 

\begin{figure}[h!]
\end{figure}

\begin{figure}[h!]
\end{figure}

\subsection{Systematic Errors Estimated from the Alternative IEMs} \label{alIEM_sys_def}

We evaluated the systematic error on each of the candidates' parameters $P$ using the results of the likelihood analysis with the alternative IEMs in Equation~\ref{eq:sys_err_weighted}. The equation can also be expressed as:
\begin{equation}
\label{eq:sys_err_weighted_exp}
 \sigma_\mathrm{sys,w}=\sqrt{(<P>_w-P_\mathrm{STD})^2+\sigma_{P,w}^2}
\end{equation}
where $P_\mathrm{STD}$ is the parameter value obtained using the standard IEM and $<P>_w$ is the weighted average of the parameter evaluated using the $M$ alternative IEMs for which we can estimate it (see Section~\ref{alIEMuse}). The weights $w_i$ used are defined in Equation \ref{eq:sys_weight}. The weighted average is:
\begin{equation}
 <P>_w=\frac{1}{\sum_i^Mw_i} \sum_i^M(w_iP_i)
\end{equation}
and the weighted standard deviation is:
\begin{equation}
 \sigma_{P,w}=\sqrt{\frac{1}{\sum_i^Mw_i} \sum_i^M w_i\left(P_i-<P>_w\right)^2} ~~.
\end{equation}
The definition of the systematic error in Equation~\ref{eq:sys_err_weighted_exp} takes into account both a possible offset of the alternative values with respect to the standard one ($<P>_w - P_\mathrm{STD}$), attributable to the differences in model-building strategy and the additional degrees of freedom in the fit, and the spread of the $P_i$ around their weighted average value, attributable to the different GALPROP input parameters of the model.

In several cases the alternative IEMs yielded parameter values $P_i$ that did not bracket the value obtained using the standard IEM, $P_\mathrm{STD}$. This can be due to the differences in how the models were built as noted in Sections~\ref{Sec:IEMSysErr} and \ref{alIEMconstr}. While in such cases there is a net displacement, we conservatively took $\sigma_P$ as an estimate of the uncertainty related to the choice of the IEM symmetrically around the value $P_\mathrm{STD}$. This posed a problem for parameters such as the flux which are naturally positive definite.  Notably, if the net displacement of the parameter is more than twice the parameter value found using the standard IEM, with linearly symmetric error bars the parameter would be formally consistent with zero, even though it is significantly different from zero for all IEMs. In those cases it is more natural to consider the error in logarithmic space. To calculate the error bars symmetrically in the natural logarithmic space, we used the following formalism. We first evaluated the sign of the displacement between the parameter value for the alternative IEMs and the standard one:
 \begin{equation}
  \sigma_\mathrm{sign}=\frac{<P>_w -P_\mathrm{STD}}{|<P>_w -P_\mathrm{STD}|}.
 \end{equation}
We then evaluated the signed shift related to the systematic error (see Equation~\ref{eq:sys_err_weighted}), that is:
 \begin{equation}
P_\mathrm{shift}=P_\mathrm{STD}+\sigma_\mathrm{sign}\cdot \sigma_\mathrm{sys}.
 \end{equation}
From this we derived the displacement in logarithmic space: 
\begin{eqnarray}
 \Delta \log_{10} P & = & |\log_{10} P_\mathrm{shift}-\log_{10} 
P_\mathrm{STD}|=\left|\log_{10} \frac{ 
P_\mathrm{shift}}{P_\mathrm{STD}}\right| = \nonumber \\
&= &\left|\log_{10} 
\frac{P_\mathrm{STD}+\sigma_\mathrm{sign}\sigma_\mathrm{sys}}{P_\mathrm{STD}} 
\right|=\left|\log_{10} \left(1+ \sigma_\mathrm{sign}\frac{\sigma_\mathrm{sys}}{P_\mathrm{STD}}\right)\right| .
\end{eqnarray}
The extrema of the error bars are therefore:
\begin{equation}
 P_{\pm}=10^{(\log_{10} P_\mathrm{STD}\pm \Delta \log_{10} P)}.
\end{equation}

\subsection{Impact of the GALPROP Input Parameters}\label{aIEMValidation}
\begin{figure}[h!]
\centering
\includegraphics[ width=0.8\columnwidth]{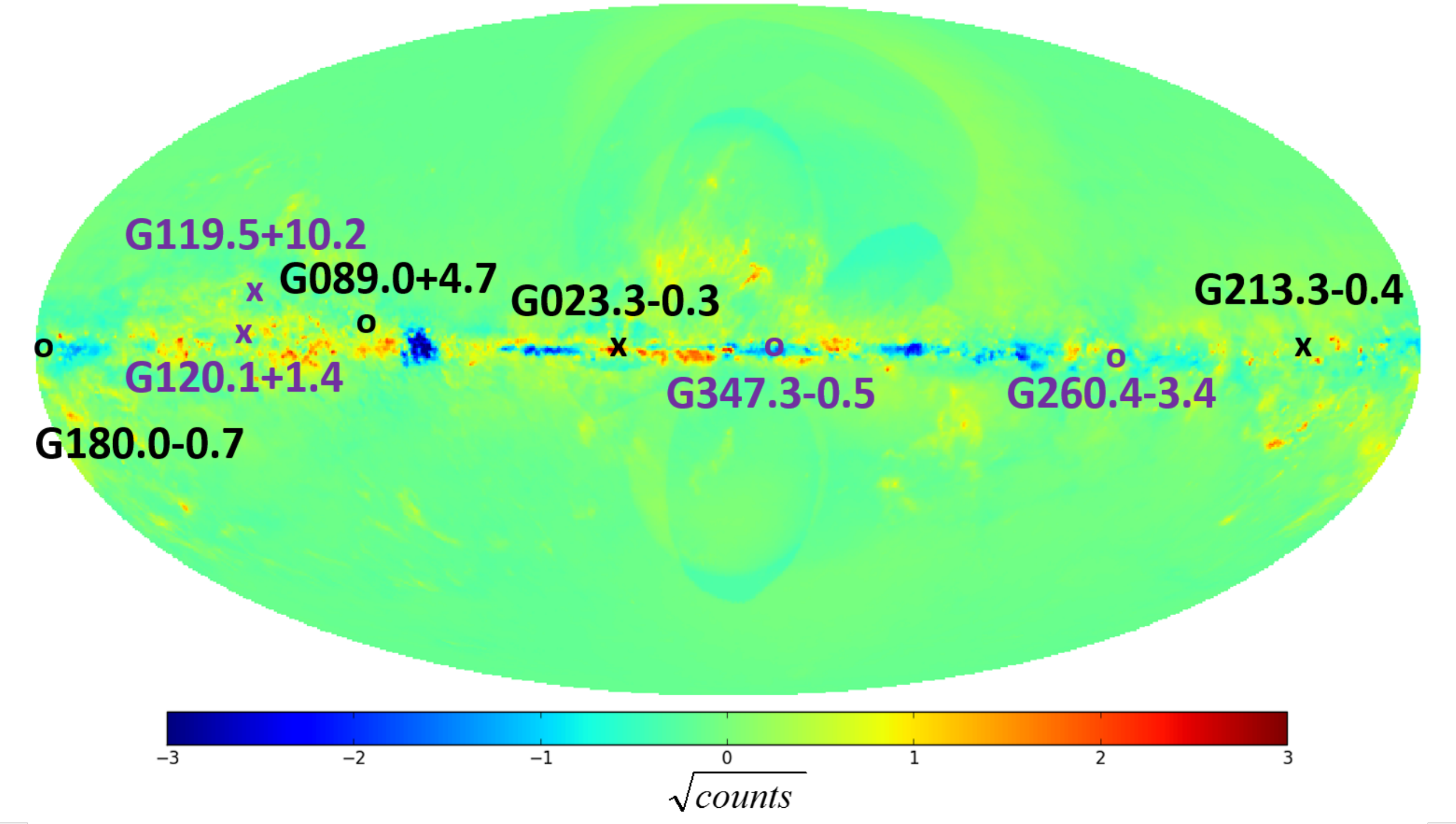} 
\caption{The eight candidate SNRs used to test the method are shown on the relative difference between the standard IEM and one of the alternative IEMs in $\sigma$ units. The selected SNRs represent all combinations of hard (purple) and soft (black), point-like (x) and extended (o) sources. The IEM intensities are converted to predicted counts for each pixel ($\sim0.052$\,deg$^2$) using $2$\,years of source class exposure in the energy band $1-10$\,GeV. The color scale shows:
   $(\mathrm{counts}_\mathrm{STD}-\mathrm{counts}_\mathrm{ALT})/\sqrt{({\mathrm{counts}_\mathrm{STD}+\mathrm{counts}_\mathrm{ALT}})/2}$,
   where $\mathrm{counts}_\mathrm{STD}$ are the predicted counts from the standard IEM and $\mathrm{counts}_\mathrm{ALT}$ are those of the alternative IEM. The alternative IEM chosen has the Lorimer distribution of CR sources, halo height of $4$\,kpc, and spin temperature $150$\,K. Further details may be found in \cite{dePalma13-AltIEMSystematics_FSymp}.}
\label{fig:snr_loc}
\end{figure}

We tested this method on a subset of eight candidate SNRs chosen to represent the range of spectral and spatial characteristics of SNRs in \g-rays and located in regions with bright or dim Galactic diffuse emission. Figure~\ref{fig:snr_loc} shows the candidate SNRs' locations on the sky, illustrating their range of Galactic longitude \citep[see also][]{dePalma13-AltIEMSystematics_FSymp}. Their locations are overlaid on a map showing the difference between the standard IEM and one of the alternative IEMs.  There are clear structures in the map that arose from the distinct model-building strategies as described in Section~\ref{alIEMconstr}. The map in Figure~\ref{fig:snr_loc} was created using models covering the full sky and therefore does not account for adjustments made to the model in the individual RoI fits as described in Section~\ref{alIEMuse}.  While the added freedom may reduce the differences between the models for an individual RoI, the freedom is not large enough to completely absorb it, and the local spatial and spectral differences between the models will remain. The differences are expected to influence the SNR's fitted parameters differently depending on the SNR's location. 

We focus in particular here on the influence of each of the three parameters varied when creating the eight alternative IEMs, described in Section~\ref{alIEMconstr}, on the fitted candidates' parameters, such as flux and PL spectral index. We then determined if any had the largest impact for all the candidates. To identify which, if any, of the three IEM parameters (CR source distribution, CR propagation halo height, and \hi{} spin temperature) had the largest impact on a candidate's fitted parameter, we adopted the following procedure. For a fitted source's parameter~$P$, e.g. flux, and a GALPROP input parameter $X = \{X_i, X_j\}$, e.g., spin temperature $T_s~=~\{150$\,K$, 10^5$\,K$\}$, let $\{P_{i}\}$ be the set of P values obtained when $X=X_i$ and $\{P_{j}\}$ the set of $P$ values obtained when $X=X_j$. If we marginalize over the other IEM input parameters by taking the average values of the two sets, $<P_{i}>$ and $<P_{j}>$, we can define the ratio $R$ as the difference between the two sets relative to their intrinsic spread:
\begin{equation}
\label{eq:ratio_aIEM}
R \equiv \frac{|<P_{i}>-<P_{j}>|}{\max(\sigma_{P_i},\sigma_{P_j})}.
\end{equation}
where $\sigma_{P_i}$,  $\sigma_{P_j}$ are the standard deviations of two sets. We divide by the maximum value of the standard deviations to conservatively account for the dispersion within each set.

A value of $R > 1$ means that changing the GALPROP input parameter $X$ from $X_i$ to $X_j$ results in a larger difference in the values determined for $P$ than the standard deviation of the samples obtained from all the combinations of other GALPROP input parameters. This implies that this GALPROP parameter $X$ has a larger influence on $P$ than any other input parameter tested. A value of $R >> 1$ for a particular GALPROP input parameter means that the effect of all the other parameters are negligible for that particular candidate. A value of $R$ between $0$ and $1$ means that changing the GALPROP input parameter results in a smaller difference in the values determined for $P$ than the standard deviation of the samples obtained from all the combinations of other GALPROP input parameters. This implies that this GALPROP parameter $X$ has a smaller influence on $P$ than any other input parameter tested. A value of $R \sim 0$ for a particular GALPROP input parameter means that the effect of this parameter is negligible for that particular candidate. If in various RoIs different GALPROP input parameters have a larger effect on the candidate's parameters than the others, we must use all the alternative IEMs to estimate the systematic error since different input parameters are more relevant in different parts of the sky.

\begin{figure}[h!]
\centering
\includegraphics[width=0.8\columnwidth]{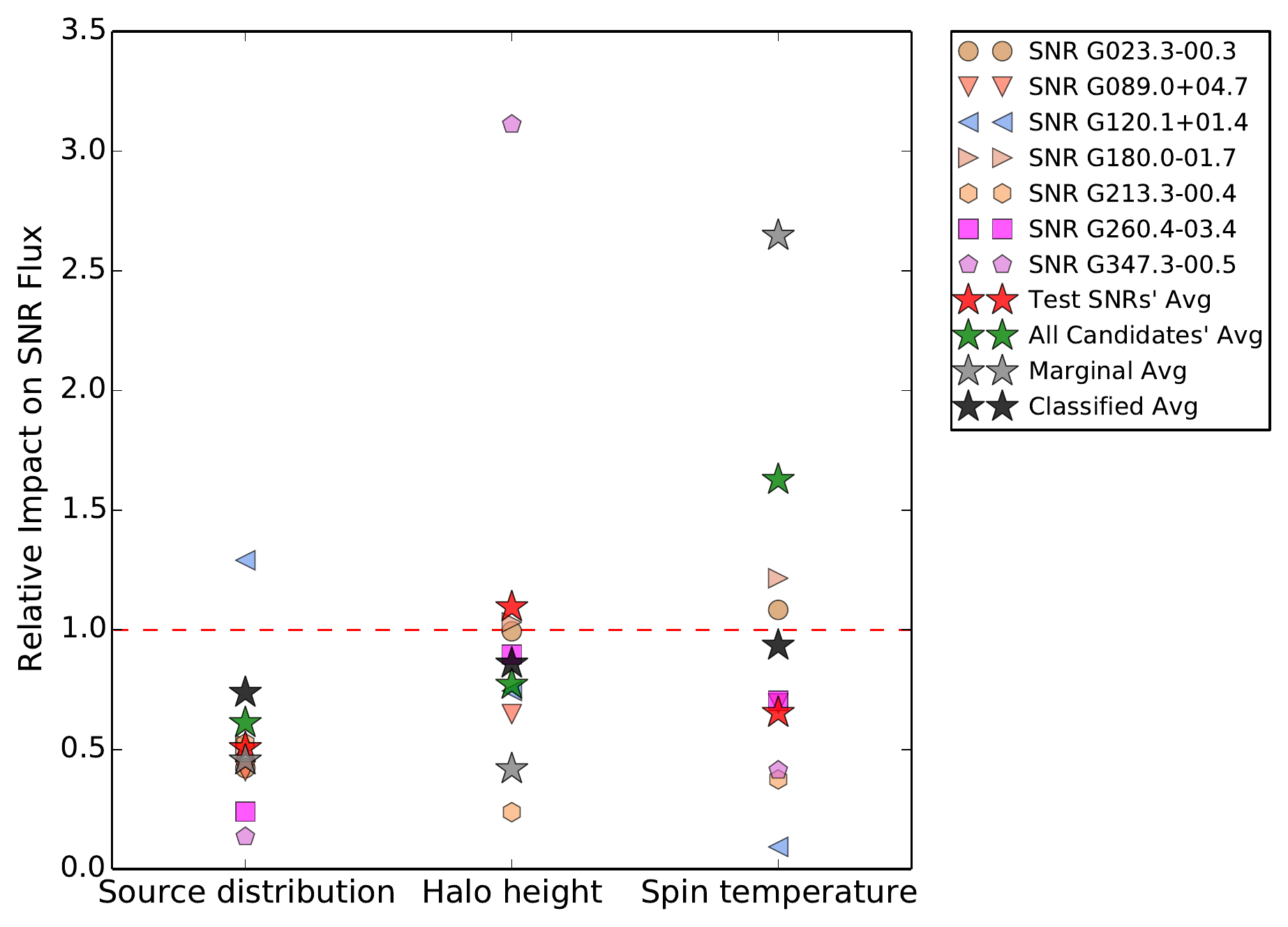} 
\caption{The impact on the candidates' fluxes for each of the alternative IEM input parameters (source distribution, halo height, and spin temperature), marginalized over the other GALPROP input parameters, shown via $R$ (Equation~\ref{eq:ratio_aIEM}). 
The stars represent the average ratio over the different candidate classes (classified, marginal, test, and all). As no alternative IEM input parameter has a ratio significantly larger than $1$ for all tested candidates, no input parameter dominates the systematic uncertainties of the fitted source parameter sufficiently to justify neglecting the others.}
\label{fig:input_param_aIEM_flux}
\end{figure}

\begin{figure}[h!]
\centering
\includegraphics[width=0.8\columnwidth]{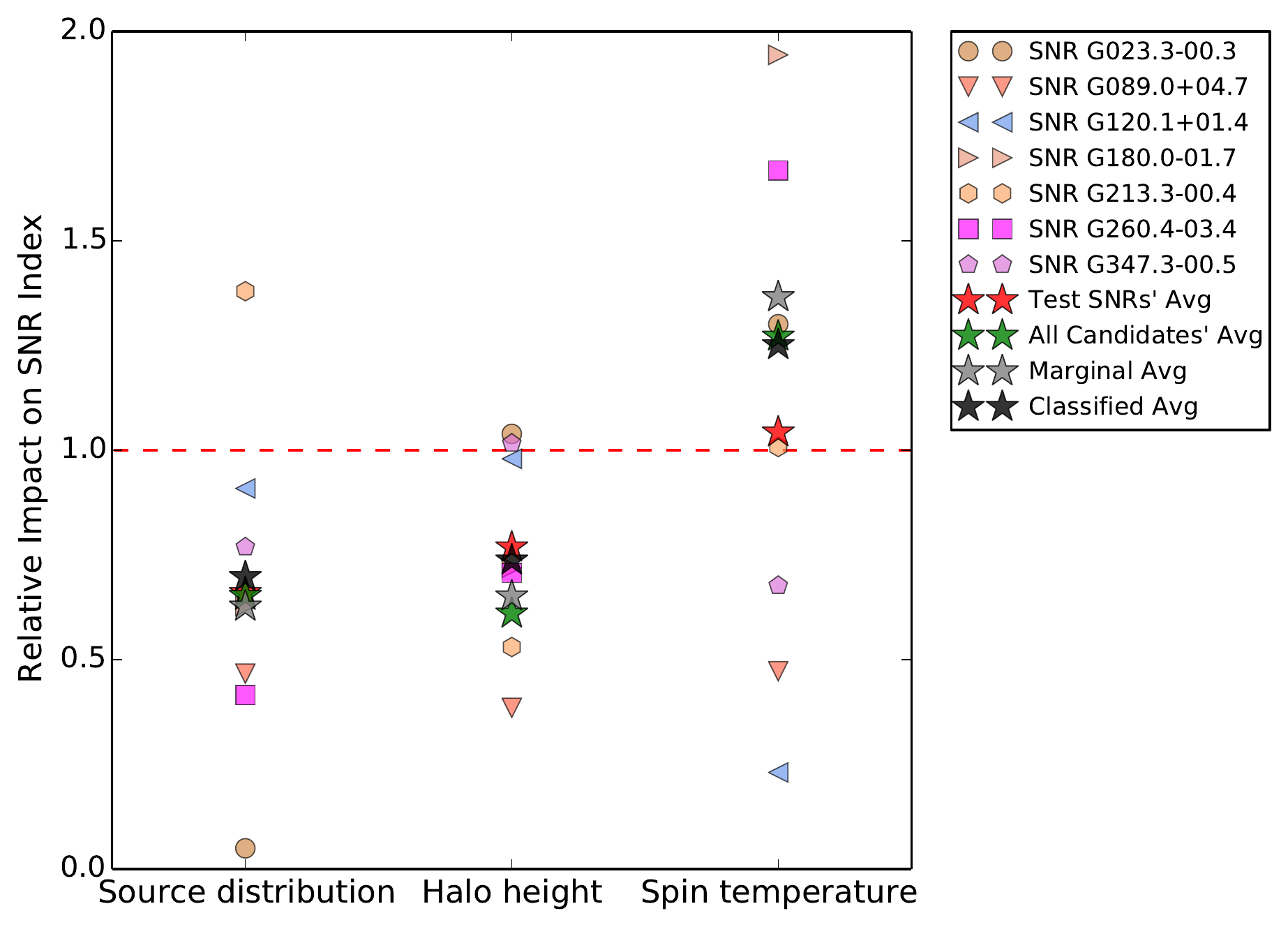} 
\caption{The impact on the candidates' PL indices for each of the alternative IEM input parameters (source distribution, halo height, and spin temperature), marginalized over the other GALPROP input parameters, shown via $R$ (Equation~\ref{eq:ratio_aIEM}). 
The stars represent the average ratio over the different candidate classes (classified, marginal, test, and all).  As no alternative IEM input parameter has a ratio significantly larger than $1$ for all tested candidates, no input parameter dominates the fitted source parameter sufficiently to justify neglecting the others. For candidates best fit with a logP model we used the value obtained from the analysis with the PL model. }
\label{fig:input_param_aIEM_Index}
\end{figure}

In Figures~\ref{fig:input_param_aIEM_flux} and \ref{fig:input_param_aIEM_Index} for the flux and PL index respectively we show the ratio $R$  for each of the alternative IEMs' input parameters for the individual SNR candidates, along with the averages for all seven test candidates, for the classified candidates, for the marginally classified candidates, and for all candidates in aggregate. The candidate associated with SNR~G120.1+1.4, being just below our threshold, was not significantly detected in our final analysis. Nevertheless it is shown in the figures for reference since it was useful for studying the effects on low significance sources. We do not however, show the candidate in the region of SNR~G119.5+10.2 as the fit in this region did not converge properly for the alternative IEMs. While the spin temperature seems to have a slightly larger effect than the other parameters, particularly for the index, none of the three GALPROP input parameters has $R$ significantly smaller or larger than $1$ for all the candidates tested such that it or the other GALPROP parameters can be neglected. Thus, as shown for the flux in Figure~\ref{fig:input_param_aIEM_flux} and the index in Figure~\ref{fig:input_param_aIEM_Index}, all three input parameters are relevant for the various candidate classes (classified, marginal, test, and all). We therefore conclude that none of the input IEM parameters dominated the systematic uncertainties of the fitted source parameters sufficiently to justify neglecting the others.


\FloatBarrier
\subsection{Alternative IEM Systematics in Context}\label{aIEMsContext}

We examined the magnitude of the systematic errors inferred from the alternative IEMs, as derived in Equation~\ref{eq:sys_err_weighted}, with respect to both the statistical error and the systematic error due to uncertainty in the effective area.

\subsubsection{Statistical and Alternative IEM Systematic Errors}\label{sec:aIEMsvStat}

\begin{figure}[h!]
  \centering
  \includegraphics[width=0.95\columnwidth]{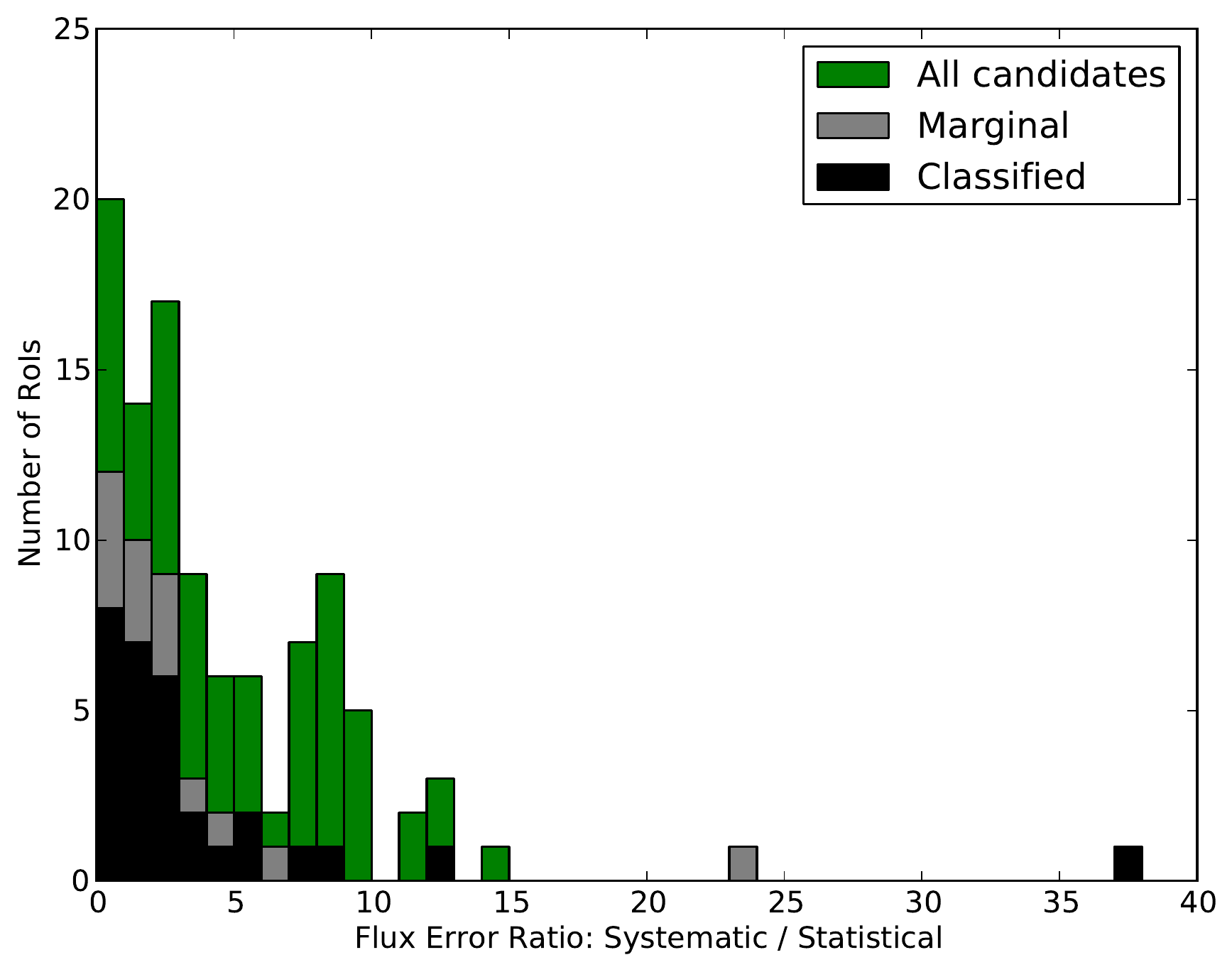} 
  \caption{ Cumulative histogram of the ratio of the systematic to statistical error on flux for all SNR regions. Classified candidates are black; marginally classified candidates are grey; and all remaining candidates are shown in green. The greatest outlier, the candidate in the region of SNR~G78.2+2.1, is discussed further in Section~\ref{sec:aIEMsvStat}.}
  \label{fig:rel_err_histo}
 \end{figure}

\begin{figure}[h!]
  \centering
  \includegraphics[width=0.95\columnwidth]{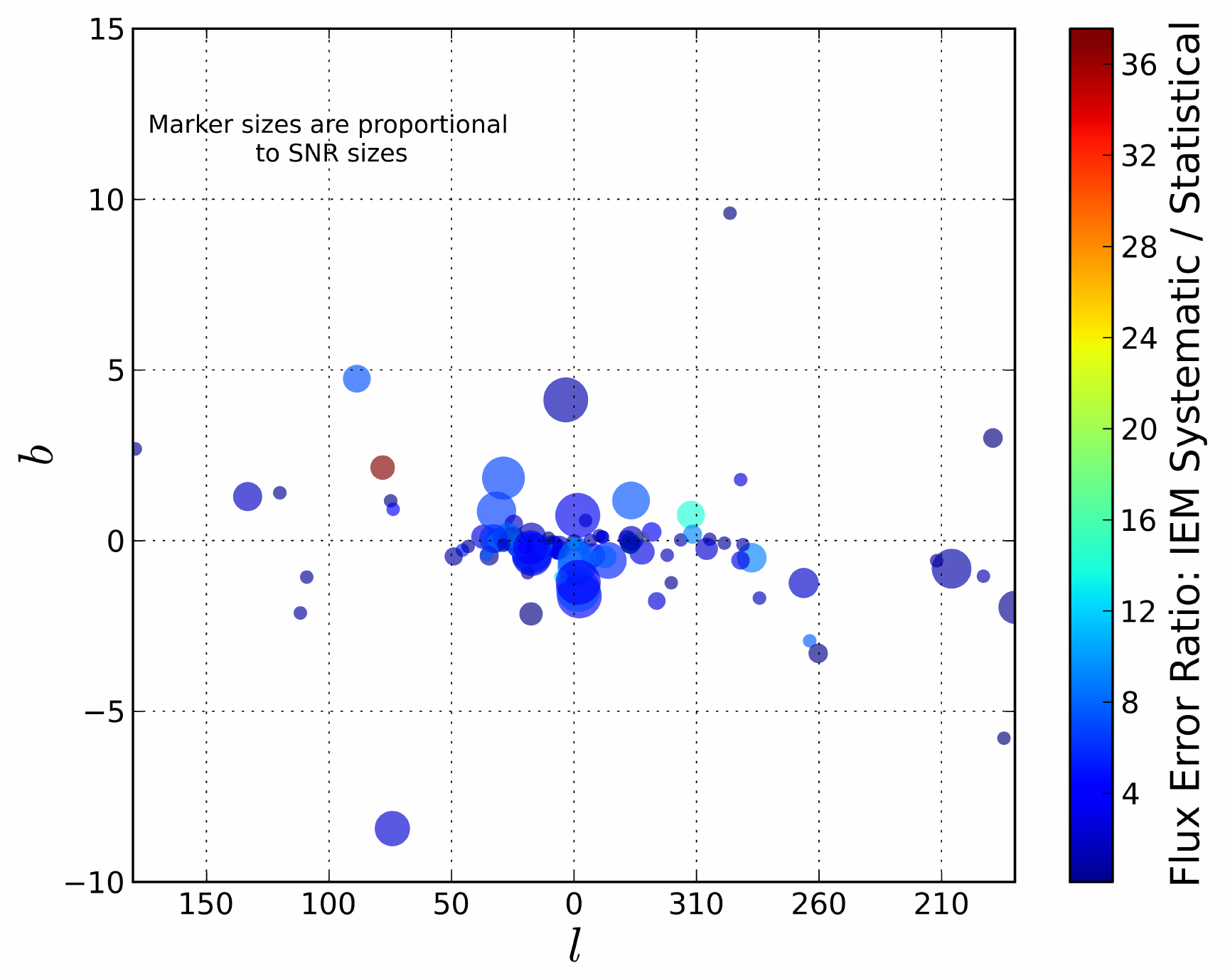} 
  \caption{The ratio of the alternative IEM systematic and statistical errors on flux for all candidates shown at their GeV positions.}
  \label{fig:rel_err_skymap}
 \end{figure}

\begin{figure}[h!]
  \centering
  \includegraphics[width=0.95\columnwidth]{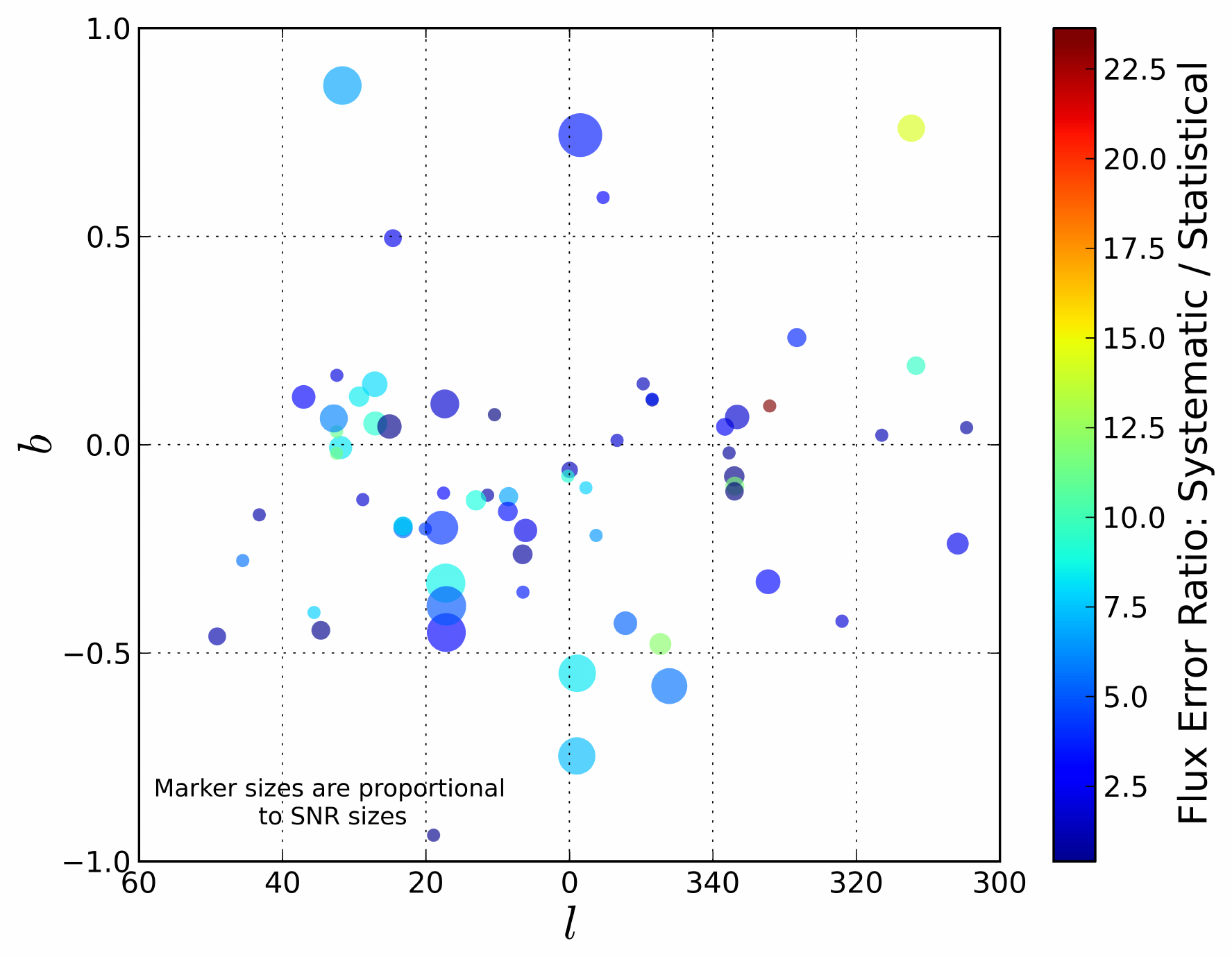} 
  \caption{Same as Figure~\ref{fig:rel_err_skymap} for the inner Galaxy only. The coordinate axes do not have the same scale.}
  \label{fig:rel_err_skymap_zoom}
 \end{figure}

Figure~\ref{fig:rel_err_histo} shows that the error on the flux due to the alternative IEM systematic error is generally the same order of magnitude as the $1\sigma$ statistical error, particularly for the \nclassifiedandmarginal{} classified and marginally classified sources. The greatest outlier, the candidate in the region of SNR~G78.2+2.1, lies in the Cygnus region, where the diffuse emission is bright, structured, and particularly difficult to model. In this region the candidate's size increases by up to six times that found with the standard IEM, leading to the large flux systematic error estimate. We also looked for any correlation between the ratio of the alternative IEM systematic flux error to the statistical error and position on the sky. Neither Figure~\ref{fig:rel_err_skymap} nor, for the inner Galaxy, Figure~\ref{fig:rel_err_skymap_zoom} show any obvious trend of relative error magnitude with sky position.

\subsubsection{Bracketing IRF and Alternative IEM Systematic Errors}\label{sec:aIEMsvbIRF}

\begin{figure}[h!]
  \centering
    \includegraphics[width=0.95\columnwidth]{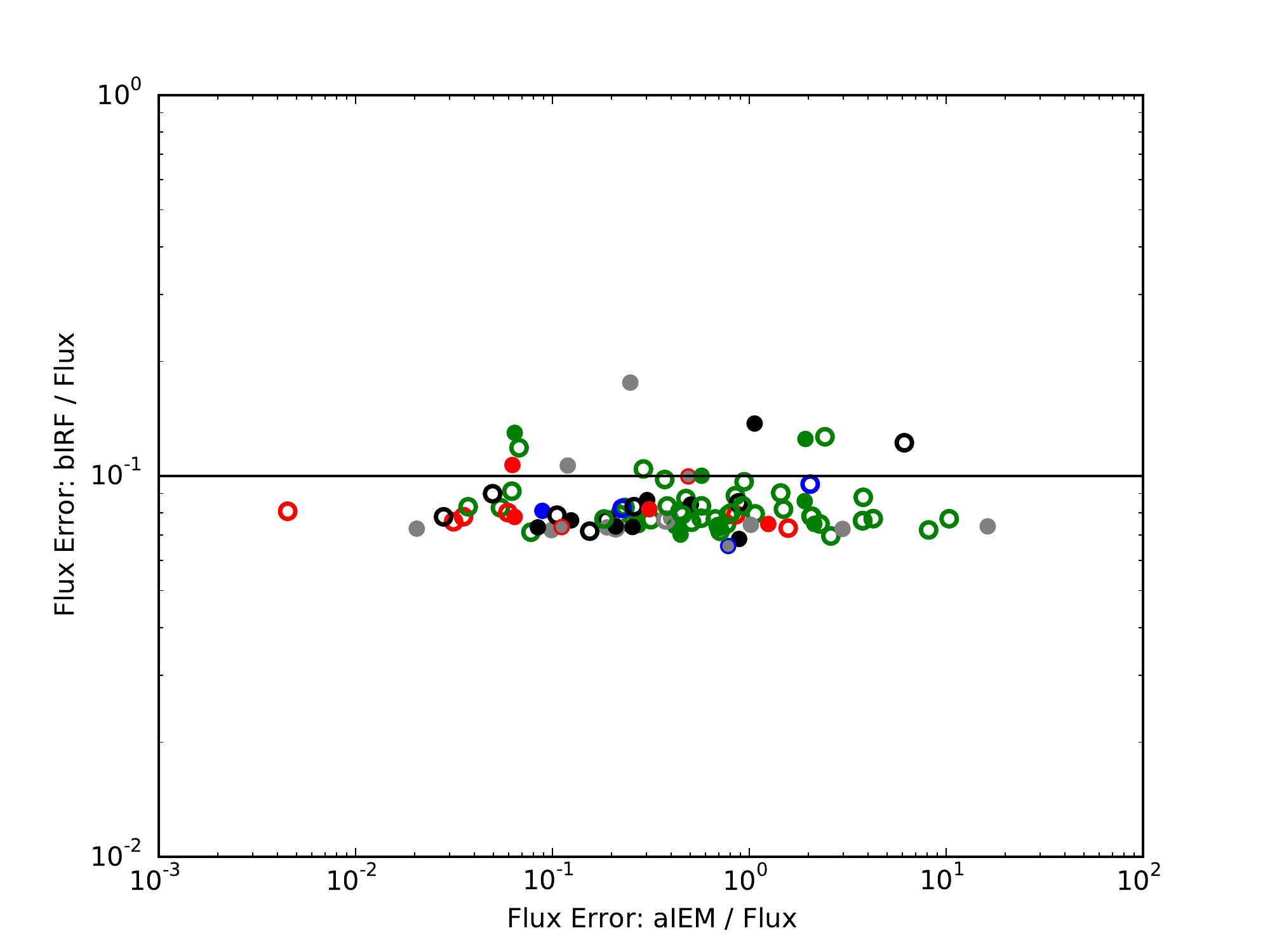} 
  \caption{Comparison of the systematic error on flux derived from the weighted alternative IEMs (labeled aIEM) and the bracketing IRFs (labeled bIRF), expressed relative to the measured flux. The line indicates a relative systematic error of $10\%$ for the bracketing IRF with respect to the flux. The relative error from the bracketing IRFs is less than $10\%$ for the majority of candidates, which span over $3$ orders of magnitude in error derived from the alternative IEMs. Markers and colors are as in Figure~\ref{fig:sysvStatFluxErrs}. } 
  \label{fig:aIEMvbIRFsRelFluxErrs}
 \end{figure}

We also investigated the relative magnitude of the alternative IEM systematic error with respect to that from the effective area derived using the bracketing IRFs (Section~\ref{Sec:EffAreaSysErr}). Figure~\ref{fig:aIEMvbIRFsRelFluxErrs} shows that the systematic error on flux estimated from the alternative IEMs tends to dominate over that due to effective area. This is true for most sources, regardless of their size, classification, or Galactic longitude. Moreover, the error propagated from the effective area tends to be $<10\%$ of the flux for the large majority of candidates, while that from the alternative IEMs ranges over $\sim3$ orders of magnitude, indicative of the diverse environments in which these candidates are found. We also observe that the candidates classified as other than SNRs, particularly when extended, tend to have somewhat larger alternative IEM errors than the classified and marginally classified candidates. This drives the increase in systematic errors observed in Figure~\ref{fig:sysvStatFluxErrs} for extended candidates classified as other,  contributing to the trend noted in Section~\ref{Sec:SysvStatErr} of extended candidates with larger systematic and statistical errors tending to be classified as other.

\begin{figure}[h!]
  \centering
  \includegraphics[width=0.95\columnwidth]{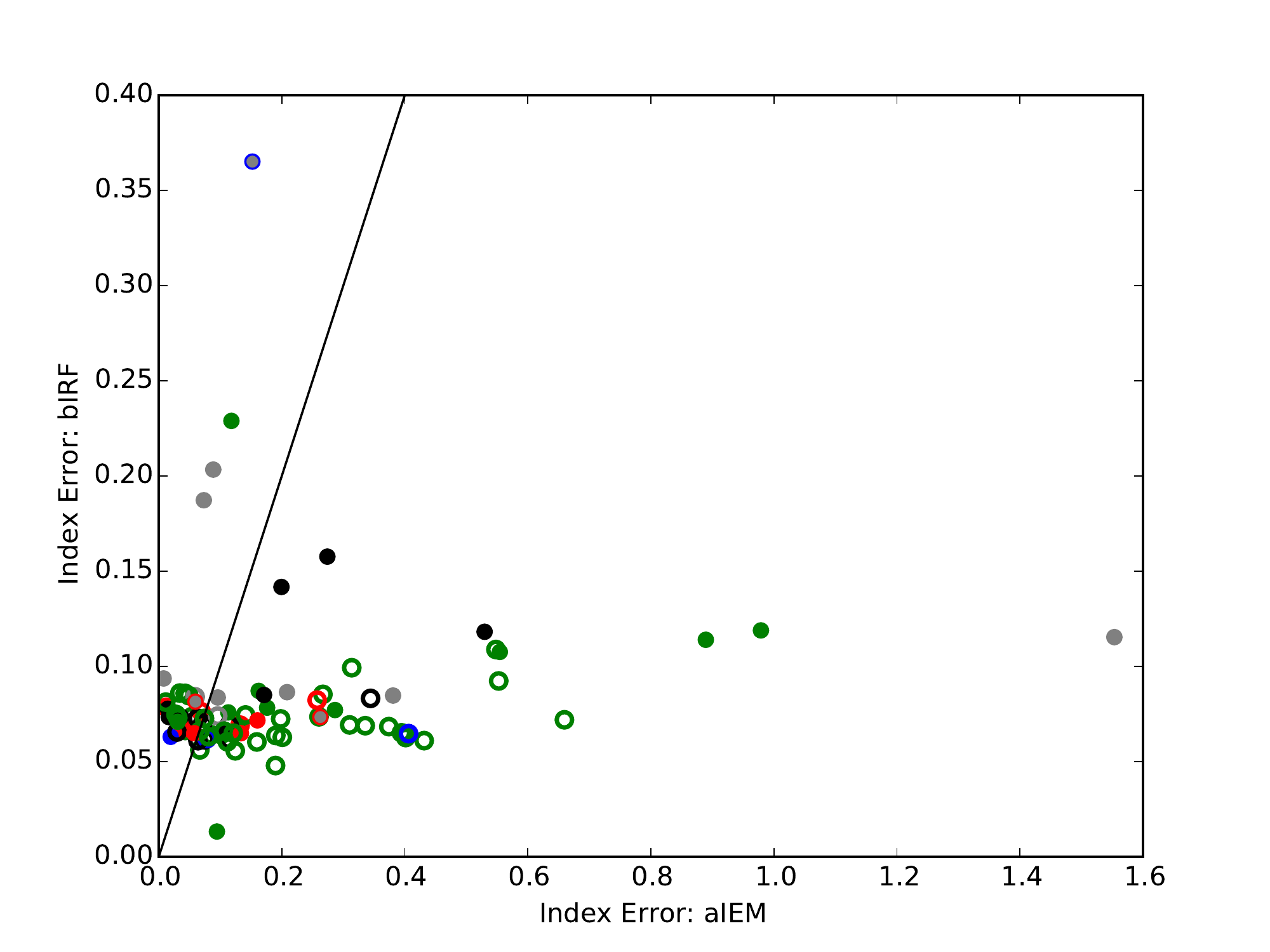} 
  \caption{Comparison of the systematic error on index derived from the weighted alternative IEMs and the bracketing IRFs. As seen by the distribution about the solid line showing equal systematic errors from the bracketing IRFs and the alternative IEMs, about half the candidates have larger alternative IEM systematic index error than bracketing IRF error and vice versa. Markers and colors are as in Figure~\ref{fig:sysvStatFluxErrs}.} 
  \label{fig:aIEMvbIRFsIndexErrs}
 \end{figure}

In contrast, Figure~\ref{fig:aIEMvbIRFsIndexErrs} shows that for the PL index the systematic errors from the bracketing IRFs dominate the alternative IEM errors for approximately half the candidates. Aside from candidates classified as other having somewhat larger systematic errors due to the alternative IEMs, systematic errors on neither flux nor index are significantly different among various subtypes: point or extended candidates, marginal or classified candidates, or young compared with interacting.

\FloatBarrier

\section{Chance Coincidence Estimation}
\label{Appen:ChanceCoinc}

For each of the associations between Green's catalog sources and \Fermi{} GeV candidates, there is some probability of a ``chance coincidence.'' This occurs if an SNR identified in other wavebands does \emph{not} emit in GeV \g-rays, but a different source or a bright region of diffuse \g-ray emission provides a photon excess in approximately the same location. If the excess overlaps sufficiently with the radio SNR (Section~\ref{Sec:Class}), the classification procedure will associate it with the SNR, leading to a false classification.

While it is not possible to identify \emph{which} of the individuals identified as classified are by chance, it is possible to determine the probability that this will occur. We used a Monte Carlo approach to calculate this false discovery rate, starting with a mock catalog that mimics the true Green's catalog in essential ways while still having non-overlapping sources (aside from edge cases described in Section~\ref{SSec:MockCatalogGen}). Running the standard analysis on the mock catalog cannot, by construction, produce a true classification, so any matches are by chance. 

\subsection{Mock Catalog Generation}\label{SSec:MockCatalogGen}

We considered four distributions as essential for generating the mock catalog: longitude coordinate, latitude coordinate, source size, and integrated IEM intensity. The last is the total amount of diffuse \g-ray emission included in the radio area of the SNR. We generated many mock catalogs and reran the analysis on the one which best matched the Green's catalog distribution for each variable.

Since the diffuse emission varies dramatically with Galactic latitude, we chose
to use a ``horizontal displacement'' method to generate the mock catalogs. Each 
source in Green's catalog was displaced a random amount in longitude proportional to its radio radius $r$ or the minimum resolvable GeV radius, as discussed in Section~\ref{Sec:SpatCoinc}. 
With an effective radius defined as 
\begin{equation}
r_e = \mathrm{max}\left(\frac{0.2^{\circ}}{3}, r\right)
\end{equation}
the allowed displacement range $d$ is defined as
\begin{equation}
3r_e \le d \le 5r_e.
\end{equation}
The simulation also randomly displaced the source toward increasing or decreasing longitude.

The simulation is designed to avoid most overlaps between displaced sources and other
Green's catalog objects, aside from the parent source which is avoided by
definition. The following steps are performed:
\begin{enumerate}
\item Generate a trial catalog of random positions
\item Check for overlaps
\begin{enumerate}
\item If there are fewer than $6$ overlaps (out of \nGalSNRs~sources), exit loop
\item If the simulation has tried to fix overlaps in the same subset of sources more than $10$ times, exit loop 
\end{enumerate}
\item Otherwise, regenerate overlapping sources, and return to step 2.
\end{enumerate}
The maximum number of overlaps was determined empirically to avoid
forcing mock sources in regions of high density of SNRs into the same location every iteration. The second loop condition prevents the simulation from getting trapped in unsolvable situations. It is unlikely
that the five allowed overlaps will lead to a coincidence, since the mock and Green's 
source must match in both size and location, but we check this in Section~\ref{ssec:mock_catalog_coincidences}.

\begin{figure}[h]
  \centering
  \includegraphics[width=0.8\columnwidth]{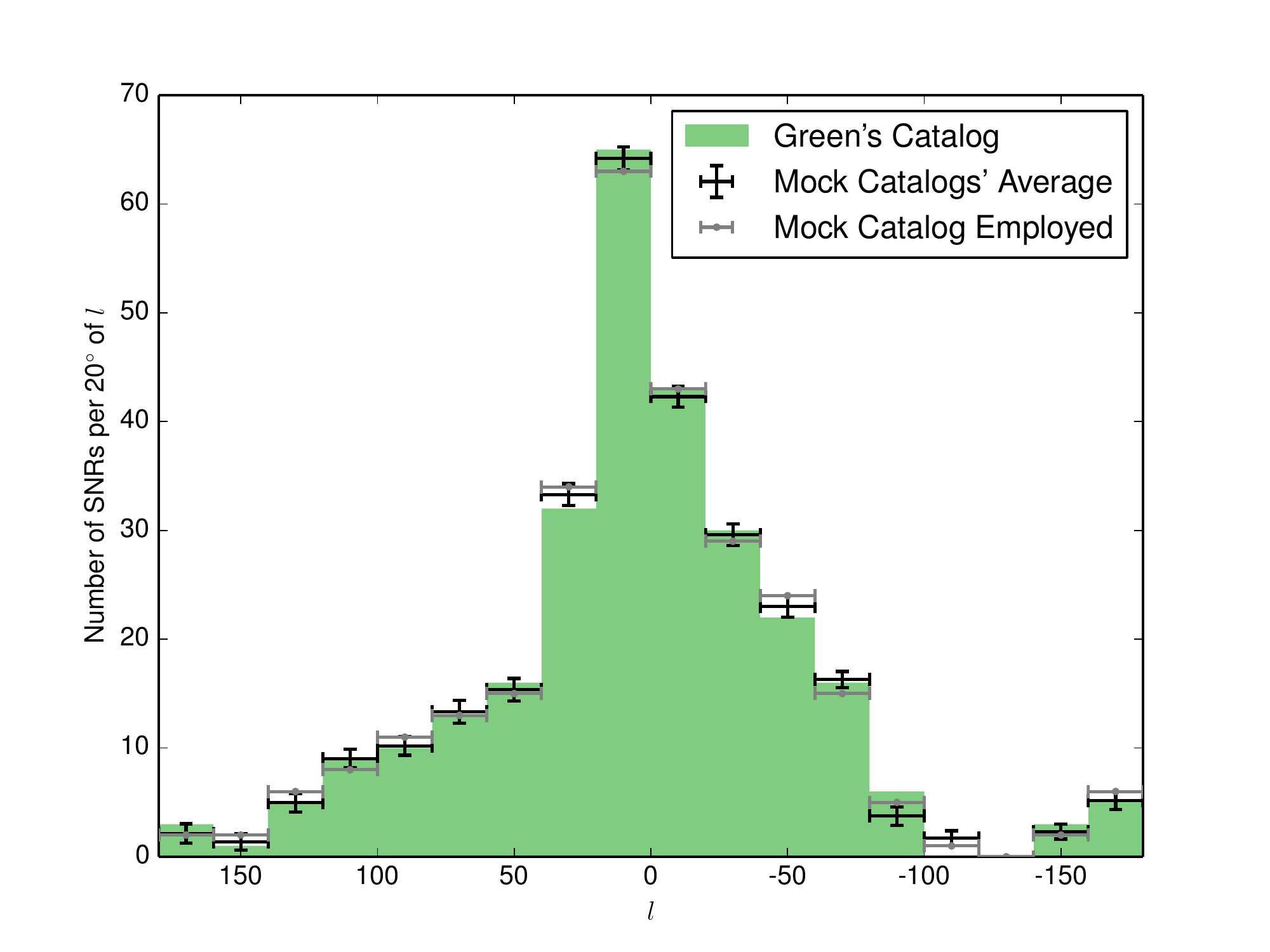} 
  \caption{The distribution of SNRs in Galactic longitude. The green bars represent the original Green's catalog, the black points are an average over $100$~realizations of the mock catalog and the grey points are from the mock catalog analyzed in this paper.}
  \label{fig:mock_l_hist}
\end{figure}

\begin{figure}[h!]
  \centering
  \includegraphics[width=0.8\columnwidth]{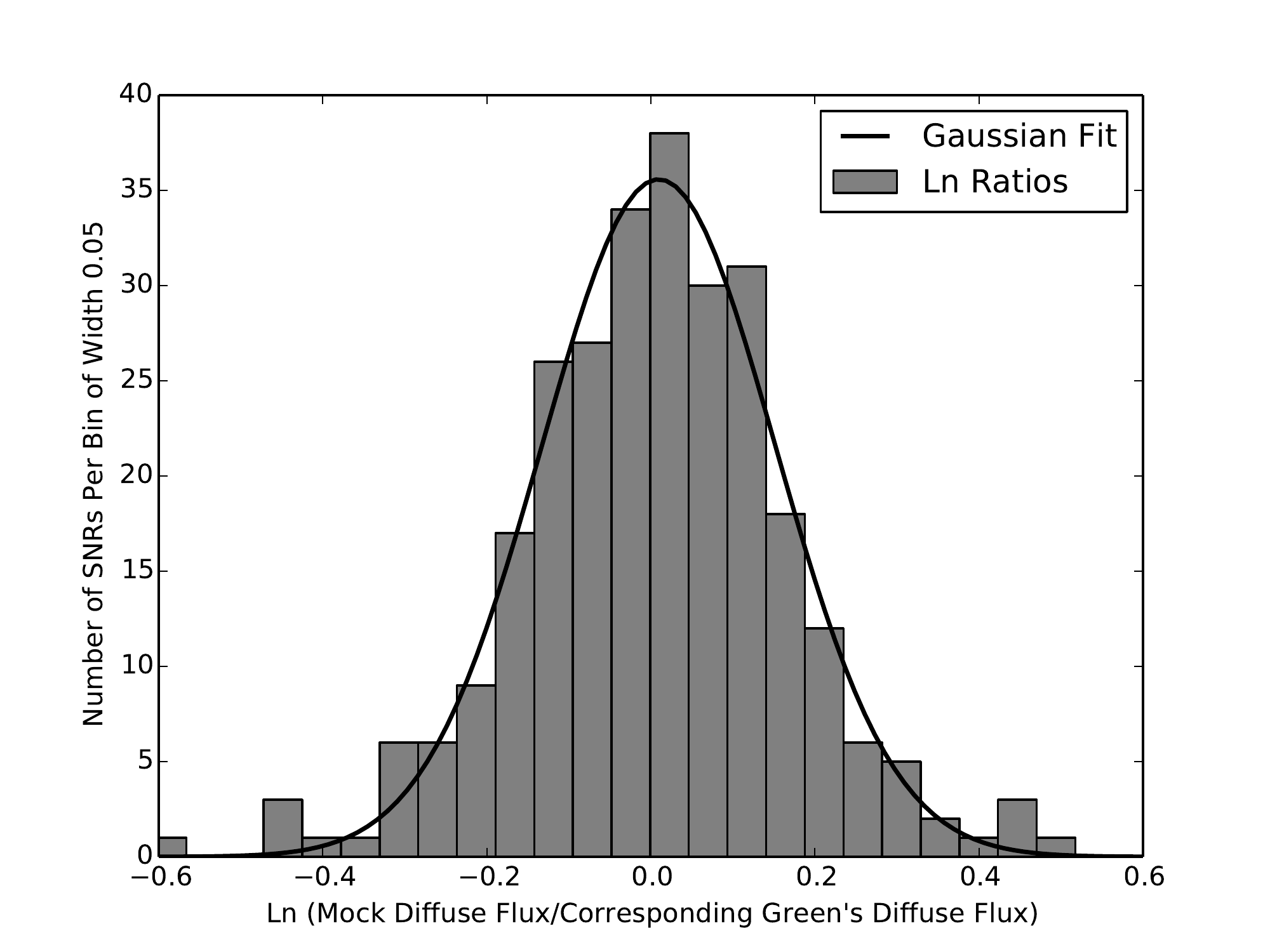} 
  \caption{The distribution of the natural logarithm of the ratio of the diffuse flux under a mock catalog source to that under the corresponding unshifted source. This is approximately a Gaussian centered at zero, which means the mock catalog accurately represents the distribution of background interstellar diffuse fluxes of the original Green's catalog.}
  \label{fig:mock_diffuse_hist}
\end{figure}

Using this method, we generated several catalogs and selected the one that best fit 
our distribution requirements, noted previously. Since the size and latitude are not modified when 
creating the mock catalog, it was only necessary to compare longitude distributions
and integrated fluxes. The longitude distributions for Green's catalog, the mock catalog analyzed, and the average of $100$ mock catalogs generated by this method, shown in Figure~\ref{fig:mock_l_hist}, match sufficiently.
Instead of directly comparing the integrated diffuse fluxes under the mock and Green's catalog sources, we plot
\begin{equation}
\mathcal{D}_i = \mathrm{ln} \left(\frac{D_{m,i}}{D_{g,i}}\right),
\end{equation}
where $D_{m,i}$ is the integrated diffuse flux under mock source $i$ and $D_{g,i}$ is the integrated diffuse under the original (unshifted) source. The variable $\mathcal{D}_i$ is very nearly normally distributed around zero, as can be seen in Figure~\ref{fig:mock_diffuse_hist}. 
The average of ${D_{m,i}}/{D_{g,i}}$ is $1.08$ 
and the dispersion ($\sigma$) is $14\%$. The marginal increase in diffuse emission under mock sources is less than the dispersion. This should not bias the study, and at worst we may marginally overestimate the number of mock sources found. 

\subsection{Mock Catalog Coincidences}\label{ssec:mock_catalog_coincidences}

Analyzing this mock catalog with the standard method produced \nmock~(\nmockclassifiedandmarginal) coincidences using the classified (marginal and classified) threshold, compared to the \nassocprobclassified~(\nclassifiedandmarginal) total number found from Green's catalog (before removing known sources and before demotions). Those sources which met at least one of the classification thresholds are shown in Table~\ref{Tab:MockCoincidences}. The coincidences tend to follow the pattern established by Green's catalog: most are clustered in the Galactic center region, with a few outliers.

The mock catalog used ended on condition 2a (Section~\ref{SSec:MockCatalogGen}) with four allowed overlaps, of which one (Mock008.7$-$00.1) was found to be associated with an SNR at the marginal classification threshold level. We allow this to remain in the results to generate conservative upper limits for the false discovery rate.

\begin{deluxetable}{lccccccc}
\tablewidth{0pt} 
\singlespace
\tabletypesize{\scriptsize}
\tablecaption{Spurious Classifications in the Mock Catalog}
\label{Tab:MockCoincidences}
\tablehead{
\colhead{} & 
\multicolumn{2}{c}{Mock} &
\multicolumn{3}{c}{Green's Catalog} &
\colhead{ } & 
\colhead{ } \\
\colhead{} & 
\colhead{$l$} &
\colhead{$b$} &
\colhead{Size} &
\colhead{$l$} &
\colhead{$b$} &
\colhead{Location} & 
\colhead{Extension} \\
\colhead{Name} & 
\colhead{[deg]} & 
\colhead{[deg]} & 
\colhead{[deg]} & 
\colhead{[deg]} & 
\colhead{[deg]} &
\colhead{ Overlap } & 
\colhead{ Overlap } \\
}
\startdata
\multicolumn{8}{c}{{{\bf Classified Mock Candidates:}}}\\
Mock020.0$-$00.2 & 20.26 & -0.17 & 0.08 & 19.98 & -0.17 & 1.00 & 0.42 \\
Mock347.3$-$00.5 & 354.08 & -0.51 & 0.50 & 347.37 & -0.51 & 0.41 & 0.84 \\
\multicolumn{8}{c}{{{\bf Marginally Classified Mock Candidates:}}}\\
Mock008.7$-$00.1 & 10.23 & -0.10 & 0.38 & 8.74 & -0.10 & 0.37 & 1.00 \\
Mock017.4$-$02.3 & 16.66 & -2.30 & 0.20 & 17.39 & -2.30 & 0.15 & 0.81\\
Mock032.0$-$04.9 & 33.54 & -4.61 & 0.50 & 31.92 & -4.61 & 0.16 & 1.00\\
Mock213.3$-$00.4 & 207.84 & -0.48 & 1.33 & 213.15 & -0.48 & 0.22 & 0.22\\
Mock296.8$-$00.3 & 297.35 & -0.34 & 0.14 & 296.88 & -0.34 & 0.63 & 0.31\\
Mock359.1$-$00.5 & 358.51 & -0.51 & 0.20 & 359.12 & -0.51 & 0.35 & 0.32 \\
Mock359.1$+$00.9 & 358.76 & 0.99 & 0.10 & 359.10 & 0.99 & 0.96 & 0.38\\ 
\enddata
\tablecomments{Mock Catalog sources passing either the classification or marginal classification threshold. The name of the source is derived from the parent source in Green's catalog.  The position of the mock source is given by the Mock $l$ and $b$ columns. The location of its parent source is given by the Green's Catalog $l$ and $b$ columns. Location and extension overlaps show the overlap between the mock candidate and radio SNR it was derived from as defined in Section \ref{Sec:SpatCoinc}. Mock008.7-00.1 overlapped a known Green's catalog source when the mock catalog generation terminated; we retain it to derive conservative constraints.}
\end{deluxetable}

\subsection{Estimating Upper Limits}\label{Sec:ULmock}

We would like to calculate a $95\%$~confidence upper bound to the chance coincidences and false discovery rates given in Section~\ref{ssec:mock_catalog_coincidences}. First, we need a distribution to model the expected number of false discoveries for a single trial. The mock catalog itself has $N=$~\nGalSNRs{}~mock SNRs, each of which has a probability $p_i \ll 1$ of returning a chance coincidence, equal to $0$ for many sources. The true chance coincidence distribution $\mathcal{C}(n)$ is the sum over each of these Bernoulli random variables.
We show below this is asymptotically similar to a binomial with $N$ trials and probability $\tilde{p} = \sum p_i /N$, denoted $B_N(n,\tilde{p})$. This amounts to assuming that all sources have the same probability of false discovery and classification as an SNR.

In order to check that this is a reasonable assumption, we consider the other extreme possibility. We define $n_{max} = \lceil (\sum p_i )/\max(p_i) \rceil$ and $p_{max} = \sum p_i /n_{max}$, where the $\lceil x \rceil$ notation denotes the ceiling function. Then we define $\tilde{\mathcal{C}}(n)$ as the distribution with $n_{max}$ random numbers having probability $p_{max}$ to equal $1$ and the remaining $N-n_{max}$ numbers having zero probability of equaling $1$. The mean probability is the same as that of the original distribution, i.e. $n_{max} p_{max} = N \tilde{p}$, but we have concentrated the probability density into a smaller number of sources each with larger probability. The $N-n_{max}$ sources with probability $0$ do not contribute, so this distribution is simply a binomial with $n_{max}$ trials and $p_{max}$ probability, i.e. $\tilde{\mathcal{C}}(n)=B_{n_{max}}(n,p_{max})$.
Using the explicit form of the binomial expression and Stirling's formula, it is possible to derive that
\begin{equation}
\begin{split}
\ln\left(\frac{B_N(n,\tilde{p})}{B_{n_{max}}(n,p_{max})}\right) = &\frac{1-n_{max}/N}{2} \left[ \frac{(n-p_{max}n_{max})^2}{n_{max}} - p_{max}\right] \\&+ O\left(\frac{1}{n_{max}}\right) + O(n_{max} p_{max}^3)
\end{split}
\end{equation}

Since the variance of the binomial distribution with $n_{max}$ trials is $n_{max} p_{max} (1-p_{max})$ and its mean is $n_{max} p_{max}$, we know that $(n-p_{max}n_{max})^2$ is of order $n_{max} p_{max} (1-p_{max})$ so $(n-p_{max}n_{max})^2/n_{max}$ is of order $p_{max}$. Thus the ratio is of order $p_{max}$, which is small, and the two distributions are asymptotically similar as long as $n_{max}$ is large and $n_{max} p_{max}^3$ is small.

However, if $p_{max}$ is not small, then the above relation is not true. Consider the case when $p_{max} = 1$, and the distribution becomes a delta function at $n=n_{max}$. In general, if $p_{max}$ is not small, the true distribution's variance will actually be \emph{smaller} than that of the approximation $B_N(n,\tilde p)$. Thus, the approximation will always provide a conservative upper limit on the error.

For the upper limit itself, from the preceding we assume that a binomial distribution $B_N(n,p)$ provides a conservative estimate of the error. Here, we have changed notation, letting $\tilde p$ become $p$. We now use a Bayesian approach to calculate the probability of a second catalog producing a number of false coincidences $n_2$ given that the first produced a number~$n_1$ among $N$ input SNRs, where $N$ is again the number of trials.

Assuming that the prior distribution of the underlying chance coincidence probability~$p$ is flat between $0$ and $1$, Bayes' theorem combined with the previous derivation gives the posterior distribution for $p$ as 
\begin{equation}
f(p \, | \, n_1) = \frac{B_N(n_1,p) }{\int_0^1 \! B_N(n_1,p) \, \mathrm{d} p}~.
\end{equation}
Using a flat prior is a conservative assumption, since it gives strong weight to large probabilities. The integral here is related to the beta function $B(x,y)$ as:
\begin{align}
\int_0^1\! B_N(n_1,p) \, \mathrm{d} p =& \int_0^1\! \binom{N}{n_1} p^{n_1}(1-p)^{N-n_1} \, \mathrm{d} p = \binom{N}{n_1} B(n_1+1,N-n_1+1) \\=& \frac{1}{N+1}
\end{align}
Thus $f(p \, | \, n_1) = (N+1) B_N(n_1,p)$.
Note that the average probability $<p>$ is
\begin{align}
<p>=& \int_0^1 \! p f(p \, | \, n_1) \, \mathrm{d} p = (N+1)\int_0^1 \! p B_N(n_1,p) \, \mathrm{d} p \\
= &(N+1)\binom{N}{n_1} B(n_1+2,N-n_1+1) \\
= &\frac{n_1+1}{N+1}
\end{align}
which is larger than $n_1/N$. For instance, if $n_1=3$ and $N=100$, $<p> \approx 0.039$.

We can identify the probability for $n$ from the second catalog to equal a value $n_2$, given that the first catalog has produced $n_1$, by marginalizing over $p$ in the binomial probability above:
\begin{align}
P(n_2) =& \int_0^1 \! f(p \, | \, n_1) B_N(n_2,p)\, \, \mathrm{d} p = \int_0^1\! (N+1)B_N(n_1, p)B_N(n_2,p) \, \mathrm{d} p \\
= & (N+1) \binom{N}{n_1} \binom{N}{n_2} B(n_1+n_2+1, 2N - n_1 - n_2 + 1) \\
= & \frac{(N+1)!}{n_1!\, (N-n_1)!}\frac{N!}{n_2!\, (N-n_2)!} \frac{(n_1+n_2)!\,(2N-n_1-n_2)!}{(2N + 1)!}
\end{align}
The large quantities involving $N$ can be expanded using Stirling's approximation, and then by assuming that $N$ is large compared with $n_1$, $n_2$, and $1$, we can derive the simple formula
\begin{equation}
P(n_2) = \frac{(n_1+n_2)!}{n_1!\,n_2!\,2^{(n_1+n_2+1)}} = \binom{n_1+n_2}{n_2} 2^{-(n_1+n_2+1)}.
\end{equation}
This gives the probability that a particular value $n_2$ is found in the second trial, given that $n_1$ was returned from the first trial and $N$ is large.

Using this formula, for the \nmock~(\nmockclassifiedandmarginal) mock coincidences passing the classified (marginal and classified) threshold, we get a 95\% confidence upper limit of \nninetyfivemock~(\nninetyfivemockclassifiedandmarginal). When divided by the \nassocprobclassified~(\nclassifiedandmarginal) total number using Green's catalog, these correspond to a $95\%$ upper limit false
discovery rate of \nninetyfivemockpercent~(\nninetyfivemockclassifiedandmarginalpercent).

\FloatBarrier

\section{Electronic Catalog Data Products}
\label{apen:SupMat}
\label{Sec:CatProds}

The results described in this First \Fermi{} Large Area Telescope Catalog of Supernova Remnants (1SC) are provided as supplemental online material. A compressed (gzip) electronic archive file (tar) called 1SC\_auxiliary\_files\_v\#\#.tgz is available at the \Fermi{} Science Support Center\footnote{\url{http://fermi.gsfc.nasa.gov/ssc/data/access/lat/1st_ SNR_catalog/}}. The archive contains a directory structure with FITS tables of the analysis results, an ASCII text version of the FITS table, individual FITS files that serve as spatial templates for the classified extended SNRs, and a DS9 region file containing the final position and extension for all GeV candidates.

Detailed column descriptions for the main FITS auxiliary file are given in Tables~\ref{Tab:FITScatalog}, \ref{Tab:FITSupperlimits}, and \ref{Tab:FITSmwinfo}. The first binary table extension, described in Table~\ref{Tab:FITScatalog}, includes the spatial and spectral results recorded in Tables~\ref{Tab:ResultsSpat} and \ref{Tab:ResultsSpec}. The second extension, described in Table~\ref{Tab:FITSupperlimits}, includes the upper limits from Table~\ref{Tab:ResultsULs}. The third extension, described in Table~\ref{Tab:FITSmwinfo}, includes the multiwavelength information presented in Tables~\ref{Tab:YoungSNRs}, \ref{Tab:InteractingSNRs}, \ref{Tab:SNRinfo}, and \ref{Tab:TeVSNRs}.

The included spatial templates are in celestial (J2000) coordinates in FITS format. They are normalized to $1$ with the background set to $0$ and are appropriate to be used in XML model files with the \Fermi{} Science Tools.

The DS9 region file includes all detected candidate sources, either as a cross if the point hypothesis is most likely or with a circle equal in size to the most likely extended disk. Colors indicate their classification. Candidates classified as SNRs are black, unless they are also identified as interacting with an MC~(red) or as young by the presence of nonthermal X-rays~(blue), as for the figures herein and described in Figure~\ref{fig:GeVFluxGeVIndex}. Marginally classified candidates are grey independent of any MW evidence for SNR-MC interaction or the SNR being young. Other detected candidate sources are likewise green.

\pagestyle{empty}
\begin{landscape}
\begin{deluxetable}{lll}
\tablewidth{0pt}
\setlength{\tabcolsep}{0.05in}
\singlespace
\tabletypesize{\scriptsize}
\tablecaption{LAT First SNR Catalog FITS Format: SNR\_CATALOG Extension \label{Tab:FITScatalog}}
\tablewidth{9.3in}
\tablehead{
\colhead{Name} & 
\colhead{Units} & 
\colhead{Description} 
}
\startdata
SNR\_Name & \nodata & Supernova remnant name \\
Classification & \nodata & Classified, Marginally Classified, Other, Not an SNR \\
RAJ2000, DEJ2000 & deg & Candidate SNR position in celestial coordinates (J2000) \\
StatUnc\_RAJ2000, StatUnc\_DEJ2000 & deg & One sigma statistical error in best-fit celestial coordinates (J2000) \\
SysUnc\_RAJ2000, SysUnc\_DEJ2000 & deg & One sigma systematic error from alternative IEMs in best-fit celestial coordinates (J2000) \\
GLON, GLAT & deg & Candidate SNR position in Galactic coordinates \\
Conf\_68\_SemiMajor	&	deg	 & Long radius of error ellipse at 68\% confidence \\
Conf\_68\_SemiMinor	&	deg	 & Short radius of error ellipse at 68\% confidence \\
Conf\_68\_PosAng	&	deg	 & Position angle of the 68\% long axis from celestial North, positive toward increasing RA (eastward) \\
TS & \nodata & Source TS \\
TS\_EXT	& \nodata & TS$_{\rm ext}$ of the best-fit extended hypothesis over a point source \\
RADIUS & deg & Best-fit radius of extended disk \\
StatUnc\_RADIUS	&	deg	 & One sigma statistical error on the best-fit disk radius \\
PosSysUnc\_RADIUS, NegSysUnc\_RADIUS	&	deg	 & One sigma systematic error on the best-fit disk radius from the alternative IEMs \\
AIEM\_FLAG1	& \nodata & Number of alternate IEMs for which a candidate’s TS became less than 9 \\
AIEM\_FLAG2	 & \nodata & Number of alternate IEMs for which the best extension hypothesis changed, between extended and point \\
AIEM\_FLAG3	 & \nodata & Number of alternate IEMs which had problems with during fitting \\
OVERLAP\_LOC	 & \nodata & Fractional overlap in the GeV candidate’s position with respect to Green’s catalog radio SNR \\
OVERLAP\_EXT & \nodata & Fractional overlap in the GeV candidate’s extension with respect to Green’s catalog radio SNR \\
SPECTRUM & \nodata & Spectral form: PL or logP \\
F1000 & ph cm$^{-2}$ s$^{-1}$ & Photon flux for 1 GeV--100 GeV \\
StatUnc\_F1000 & ph cm$^{-2}$ s$^{-1}$ & One sigma statistical uncertainty for photon flux \\
PosSysUnc\_F1000,NegSysUnc\_F1000	& ph cm$^{-2}$ s$^{-1}$ & One sigma systematic uncertainty for photon flux in the positive and negative directions \\
INDEX	& \nodata & Photon index for PL model; alpha for logP model \\
StatUnc\_INDEX & \nodata & 	One sigma statistical uncertainty for photon index \\
PosSysUnc\_INDEX, NegSysUnc\_INDEX	& \nodata & One sigma systematic uncertainty for photon index in the positive and negative directions \\
TS\_CURVE	& \nodata & TS of logP model over simple PL \\
BETA	& \nodata & logP beta \\
StatUnc\_BETA	& \nodata & One sigma statistical uncertainty for logP beta \\
PosSysUnc\_BETA,NegSysUnc\_BETA	& \nodata & One sigma systematic uncertainty for logP beta in the positive and negative directions \\
BEST\_HYP	& \nodata & Best hypothesis from the final gtlike fit: Disk, Neardisk or Point
\enddata
\end{deluxetable}
\end{landscape}

\pagestyle{empty}
\begin{landscape}
\begin{deluxetable}{lll}
\tablewidth{0pt}
\singlespace
\tabletypesize{\scriptsize}
\tablecaption{LAT First SNR Catalog FITS Format: SNR\_UPPER\_LIMITS Extension\label{Tab:FITSupperlimits}}
\tablehead{
\colhead{Name} & 
\colhead{Units} & 
\colhead{Description} 
}
\startdata
SNR\_Name & \nodata & Supernova remnant name \\
UL95\_I20, UL99\_I20 & ph cm$^{-2}$ s$^{-1}$ & 95\% and 99\% confidence level upper limit assuming an index of 2.0\\
UL95\_I25, UL99\_I25 & ph cm$^{-2}$ s$^{-1}$ & 95\% and 99\% confidence level upper limit assuming an index of 2.5
\enddata
\end{deluxetable}
\end{landscape}

\pagestyle{empty}
\begin{landscape}
\begin{deluxetable}{lll}
\tablewidth{0pt}
\singlespace
\tabletypesize{\scriptsize}
\tablecaption{LAT First SNR Catalog FITS Format: MW\_INFO Extension\label{Tab:FITSmwinfo}}
\tablehead{
\colhead{Name} & 
\colhead{Units} & 
\colhead{Description} 
}
\startdata
SNR\_Name & \nodata & Supernova remnant name. \\
RADIO\_RAJ2000 & deg & Right Ascension (J2000) of radio centroid \\
RADIO\_DEJ2000 & deg & Declination (J2000) of radio centroid \\
RADIO\_RADIUS & deg & Radius of the radio remnant \\
RADIO\_GLON & deg & Galactic longitude radio centroid \\
RADIO\_GLAT & deg & Galactic latitude of radio centroid \\
Is\_Xray\_Sync & \nodata & Is SNR X-ray synchrotron emitting? \\
Bib\_Xray\_Sync & \nodata & Bibcode for X-ray synchrotron reference\\
Is\_Interacting & \nodata & Is SNR interacting with molecular clouds?\\
Interacting\_Evidence & \nodata & Is SNR interacting with molecular clouds? OH, LB, or H2 as in Table~\ref{Tab:InteractingSNRs}\\
Bib\_Interacting & \nodata & ADS bibcode for Interacting reference\\
Distance & kpc & Distance to SNR in kpc\\
NearUnc\_Distance\, FarUnc\_Distance & kpc & Distance uncertainty toward and away from the Earth.\\
Distance\_Method & \nodata & Method used for distance determination: HI, Maser, CO, PSR, PM, NH, as in Table~\ref{Tab:SNRinfo}\\
Bib\_Distance & \nodata & Reference for distance measurement\\
TeV\_INDEX & \nodata & TeV index\\
Unc\_TeV\_INDEX & \nodata & TeV index uncertainty\\
Bib\_TeV & \nodata & Reference for TeV index\\
\enddata
\end{deluxetable}
\end{landscape}

\newpage


\begin{thebibliography}{}
\expandafter\ifx\csname natexlab\endcsname\relax\def\natexlab#1{#1}\fi

\bibitem[{{Abdo} {et~al.}(2010{\natexlab{a}}){Abdo}, {Ackermann}, {Ajello},
  {Allafort}, {Asano}, {Baldini}, {Ballet}, {Barbiellini}, {Baring},
  {Bastieri}, {Bechtol}, {Bellazzini}, {Berenji}, {Blandford}, {Bloom},
  {Bonamente}, {Borgland}, {Bregeon}, {Brez}, {Brigida}, {Bruel}, {Buson},
  {Caliandro}, {Cameron}, {Camilo}, {Caraveo}, {Carrigan}, {Casandjian},
  {Cecchi}, {{\c C}elik}, {Chekhtman}, {Cheung}, {Chiang}, {Ciprini}, {Claus},
  {Cohen-Tanugi}, {Conrad}, {den Hartog}, {Dermer}, {de Luca}, {de Palma},
  {Dormody}, {Silva}, {Drell}, {Dubois}, {Dumora}, {Farnier}, {Favuzzi},
  {Fegan}, {Ferrara}, {Focke}, {Frailis}, {Fukazawa}, {Funk}, {Fusco},
  {Gargano}, {Gehrels}, {Germani}, {Giglietto}, {Giordano}, {Glanzman},
  {Godfrey}, {Gotthelf}, {Grenier}, {Grondin}, {Grove}, {Guillemot}, {Guiriec},
  {Hanabata}, {Harding}, {Hays}, {Hobbs}, {Horan}, {Hughes}, {J{\'o}hannesson},
  {Johnson}, {Johnson}, {Johnson}, {Johnston}, {Kamae}, {Kanai}, {Kanbach},
  {Katagiri}, {Kataoka}, {Kawai}, {Keith}, {Kerr}, {Kn{\"o}dlseder}, {Kuss},
  {Lande}, {Latronico}, {Lemoine-Goumard}, {Llena Garde}, {Longo}, {Loparco},
  {Lott}, {Lovellette}, {Lubrano}, {Makeev}, {Manchester}, {Marelli},
  {Mazziotta}, {McEnery}, {Michelson}, {Mitthumsiri}, {Mizuno}, {Moiseev},
  {Monte}, {Monzani}, {Morselli}, {Moskalenko}, {Murgia}, {Nakamori}, {Nolan},
  {Norris}, {Nuss}, {Ohno}, {Ohsugi}, {Omodei}, {Orlando}, {Ormes}, {Paneque},
  {Panetta}, {Parent}, {Pelassa}, {Pepe}, {Pesce-Rollins}, {Piron}, {Porter},
  {Rain{\`o}}, {Rando}, {Razzano}, {Rea}, {Reimer}, {Reimer}, {Reposeur},
  {Rodriguez}, {Romani}, {Roth}, {Ryde}, {Sadrozinski}, {Sander}, {Saz
  Parkinson}, {Sgr{\`o}}, {Siskind}, {Smith}, {Smith}, {Spandre}, {Spinelli},
  {Starck}, {Strickman}, {Suson}, {Takahashi}, {Takahashi}, {Tanaka}, {Thayer},
  {Thayer}, {Thompson}, {Thorsett}, {Tibaldo}, {Torres}, {Tosti}, {Tramacere},
  {Uchiyama}, {Usher}, {Vasileiou}, {Venter}, {Vilchez}, {Vitale}, {Waite},
  {Wang}, {Weltevrede}, {Winer}, {Wood}, {Yang}, {Ylinen}, {Ziegler}, {Fermi
  LAT Collaboration}, \& {Pulsar Timing Consortium}}]{Abdo10-msh15--52}
{Abdo}, A.~A., {Ackermann}, M., {Ajello}, M., {et~al.} 2010{\natexlab{a}},
  \apj, 714, 927

\bibitem[{{Abdo} {et~al.}(2010{\natexlab{b}}){Abdo}, {Ackermann}, {Ajello},
  {Atwood}, {Axelsson}, {Baldini}, {Ballet}, {Barbiellini}, {Baring},
  {Bastieri}, {Bechtol}, {Bellazzini}, {Berenji}, {Blandford}, {Bloom},
  {Bonamente}, {Borgland}, {Bregeon}, {Brez}, {Brigida}, {Bruel}, {Burnett},
  {Caliandro}, {Cameron}, {Camilo}, {Caraveo}, {Casandjian}, {Cecchi}, {{\c
  C}elik}, {Chekhtman}, {Cheung}, {Chiang}, {Ciprini}, {Claus}, {Cognard},
  {Cohen-Tanugi}, {Cominsky}, {Conrad}, {Dermer}, {de Angelis}, {de Luca}, {de
  Palma}, {Digel}, {Silva}, {Drell}, {Dubois}, {Dumora}, {Espinoza}, {Farnier},
  {Favuzzi}, {Fegan}, {Ferrara}, {Focke}, {Frailis}, {Freire}, {Fukazawa},
  {Funk}, {Fusco}, {Gargano}, {Gasparrini}, {Gehrels}, {Germani}, {Giavitto},
  {Giebels}, {Giglietto}, {Giordano}, {Glanzman}, {Godfrey}, {Grenier},
  {Grondin}, {Grove}, {Guillemot}, {Guiriec}, {Hanabata}, {Harding},
  {Hayashida}, {Hays}, {Hughes}, {J{\'o}hannesson}, {Johnson}, {Johnson},
  {Johnson}, {Johnson}, {Johnston}, {Kamae}, {Katagiri}, {Kataoka}, {Kawai},
  {Kerr}, {Kn{\"o}dlseder}, {Kocian}, {Kramer}, {Kuehn}, {Kuss}, {Lande},
  {Latronico}, {Lee}, {Lemoine-Goumard}, {Longo}, {Loparco}, {Lott},
  {Lovellette}, {Lubrano}, {Lyne}, {Makeev}, {Marelli}, {Mazziotta}, {McEnery},
  {Meurer}, {Michelson}, {Mitthumsiri}, {Mizuno}, {Moiseev}, {Monte},
  {Monzani}, {Moretti}, {Morselli}, {Moskalenko}, {Murgia}, {Nakamori},
  {Nolan}, {Norris}, {Noutsos}, {Nuss}, {Ohsugi}, {Omodei}, {Orlando}, {Ormes},
  {Ozaki}, {Paneque}, {Panetta}, {Parent}, {Pelassa}, {Pepe}, {Pesce-Rollins},
  {Pierbattista}, {Piron}, {Porter}, {Rain{\`o}}, {Rando}, {Ray}, {Razzano},
  {Reimer}, {Reimer}, {Reposeur}, {Ritz}, {Rochester}, {Rodriguez}, {Romani},
  {Roth}, {Ryde}, {Sadrozinski}, {Sanchez}, {Sander}, {Saz Parkinson},
  {Scargle}, {Sgr{\`o}}, {Siskind}, {Smith}, {Smith}, {Spandre}, {Spinelli},
  {Stappers}, {Strickman}, {Suson}, {Tajima}, {Takahashi}, {Tanaka}, {Thayer},
  {Thayer}, {Theureau}, {Thompson}, {Thorsett}, {Tibaldo}, {Torres}, {Tosti},
  {Tramacere}, {Uchiyama}, {Usher}, {Van Etten}, {Vasileiou}, {Vilchez},
  {Vitale}, {Waite}, {Wallace}, {Wang}, {Watters}, {Weltevrede}, {Winer},
  {Wood}, {Ylinen}, \& {Ziegler}}]{2010ApJ...708.1254A}
---. 2010{\natexlab{b}}, \apj, 708, 1254

\bibitem[{{Abdo} {et~al.}(2010{\natexlab{c}}){Abdo}, {Ackermann}, {Ajello},
  {Allafort}, {Baldini}, {Ballet}, {Barbiellini}, {Bastieri}, {Bechtol},
  {Bellazzini}, {Berenji}, {Blandford}, {Bloom}, {Bonamente}, {Borgland},
  {Bouvier}, {Brandt}, {Bregeon}, {Brigida}, {Bruel}, {Buehler}, {Buson},
  {Caliandro}, {Cameron}, {Caraveo}, {Carrigan}, {Casandjian}, {Cecchi}, {{\c
  C}elik}, {Chekhtman}, {Chiang}, {Ciprini}, {Claus}, {Cohen-Tanugi}, {Conrad},
  {Dermer}, {de Palma}, {Silva}, {Drell}, {Dubois}, {Dumora}, {Farnier},
  {Favuzzi}, {Fegan}, {Fukazawa}, {Fukui}, {Funk}, {Fusco}, {Gargano},
  {Gehrels}, {Germani}, {Giglietto}, {Giordano}, {Glanzman}, {Godfrey},
  {Grenier}, {Grove}, {Guiriec}, {Hadasch}, {Hanabata}, {Harding}, {Hays},
  {Horan}, {Hughes}, {J{\'o}hannesson}, {Johnson}, {Johnson}, {Kamae},
  {Katagiri}, {Kataoka}, {Kn{\"o}dlseder}, {Kuss}, {Lande}, {Latronico}, {Lee},
  {Lemoine-Goumard}, {Llena Garde}, {Longo}, {Loparco}, {Lovellette},
  {Lubrano}, {Makeev}, {Mazziotta}, {Michelson}, {Mitthumsiri}, {Mizuno},
  {Moiseev}, {Monte}, {Monzani}, {Morselli}, {Moskalenko}, {Murgia},
  {Nakamori}, {Nolan}, {Norris}, {Nuss}, {Ohno}, {Ohsugi}, {Omodei}, {Orlando},
  {Ormes}, {Ozaki}, {Panetta}, {Parent}, {Pelassa}, {Pepe}, {Pesce-Rollins},
  {Piron}, {Porter}, {Rain{\`o}}, {Rando}, {Razzano}, {Reimer}, {Reimer},
  {Reposeur}, {Rodriguez}, {Roth}, {Sadrozinski}, {Sander}, {Saz Parkinson},
  {Sgr{\`o}}, {Siskind}, {Smith}, {Smith}, {Spandre}, {Spinelli}, {Strickman},
  {Suson}, {Tajima}, {Takahashi}, {Takahashi}, {Tanaka}, {Thayer}, {Thayer},
  {Thompson}, {Tibaldo}, {Tibolla}, {Torres}, {Tosti}, {Uchiyama}, {Uehara},
  {Usher}, {Vasileiou}, {Vilchez}, {Vitale}, {Waite}, {Wang}, {Winer}, {Wood},
  {Yamamoto}, {Yamazaki}, {Yang}, {Ylinen}, \& {Ziegler}}]{Abdo10-W28}
---. 2010{\natexlab{c}}, \apj, 718, 348

\bibitem[{{Abdo} {et~al.}(2010{\natexlab{d}}){Abdo}, {Ackermann}, {Ajello},
  {Allafort}, {Baldini}, {Ballet}, {Barbiellini}, {Baring}, {Bastieri},
  {Baughman}, {Bechtol}, {Bellazzini}, {Berenji}, {Blandford}, {Bloom},
  {Bonamente}, {Borgland}, {Bregeon}, {Brez}, {Brigida}, {Bruel}, {Buehler},
  {Burnett}, {Busetto}, {Caliandro}, {Cameron}, {Caraveo}, {Casandjian},
  {Cecchi}, {{\c C}elik}, {Charles}, {Chaty}, {Chekhtman}, {Cheung}, {Chiang},
  {Cillis}, {Ciprini}, {Claus}, {Cohen-Tanugi}, {Conrad}, {Corbel}, {de Palma},
  {Digel}, {Dormody}, {Silva}, {Drell}, {Dubois}, {Dumora}, {Edmonds},
  {Farnier}, {Favuzzi}, {Fegan}, {Ferrara}, {Focke}, {Fortin}, {Frailis},
  {Fukazawa}, {Funk}, {Fusco}, {Gargano}, {Gasparrini}, {Gehrels}, {Germani},
  {Giavitto}, {Giglietto}, {Giordano}, {Glanzman}, {Godfrey}, {Grenier},
  {Grondin}, {Grove}, {Guillemot}, {Guiriec}, {Hanabata}, {Hays}, {Harding},
  {Hayashida}, {Horan}, {Hughes}, {Jackson}, {Johnson}, {Johnson}, {Johnson},
  {Kamae}, {Katagiri}, {Kataoka}, {Kawai}, {Kerr}, {Kn{\"o}dlseder}, {Kuss},
  {Lande}, {Latronico}, {Lemoine-Goumard}, {Longo}, {Loparco}, {Lott},
  {Lovellette}, {Lubrano}, {Makeev}, {Mazziotta}, {Meurer}, {Michelson},
  {Mitthumsiri}, {Mizuno}, {Monte}, {Monzani}, {Morselli}, {Moskalenko},
  {Murgia}, {Nakamori}, {Nolan}, {Norris}, {Nuss}, {Ohsugi}, {Okumura},
  {Omodei}, {Orlando}, {Ormes}, {Paneque}, {Panetta}, {Pelassa}, {Pepe},
  {Pesce-Rollins}, {Piron}, {Pohl}, {Porter}, {Rain{\`o}}, {Rando}, {Reimer},
  {Reimer}, {Reposeur}, {Ritz}, {Rodriguez}, {Romani}, {Roth}, {Sadrozinski},
  {Sander}, {Saz Parkinson}, {Scargle}, {Sgr{\`o}}, {Siskind}, {Smith},
  {Smith}, {Spinelli}, {Strickman}, {Suson}, {Tajima}, {Takahashi}, {Tanaka},
  {Thayer}, {Thayer}, {Thompson}, {Thorsett}, {Tibaldo}, {Tibolla}, {Torres},
  {Tosti}, {Tramacere}, {Uchiyama}, {Usher}, {Van Etten}, {Vasileiou},
  {Venter}, {Vilchez}, {Vitale}, {Waite}, {Wang}, {Winer}, {Wood}, {Yamazaki},
  {Ylinen}, \& {Ziegler}}]{Abdo10-CasA}
---. 2010{\natexlab{d}}, \apjl, 710, L92

\bibitem[{{Abdo} {et~al.}(2010{\natexlab{e}}){Abdo}, {Ackermann}, {Ajello},
  {Baldini}, {Ballet}, {Barbiellini}, {Bastieri}, {Bechtol}, {Bellazzini},
  {Bloom}, {Bonamente}, {Borgland}, {Bouvier}, {Bregeon}, {Brez}, {Brigida},
  {Bruel}, {Buehler}, {Buson}, {Caliandro}, {Cameron}, {Caraveo}, {Casandjian},
  {Cecchi}, {{\c C}elik}, {Cheung}, {Chiang}, {Ciprini}, {Claus},
  {Cohen-Tanugi}, {Conrad}, {Dermer}, {de Palma}, {Digel}, {Silva}, {Drell},
  {Dumora}, {Favuzzi}, {Funk}, {Fusco}, {Gargano}, {Gehrels}, {Giglietto},
  {Giordano}, {Giroletti}, {Glanzman}, {Godfrey}, {Grenier}, {Grondin},
  {Grove}, {Guillemot}, {Guiriec}, {Hadasch}, {Hanabata}, {Harding},
  {Hayashida}, {Hays}, {Horan}, {Hughes}, {Jackson}, {J{\'o}hannesson},
  {Johnson}, {Johnson}, {Kamae}, {Katagiri}, {Kataoka}, {Katsuta},
  {Kn{\"o}dlseder}, {Kuss}, {Lande}, {Latronico}, {Lee}, {Lemoine-Goumard},
  {Longo}, {Loparco}, {Lovellette}, {Lubrano}, {Makeev}, {Mazziotta}, {Mizuno},
  {Monte}, {Monzani}, {Morselli}, {Moskalenko}, {Murgia}, {Naumann-Godo},
  {Nolan}, {Norris}, {Nuss}, {Ohsugi}, {Okumura}, {Omodei}, {Orlando}, {Ormes},
  {Pelassa}, {Pepe}, {Pesce-Rollins}, {Piron}, {Rain{\`o}}, {Rando}, {Razzano},
  {Reimer}, {Reimer}, {Reposeur}, {Ripken}, {Roth}, {Sadrozinski}, {Sander},
  {Saz Parkinson}, {Sgr{\`o}}, {Siskind}, {Smith}, {Smith}, {Spinelli},
  {Strickman}, {Suson}, {Tajima}, {Takahashi}, {Takahashi}, {Tanaka},
  {Tibaldo}, {Tibolla}, {Torres}, {Tosti}, {Tramacere}, {Uchiyama}, {Usher},
  {Vandenbroucke}, {Vasileiou}, {Vitale}, {Waite}, {Wang}, {Winer}, {Wood},
  {Ylinen}, \& {Ziegler}}]{Abdo10-W49B}
---. 2010{\natexlab{e}}, \apj, 722, 1303

\bibitem[{{Abdo} {et~al.}(2010{\natexlab{f}}){Abdo}, {Ackermann}, {Ajello},
  {Baldini}, {Ballet}, {Barbiellini}, {Baring}, {Bastieri}, {Baughman},
  {Bechtol}, {Bellazzini}, {Berenji}, {Blandford}, {Bloom}, {Bonamente},
  {Borgland}, {Bregeon}, {Brez}, {Brigida}, {Bruel}, {Burnett}, {Buson},
  {Caliandro}, {Cameron}, {Caraveo}, {Casandjian}, {Cecchi}, {{\c C}elik},
  {Chekhtman}, {Cheung}, {Chiang}, {Ciprini}, {Claus}, {Cognard},
  {Cohen-Tanugi}, {Cominsky}, {Conrad}, {Cutini}, {Dermer}, {de Angelis}, {de
  Palma}, {Digel}, {do Couto e Silva}, {Drell}, {Dubois}, {Dumora}, {Espinoza},
  {Farnier}, {Favuzzi}, {Fegan}, {Focke}, {Fortin}, {Frailis}, {Fukazawa},
  {Funk}, {Fusco}, {Gargano}, {Gasparrini}, {Gehrels}, {Germani}, {Giavitto},
  {Giebels}, {Giglietto}, {Giordano}, {Glanzman}, {Godfrey}, {Grenier},
  {Grondin}, {Grove}, {Guillemot}, {Guiriec}, {Hanabata}, {Harding},
  {Hayashida}, {Hays}, {Hughes}, {Jackson}, {J{\'o}hannesson}, {Johnson},
  {Johnson}, {Johnson}, {Kamae}, {Katagiri}, {Kataoka}, {Katsuta}, {Kawai},
  {Kerr}, {Kn{\"o}dlseder}, {Kocian}, {Kramer}, {Kuss}, {Lande}, {Latronico},
  {Lemoine-Goumard}, {Longo}, {Loparco}, {Lott}, {Lovellette}, {Lubrano},
  {Lyne}, {Madejski}, {Makeev}, {Mazziotta}, {McEnery}, {Meurer}, {Michelson},
  {Mitthumsiri}, {Mizuno}, {Monte}, {Monzani}, {Morselli}, {Moskalenko},
  {Murgia}, {Nakamori}, {Nolan}, {Norris}, {Noutsos}, {Nuss}, {Ohsugi},
  {Omodei}, {Orlando}, {Ormes}, {Paneque}, {Parent}, {Pelassa}, {Pepe},
  {Pesce-Rollins}, {Piron}, {Porter}, {Rain{\`o}}, {Rando}, {Razzano},
  {Reimer}, {Reimer}, {Reposeur}, {Rochester}, {Rodriguez}, {Romani}, {Roth},
  {Ryde}, {Sadrozinski}, {Sanchez}, {Sander}, {Parkinson}, {Scargle},
  {Sgr{\`o}}, {Siskind}, {Smith}, {Smith}, {Spandre}, {Spinelli}, {Stappers},
  {Stecker}, {Strickman}, {Suson}, {Tajima}, {Takahashi}, {Takahashi},
  {Tanaka}, {Thayer}, {Thayer}, {Theureau}, {Thompson}, {Tibaldo}, {Tibolla},
  {Torres}, {Tosti}, {Tramacere}, {Uchiyama}, {Usher}, {Vasileiou}, {Venter},
  {Vilchez}, {Vitale}, {Waite}, {Wang}, {Winer}, {Wood}, {Yamazaki}, {Ylinen},
  \& {Ziegler}}]{Abdo10-W44}
---. 2010{\natexlab{f}}, Science, 327, 1103

\bibitem[{{Abdo} {et~al.}(2010{\natexlab{g}}){Abdo}, {Ackermann}, {Ajello},
  {Baldini}, {Ballet}, {Barbiellini}, {Bastieri}, {Baughman}, {Bechtol},
  {Bellazzini}, {Berenji}, {Blandford}, {Bloom}, {Bonamente}, {Borgland},
  {Bregeon}, {Brez}, {Brigida}, {Bruel}, {Burnett}, {Buson}, {Caliandro},
  {Cameron}, {Caraveo}, {Casandjian}, {Cecchi}, {{\c C}elik}, {Chekhtman},
  {Cheung}, {Chiang}, {Cillis}, {Ciprini}, {Claus}, {Cohen-Tanugi}, {Cominsky},
  {Conrad}, {Cutini}, {Dermer}, {de Angelis}, {de Palma}, {Silva}, {Drell},
  {Drlica-Wagner}, {Dubois}, {Dumora}, {Farnier}, {Favuzzi}, {Fegan}, {Focke},
  {Fortin}, {Frailis}, {Fukazawa}, {Funk}, {Fusco}, {Gargano}, {Gasparrini},
  {Gehrels}, {Germani}, {Giavitto}, {Giebels}, {Giglietto}, {Giordano},
  {Glanzman}, {Godfrey}, {Grenier}, {Grondin}, {Grove}, {Guillemot}, {Guiriec},
  {Hanabata}, {Harding}, {Hayashida}, {Hughes}, {Jackson}, {J{\'o}hannesson},
  {Johnson}, {Johnson}, {Johnson}, {Kamae}, {Katagiri}, {Kataoka}, {Kawai},
  {Kerr}, {Kn{\"o}dlseder}, {Kocian}, {Kuss}, {Lande}, {Latronico}, {Lee},
  {Lemoine-Goumard}, {Longo}, {Loparco}, {Lott}, {Lovellette}, {Lubrano},
  {Madejski}, {Makeev}, {Mazziotta}, {Meurer}, {Michelson}, {Mitthumsiri},
  {Moiseev}, {Monte}, {Monzani}, {Morselli}, {Moskalenko}, {Murgia},
  {Nakamori}, {Nolan}, {Norris}, {Nuss}, {Ohsugi}, {Orlando}, {Ormes}, {Ozaki},
  {Paneque}, {Panetta}, {Parent}, {Pelassa}, {Pepe}, {Pesce-Rollins}, {Piron},
  {Porter}, {Rain{\`o}}, {Rando}, {Razzano}, {Reimer}, {Reimer}, {Reposeur},
  {Rochester}, {Rodriguez}, {Romani}, {Roth}, {Ryde}, {Sadrozinski}, {Sanchez},
  {Sander}, {Saz Parkinson}, {Scargle}, {Sgr{\`o}}, {Siskind}, {Smith},
  {Smith}, {Spandre}, {Spinelli}, {Strickman}, {Strong}, {Suson}, {Tajima},
  {Takahashi}, {Takahashi}, {Tanaka}, {Thayer}, {Thayer}, {Thompson},
  {Tibaldo}, {Torres}, {Tosti}, {Tramacere}, {Uchiyama}, {Usher}, {Van Etten},
  {Vasileiou}, {Venter}, {Vilchez}, {Vitale}, {Waite}, {Wang}, {Winer}, {Wood},
  {Ylinen}, \& {Ziegler}}]{Abdo10-IC443}
---. 2010{\natexlab{g}}, \apj, 712, 459

\bibitem[{{Abdo} {et~al.}(2011){Abdo}, {Ackermann}, {Ajello}, {Allafort},
  {Baldini}, {Ballet}, {Barbiellini}, {Baring}, {Bastieri}, {Bellazzini},
  {Berenji}, {Blandford}, {Bloom}, {Bonamente}, {Borgland}, {Bouvier},
  {Brandt}, {Bregeon}, {Brigida}, {Bruel}, {Buehler}, {Buson}, {Caliandro},
  {Cameron}, {Caraveo}, {Casandjian}, {Cecchi}, {Chaty}, {Chekhtman}, {Cheung},
  {Chiang}, {Cillis}, {Ciprini}, {Claus}, {Cohen-Tanugi}, {Conrad}, {Corbel},
  {Cutini}, {de Angelis}, {de Palma}, {Dermer}, {Digel}, {Silva}, {Drell},
  {Drlica-Wagner}, {Dubois}, {Dumora}, {Favuzzi}, {Ferrara}, {Fortin},
  {Frailis}, {Fukazawa}, {Fukui}, {Funk}, {Fusco}, {Gargano}, {Gasparrini},
  {Gehrels}, {Germani}, {Giglietto}, {Giordano}, {Giroletti}, {Glanzman},
  {Godfrey}, {Grenier}, {Grondin}, {Guiriec}, {Hadasch}, {Hanabata}, {Harding},
  {Hayashida}, {Hayashi}, {Hays}, {Horan}, {Jackson}, {J{\'o}hannesson},
  {Johnson}, {Kamae}, {Katagiri}, {Kataoka}, {Kerr}, {Kn{\"o}dlseder}, {Kuss},
  {Lande}, {Latronico}, {Lee}, {Lemoine-Goumard}, {Longo}, {Loparco},
  {Lovellette}, {Lubrano}, {Madejski}, {Makeev}, {Mazziotta}, {McEnery},
  {Michelson}, {Mignani}, {Mitthumsiri}, {Mizuno}, {Moiseev}, {Monte},
  {Monzani}, {Morselli}, {Moskalenko}, {Murgia}, {Naumann-Godo}, {Nolan},
  {Norris}, {Nuss}, {Ohsugi}, {Okumura}, {Orlando}, {Ormes}, {Paneque},
  {Parent}, {Pelassa}, {Pesce-Rollins}, {Pierbattista}, {Piron}, {Pohl},
  {Porter}, {Rain{\`o}}, {Rando}, {Razzano}, {Reimer}, {Reposeur}, {Ritz},
  {Romani}, {Roth}, {Sadrozinski}, {Saz Parkinson}, {Sgr{\`o}}, {Smith},
  {Smith}, {Spandre}, {Spinelli}, {Strickman}, {Tajima}, {Takahashi},
  {Takahashi}, {Tanaka}, {Thayer}, {Thayer}, {Thompson}, {Tibaldo}, {Tibolla},
  {Torres}, {Tosti}, {Tramacere}, {Troja}, {Uchiyama}, {Vandenbroucke},
  {Vasileiou}, {Vianello}, {Vilchez}, {Vitale}, {Waite}, {Wang}, {Winer},
  {Wood}, {Yamamoto}, {Yamazaki}, {Yang}, \& {Ziegler}}]{Abdo11-RXJ1713}
---. 2011, \apj, 734, 28

\bibitem[{{Abdo} {et~al.}(2013){Abdo}, {Ajello}, {Allafort}, {Baldini},
  {Ballet}, {Barbiellini}, {Baring}, {Bastieri}, {Belfiore}, {Bellazzini}, \&
  et~al.}]{Abdo13-2PC}
{Abdo}, A.~A., {Ajello}, M., {Allafort}, A., {et~al.} 2013, \apjs, 208, 17

\bibitem[{{Acciari} {et~al.}(2009){Acciari}, {Aliu}, {Arlen}, {Aune},
  {Bautista}, {Beilicke}, {Benbow}, {Bradbury}, {Buckley}, {Bugaev}, {Butt},
  {Byrum}, {Cannon}, {Celik}, {Cesarini}, {Chow}, {Ciupik}, {Cogan}, {Colin},
  {Cui}, {Daniel}, {Dickherber}, {Duke}, {Dwarkadas}, {Ergin}, {Fegan},
  {Finley}, {Finnegan}, {Fortin}, {Fortson}, {Furniss}, {Gall}, {Gibbs},
  {Gillanders}, {Godambe}, {Grube}, {Guenette}, {Gyuk}, {Hanna}, {Hays},
  {Holder}, {Horan}, {Hui}, {Humensky}, {Imran}, {Kaaret}, {Karlsson},
  {Kertzman}, {Kieda}, {Kildea}, {Konopelko}, {Krawczynski}, {Krennrich},
  {Lang}, {LeBohec}, {Maier}, {McCann}, {McCutcheon}, {Millis}, {Moriarty},
  {Ong}, {Otte}, {Pandel}, {Perkins}, {Pohl}, {Quinn}, {Ragan}, {Reyes},
  {Reynolds}, {Roache}, {Rose}, {Schroedter}, {Sembroski}, {Smith}, {Steele},
  {Swordy}, {Theiling}, {Toner}, {Valcarcel}, {Varlotta}, {Vassiliev},
  {Vincent}, {Wagner}, {Wakely}, {Ward}, {Weekes}, {Weinstein}, {Weisgarber},
  {Williams}, {Wissel}, {Wood}, \& {Zitzer}}]{2009ApJ...698L.133A}
{Acciari}, V.~A., {Aliu}, E., {Arlen}, T., {et~al.} 2009, \apjl, 698, L133

\bibitem[{{Acciari} {et~al.}(2010){Acciari}, {Aliu}, {Arlen}, {Aune},
  {Bautista}, {Beilicke}, {Benbow}, {Boltuch}, {Bradbury}, {Buckley}, {Bugaev},
  {Butt}, {Byrum}, {Cannon}, {Cesarini}, {Chow}, {Ciupik}, {Cogan}, {Cui},
  {Dickherber}, {Duke}, {Ergin}, {Fegan}, {Finley}, {Finnegan}, {Fortin},
  {Fortson}, {Furniss}, {Galante}, {Gall}, {Gillanders}, {Grube}, {Guenette},
  {Gyuk}, {Hanna}, {Holder}, {Huang}, {Hui}, {Humensky}, {Kaaret}, {Karlsson},
  {Kertzman}, {Kieda}, {Konopelko}, {Krawczynski}, {Krennrich}, {Lang},
  {LeBohec}, {Maier}, {McArthur}, {McCann}, {McCutcheon}, {Millis}, {Moriarty},
  {Ong}, {Pandel}, {Perkins}, {Pohl}, {Quinn}, {Ragan}, {Reynolds}, {Roache},
  {Rose}, {Schroedter}, {Sembroski}, {Smith}, {Smith}, {Steele}, {Swordy},
  {Theiling}, {Thibadeau}, {Varlotta}, {Vassiliev}, {Vincent}, {Wagner},
  {Wakely}, {Ward}, {Weekes}, {Weinstein}, {Weisgarber}, {Wissel}, {Wood}, \&
  {VERITAS Collaboration}}]{2010ApJ...714..163A}
---. 2010, \apj, 714, 163

\bibitem[{{Acero} {et~al.}(2009){Acero}, {Ballet}, {Decourchelle},
  {Lemoine-Goumard}, {Ortega}, {Giacani}, {Dubner}, \&
  {Cassam-Chena{\"i}}}]{2009AA...505..157A}
{Acero}, F., {Ballet}, J., {Decourchelle}, A., {et~al.} 2009, \aap, 505, 157

\bibitem[{{Acero} {et~al.}(2013){Acero}, {Ackermann}, {Ajello}, {Allafort},
  {Baldini}, {Ballet}, {Barbiellini}, {Bastieri}, {Bechtol}, {Bellazzini},
  {Blandford}, {Bloom}, {Bonamente}, {Bottacini}, {Brandt}, {Bregeon},
  {Brigida}, {Bruel}, {Buehler}, {Buson}, {Caliandro}, {Cameron}, {Caraveo},
  {Cecchi}, {Charles}, {Chaves}, {Chekhtman}, {Chiang}, {Chiaro}, {Ciprini},
  {Claus}, {Cohen-Tanugi}, {Conrad}, {Cutini}, {Dalton}, {D'Ammando}, {de
  Palma}, {Dermer}, {Di Venere}, {Silva}, {Drell}, {Drlica-Wagner}, {Falletti},
  {Favuzzi}, {Fegan}, {Ferrara}, {Focke}, {Franckowiak}, {Fukazawa}, {Funk},
  {Fusco}, {Gargano}, {Gasparrini}, {Giglietto}, {Giordano}, {Giroletti},
  {Glanzman}, {Godfrey}, {Gr{\'e}goire}, {Grenier}, {Grondin}, {Grove},
  {Guiriec}, {Hadasch}, {Hanabata}, {Harding}, {Hayashida}, {Hayashi}, {Hays},
  {Hewitt}, {Hill}, {Horan}, {Hou}, {Hughes}, {Inoue}, {Jackson}, {Jogler},
  {J{\'o}hannesson}, {Johnson}, {Kamae}, {Kawano}, {Kerr}, {Kn{\"o}dlseder},
  {Kuss}, {Lande}, {Larsson}, {Latronico}, {Lemoine-Goumard}, {Longo},
  {Loparco}, {Lovellette}, {Lubrano}, {Marelli}, {Massaro}, {Mayer},
  {Mazziotta}, {McEnery}, {Mehault}, {Michelson}, {Mitthumsiri}, {Mizuno},
  {Monte}, {Monzani}, {Morselli}, {Moskalenko}, {Murgia}, {Nakamori}, {Nemmen},
  {Nuss}, {Ohsugi}, {Okumura}, {Orienti}, {Orlando}, {Ormes}, {Paneque},
  {Panetta}, {Perkins}, {Pesce-Rollins}, {Piron}, {Pivato}, {Porter},
  {Rain{\`o}}, {Rando}, {Razzano}, {Reimer}, {Reimer}, {Reposeur}, {Ritz},
  {Roth}, {Rousseau}, {Saz Parkinson}, {Schulz}, {Sgr{\`o}}, {Siskind},
  {Smith}, {Spandre}, {Spinelli}, {Suson}, {Takahashi}, {Takeuchi}, {Thayer},
  {Thayer}, {Thompson}, {Tibaldo}, {Tibolla}, {Tinivella}, {Torres}, {Tosti},
  {Troja}, {Uchiyama}, {Vandenbroucke}, {Vasileiou}, {Vianello}, {Vitale},
  {Werner}, {Winer}, {Wood}, \& {Yang}}]{Acero13-TevPwnCat}
{Acero}, F., {Ackermann}, M., {Ajello}, M., {et~al.} 2013, \apj, 773, 77

\bibitem[{{Acero} {et~al.}(2015){Acero}, {Ackermann}, {Ajello}, {Albert},
  {Atwood}, {Axelsson}, {Baldini}, {Ballet}, {Barbiellini}, {Bastieri},
  {Belfiore}, {Bellazzini}, {Bissaldi}, {Blandford}, {Bloom}, {Bogart},
  {Bonino}, {Bottacini}, {Bregeon}, {Britto}, {Bruel}, {Buehler}, {Burnett},
  {Buson}, {Caliandro}, {Cameron}, {Caputo}, {Caragiulo}, {Caraveo},
  {Casandjian}, {Cavazzuti}, {Charles}, {Chaves}, {Chekhtman}, {Cheung},
  {Chiang}, {Chiaro}, {Ciprini}, {Claus}, {Cohen-Tanugi}, {Cominsky}, {Conrad},
  {Cutini}, {D'Ammando}, {de Angelis}, {DeKlotz}, {de Palma}, {Desiante},
  {Digel}, {Di Venere}, {Drell}, {Dubois}, {Dumora}, {Favuzzi}, {Fegan},
  {Ferrara}, {Finke}, {Franckowiak}, {Fukazawa}, {Funk}, {Fusco}, {Gargano},
  {Gasparrini}, {Giebels}, {Giglietto}, {Giommi}, {Giordano}, {Giroletti},
  {Glanzman}, {Godfrey}, {Grenier}, {Grondin}, {Grove}, {Guillemot}, {Guiriec},
  {Hadasch}, {Harding}, {Hays}, {Hewitt}, {Hill}, {Horan}, {Iafrate}, {Jogler},
  {J{\'o}hannesson}, {Johnson}, {Johnson}, {Johnson}, {Johnson}, {Kamae},
  {Kataoka}, {Katsuta}, {Kuss}, {La Mura}, {Landriu}, {Larsson}, {Latronico},
  {Lemoine-Goumard}, {Li}, {Li}, {Longo}, {Loparco}, {Lott}, {Lovellette},
  {Lubrano}, {Madejski}, {Massaro}, {Mayer}, {Mazziotta}, {McEnery},
  {Michelson}, {Mirabal}, {Mizuno}, {Moiseev}, {Mongelli}, {Monzani},
  {Morselli}, {Moskalenko}, {Murgia}, {Nuss}, {Ohno}, {Ohsugi}, {Omodei},
  {Orienti}, {Orlando}, {Ormes}, {Paneque}, {Panetta}, {Perkins},
  {Pesce-Rollins}, {Piron}, {Pivato}, {Porter}, {Racusin}, {Rando}, {Razzano},
  {Razzaque}, {Reimer}, {Reimer}, {Reposeur}, {Rochester}, {Romani},
  {Salvetti}, {S{\'a}nchez-Conde}, {Saz Parkinson}, {Schulz}, {Siskind},
  {Smith}, {Spada}, {Spandre}, {Spinelli}, {Stephens}, {Strong}, {Suson},
  {Takahashi}, {Takahashi}, {Tanaka}, {Thayer}, {Thayer}, {Thompson},
  {Tibaldo}, {Tibolla}, {Torres}, {Torresi}, {Tosti}, {Troja}, {Van Klaveren},
  {Vianello}, {Winer}, {Wood}, {Wood}, {Zimmer}, \& {Fermi-LAT
  Collaboration}}]{Acero15-3FGL}
---. 2015, \apjs, 218, 23

\bibitem[{{Ackermann} {et~al.}(2012{\natexlab{a}}){Ackermann}, {Ajello},
  {Allafort}, {Antolini}, {Baldini}, {Ballet}, {Barbiellini}, {Bastieri},
  {Bellazzini}, {Berenji}, {Blandford}, {Bloom}, {Bonamente}, {Borgland},
  {Bouvier}, {Brandt}, {Bregeon}, {Brigida}, {Bruel}, {Buehler}, {Burnett},
  {Buson}, {Caliandro}, {Cameron}, {Caraveo}, {Casandjian}, {Cavazzuti},
  {Cecchi}, {{\c C}elik}, {Charles}, {Chekhtman}, {Chen}, {Cheung}, {Chiang},
  {Ciprini}, {Claus}, {Cohen-Tanugi}, {Conrad}, {Cutini}, {de Angelis},
  {DeCesar}, {De Luca}, {de Palma}, {Dermer}, {Silva}, {Drell},
  {Drlica-Wagner}, {Dubois}, {Enoto}, {Favuzzi}, {Fegan}, {Ferrara}, {Focke},
  {Fortin}, {Fukazawa}, {Funk}, {Fusco}, {Gargano}, {Gasparrini}, {Gehrels},
  {Germani}, {Giglietto}, {Giordano}, {Giroletti}, {Glanzman}, {Godfrey},
  {Grenier}, {Grondin}, {Grove}, {Guillemot}, {Guiriec}, {Gustafsson},
  {Hadasch}, {Hanabata}, {Harding}, {Hayashida}, {Hays}, {Healey}, {Hill},
  {Horan}, {Hou}, {J{\'o}hannesson}, {Johnson}, {Johnson}, {Kamae}, {Katagiri},
  {Kataoka}, {Kerr}, {Kn{\"o}dlseder}, {Kuss}, {Lande}, {Latronico}, {Lee},
  {Lemoine-Goumard}, {Longo}, {Loparco}, {Lott}, {Lovellette}, {Lubrano},
  {Madejski}, {Mazziotta}, {McEnery}, {Mehault}, {Michelson}, {Mignani},
  {Mitthumsiri}, {Mizuno}, {Monte}, {Monzani}, {Morselli}, {Moskalenko},
  {Murgia}, {Nakamori}, {Naumann-Godo}, {Nolan}, {Norris}, {Nuss}, {Ohsugi},
  {Okumura}, {Omodei}, {Orlando}, {Ormes}, {Ozaki}, {Paneque}, {Panetta},
  {Parent}, {Pelassa}, {Pesce-Rollins}, {Pierbattista}, {Piron}, {Pivato},
  {Porter}, {Rain{\`o}}, {Rando}, {Ray}, {Razzano}, {Reimer}, {Reimer},
  {Reposeur}, {Romani}, {Sadrozinski}, {Salvetti}, {Saz Parkinson}, {Schalk},
  {Sgr{\`o}}, {Shaw}, {Siskind}, {Smith}, {Spandre}, {Spinelli}, {Suson},
  {Takahashi}, {Tanaka}, {Thayer}, {Thayer}, {Thompson}, {Tibaldo}, {Tibolla},
  {Torres}, {Tosti}, {Tramacere}, {Troja}, {Usher}, {Vandenbroucke},
  {Vasileiou}, {Vianello}, {Vilchez}, {Vitale}, {Waite}, {Wallace}, {Wang},
  {Winer}, {Wolff}, {Wood}, {Wood}, {Yang}, \&
  {Zimmer}}]{Ackermann12_1FGLUnassoc}
{Ackermann}, M., {Ajello}, M., {Allafort}, A., {et~al.} 2012{\natexlab{a}},
  \apj, 753, 83

\bibitem[{{Ackermann} {et~al.}(2012{\natexlab{b}}){Ackermann}, {Ajello},
  {Atwood}, {Baldini}, {Ballet}, {Barbiellini}, {Bastieri}, {Bechtol},
  {Bellazzini}, {Berenji}, {Blandford}, {Bloom}, {Bonamente}, {Borgland},
  {Brandt}, {Bregeon}, {Brigida}, {Bruel}, {Buehler}, {Buson}, {Caliandro},
  {Cameron}, {Caraveo}, {Cavazzuti}, {Cecchi}, {Charles}, {Chekhtman},
  {Chiang}, {Ciprini}, {Claus}, {Cohen-Tanugi}, {Conrad}, {Cutini}, {de
  Angelis}, {de Palma}, {Dermer}, {Digel}, {Silva}, {Drell}, {Drlica-Wagner},
  {Falletti}, {Favuzzi}, {Fegan}, {Ferrara}, {Focke}, {Fortin}, {Fukazawa},
  {Funk}, {Fusco}, {Gaggero}, {Gargano}, {Germani}, {Giglietto}, {Giordano},
  {Giroletti}, {Glanzman}, {Godfrey}, {Grove}, {Guiriec}, {Gustafsson},
  {Hadasch}, {Hanabata}, {Harding}, {Hayashida}, {Hays}, {Horan}, {Hou},
  {Hughes}, {J{\'o}hannesson}, {Johnson}, {Johnson}, {Kamae}, {Katagiri},
  {Kataoka}, {Kn{\"o}dlseder}, {Kuss}, {Lande}, {Latronico}, {Lee},
  {Lemoine-Goumard}, {Longo}, {Loparco}, {Lott}, {Lovellette}, {Lubrano},
  {Mazziotta}, {McEnery}, {Michelson}, {Mitthumsiri}, {Mizuno}, {Monte},
  {Monzani}, {Morselli}, {Moskalenko}, {Murgia}, {Naumann-Godo}, {Norris},
  {Nuss}, {Ohsugi}, {Okumura}, {Omodei}, {Orlando}, {Ormes}, {Paneque},
  {Panetta}, {Parent}, {Pesce-Rollins}, {Pierbattista}, {Piron}, {Pivato},
  {Porter}, {Rain{\`o}}, {Rando}, {Razzano}, {Razzaque}, {Reimer}, {Reimer},
  {Sadrozinski}, {Sgr{\`o}}, {Siskind}, {Spandre}, {Spinelli}, {Strong},
  {Suson}, {Takahashi}, {Tanaka}, {Thayer}, {Thayer}, {Thompson}, {Tibaldo},
  {Tinivella}, {Torres}, {Tosti}, {Troja}, {Usher}, {Vandenbroucke},
  {Vasileiou}, {Vianello}, {Vitale}, {Waite}, {Wang}, {Winer}, {Wood}, {Wood},
  {Yang}, {Ziegler}, \& {Zimmer}}]{Ackermann12-aIEMs}
{Ackermann}, M., {Ajello}, M., {Atwood}, W.~B., {et~al.} 2012{\natexlab{b}},
  \apj, 750, 3

\bibitem[{{Ackermann} {et~al.}(2012{\natexlab{c}}){Ackermann}, {Ajello},
  {Allafort}, {Baldini}, {Ballet}, {Bastieri}, {Bechtol}, {Bellazzini},
  {Berenji}, {Bloom}, {Bonamente}, {Borgland}, {Bouvier}, {Bregeon}, {Brigida},
  {Bruel}, {Buehler}, {Buson}, {Caliandro}, {Cameron}, {Caraveo}, {Casandjian},
  {Cecchi}, {Charles}, {Chekhtman}, {Cheung}, {Chiang}, {Cillis}, {Ciprini},
  {Claus}, {Cohen-Tanugi}, {Conrad}, {Cutini}, {de Palma}, {Dermer}, {Digel},
  {Silva}, {Drell}, {Drlica-Wagner}, {Favuzzi}, {Fegan}, {Fortin}, {Fukazawa},
  {Funk}, {Fusco}, {Gargano}, {Gasparrini}, {Germani}, {Giglietto}, {Giordano},
  {Glanzman}, {Godfrey}, {Grenier}, {Guiriec}, {Gustafsson}, {Hadasch},
  {Hayashida}, {Hays}, {Hughes}, {J{\'o}hannesson}, {Johnson}, {Kamae},
  {Katagiri}, {Kataoka}, {Kn{\"o}dlseder}, {Kuss}, {Lande}, {Longo}, {Loparco},
  {Lott}, {Lovellette}, {Lubrano}, {Madejski}, {Martin}, {Mazziotta},
  {McEnery}, {Michelson}, {Mizuno}, {Monte}, {Monzani}, {Morselli},
  {Moskalenko}, {Murgia}, {Nishino}, {Norris}, {Nuss}, {Ohno}, {Ohsugi},
  {Okumura}, {Omodei}, {Orlando}, {Ozaki}, {Parent}, {Persic}, {Pesce-Rollins},
  {Petrosian}, {Pierbattista}, {Piron}, {Pivato}, {Porter}, {Rain{\`o}},
  {Rando}, {Razzano}, {Reimer}, {Reimer}, {Ritz}, {Roth}, {Sbarra}, {Sgr{\`o}},
  {Siskind}, {Spandre}, {Spinelli}, {Stawarz}, {Strong}, {Takahashi}, {Tanaka},
  {Thayer}, {Tibaldo}, {Tinivella}, {Torres}, {Tosti}, {Troja}, {Uchiyama},
  {Vandenbroucke}, {Vianello}, {Vitale}, {Waite}, {Wood}, \&
  {Yang}}]{Ackermann12-SFGs}
{Ackermann}, M., {Ajello}, M., {Allafort}, A., {et~al.} 2012{\natexlab{c}},
  \apj, 755, 164

\bibitem[{{Ackermann} {et~al.}(2012{\natexlab{d}}){Ackermann}, {Ajello},
  {Ballet}, {Barbiellini}, {Bastieri}, {Belfiore}, {Bellazzini}, {Berenji},
  {Blandford}, {Bloom}, {Bonamente}, {Borgland}, {Bregeon}, {Brigida}, {Bruel},
  {Buehler}, {Buson}, {Caliandro}, {Cameron}, {Caraveo}, {Cavazzuti}, {Cecchi},
  {{\c C}elik}, {Charles}, {Chaty}, {Chekhtman}, {Cheung}, {Chiang}, {Ciprini},
  {}, {Claus}, {Cohen-Tanugi}, {Corbel}, {Corbet}, {Cutini}, {de Luca}, {den
  Hartog}, {de Palma}, {Dermer}, {Digel}, {do Couto e Silva}, {Donato},
  {Drell}, {Drlica-Wagner}, {Dubois}, {Dubus}, {Favuzzi}, {Fegan}, {Ferrara},
  {Focke}, {Fortin}, {Fukazawa}, {Funk}, {Fusco}, {Gargano}, {Gasparrini},
  {Gehrels}, {Germani}, {Giglietto}, {Giordano}, {Giroletti}, {Glanzman},
  {Godfrey}, {Grenier}, {Grove}, {Guiriec}, {Hadasch}, {Hanabata}, {Harding},
  {Hayashida}, {Hays}, {Hill}, {Hughes}, {J{\'o}hannesson}, {Johnson},
  {Johnson}, {Kamae}, {Katagiri}, {Kataoka}, {Kerr}, {Kn{\"o}dlseder}, {Kuss},
  {Lande}, {Longo}, {Loparco}, {Lovellette}, {Lubrano}, {Mazziotta}, {McEnery},
  {Michelson}, {Mitthumsiri}, {Mizuno}, {Monte}, {Monzani}, {Morselli},
  {Moskalenko}, {Murgia}, {Nakamori}, {Naumann-Godo}, {Norris}, {Nuss}, {Ohno},
  {Ohsugi}, {Okumura}, {Omodei}, {Orlando}, {Ozaki}, {Paneque}, {Parent},
  {Pesce-Rollins}, {Pierbattista}, {Piron}, {Pivato}, {Porter}, {Rain{\`o}},
  {Rando}, {Razzano}, {Reimer}, {Reimer}, {Ritz}, {Romani}, {Roth}, {Saz
  Parkinson}, {Sgr{\`o}}, {Siskind}, {Spandre}, {Spinelli}, {Suson},
  {Takahashi}, {Tanaka}, {Thayer}, {Thayer}, {Thompson}, {Tibaldo},
  {Tinivella}, {Torres}, {Tosti}, {Troja}, {Uchiyama}, {Usher},
  {Vandenbroucke}, {Vianello}, {Vitale}, {Waite}, {Winer}, {Wood}, {Wood},
  {Yang}, {Zimmer}, {Coe}, {Di Mille}, {Edwards}, {Filipovi{\'c}}, {Payne},
  {Stevens}, \& {Torres}}]{Abdo12-1FGLJ1018}
{Ackermann}, M., {Ajello}, M., {Ballet}, J., {et~al.} 2012{\natexlab{d}},
  Science, 335, 189

\bibitem[{{Ackermann} {et~al.}(2012{\natexlab{e}}){Ackermann}, {Ajello},
  {Allafort}, {Baldini}, {Ballet}, {Barbiellini}, {Bastieri}, {Belfiore},
  {Bellazzini}, {Berenji}, {Blandford}, {Bloom}, {Bonamente}, {Borgland},
  {Bottacini}, {Bregeon}, {Brigida}, {Bruel}, {Buehler}, {Buson}, {Caliandro},
  {Cameron}, {Caraveo}, {Casandjian}, {Cecchi}, {Chekhtman}, {Ciprini},
  {Claus}, {Cohen-Tanugi}, {de Angelis}, {de Palma}, {Dermer}, {Silva},
  {Drell}, {Dumora}, {Favuzzi}, {Fegan}, {Focke}, {Fortin}, {Fukazawa},
  {Fusco}, {Gargano}, {Germani}, {Giglietto}, {Giordano}, {Giroletti},
  {Glanzman}, {Godfrey}, {Grenier}, {Guillemot}, {Guiriec}, {Hadasch},
  {Hanabata}, {Harding}, {Hayashida}, {Hayashi}, {Hays}, {J{\'o}hannesson},
  {Johnson}, {Kamae}, {Katagiri}, {Kataoka}, {Kerr}, {Kn{\"o}dlseder}, {Kuss},
  {Lande}, {Latronico}, {Lee}, {Longo}, {Loparco}, {Lott}, {Lovellette},
  {Lubrano}, {Martin}, {Mazziotta}, {McEnery}, {Mehault}, {Michelson},
  {Mitthumsiri}, {Mizuno}, {Monte}, {Monzani}, {Morselli}, {Moskalenko},
  {Murgia}, {Naumann-Godo}, {Nolan}, {Norris}, {Nuss}, {Ohsugi}, {Okumura},
  {Omodei}, {Orlando}, {Ormes}, {Ozaki}, {Paneque}, {Parent}, {Pesce-Rollins},
  {Pierbattista}, {Piron}, {Porter}, {Rain{\`o}}, {Rando}, {Razzano}, {Reimer},
  {Reposeur}, {Ritz}, {Saz Parkinson}, {Sgr{\`o}}, {Siskind}, {Smith},
  {Spinelli}, {Strong}, {Takahashi}, {Tanaka}, {Thayer}, {Thayer}, {Thompson},
  {Tibaldo}, {Torres}, {Tosti}, {Tramacere}, {Troja}, {Uchiyama},
  {Vandenbroucke}, {Vasileiou}, {Vianello}, {Vitale}, {Waite}, {Wang}, {Winer},
  {Wood}, {Yang}, {Zimmer}, \& {Bontemps}}]{Ackermann12-CygnusCR}
{Ackermann}, M., {Ajello}, M., {Allafort}, A., {et~al.} 2012{\natexlab{e}},
  \aap, 538, A71

\bibitem[{{Ackermann} {et~al.}(2012{\natexlab{f}}){Ackermann}, {Ajello},
  {Albert}, {Allafort}, {Atwood}, {Axelsson}, {Baldini}, {Ballet},
  {Barbiellini}, {Bastieri}, {Bechtol}, {Bellazzini}, {Bissaldi}, {Blandford},
  {Bloom}, {Bogart}, {Bonamente}, {Borgland}, {Bottacini}, {Bouvier}, {Brandt},
  {Bregeon}, {Brigida}, {Bruel}, {Buehler}, {Burnett}, {Buson}, {Caliandro},
  {Cameron}, {Caraveo}, {Casandjian}, {Cavazzuti}, {Cecchi}, {{\c C}elik},
  {Charles}, {Chaves}, {Chekhtman}, {Cheung}, {Chiang}, {Ciprini}, {Claus},
  {Cohen-Tanugi}, {Conrad}, {Corbet}, {Cutini}, {D'Ammando}, {Davis}, {de
  Angelis}, {DeKlotz}, {de Palma}, {Dermer}, {Digel}, {Silva}, {Drell},
  {Drlica-Wagner}, {Dubois}, {Favuzzi}, {Fegan}, {Ferrara}, {Focke}, {Fortin},
  {Fukazawa}, {Funk}, {Fusco}, {Gargano}, {Gasparrini}, {Gehrels}, {Giebels},
  {Giglietto}, {Giordano}, {Giroletti}, {Glanzman}, {Godfrey}, {Grenier},
  {Grove}, {Guiriec}, {Hadasch}, {Hayashida}, {Hays}, {Horan}, {Hou}, {Hughes},
  {Jackson}, {Jogler}, {J{\'o}hannesson}, {Johnson}, {Johnson}, {Johnson},
  {Kamae}, {Katagiri}, {Kataoka}, {Kerr}, {Kn{\"o}dlseder}, {Kuss}, {Lande},
  {Larsson}, {Latronico}, {Lavalley}, {Lemoine-Goumard}, {Longo}, {Loparco},
  {Lott}, {Lovellette}, {Lubrano}, {Mazziotta}, {McConville}, {McEnery},
  {Mehault}, {Michelson}, {Mitthumsiri}, {Mizuno}, {Moiseev}, {Monte},
  {Monzani}, {Morselli}, {Moskalenko}, {Murgia}, {Naumann-Godo}, {Nemmen},
  {Nishino}, {Norris}, {Nuss}, {Ohno}, {Ohsugi}, {Okumura}, {Omodei},
  {Orienti}, {Orlando}, {Ormes}, {Paneque}, {Panetta}, {Perkins},
  {Pesce-Rollins}, {Pierbattista}, {Piron}, {Pivato}, {Porter}, {Racusin},
  {Rain{\`o}}, {Rando}, {Razzano}, {Razzaque}, {Reimer}, {Reimer}, {Reposeur},
  {Reyes}, {Ritz}, {Rochester}, {Romoli}, {Roth}, {Sadrozinski}, {Sanchez},
  {Saz Parkinson}, {Sbarra}, {Scargle}, {Sgr{\`o}}, {Siegal-Gaskins},
  {Siskind}, {Spandre}, {Spinelli}, {Stephens}, {Suson}, {Tajima}, {Takahashi},
  {Tanaka}, {Thayer}, {Thayer}, {Thompson}, {Tibaldo}, {Tinivella}, {Tosti},
  {Troja}, {Usher}, {Vandenbroucke}, {Van Klaveren}, {Vasileiou}, {Vianello},
  {Vitale}, {Waite}, {Wallace}, {Winer}, {Wood}, {Wood}, {Wood}, {Yang}, \&
  {Zimmer}}]{LAT-instrument-paper}
{Ackermann}, M., {Ajello}, M., {Albert}, A., {et~al.} 2012{\natexlab{f}},
  \apjs, 203, 4

\bibitem[{{Ackermann} {et~al.}(2013{\natexlab{a}}){Ackermann}, {Ajello},
  {Allafort}, {Baldini}, {Ballet}, {Barbiellini}, {Baring}, {Bastieri},
  {Bechtol}, {Bellazzini}, {Blandford}, {Bloom}, {Bonamente}, {Borgland},
  {Bottacini}, {Brandt}, {Bregeon}, {Brigida}, {Bruel}, {Buehler}, {Busetto},
  {Buson}, {Caliandro}, {Cameron}, {Caraveo}, {Casandjian}, {Cecchi}, {{\c
  C}elik}, {Charles}, {Chaty}, {Chaves}, {Chekhtman}, {Cheung}, {Chiang},
  {Chiaro}, {Cillis}, {Ciprini}, {Claus}, {Cohen-Tanugi}, {Cominsky}, {Conrad},
  {Corbel}, {Cutini}, {D'Ammando}, {de Angelis}, {de Palma}, {Dermer}, {do
  Couto e Silva}, {Drell}, {Drlica-Wagner}, {Falletti}, {Favuzzi}, {Ferrara},
  {Franckowiak}, {Fukazawa}, {Funk}, {Fusco}, {Gargano}, {Germani},
  {Giglietto}, {Giommi}, {Giordano}, {Giroletti}, {Glanzman}, {Godfrey},
  {Grenier}, {Grondin}, {Grove}, {Guiriec}, {Hadasch}, {Hanabata}, {Harding},
  {Hayashida}, {Hayashi}, {Hays}, {Hewitt}, {Hill}, {Hughes}, {Jackson},
  {Jogler}, {J{\'o}hannesson}, {Johnson}, {Kamae}, {Kataoka}, {Katsuta},
  {Kn{\"o}dlseder}, {Kuss}, {Lande}, {Larsson}, {Latronico}, {Lemoine-Goumard},
  {Longo}, {Loparco}, {Lovellette}, {Lubrano}, {Madejski}, {Massaro}, {Mayer},
  {Mazziotta}, {McEnery}, {Mehault}, {Michelson}, {Mignani}, {Mitthumsiri},
  {Mizuno}, {Moiseev}, {Monzani}, {Morselli}, {Moskalenko}, {Murgia},
  {Nakamori}, {Nemmen}, {Nuss}, {Ohno}, {Ohsugi}, {Omodei}, {Orienti},
  {Orlando}, {Ormes}, {Paneque}, {Perkins}, {Pesce-Rollins}, {Piron}, {Pivato},
  {Rain{\`o}}, {Rando}, {Razzano}, {Razzaque}, {Reimer}, {Reimer}, {Ritz},
  {Romoli}, {S{\'a}nchez-Conde}, {Schulz}, {Sgr{\`o}}, {Simeon}, {Siskind},
  {Smith}, {Spandre}, {Spinelli}, {Stecker}, {Strong}, {Suson}, {Tajima},
  {Takahashi}, {Takahashi}, {Tanaka}, {Thayer}, {Thayer}, {Thompson},
  {Thorsett}, {Tibaldo}, {Tibolla}, {Tinivella}, {Troja}, {Uchiyama}, {Usher},
  {Vandenbroucke}, {Vasileiou}, {Vianello}, {Vitale}, {Waite}, {Werner},
  {Winer}, {Wood}, {Wood}, {Yamazaki}, {Yang}, \&
  {Zimmer}}]{Ackermann13-pionBump}
{Ackermann}, M., {Ajello}, M., {Allafort}, A., {et~al.} 2013{\natexlab{a}},
  Science, 339, 807

\bibitem[{{Ackermann} {et~al.}(2013{\natexlab{b}}){Ackermann}, {Ajello},
  {Albert}, {Allafort}, {Antolini}, {Baldini}, {Ballet}, {Barbiellini},
  {Bastieri}, {Bechtol}, {Bellazzini}, {Blandford}, {Bloom}, {Bonamente},
  {Bottacini}, {Bouvier}, {Brandt}, {Bregeon}, {Brigida}, {Bruel}, {Buehler},
  {Buson}, {Caliandro}, {Cameron}, {Caraveo}, {Cavazzuti}, {Cecchi}, {Charles},
  {Chekhtman}, {Cheung}, {Chiang}, {Chiaro}, {Ciprini}, {Claus},
  {Cohen-Tanugi}, {Conrad}, {Cutini}, {Dalton}, {D'Ammando}, {de Angelis}, {de
  Palma}, {Dermer}, {Di Venere}, {Drell}, {Drlica-Wagner}, {Favuzzi}, {Fegan},
  {Ferrara}, {Focke}, {Franckowiak}, {Fukazawa}, {Funk}, {Fusco}, {Gargano},
  {Gasparrini}, {Germani}, {Giglietto}, {Giordano}, {Giroletti}, {Glanzman},
  {Godfrey}, {Grenier}, {Grondin}, {Grove}, {Guiriec}, {Hadasch}, {Hanabata},
  {Harding}, {Hayashida}, {Hays}, {Hewitt}, {Hill}, {Horan}, {Hou}, {Hughes},
  {Inoue}, {Jackson}, {Jogler}, {J{\'o}hannesson}, {Johnson}, {Kamae},
  {Kataoka}, {Kawano}, {Kn{\"o}dlseder}, {Kuss}, {Lande}, {Larsson},
  {Latronico}, {Lemoine-Goumard}, {Longo}, {Loparco}, {Lott}, {Lovellette},
  {Lubrano}, {Mayer}, {Mazziotta}, {McEnery}, {Michelson}, {Mitthumsiri},
  {Mizuno}, {Monte}, {Monzani}, {Morselli}, {Moskalenko}, {Murgia}, {Nemmen},
  {Nuss}, {Ohsugi}, {Okumura}, {Omodei}, {Orienti}, {Orlando}, {Ormes},
  {Paneque}, {Panetta}, {Perkins}, {Pesce-Rollins}, {Piron}, {Pivato},
  {Porter}, {Rain{\`o}}, {Rando}, {Razzano}, {Reimer}, {Reimer}, {Romoli},
  {Roth}, {S{\'a}nchez-Conde}, {Scargle}, {Schulz}, {Sgr{\`o}}, {Siskind},
  {Spandre}, {Spinelli}, {Suson}, {Takahashi}, {Takeuchi}, {Thayer}, {Thayer},
  {Thompson}, {Tibaldo}, {Tinivella}, {Torres}, {Tosti}, {Troja}, {Tronconi},
  {Usher}, {Vandenbroucke}, {Vasileiou}, {Vianello}, {Vitale}, {Winer}, {Wood},
  {Wood}, \& {Yang}}]{Ackermann13-FAVA}
{Ackermann}, M., {Ajello}, M., {Albert}, A., {et~al.} 2013{\natexlab{b}}, \apj,
  771, 57

\bibitem[{{Aharonian} {et~al.}(2006){Aharonian}, {Akhperjanian}, {Bazer-Bachi},
  {Beilicke}, {Benbow}, {Berge}, {Bernl{\"o}hr}, {Boisson}, {Bolz}, {Borrel},
  {Braun}, {Breitling}, {Brown}, {Chadwick}, {Chounet}, {Cornils},
  {Costamante}, {Degrange}, {Dickinson}, {Djannati-Ata{\"i}}, {Drury}, {Dubus},
  {Emmanoulopoulos}, {Espigat}, {Feinstein}, {Fontaine}, {Fuchs}, {Funk},
  {Gallant}, {Giebels}, {Gillessen}, {Glicenstein}, {Goret}, {Hadjichristidis},
  {Hauser}, {Heinzelmann}, {Henri}, {Hermann}, {Hinton}, {Hofmann}, {Holleran},
  {Horns}, {Jacholkowska}, {de Jager}, {Kh{\'e}lifi}, {Komin}, {Konopelko},
  {Latham}, {Le Gallou}, {Lemi{\`e}re}, {Lemoine-Goumard}, {Leroy}, {Lohse},
  {Martin}, {Martineau-Huynh}, {Marcowith}, {Masterson}, {McComb}, {de
  Naurois}, {Nolan}, {Noutsos}, {Orford}, {Osborne}, {Ouchrif}, {Panter},
  {Pelletier}, {Pita}, {P{\"u}hlhofer}, {Punch}, {Raubenheimer}, {Raue},
  {Raux}, {Rayner}, {Reimer}, {Reimer}, {Ripken}, {Rob}, {Rolland}, {Rowell},
  {Sahakian}, {Saug{\'e}}, {Schlenker}, {Schlickeiser}, {Schuster}, {Schwanke},
  {Siewert}, {Sol}, {Spangler}, {Steenkamp}, {Stegmann}, {Tavernet}, {Terrier},
  {Th{\'e}oret}, {Tluczykont}, {Vasileiadis}, {Venter}, {Vincent}, {V{\"o}lk},
  \& {Wagner}}]{2006ApJ...636..777A}
{Aharonian}, F., {Akhperjanian}, A.~G., {Bazer-Bachi}, A.~R., {et~al.} 2006,
  \apj, 636, 777

\bibitem[{{Aharonian} {et~al.}(2007{\natexlab{a}}){Aharonian}, {Akhperjanian},
  {Bazer-Bachi}, {Beilicke}, {Benbow}, {Berge}, {Bernl{\"o}hr}, {Boisson},
  {Bolz}, {Borrel}, {Braun}, {Brown}, {B{\"u}hler}, {B{\"u}sching}, {Carrigan},
  {Chadwick}, {Chounet}, {Coignet}, {Cornils}, {Costamante}, {Degrange},
  {Dickinson}, {Djannati-Ata{\"i}}, {Drury}, {Dubus}, {Egberts},
  {Emmanoulopoulos}, {Espigat}, {Feinstein}, {Ferrero}, {Fiasson}, {Filipovic},
  {Fontaine}, {Fukui}, {Funk}, {Funk}, {F{\"u}{\ss}ling}, {Gallant}, {Giebels},
  {Glicenstein}, {Goret}, {Hadjichristidis}, {Hauser}, {Hauser}, {Heinzelmann},
  {Henri}, {Hermann}, {Hinton}, {Hiraga}, {Hoffmann}, {Hofmann}, {Holleran},
  {Hoppe}, {Horns}, {Ishisaki}, {Jacholkowska}, {de Jager}, {Kendziorra},
  {Kerschhaggl}, {Kh{\'e}lifi}, {Komin}, {Konopelko}, {Kosack}, {Lamanna},
  {Latham}, {Le Gallou}, {Lemi{\`e}re}, {Lemoine-Goumard}, {Lohse}, {Martin},
  {Martineau-Huynh}, {Marcowith}, {Masterson}, {Maurin}, {McComb}, {Moulin},
  {Moriguchi}, {de Naurois}, {Nedbal}, {Nolan}, {Noutsos}, {Orford}, {Osborne},
  {Ouchrif}, {Panter}, {Pelletier}, {Pita}, {P{\"u}hlhofer}, {Punch},
  {Ranchon}, {Raubenheimer}, {Raue}, {Rayner}, {Reimer}, {Ripken}, {Rob},
  {Rolland}, {Rosier-Lees}, {Rowell}, {Sahakian}, {Santangelo}, {Saug{\'e}},
  {Schlenker}, {Schlickeiser}, {Schr{\"o}der}, {Schwanke}, {Schwarzburg},
  {Schwemmer}, {Shalchi}, {Sol}, {Spangler}, {Spanier}, {Steenkamp},
  {Stegmann}, {Superina}, {Tam}, {Tavernet}, {Terrier}, {Tluczykont}, {van
  Eldik}, {Vasileiadis}, {Venter}, {Vialle}, {Vincent}, {V{\"o}lk}, {Wagner},
  \& {Ward}}]{2007ApJ...661..236A}
---. 2007{\natexlab{a}}, \apj, 661, 236

\bibitem[{{Aharonian} {et~al.}(2007{\natexlab{b}}){Aharonian}, {Akhperjanian},
  {Bazer-Bachi}, {Beilicke}, {Benbow}, {Berge}, {Bernl{\"o}hr}, {Boisson},
  {Bolz}, {Borrel}, {Braun}, {Brion}, {Brown}, {B{\"u}hler}, {B{\"u}sching},
  {Carrigan}, {Chadwick}, {Chounet}, {Coignet}, {Cornils}, {Costamante},
  {Degrange}, {Dickinson}, {Djannati-Ata{\"i}}, {O'C.~Drury}, {Dubus},
  {Egberts}, {Emmanoulopoulos}, {Espigat}, {Feinstein}, {Ferrero}, {Fiasson},
  {Fontaine}, {Funk}, {Funk}, {F{\"u}{\ss}ling}, {Gallant}, {Giebels},
  {Glicenstein}, {Gl{\"u}ck}, {Goret}, {Hadjichristidis}, {Hauser}, {Hauser},
  {Heinzelmann}, {Henri}, {Hermann}, {Hinton}, {Hoffmann}, {Hofmann},
  {Holleran}, {Hoppe}, {Horns}, {Jacholkowska}, {de Jager}, {Kendziorra},
  {Kerschhaggl}, {Kh{\'e}lifi}, {Komin}, {Konopelko}, {Kosack}, {Lamanna},
  {Latham}, {Le Gallou}, {Lemi{\`e}re}, {Lemoine-Goumard}, {Lohse}, {Martin},
  {Martineau-Huynh}, {Marcowith}, {Masterson}, {Maurin}, {McComb}, {Moulin},
  {de Naurois}, {Nedbal}, {Nolan}, {Noutsos}, {Olive}, {Orford}, {Osborne},
  {Panter}, {Pelletier}, {Pita}, {P{\"u}hlhofer}, {Punch}, {Ranchon},
  {Raubenheimer}, {Raue}, {Rayner}, {Reimer}, {Reimer}, {Ripken}, {Rob},
  {Rolland}, {Rosier-Lees}, {Rowell}, {Sahakian}, {Santangelo}, {Saug{\'e}},
  {Schlenker}, {Schlickeiser}, {Schr{\"o}der}, {Schwanke}, {Schwarzburg},
  {Schwemmer}, {Shalchi}, {Sol}, {Spangler}, {Spanier}, {Steenkamp},
  {Stegmann}, {Superina}, {Tam}, {Tavernet}, {Terrier}, {Tluczykont}, {van
  Eldik}, {Vasileiadis}, {Venter}, {Vialle}, {Vincent}, {V{\"o}lk}, {Wagner},
  \& {Ward}}]{2007AA...464..235A}
---. 2007{\natexlab{b}}, \aap, 464, 235

\bibitem[{{Aharonian} {et~al.}(2008{\natexlab{a}}){Aharonian}, {Akhperjanian},
  {Barres de Almeida}, {Bazer-Bachi}, {Behera}, {Beilicke}, {Benbow},
  {Bernl{\"o}hr}, {Boisson}, {Borrel}, {Braun}, {Brion}, {Brucker},
  {B{\"u}hler}, {Bulik}, {B{\"u}sching}, {Boutelier}, {Carrigan}, {Chadwick},
  {Chaves}, {Chounet}, {Clapson}, {Coignet}, {Cornils}, {Costamante}, {Dalton},
  {Degrange}, {Dickinson}, {Djannati-Ata{\"i}}, {Domainko}, {O'C.~Drury},
  {Dubois}, {Dubus}, {Dyks}, {Egberts}, {Emmanoulopoulos}, {Espigat},
  {Farnier}, {Feinstein}, {Fiasson}, {F{\"o}rster}, {Fontaine}, {Funk},
  {F{\"u}{\ss}ling}, {Gabici}, {Gallant}, {Giebels}, {Glicenstein},
  {Gl{\"u}ck}, {Goret}, {Hadjichristidis}, {Hauser}, {Hauser}, {Heinzelmann},
  {Henri}, {Hermann}, {Hinton}, {Hoffmann}, {Hofmann}, {Holleran}, {Hoppe},
  {Horns}, {Jacholkowska}, {de Jager}, {Jung}, {Katarzy{\'n}ski}, {Kaufmann},
  {Kendziorra}, {Kerschhaggl}, {Khangulyan}, {Kh{\'e}lifi}, {Keogh}, {Komin},
  {Kosack}, {Lamanna}, {Latham}, {Lemoine-Goumard}, {Lenain}, {Lohse},
  {Martin}, {Martineau-Huynh}, {Marcowith}, {Masterson}, {Maurin}, {McComb},
  {Moderski}, {Moulin}, {Nakajima}, {Naumann-Godo}, {de Naurois}, {Nedbal},
  {Nekrassov}, {Nolan}, {Ohm}, {Olive}, {de O{\~n}a Wilhelmi}, {Orford},
  {Osborne}, {Ostrowski}, {Panter}, {Pedaletti}, {Pelletier}, {Petrucci},
  {Pita}, {P{\"u}hlhofer}, {Punch}, {Quirrenbach}, {Raubenheimer}, {Raue},
  {Rayner}, {Reimer}, {Renaud}, {Rieger}, {Ripken}, {Rob}, {Rosier-Lees},
  {Rowell}, {Rudak}, {Ruppel}, {Sahakian}, {Santangelo}, {Schlickeiser},
  {Sch{\"o}ck}, {Schr{\"o}der}, {Schwanke}, {Schwarzburg}, {Schwemmer},
  {Shalchi}, {Skilton}, {Sol}, {Spangler}, {Stawarz}, {Steenkamp}, {Stegmann},
  {Superina}, {Tam}, {Tavernet}, {Terrier}, {Tibolla}, {van Eldik},
  {Vasileiadis}, {Venter}, {Vialle}, {Vincent}, {Vivier}, {V{\"o}lk}, {Volpe},
  {Wagner}, {Ward}, {Zdziarski}, \& {Zech}}]{2008AA...490..685A}
{Aharonian}, F., {Akhperjanian}, A.~G., {Barres de Almeida}, U., {et~al.}
  2008{\natexlab{a}}, \aap, 490, 685

\bibitem[{{Aharonian} {et~al.}(2008{\natexlab{b}}){Aharonian}, {Akhperjanian},
  {Bazer-Bachi}, {Behera}, {Beilicke}, {Benbow}, {Berge}, {Bernl{\"o}hr},
  {Boisson}, {Bolz}, {Borrel}, {Braun}, {Brion}, {Brown}, {B{\"u}hler},
  {Bulik}, {B{\"u}sching}, {Boutelier}, {Carrigan}, {Chadwick}, {Chounet},
  {Clapson}, {Coignet}, {Cornils}, {Costamante}, {Degrange}, {Dickinson},
  {Djannati-Ata{\"i}}, {Domainko}, {O'C.~Drury}, {Dubus}, {Dyks}, {Egberts},
  {Emmanoulopoulos}, {Espigat}, {Farnier}, {Feinstein}, {Fiasson},
  {F{\"o}rster}, {Fontaine}, {Fukui}, {Funk}, {Funk}, {F{\"u}{\ss}ling},
  {Gallant}, {Giebels}, {Glicenstein}, {Gl{\"u}ck}, {Goret}, {Hadjichristidis},
  {Hauser}, {Hauser}, {Heinzelmann}, {Henri}, {Hermann}, {Hinton}, {Hoffmann},
  {Hofmann}, {Holleran}, {Hoppe}, {Horns}, {Jacholkowska}, {de Jager},
  {Kendziorra}, {Kerschhaggl}, {Kh{\'e}lifi}, {Komin}, {Kosack}, {Lamanna},
  {Latham}, {Le Gallou}, {Lemi{\`e}re}, {Lemoine-Goumard}, {Lenain}, {Lohse},
  {Martin}, {Martineau-Huynh}, {Marcowith}, {Masterson}, {Maurin}, {McComb},
  {Moderski}, {Moriguchi}, {Moulin}, {de Naurois}, {Nedbal}, {Nolan}, {Olive},
  {Orford}, {Osborne}, {Ostrowski}, {Panter}, {Pedaletti}, {Pelletier},
  {Petrucci}, {Pita}, {P{\"u}hlhofer}, {Punch}, {Ranchon}, {Raubenheimer},
  {Raue}, {Rayner}, {Reimer}, {Renaud}, {Ripken}, {Rob}, {Rolland},
  {Rosier-Lees}, {Rowell}, {Rudak}, {Ruppel}, {Sahakian}, {Santangelo},
  {Saug{\'e}}, {Schlenker}, {Schlickeiser}, {Schr{\"o}der}, {Schwanke},
  {Schwarzburg}, {Schwemmer}, {Shalchi}, {Sol}, {Spangler}, {Stawarz},
  {Steenkamp}, {Stegmann}, {Superina}, {Takeuchi}, {Tam}, {Tavernet},
  {Terrier}, {van Eldik}, {Vasileiadis}, {Venter}, {Vialle}, {Vincent},
  {Vivier}, {V{\"o}lk}, {Volpe}, {Wagner}, \& {Ward}}]{Aharonian08-w28}
{Aharonian}, F., {Akhperjanian}, A.~G., {Bazer-Bachi}, A.~R., {et~al.}
  2008{\natexlab{b}}, \aap, 481, 401

\bibitem[{{Aharonian} {et~al.}(2008{\natexlab{c}}){Aharonian}, {Akhperjanian},
  {Barres de Almeida}, {Bazer-Bachi}, {Behera}, {Beilicke}, {Benbow}, {Berge},
  {Bernl{\"o}hr}, {Boisson}, {Bolz}, {Borrel}, {Braun}, {Brion}, {Brucker},
  {B{\"u}hler}, {Bulik}, {B{\"u}sching}, {Boutelier}, {Carrigan}, {Chadwick},
  {Chounet}, {Clapson}, {Coignet}, {Cornils}, {Costamante}, {Dalton},
  {Degrange}, {Dickinson}, {Djannati-Ata{\"i}}, {Domainko}, {O'C.~Drury},
  {Dubois}, {Dubus}, {Dyks}, {Egberts}, {Emmanoulopoulos}, {Espigat},
  {Farnier}, {Feinstein}, {Fiasson}, {F{\"o}rster}, {Fontaine},
  {F{\"u}{\ss}ling}, {Gallant}, {Giebels}, {Glicenstein}, {Gl{\"u}ck}, {Goret},
  {Hadjichristidis}, {Hauser}, {Hauser}, {Heinzelmann}, {Henri}, {Hermann},
  {Hinton}, {Hoffmann}, {Hofmann}, {Holleran}, {Hoppe}, {Horns},
  {Jacholkowska}, {de Jager}, {Jung}, {Katarzy{\'n}ski}, {Kendziorra},
  {Kerschhaggl}, {Kh{\'e}lifi}, {Keogh}, {Komin}, {Kosack}, {Lamanna},
  {Latham}, {Lemoine-Goumard}, {Lenain}, {Lohse}, {Martin}, {Martineau-Huynh},
  {Marcowith}, {Masterson}, {Maurin}, {McComb}, {Moderski}, {Moulin},
  {Naumann-Godo}, {de Naurois}, {Nedbal}, {Nekrassov}, {Nolan}, {Ohm}, {Olive},
  {de O{\~n}a Wilhelmi}, {Orford}, {Osborne}, {Ostrowski}, {Panter},
  {Pedaletti}, {Pelletier}, {Petrucci}, {Pita}, {P{\"u}hlhofer}, {Punch},
  {Quirrenbach}, {Raubenheimer}, {Raue}, {Rayner}, {Renaud}, {Ripken}, {Rob},
  {Rosier-Lees}, {Rowell}, {Rudak}, {Ruppel}, {Sahakian}, {Santangelo},
  {Schlickeiser}, {Sch{\"o}ck}, {Schr{\"o}der}, {Schwanke}, {Schwarzburg},
  {Schwemmer}, {Shalchi}, {Sol}, {Spangler}, {Stawarz}, {Steenkamp},
  {Stegmann}, {Superina}, {Tam}, {Tavernet}, {Terrier}, {van Eldik},
  {Vasileiadis}, {Venter}, {Vialle}, {Vincent}, {Vivier}, {V{\"o}lk}, {Volpe},
  {Wagner}, {Ward}, {Zdziarski}, \& {Zech}}]{2008AA...488..219A}
{Aharonian}, F., {Akhperjanian}, A.~G., {Barres de Almeida}, U., {et~al.}
  2008{\natexlab{c}}, \aap, 488, 219

\bibitem[{{Aharonian} {et~al.}(2008{\natexlab{d}}){Aharonian}, {Akhperjanian},
  {Barres de Almeida}, {Bazer-Bachi}, {Behera}, {Beilicke}, {Benbow},
  {Bernl{\"o}hr}, {Boisson}, {Bolz}, {Borrel}, {Braun}, {Brion}, {Brown},
  {B{\"u}hler}, {Bulik}, {B{\"u}sching}, {Boutelier}, {Carrigan}, {Chadwick},
  {Chounet}, {Clapson}, {Coignet}, {Cornils}, {Costamante}, {Dalton},
  {Degrange}, {Dickinson}, {Djannati-Ata{\"i}}, {Domainko}, {Drury}, {Dubois},
  {Dubus}, {Dyks}, {Egberts}, {Emmanoulopoulos}, {Espigat}, {Farnier},
  {Feinstein}, {Fiasson}, {F{\"o}rster}, {Fontaine}, {Funk}, {F{\"u}{\ss}ling},
  {Gallant}, {Giebels}, {Glicenstein}, {Gl{\"u}ck}, {Goret}, {Hadjichristidis},
  {Hauser}, {Hauser}, {Heinzelmann}, {Henri}, {Hermann}, {Hinton}, {Hoffmann},
  {Hofmann}, {Holleran}, {Hoppe}, {Horns}, {Jacholkowska}, {de Jager}, {Jung},
  {Katarzy{\'n}ski}, {Kendziorra}, {Kerschhaggl}, {Kh{\'e}lifi}, {Keogh},
  {Komin}, {Kosack}, {Lamanna}, {Latham}, {Lemi{\`e}re}, {Lemoine-Goumard},
  {Lenain}, {Lohse}, {Martin}, {Martineau-Huynh}, {Marcowith}, {Masterson},
  {Maurin}, {Maurin}, {McComb}, {Moderski}, {Moulin}, {de Naurois}, {Nedbal},
  {Nolan}, {Ohm}, {Olive}, {de O{\~n}a Wilhelmi}, {Orford}, {Osborne},
  {Ostrowski}, {Panter}, {Pedaletti}, {Pelletier}, {Petrucci}, {Pita},
  {P{\"u}hlhofer}, {Punch}, {Ranchon}, {Raubenheimer}, {Raue}, {Rayner},
  {Renaud}, {Ripken}, {Rob}, {Rolland}, {Rosier-Lees}, {Rowell}, {Rudak},
  {Ruppel}, {Sahakian}, {Santangelo}, {Schlickeiser}, {Sch{\"o}ck},
  {Schr{\"o}der}, {Schwanke}, {Schwarzburg}, {Schwemmer}, {Shalchi}, {Sol},
  {Spangler}, {Stawarz}, {Steenkamp}, {Stegmann}, {Superina}, {Tam},
  {Tavernet}, {Terrier}, {van Eldik}, {Vasileiadis}, {Venter}, {Vialle},
  {Vincent}, {Vivier}, {V{\"o}lk}, {Volpe}, {Wagner}, {Ward}, {Zdziarski}, \&
  {Zech}}]{Aharonian08-HESSJ1858+020}
---. 2008{\natexlab{d}}, \aap, 477, 353

\bibitem[{{Albert} {et~al.}(2007){Albert}, {Aliu}, {Anderhub}, {Antoranz},
  {Armada}, {Baixeras}, {Barrio}, {Bartko}, {Bastieri}, {Becker}, {Bednarek},
  {Berger}, {Bigongiari}, {Biland}, {Bock}, {Bordas}, {Bosch-Ramon}, {Bretz},
  {Britvitch}, {Camara}, {Carmona}, {Chilingarian}, {Coarasa}, {Commichau},
  {Contreras}, {Cortina}, {Costado}, {Curtef}, {Danielyan}, {Dazzi}, {de
  Angelis}, {Delgado}, {de Los Reyes}, {de Lotto}, {Domingo-Santamar{\'{\i}}a},
  {Dorner}, {Doro}, {Errando}, {Fagiolini}, {Ferenc}, {Fern{\'a}ndez}, {Firpo},
  {Flix}, {Fonseca}, {Font}, {Fuchs}, {Galante}, {Garc{\'{\i}}a-L{\'o}pez},
  {Garczarczyk}, {Gaug}, {Giller}, {Goebel}, {Hakobyan}, {Hayashida},
  {Hengstebeck}, {Herrero}, {H{\"o}hne}, {Hose}, {Hsu}, {Jacon}, {Jogler},
  {Kosyra}, {Kranich}, {Kritzer}, {Laille}, {Lindfors}, {Lombardi}, {Longo},
  {L{\'o}pez}, {L{\'o}pez}, {Lorenz}, {Majumdar}, {Maneva}, {Mannheim},
  {Mansutti}, {Mariotti}, {Mart{\'{\i}}nez}, {Mazin}, {Merck}, {Meucci},
  {Meyer}, {Miranda}, {Mirzoyan}, {Mizobuchi}, {Moralejo}, {Nilsson},
  {Ninkovic}, {O{\~n}a-Wilhelmi}, {Otte}, {Oya}, {Paneque}, {Panniello},
  {Paoletti}, {Paredes}, {Pasanen}, {Pascoli}, {Pauss}, {Pegna}, {Persic},
  {Peruzzo}, {Piccioli}, {Poller}, {Puchades}, {Prandini}, {Raymers}, {Rhode},
  {Rib{\'o}}, {Rico}, {Rissi}, {Robert}, {R{\"u}gamer}, {Saggion},
  {S{\'a}nchez}, {Sartori}, {Scalzotto}, {Scapin}, {Schmitt}, {Schweizer},
  {Shayduk}, {Shinozaki}, {Shore}, {Sidro}, {Sillanp{\"a}{\"a}}, {Sobczynska},
  {Stamerra}, {Stark}, {Takalo}, {Temnikov}, {Tescaro}, {Teshima}, {Tonello},
  {Torres}, {Turini}, {Vankov}, {Vitale}, {Wagner}, {Wibig}, {Wittek},
  {Zandanel}, {Zanin}, \& {Zapatero}}]{2007AA...474..937A}
{Albert}, J., {Aliu}, E., {Anderhub}, H., {et~al.} 2007, \aap, 474, 937

\bibitem[{{Araya}(2013)}]{Araya13-G295.6+10.0}
{Araya}, M. 2013, \mnras, 434, 2202

\bibitem[{{Arzoumanian} {et~al.}(2011){Arzoumanian}, {Gotthelf}, {Ransom},
  {Safi-Harb}, {Kothes}, \& {Landecker}}]{2011ApJ...739...39A}
{Arzoumanian}, Z., {Gotthelf}, E.~V., {Ransom}, S.~M., {et~al.} 2011, \apj,
  739, 39

\bibitem[{{Atwood} {et~al.}(2013){Atwood}, {Albert}, {Baldini}, {Tinivella},
  {Bregeon}, {Pesce-Rollins}, {Sgr{\`o}}, {Bruel}, {Charles}, {Drlica-Wagner},
  {Franckowiak}, {Jogler}, {Rochester}, {Usher}, {Wood}, {Cohen-Tanugi}, \&
  {S.~Zimmer for the Fermi-LAT Collaboration}}]{Atwood13-pass8}
{Atwood}, W., {Albert}, A., {Baldini}, L., {et~al.} 2013, $4^{th}$ Fermi
  Symposium Proceedings, arXiv:1303.3514

\bibitem[{{Atwood} {et~al.}(2009){Atwood}, {Abdo}, {Ackermann}, {Althouse},
  {Anderson}, {Axelsson}, {Baldini}, {Ballet}, {Band}, {Barbiellini}, \&
  et~al.}]{Atwood09-FermiLAT}
{Atwood}, W.~B., {Abdo}, A.~A., {Ackermann}, M., {et~al.} 2009, \apj, 697, 1071

\bibitem[{{Bamba} {et~al.}(2001){Bamba}, {Ueno}, {Koyama}, \&
  {Yamauchi}}]{2001PASJ...53L..21B}
{Bamba}, A., {Ueno}, M., {Koyama}, K., \& {Yamauchi}, S. 2001, \pasj, 53, L21

\bibitem[{{Bamba} {et~al.}(2005{\natexlab{a}}){Bamba}, {Yamazaki}, \&
  {Hiraga}}]{2005ApJ...632..294B}
{Bamba}, A., {Yamazaki}, R., \& {Hiraga}, J.~S. 2005{\natexlab{a}}, \apj, 632,
  294

\bibitem[{{Bamba} {et~al.}(2005{\natexlab{b}}){Bamba}, {Yamazaki}, {Yoshida},
  {Terasawa}, \& {Koyama}}]{2005ApJ...621..793B}
{Bamba}, A., {Yamazaki}, R., {Yoshida}, T., {Terasawa}, T., \& {Koyama}, K.
  2005{\natexlab{b}}, \apj, 621, 793

\bibitem[{{Bamba} {et~al.}(2008){Bamba}, {Fukazawa}, {Hiraga}, {Hughes},
  {Katagiri}, {Kokubun}, {Koyama}, {Miyata}, {Mizuno}, {Mori}, {Nakajima},
  {Ozaki}, {Petre}, {Takahashi}, {Takahashi}, {Tanaka}, {Terada}, {Uchiyama},
  {Watanabe}, \& {Yamaguchi}}]{2008PASJ...60S.153B}
{Bamba}, A., {Fukazawa}, Y., {Hiraga}, J.~S., {et~al.} 2008, \pasj, 60, 153

\bibitem[{{Bamba} {et~al.}(2012){Bamba}, {P{\"u}hlhofer}, {Acero}, {Klochkov},
  {Tian}, {Yamazaki}, {Li}, {Horns}, {Kosack}, \&
  {Komin}}]{2012ApJ...756..149B}
{Bamba}, A., {P{\"u}hlhofer}, G., {Acero}, F., {et~al.} 2012, \apj, 756, 149

\bibitem[{{Becker} \& {Helfand}(1987)}]{1987ApJ...316..660B}
{Becker}, R.~H., \& {Helfand}, D.~J. 1987, \apj, 316, 660

\bibitem[{{Becker} {et~al.}(1985){Becker}, {Markert}, \&
  {Donahue}}]{1985ApJ...296..461B}
{Becker}, R.~H., {Markert}, T., \& {Donahue}, M. 1985, \apj, 296, 461

\bibitem[{{Bell}(2004)}]{Bell04-Emax}
{Bell}, A.~R. 2004, \mnras, 353, 550

\bibitem[{{Bell}(2015)}]{Bell14_CRsFromSNe}
---. 2015, \mnras, 447, 2224

\bibitem[{{Bell} \& {Lucek}(2001)}]{Bell01-Emax}
{Bell}, A.~R., \& {Lucek}, S.~G. 2001, \mnras, 321, 433

\bibitem[{{Bell} {et~al.}(2013){Bell}, {Schure}, {Reville}, \&
  {Giacinti}}]{Bell13-Emax}
{Bell}, A.~R., {Schure}, K.~M., {Reville}, B., \& {Giacinti}, G. 2013, \mnras,
  431, 415

\bibitem[{{Blair} {et~al.}(2009){Blair}, {Sankrit}, {Torres}, {Chayer}, \&
  {Danforth}}]{2009ApJ...692..335B}
{Blair}, W.~P., {Sankrit}, R., {Torres}, S.~I., {Chayer}, P., \& {Danforth},
  C.~W. 2009, \apj, 692, 335

\bibitem[{{Blasi} {et~al.}(2007){Blasi}, {Amato}, \& {Caprioli}}]{Blasi07-Emax}
{Blasi}, P., {Amato}, E., \& {Caprioli}, D. 2007, \mnras, 375, 1471

\bibitem[{{Boumis} {et~al.}(2008){Boumis}, {Alikakos}, {Christopoulou},
  {Mavromatakis}, {Xilouris}, \& {Goudis}}]{2008AA...481..705B}
{Boumis}, P., {Alikakos}, J., {Christopoulou}, P.~E., {et~al.} 2008, \aap, 481,
  705

\bibitem[{{Brandt} {et~al.}(2013){Brandt}, {Acero}, {de Palma}, {Hewitt},
  {Renaud}, \& {for the Fermi LAT Collaboration}}]{Brandt13-snrCatCrConst_ICRC}
{Brandt}, T.~J., {Acero}, F., {de Palma}, F., {et~al.} 2013, $4^{th}$ Fermi
  Symposium Proceedings, arXiv:1307.6571

\bibitem[{{Brandt} \& {Fermi-LAT Collaboration}(2013)}]{Brandt13-CTB37A}
{Brandt}, T.~J., \& {Fermi-LAT Collaboration}. 2013, Advances in Space
  Research, 51, 247

\bibitem[{{Brogan} {et~al.}(2006){Brogan}, {Gelfand}, {Gaensler}, {Kassim}, \&
  {Lazio}}]{Brogan06}
{Brogan}, C.~L., {Gelfand}, J.~D., {Gaensler}, B.~M., {Kassim}, N.~E., \&
  {Lazio}, T.~J.~W. 2006, \apjl, 639, L25

\bibitem[{{Brogan} \& {Troland}(2001)}]{2001ApJ...550..799B}
{Brogan}, C.~L., \& {Troland}, T.~H. 2001, \apj, 550, 799

\bibitem[{{Brun} {et~al.}(2011){Brun}, {de Naurois}, {Hofmann}, {Carrigan},
  {Djannati-Ata{\"i}}, {Ohm}, \& {for the
  H.~E.~S.~S.~Collaboration}}]{2011arXiv1104.5003B}
{Brun}, F., {de Naurois}, M., {Hofmann}, W., {et~al.} 2011, in 25th Texas
  Symposium on Relativistic Astrophysics, 196

\bibitem[{{Byun} {et~al.}(2006){Byun}, {Koo}, {Tatematsu}, \&
  {Sunada}}]{2006ApJ...637..283B}
{Byun}, D.-Y., {Koo}, B.-C., {Tatematsu}, K., \& {Sunada}, K. 2006, \apj, 637,
  283

\bibitem[{{Camilo} {et~al.}(2000){Camilo}, {Kaspi}, {Lyne}, {Manchester},
  {Bell}, {D'Amico}, {McKay}, \& {Crawford}}]{2000ApJ...541..367C}
{Camilo}, F., {Kaspi}, V.~M., {Lyne}, A.~G., {et~al.} 2000, \apj, 541, 367

\bibitem[{{Camilo} {et~al.}(2009{\natexlab{a}}){Camilo}, {Ng}, {Gaensler},
  {Ransom}, {Chatterjee}, {Reynolds}, \& {Sarkissian}}]{2009ApJ...703L..55C}
{Camilo}, F., {Ng}, C.-Y., {Gaensler}, B.~M., {et~al.} 2009{\natexlab{a}},
  \apjl, 703, L55

\bibitem[{{Camilo} {et~al.}(2009{\natexlab{b}}){Camilo}, {Ransom}, {Gaensler},
  \& {Lorimer}}]{2009ApJ...700L..34C}
{Camilo}, F., {Ransom}, S.~M., {Gaensler}, B.~M., \& {Lorimer}, D.~R.
  2009{\natexlab{b}}, \apjl, 700, L34

\bibitem[{{Camilo} {et~al.}(2006){Camilo}, {Ransom}, {Gaensler}, {Slane},
  {Lorimer}, {Reynolds}, {Manchester}, \& {Murray}}]{2006ApJ...637..456C}
{Camilo}, F., {Ransom}, S.~M., {Gaensler}, B.~M., {et~al.} 2006, \apj, 637, 456

\bibitem[{{Caprioli}(2011)}]{2011JCAP...05..026C}
{Caprioli}, D. 2011, \jcap, 5, 26

\bibitem[{{Caraveo} {et~al.}(2001){Caraveo}, {De Luca}, {Mignani}, \&
  {Bignami}}]{2001ApJ...561..930C}
{Caraveo}, P.~A., {De Luca}, A., {Mignani}, R.~P., \& {Bignami}, G.~F. 2001,
  \apj, 561, 930

\bibitem[{{Carmona}(2011)}]{2011ICRC....7..115C}
{Carmona}, E. 2011, International Cosmic Ray Conference, 7, 115

\bibitem[{{Carrigan} {et~al.}(2013){Carrigan}, {Brun}, {Chaves}, {Deil},
  {Donath}, {Gast}, {Marandon}, {Renaud}, \& {for the
  H.~E.~S.~S.~collaboration}}]{Carrigan13-HESSgpsurvey}
{Carrigan}, S., {Brun}, F., {Chaves}, R.~C.~G., {et~al.} 2013, ICRC,
  arXiv:1307.4690

\bibitem[{{Casandjian} {et~al.}(2009){Casandjian}, {Grenier}, \& {for the Fermi
  Large Area Telescope Collaboration}}]{Casandjian09-LoopI}
{Casandjian}, J.-M., {Grenier}, I., \& {for the Fermi Large Area Telescope
  Collaboration}. 2009, Fermi Symposium Proceedings, arXiv:0912.3478

\bibitem[{{Case} \& {Bhattacharya}(1998)}]{Case98-SigmaDSNR}
{Case}, G.~L., \& {Bhattacharya}, D. 1998, \apj, 504, 761

\bibitem[{{Castelletti} {et~al.}(2013){Castelletti}, {Supan}, {Dubner},
  {Joshi}, \& {Surnis}}]{2013AA...557L..15C}
{Castelletti}, G., {Supan}, L., {Dubner}, G., {Joshi}, B.~C., \& {Surnis},
  M.~P. 2013, \aap, 557, L15

\bibitem[{{Castro} \& {Slane}(2010)}]{Castro10-interactMCs}
{Castro}, D., \& {Slane}, P. 2010, \apj, 717, 372

\bibitem[{{Castro} {et~al.}(2013){Castro}, {Slane}, {Carlton}, \&
  {Figueroa-Feliciano}}]{Castro13-3LatSnrs}
{Castro}, D., {Slane}, P., {Carlton}, A., \& {Figueroa-Feliciano}, E. 2013,
  \apj, 774, 36

\bibitem[{{Caswell}(1985)}]{1985AJ.....90.1224C}
{Caswell}, J.~L. 1985, \aj, 90, 1224

\bibitem[{{Caswell} {et~al.}(2004){Caswell}, {McClure-Griffiths}, \&
  {Cheung}}]{2004MNRAS.352.1405C}
{Caswell}, J.~L., {McClure-Griffiths}, N.~M., \& {Cheung}, M.~C.~M. 2004,
  \mnras, 352, 1405

\bibitem[{{Caswell} {et~al.}(1975){Caswell}, {Murray}, {Roger}, {Cole}, \&
  {Cooke}}]{1975AA....45..239C}
{Caswell}, J.~L., {Murray}, J.~D., {Roger}, R.~S., {Cole}, D.~J., \& {Cooke},
  D.~J. 1975, \aap, 45, 239

\bibitem[{{Chatterjee} {et~al.}(2009){Chatterjee}, {Brisken}, {Vlemmings},
  {Goss}, {Lazio}, {Cordes}, {Thorsett}, {Fomalont}, {Lyne}, \&
  {Kramer}}]{2009ApJ...698..250C}
{Chatterjee}, S., {Brisken}, W.~F., {Vlemmings}, W.~H.~T., {et~al.} 2009, \apj,
  698, 250

\bibitem[{{Claussen} {et~al.}(1999){Claussen}, {Goss}, {Frail}, \&
  {Desai}}]{1999ApJ...522..349C}
{Claussen}, M.~J., {Goss}, W.~M., {Frail}, D.~A., \& {Desai}, K. 1999, \apj,
  522, 349

\bibitem[{{Combi} {et~al.}(2006){Combi}, {Albacete Colombo}, {Romero}, \&
  {Benaglia}}]{2006ApJ...653L..41C}
{Combi}, J.~A., {Albacete Colombo}, J.~F., {Romero}, G.~E., \& {Benaglia}, P.
  2006, \apjl, 653, L41

\bibitem[{{Combi} {et~al.}(2005){Combi}, {Benaglia}, {Romero}, \&
  {Sugizaki}}]{2005AA...431L...9C}
{Combi}, J.~A., {Benaglia}, P., {Romero}, G.~E., \& {Sugizaki}, M. 2005, \aap,
  431, L9

\bibitem[{{Combi} {et~al.}(2010){Combi}, {Albacete Colombo},
  {S{\'a}nchez-Ayaso}, {Romero}, {Mart{\'{\i}}}, {Luque-Escamilla},
  {Mu{\~n}oz-Arjonilla}, {S{\'a}nchez-Sutil}, \&
  {L{\'o}pez-Santiago}}]{2010AA...523A..76C}
{Combi}, J.~A., {Albacete Colombo}, J.~F., {S{\'a}nchez-Ayaso}, E., {et~al.}
  2010, \aap, 523, A76

\bibitem[{{Cristofari} {et~al.}(2013){Cristofari}, {Gabici}, {Casanova},
  {Terrier}, \& {Parizot}}]{Cristofari13-SNRpop}
{Cristofari}, P., {Gabici}, S., {Casanova}, S., {Terrier}, R., \& {Parizot}, E.
  2013, \mnras, 434, 2748

\bibitem[{{De Horta} {et~al.}(2013){De Horta}, {Collier}, {Filipovi{\'c}},
  {Crawford}, {Uro{\v s}evi{\'c}}, {Stootman}, \& {Tothill}}]{DeHorta13}
{De Horta}, A.~Y., {Collier}, J.~D., {Filipovi{\'c}}, M.~D., {et~al.} 2013,
  \mnras, 428, 1980

\bibitem[{{de Palma} {et~al.}(2013){de Palma}, {Brandt}, {Johannesson},
  {Tibaldo}, \& {for the Fermi LAT
  collaboration}}]{dePalma13-AltIEMSystematics_FSymp}
{de Palma}, F., {Brandt}, T.~J., {Johannesson}, G., {Tibaldo}, L., \& {for the
  Fermi LAT collaboration}. 2013, $4^{th}$ Fermi Symposium Proceedings,
  arXiv:1304.1395

\bibitem[{{Dermer} \& {Powale}(2013)}]{Dermer13-CRsSNRs}
{Dermer}, C.~D., \& {Powale}, G. 2013, \aap, 553, A34

\bibitem[{{Dodson} {et~al.}(2003){Dodson}, {Legge}, {Reynolds}, \&
  {McCulloch}}]{2003ApJ...596.1137D}
{Dodson}, R., {Legge}, D., {Reynolds}, J.~E., \& {McCulloch}, P.~M. 2003, \apj,
  596, 1137

\bibitem[{{Drury} {et~al.}(1994){Drury}, {Aharonian}, \&
  {Voelk}}]{Drury94-GraySNRsCRs}
{Drury}, L.~O., {Aharonian}, F.~A., \& {Voelk}, H.~J. 1994, \aap, 287, 959

\bibitem[{{Dubner} {et~al.}(1999){Dubner}, {Giacani}, {Reynoso}, {Goss},
  {Roth}, \& {Green}}]{1999AJ....118..930D}
{Dubner}, G., {Giacani}, E., {Reynoso}, E., {et~al.} 1999, \aj, 118, 930

\bibitem[{{Dubner} {et~al.}(2004){Dubner}, {Giacani}, {Reynoso}, \&
  {Par{\'o}n}}]{2004AA...426..201D}
{Dubner}, G., {Giacani}, E., {Reynoso}, E., \& {Par{\'o}n}, S. 2004, \aap, 426,
  201

\bibitem[{{Dubner} \& {Arnal}(1988)}]{1988AAS...75..363D}
{Dubner}, G.~M., \& {Arnal}, E.~M. 1988, \aaps, 75, 363

\bibitem[{{Dubus} {et~al.}(2013){Dubus}, {Contreras}, {Funk}, {Gallant},
  {Hassan}, {Hinton}, {Inoue}, {Kn{\"o}dlseder}, {Martin}, {Mirabal}, {de
  Naurois}, {Renaud}, \& {CTA Consortium}}]{Dubus13_CTA_GalPlane}
{Dubus}, G., {Contreras}, J.~L., {Funk}, S., {et~al.} 2013, Astroparticle
  Physics, 43, 317

\bibitem[{{Eger} {et~al.}(2011){Eger}, {Rowell}, {Kawamura}, {Fukui},
  {Rolland}, \& {Stegmann}}]{2011AA...526A..82E}
{Eger}, P., {Rowell}, G., {Kawamura}, A., {et~al.} 2011, \aap, 526, A82

\bibitem[{{Eriksen} {et~al.}(2011){Eriksen}, {Hughes}, {Badenes}, {Fesen},
  {Ghavamian}, {Moffett}, {Plucinksy}, {Rakowski}, {Reynoso}, \&
  {Slane}}]{2011ApJ...728L..28E}
{Eriksen}, K.~A., {Hughes}, J.~P., {Badenes}, C., {et~al.} 2011, \apjl, 728,
  L28

\bibitem[{{Esposito} {et~al.}(1996){Esposito}, {Hunter}, {Kanbach}, \&
  {Sreekumar}}]{Esposito96}
{Esposito}, J.~A., {Hunter}, S.~D., {Kanbach}, G., \& {Sreekumar}, P. 1996,
  \apj, 461, 820

\bibitem[{{Ferrand} \& {Safi-Harb}(2012)}]{Ferrand12-XraySNRs}
{Ferrand}, G., \& {Safi-Harb}, S. 2012, Advances in Space Research, 49, 1313

\bibitem[{{Fesen} {et~al.}(1985){Fesen}, {Blair}, \&
  {Kirshner}}]{1985ApJ...292...29F}
{Fesen}, R.~A., {Blair}, W.~P., \& {Kirshner}, R.~P. 1985, \apj, 292, 29

\bibitem[{{Frail} {et~al.}(2013){Frail}, {Claussen}, \&
  {M{\'e}hault}}]{2013ApJ...773L..19F}
{Frail}, D.~A., {Claussen}, M.~J., \& {M{\'e}hault}, J. 2013, \apjl, 773, L19

\bibitem[{{Frail} \& {Clifton}(1989)}]{1989ApJ...336..854F}
{Frail}, D.~A., \& {Clifton}, T.~R. 1989, \apj, 336, 854

\bibitem[{{Frail} {et~al.}(1996){Frail}, {Goss}, {Reynoso}, {Giacani}, {Green},
  \& {Otrupcek}}]{1996AJ....111.1651F}
{Frail}, D.~A., {Goss}, W.~M., {Reynoso}, E.~M., {et~al.} 1996, \aj, 111, 1651

\bibitem[{{Funk} {et~al.}(2008){Funk}, {Reimer}, {Torres}, \&
  {Hinton}}]{Funk08_GeVTeVGalactic}
{Funk}, S., {Reimer}, O., {Torres}, D.~F., \& {Hinton}, J.~A. 2008, \apj, 679,
  1299

\bibitem[{{Gabici} {et~al.}(2009){Gabici}, {Aharonian}, \&
  {Casanova}}]{2009MNRAS.396.1629G}
{Gabici}, S., {Aharonian}, F.~A., \& {Casanova}, S. 2009, \mnras, 396, 1629

\bibitem[{{Gaensler} {et~al.}(1999){Gaensler}, {Brazier}, {Manchester},
  {Johnston}, \& {Green}}]{1999MNRAS.305..724G}
{Gaensler}, B.~M., {Brazier}, K.~T.~S., {Manchester}, R.~N., {Johnston}, S., \&
  {Green}, A.~J. 1999, \mnras, 305, 724

\bibitem[{{Gaensler} {et~al.}(2003){Gaensler}, {Fogel}, {Slane}, {Miller},
  {Wijnands}, {Eikenberry}, \& {Lewin}}]{2003ApJ...594L..35G}
{Gaensler}, B.~M., {Fogel}, J.~K.~J., {Slane}, P.~O., {et~al.} 2003, \apjl,
  594, L35

\bibitem[{{Gaensler} {et~al.}(2008{\natexlab{a}}){Gaensler}, {Madsen},
  {Chatterjee}, \& {Mao}}]{Gaensler08-HIImodel}
{Gaensler}, B.~M., {Madsen}, G.~J., {Chatterjee}, S., \& {Mao}, S.~A.
  2008{\natexlab{a}}, \pasa, 25, 184

\bibitem[{{Gaensler} \& {Wallace}(2003)}]{2003ApJ...594..326G}
{Gaensler}, B.~M., \& {Wallace}, B.~J. 2003, \apj, 594, 326

\bibitem[{{Gaensler} {et~al.}(2008{\natexlab{b}}){Gaensler}, {Tanna}, {Slane},
  {Brogan}, {Gelfand}, {McClure-Griffiths}, {Camilo}, {Ng}, \&
  {Miller}}]{2008ApJ...680L..37G}
{Gaensler}, B.~M., {Tanna}, A., {Slane}, P.~O., {et~al.} 2008{\natexlab{b}},
  \apjl, 680, L37

\bibitem[{{Gaisser} {et~al.}(1998){Gaisser}, {Protheroe}, \&
  {Stanev}}]{Gaisser98}
{Gaisser}, T.~K., {Protheroe}, R.~J., \& {Stanev}, T. 1998, \apj, 492, 219

\bibitem[{{Gerardy} \& {Fesen}(2007)}]{2007MNRAS.376..929G}
{Gerardy}, C.~L., \& {Fesen}, R.~A. 2007, \mnras, 376, 929

\bibitem[{{Giacani} {et~al.}(2009){Giacani}, {Smith}, {Dubner}, {Loiseau},
  {Castelletti}, \& {Paron}}]{2009AA...507..841G}
{Giacani}, E., {Smith}, M.~J.~S., {Dubner}, G., {et~al.} 2009, \aap, 507, 841

\bibitem[{{Giacani} {et~al.}(2000){Giacani}, {Dubner}, {Green}, {Goss}, \&
  {Gaensler}}]{2000AJ....119..281G}
{Giacani}, E.~B., {Dubner}, G.~M., {Green}, A.~J., {Goss}, W.~M., \&
  {Gaensler}, B.~M. 2000, \aj, 119, 281

\bibitem[{{Giordano} {et~al.}(2012){Giordano}, {Naumann-Godo}, {Ballet},
  {Bechtol}, {Funk}, {Lande}, {Mazziotta}, {Rain{\`o}}, {Tanaka}, {Tibolla}, \&
  {Uchiyama}}]{Giordano12-Tycho}
{Giordano}, F., {Naumann-Godo}, M., {Ballet}, J., {et~al.} 2012, \apjl, 744, L2

\bibitem[{{Giuliani} {et~al.}(2011){Giuliani}, {Cardillo}, {Tavani}, {Fukui},
  {Yoshiike}, {Torii}, {Dubner}, {Castelletti}, {Barbiellini}, {Bulgarelli},
  {Caraveo}, {Costa}, {Cattaneo}, {Chen}, {Contessi}, {Del Monte},
  {Donnarumma}, {Evangelista}, {Feroci}, {Gianotti}, {Lazzarotto}, {Lucarelli},
  {Longo}, {Marisaldi}, {Mereghetti}, {Pacciani}, {Pellizzoni}, {Piano},
  {Picozza}, {Pittori}, {Pucella}, {Rapisarda}, {Rappoldi}, {Sabatini},
  {Soffitta}, {Striani}, {Trifoglio}, {Trois}, {Vercellone}, {Verrecchia},
  {Vittorini}, {Colafrancesco}, {Giommi}, \& {Bignami}}]{Giuliani11-W44agile}
{Giuliani}, A., {Cardillo}, M., {Tavani}, M., {et~al.} 2011, \apjl, 742, L30

\bibitem[{{Green} {et~al.}(1997){Green}, {Frail}, {Goss}, \&
  {Otrupcek}}]{Green97}
{Green}, A.~J., {Frail}, D.~A., {Goss}, W.~M., \& {Otrupcek}, R. 1997, \aj,
  114, 2058

\bibitem[{{Green}(1991)}]{Green91}
{Green}, D.~A. 1991, \pasp, 103, 209

\bibitem[{{Green}(2004)}]{Green04}
---. 2004, Bulletin of the Astronomical Society of India, 32, 335

\bibitem[{{Green}(2009{\natexlab{a}})}]{Green09-GreensCat}
---. 2009{\natexlab{a}}, Bulletin of the Astronomical Society of India, 37, 45

\bibitem[{{Green}(2009{\natexlab{b}})}]{Green09-g35.6}
---. 2009{\natexlab{b}}, \mnras, 399, 177

\bibitem[{{Green}(2012)}]{Green12-distances}
{Green}, D.~A. 2012, in American Institute of Physics Conference Series, Vol.
  1505, American Institute of Physics Conference Series, ed. F.~A. {Aharonian},
  W.~{Hofmann}, \& F.~M. {Rieger}, 5--12

\bibitem[{{Green}(2014)}]{Green14-GreensCat}
---. 2014, Bulletin of the Astronomical Society of India, 42, 47

\bibitem[{{Green} {et~al.}(1988){Green}, {Gull}, {Tan}, \&
  {Simon}}]{1988MNRAS.231..735G}
{Green}, D.~A., {Gull}, S.~F., {Tan}, S.~M., \& {Simon}, A.~J.~B. 1988, \mnras,
  231, 735

\bibitem[{{Grondin} {et~al.}(2013){Grondin}, {Romani}, {Lemoine-Goumard},
  {Guillemot}, {Harding}, \& {Reposeur}}]{Grondin2013}
{Grondin}, M.-H., {Romani}, R.~W., {Lemoine-Goumard}, M., {et~al.} 2013, \apj,
  774, 110

\bibitem[{{H.~E.~S.~S.~Collaboration}
  {et~al.}(2012){H.~E.~S.~S.~Collaboration}, {Abramowski}, {Acero},
  {Aharonian}, {Akhperjanian}, {Anton}, {Balzer}, {Barnacka}, {Becherini},
  {Becker}, {Bernl{\"o}h}, {Birsin}, {Biteau}, {Bochow}, {Boisson}, {Bolmont},
  {Bordas}, {Brucker}, {Brun}, {Brun}, {Bulik}, {B{\"u}sching}, {Carrigan},
  {Casanova}, {Cerruti}, {Chadwick}, {Charbonnier}, {Chaves}, {Cheesebrough},
  {Cologna}, {Conrad}, {Dalton}, {Daniel}, {Davids}, {Degrange}, {Deil},
  {Dickinson}, {Djannati-Ata{\"i}}, {Domainko}, {Drury}, {Dubus}, {Dutson},
  {Dyks}, {Dyrda}, {Egberts}, {Eger}, {Espigat}, {Fallon}, {Fegan},
  {Feinstein}, {Fernandes}, {Fiasson}, {Fontaine}, {F{\"o}rster},
  {F{\"u}{\ss}ling}, {Gallant}, {Gast}, {G{\'e}rard}, {Gerbig}, {Giebels},
  {Glicenstein}, {Gl{\"u}ck}, {G{\"o}ring}, {H{\"a}ffner}, {Hague}, {Hahn},
  {Hampf}, {Harris}, {Hauser}, {Heinz}, {Heinzelmann}, {Henri}, {Hermann},
  {Hillert}, {Hinton}, {Hofmann}, {Hofverberg}, {Holler}, {Horns},
  {Jacholkowska}, {de Jager}, {Jahn}, {Jamrozy}, {Jung}, {Kastendieck},
  {Katarzy{\'n}ski}, {Katz}, {Kaufmann}, {Keogh}, {Kh{\'e}lifi}, {Klochkov},
  {Klu{\.z}niak}, {Kneiske}, {Komin}, {Kosack}, {Kossakowski}, {Krayzel},
  {Laffon}, {Lamanna}, {Lenain}, {Lennarz}, {Lohse}, {Lopatin}, {Lu},
  {Marandon}, {Marcowith}, {Masbou}, {Maxted}, {Mayer}, {McComb}, {Medina},
  {M{\'e}hault}, {Moderski}, {Mohamed}, {Moulin}, {Naumann}, {Naumann-Godo},
  {de Naurois}, {Nedbal}, {Nekrassov}, {Nguyen}, {Nicholas}, {Niemiec},
  {Nolan}, {Ohm}, {de O{\~n}a Wilhelmi}, {Opitz}, {Ostrowski}, {Oya}, {Panter},
  {Paz Arribas}, {Pekeur}, {Pelletier}, {Perez}, {Petrucci}, {Peyaud}, {Pita},
  {P{\"u}hlhofer}, {Punch}, {Quirrenbach}, {Raue}, {Rayner}, {Reimer},
  {Reimer}, {Renaud}, {de los Reyes}, {Rieger}, {Ripken}, {Rob}, {Rosier-Lees},
  {Rowell}, {Rudak}, {Rulten}, {Sahakian}, {Sanchez}, {Santangelo},
  {Schlickeiser}, {Schulz}, {Schwanke}, {Schwarzburg}, {Schwemmer}, {Sheidaei},
  {Skilton}, {Sol}, {Spengler}, {Stawarz}, {Steenkamp}, {Stegmann}, {Stinzing},
  {Stycz}, {Sushch}, {Szostek}, {Tavernet}, {Terrier}, {Tluczykont},
  {Valerius}, {van Eldik}, {Vasileiadis}, {Venter}, {Viana}, {Vincent},
  {V{\"o}lk}, {Volpe}, {Vorobiov}, {Vorster}, {Wagner}, {Ward}, {White},
  {Wierzcholska}, {Zacharias}, {Zajczyk}, {Zdziarski}, {Zech}, \&
  {Zechlin}}]{HESS-J1018}
{H.~E.~S.~S.~Collaboration}, {Abramowski}, A., {Acero}, F., {et~al.} 2012,
  \aap, 541, A5

\bibitem[{{H.~E.~S.~S.~Collaboration}
  {et~al.}(2015){H.~E.~S.~S.~Collaboration}, {Abramowski}, {Aharonian}, {Ait
  Benkhali}, {Akhperjanian}, {Ang{\"u}ner}, {Anton}, {Backes}, {Balenderan},
  {Balzer}, \& et~al.}]{HESS2015-J1834-087}
{H.~E.~S.~S.~Collaboration}, {Abramowski}, A., {Aharonian}, F., {et~al.} 2015,
  \aap, 574, A27

\bibitem[{{Hailey} \& {Craig}(1994)}]{1994ApJ...434..635H}
{Hailey}, C.~J., \& {Craig}, W.~W. 1994, \apj, 434, 635

\bibitem[{{Halpern} {et~al.}(2012){Halpern}, {Gotthelf}, \&
  {Camilo}}]{2012ApJ...753L..14H}
{Halpern}, J.~P., {Gotthelf}, E.~V., \& {Camilo}, F. 2012, \apjl, 753, L14

\bibitem[{{Hanabata} {et~al.}(2014){Hanabata}, {Katagiri}, {Hewitt}, {Ballet},
  {Fukazawa}, {Fukui}, {Hayakawa}, {Lemoine-Goumard}, {Pedaletti}, {Strong},
  {Torres}, \& {Yamazaki}}]{Hanabata14}
{Hanabata}, Y., {Katagiri}, H., {Hewitt}, J.~W., {et~al.} 2014, \apj, 786, 145

\bibitem[{{Harrus} {et~al.}(1998){Harrus}, {Hughes}, \&
  {Slane}}]{1998ApJ...499..273H}
{Harrus}, I.~M., {Hughes}, J.~P., \& {Slane}, P.~O. 1998, \apj, 499, 273

\bibitem[{{Hayato} {et~al.}(2010){Hayato}, {Yamaguchi}, {Tamagawa}, {Katsuda},
  {Hwang}, {Hughes}, {Ozawa}, {Bamba}, {Kinugasa}, {Terada}, {Furuzawa},
  {Kunieda}, \& {Makishima}}]{2010ApJ...725..894H}
{Hayato}, A., {Yamaguchi}, H., {Tamagawa}, T., {et~al.} 2010, \apj, 725, 894

\bibitem[{{Helder} \& {Vink}(2008)}]{2008ApJ...686.1094H}
{Helder}, E.~A., \& {Vink}, J. 2008, \apj, 686, 1094

\bibitem[{{Helder} {et~al.}(2012){Helder}, {Vink}, {Bykov}, {Ohira}, {Raymond},
  \& {Terrier}}]{Helder12}
{Helder}, E.~A., {Vink}, J., {Bykov}, A.~M., {et~al.} 2012, \ssr, 173, 369

\bibitem[{{Helene}(1983)}]{Helene83}
{Helene}, O. 1983, Nuclear Instruments and Methods in Physics Research, 212,
  319

\bibitem[{{Hewitt} {et~al.}(2013){Hewitt}, {Acero}, {Brandt}, {Cohen}, {de
  Palma}, {Giordano}, \& {for the Fermi LAT
  Collaboration}}]{Hewitt13-snrCat_ICRC}
{Hewitt}, J.~W., {Acero}, F., {Brandt}, T.~J., {et~al.} 2013, ICRC,
  arXiv:1307.6570

\bibitem[{{Hewitt} {et~al.}(2009{\natexlab{a}}){Hewitt}, {Rho}, {Andersen}, \&
  {Reach}}]{2009ApJ...694.1266H}
{Hewitt}, J.~W., {Rho}, J., {Andersen}, M., \& {Reach}, W.~T.
  2009{\natexlab{a}}, \apj, 694, 1266

\bibitem[{{Hewitt} \& {Yusef-Zadeh}(2009)}]{2009ApJ...694L..16H}
{Hewitt}, J.~W., \& {Yusef-Zadeh}, F. 2009, \apjl, 694, L16

\bibitem[{{Hewitt} {et~al.}(2008){Hewitt}, {Yusef-Zadeh}, \&
  {Wardle}}]{2008ApJ...683..189H}
{Hewitt}, J.~W., {Yusef-Zadeh}, F., \& {Wardle}, M. 2008, \apj, 683, 189

\bibitem[{{Hewitt} {et~al.}(2009{\natexlab{b}}){Hewitt}, {Yusef-Zadeh}, \&
  {Wardle}}]{2009ApJ...706L.270H}
---. 2009{\natexlab{b}}, \apjl, 706, L270

\bibitem[{{Hewitt} {et~al.}(2006){Hewitt}, {Yusef-Zadeh}, {Wardle}, {Roberts},
  \& {Kassim}}]{2006ApJ...652.1288H}
{Hewitt}, J.~W., {Yusef-Zadeh}, F., {Wardle}, M., {Roberts}, D.~A., \&
  {Kassim}, N.~E. 2006, \apj, 652, 1288

\bibitem[{{Hofverberg} {et~al.}(2010){Hofverberg}, {Chaves}, {Fiasson},
  {Kosack}, {M{\'e}hault}, {de On{\~a} Wilhelmi}, \&
  {H.E.S.S.~Collaboration}}]{2011arXiv1104.5119H}
{Hofverberg}, P., {Chaves}, R.~C.~G., {Fiasson}, A., {et~al.} 2010, in 25th
  Texas Symposium on Relativistic Astrophysics, 196

\bibitem[{{Hui} {et~al.}(2012){Hui}, {Seo}, {Huang}, {Trepl}, {Woo}, {Lu},
  {Kong}, \& {Walter}}]{Hui12}
{Hui}, C.~Y., {Seo}, K.~A., {Huang}, R.~H.~H., {et~al.} 2012, \apj, 750, 7

\bibitem[{{Jiang} {et~al.}(2010){Jiang}, {Chen}, {Wang}, {Su}, {Zhou},
  {Safi-Harb}, \& {DeLaney}}]{2010ApJ...712.1147J}
{Jiang}, B., {Chen}, Y., {Wang}, J., {et~al.} 2010, \apj, 712, 1147

\bibitem[{{Johanson} \& {Kerton}(2009)}]{2009AJ....138.1615J}
{Johanson}, A.~K., \& {Kerton}, C.~R. 2009, \aj, 138, 1615

\bibitem[{{Junkes} {et~al.}(1992){Junkes}, {Fuerst}, \&
  {Reich}}]{1992AAS...96....1J}
{Junkes}, N., {Fuerst}, E., \& {Reich}, W. 1992, \aaps, 96, 1

\bibitem[{{Kachelriess} {et~al.}(2014){Kachelriess}, {Moskalenko}, \&
  {Ostapchenko}}]{Kachelriess14-EnhancementFactor}
{Kachelriess}, M., {Moskalenko}, I.~V., \& {Ostapchenko}, S.~S. 2014, \apj,
  789, 136

\bibitem[{{Kamae} {et~al.}(2006){Kamae}, {Karlsson}, {Mizuno}, {Abe}, \&
  {Koi}}]{Kamae06-PPinteractions}
{Kamae}, T., {Karlsson}, N., {Mizuno}, T., {Abe}, T., \& {Koi}, T. 2006, \apj,
  647, 692

\bibitem[{{Kassim} {et~al.}(1993){Kassim}, {Hertz}, \&
  {Weiler}}]{1993ApJ...419..733K}
{Kassim}, N.~E., {Hertz}, P., \& {Weiler}, K.~W. 1993, \apj, 419, 733

\bibitem[{{Kassim} \& {Weiler}(1990)}]{1990Natur.343..146K}
{Kassim}, N.~E., \& {Weiler}, K.~W. 1990, \nat, 343, 146

\bibitem[{{Katagiri} {et~al.}(2011){Katagiri}, {Tibaldo}, {Ballet}, {Giordano},
  {Grenier}, {Porter}, {Roth}, {Tibolla}, {Uchiyama}, \&
  {Yamazaki}}]{Katagiri11-CygLoop}
{Katagiri}, H., {Tibaldo}, L., {Ballet}, J., {et~al.} 2011, \apj, 741, 44

\bibitem[{{Katsuda} {et~al.}(2008){Katsuda}, {Tsunemi}, \&
  {Mori}}]{2008ApJ...678L..35K}
{Katsuda}, S., {Tsunemi}, H., \& {Mori}, K. 2008, \apjl, 678, L35

\bibitem[{{Katsuta} {et~al.}(2012){Katsuta}, {Uchiyama}, {Tanaka}, {Tajima},
  {Bechtol}, {Funk}, {Lande}, {Ballet}, {Hanabata}, {Lemoine-Goumard}, \&
  {Takahashi}}]{Katsuta12-S147}
{Katsuta}, J., {Uchiyama}, Y., {Tanaka}, T., {et~al.} 2012, \apj, 752, 135

\bibitem[{{Kerr} \& {Lynden-Bell}(1986)}]{1986MNRAS.221.1023K}
{Kerr}, F.~J., \& {Lynden-Bell}, D. 1986, \mnras, 221, 1023

\bibitem[{{Kerr}(2010)}]{Kerr10-pointlike}
{Kerr}, M. 2010, PhD thesis, University of Washington

\bibitem[{{Kim} {et~al.}(2010){Kim}, {Min}, {Seon}, {Han}, \&
  {Edelstein}}]{2010ApJ...709..823K}
{Kim}, I.-J., {Min}, K.-W., {Seon}, K.-I., {Han}, W., \& {Edelstein}, J. 2010,
  \apj, 709, 823

\bibitem[{{Koo} \& {Moon}(1997)}]{1997ApJ...475..194K}
{Koo}, B.-C., \& {Moon}, D.-S. 1997, \apj, 475, 194

\bibitem[{{Koralesky} {et~al.}(1998){Koralesky}, {Frail}, {Goss}, {Claussen},
  \& {Green}}]{1998AJ....116.1323K}
{Koralesky}, B., {Frail}, D.~A., {Goss}, W.~M., {Claussen}, M.~J., \& {Green},
  A.~J. 1998, \aj, 116, 1323

\bibitem[{{Kothes} \& {Foster}(2012)}]{2012ApJ...746L...4K}
{Kothes}, R., \& {Foster}, T. 2012, \apjl, 746, L4

\bibitem[{{Kothes} {et~al.}(2003){Kothes}, {Reich}, {Foster}, \&
  {Byun}}]{2003ApJ...588..852K}
{Kothes}, R., {Reich}, W., {Foster}, T., \& {Byun}, D.-Y. 2003, \apj, 588, 852

\bibitem[{{Kothes} {et~al.}(2001){Kothes}, {Uyaniker}, \&
  {Pineault}}]{2001ApJ...560..236K}
{Kothes}, R., {Uyaniker}, B., \& {Pineault}, S. 2001, \apj, 560, 236

\bibitem[{{Krause} {et~al.}(2008){Krause}, {Tanaka}, {Usuda}, {Hattori},
  {Goto}, {Birkmann}, \& {Nomoto}}]{2008Natur.456..617K}
{Krause}, O., {Tanaka}, M., {Usuda}, T., {et~al.} 2008, \nat, 456, 617

\bibitem[{{Ladouceur} \& {Pineault}(2008)}]{2008AA...490..197L}
{Ladouceur}, Y., \& {Pineault}, S. 2008, \aap, 490, 197

\bibitem[{{Lagage} \& {Cesarsky}(1983)}]{Lagage83-Emax}
{Lagage}, P.~O., \& {Cesarsky}, C.~J. 1983, \aap, 125, 249

\bibitem[{{Lande} {et~al.}(2012){Lande}, {Ackermann}, {Allafort}, {Ballet},
  {Bechtol}, {Burnett}, {Cohen-Tanugi}, {Drlica-Wagner}, {Funk}, {Giordano},
  {Grondin}, {Kerr}, \& {Lemoine-Goumard}}]{Lande12-extSrcSearch}
{Lande}, J., {Ackermann}, M., {Allafort}, A., {et~al.} 2012, \apj, 756, 5

\bibitem[{{Landecker} {et~al.}(1989){Landecker}, {Pineault}, {Routledge}, \&
  {Vaneldik}}]{1989MNRAS.237..277L}
{Landecker}, T.~L., {Pineault}, S., {Routledge}, D., \& {Vaneldik}, J.~F. 1989,
  \mnras, 237, 277

\bibitem[{{Lang} {et~al.}(2010){Lang}, {Goss}, {Cyganowski}, \&
  {Clubb}}]{2010ApJS..191..275L}
{Lang}, C.~C., {Goss}, W.~M., {Cyganowski}, C., \& {Clubb}, K.~I. 2010, \apjs,
  191, 275

\bibitem[{{Lazendic} {et~al.}(2004){Lazendic}, {Wardle}, {Burton},
  {Yusef-Zadeh}, {Green}, \& {Whiteoak}}]{2004MNRAS.354..393L}
{Lazendic}, J.~S., {Wardle}, M., {Burton}, M.~G., {et~al.} 2004, \mnras, 354,
  393

\bibitem[{{Leahy} {et~al.}(2014){Leahy}, {Green}, \&
  {Tian}}]{2014MNRAS.438.1813L}
{Leahy}, D., {Green}, K., \& {Tian}, W. 2014, \mnras, 438, 1813

\bibitem[{{Leahy} \& {Tian}(2006)}]{2006AA...451..251L}
{Leahy}, D., \& {Tian}, W. 2006, \aap, 451, 251

\bibitem[{{Leahy} {et~al.}(2013){Leahy}, {Green}, \&
  {Ranasinghe}}]{2013MNRAS.436..968L}
{Leahy}, D.~A., {Green}, K., \& {Ranasinghe}, S. 2013, \mnras, 436, 968

\bibitem[{{Leahy} \& {Ranasinghe}(2012)}]{2012MNRAS.423..718L}
{Leahy}, D.~A., \& {Ranasinghe}, S. 2012, \mnras, 423, 718

\bibitem[{{Leahy} \& {Roger}(1991)}]{1991AJ....101.1033L}
{Leahy}, D.~A., \& {Roger}, R.~S. 1991, \aj, 101, 1033

\bibitem[{{Leahy} {et~al.}(2008){Leahy}, {Tian}, \&
  {Wang}}]{2008AJ....136.1477L}
{Leahy}, D.~A., {Tian}, W., \& {Wang}, Q.~D. 2008, \aj, 136, 1477

\bibitem[{{Leahy} \& {Tian}(2007)}]{2007AA...461.1013L}
{Leahy}, D.~A., \& {Tian}, W.~W. 2007, \aap, 461, 1013

\bibitem[{{Leahy} \& {Tian}(2008{\natexlab{a}})}]{2008AA...480L..25L}
---. 2008{\natexlab{a}}, \aap, 480, L25

\bibitem[{{Leahy} \& {Tian}(2008{\natexlab{b}})}]{2008AJ....135..167L}
---. 2008{\natexlab{b}}, \aj, 135, 167

\bibitem[{{Lee} {et~al.}(2014){Lee}, {Patnaude}, {Ellison}, {Nagataki}, \&
  {Slane}}]{LeeEllison}
{Lee}, S.-H., {Patnaude}, D.~J., {Ellison}, D.~C., {Nagataki}, S., \& {Slane},
  P.~O. 2014, \apj, 791, 97

\bibitem[{{Lemiere} {et~al.}(2009){Lemiere}, {Slane}, {Gaensler}, \&
  {Murray}}]{2009ApJ...706.1269L}
{Lemiere}, A., {Slane}, P., {Gaensler}, B.~M., \& {Murray}, S. 2009, \apj, 706,
  1269

\bibitem[{{Lemoine-Goumard} {et~al.}(2012){Lemoine-Goumard}, {Renaud}, {Vink},
  {Allen}, {Bamba}, {Giordano}, \& {Uchiyama}}]{2012AA...545A..28L}
{Lemoine-Goumard}, M., {Renaud}, M., {Vink}, J., {et~al.} 2012, \aap, 545, A28

\bibitem[{{Li} {et~al.}(2011){Li}, {Chornock}, {Leaman}, {Filippenko},
  {Poznanski}, {Wang}, {Ganeshalingam}, \& {Mannucci}}]{Li11-SNrate}
{Li}, W., {Chornock}, R., {Leaman}, J., {et~al.} 2011, \mnras, 412, 1473

\bibitem[{{Livingstone} {et~al.}(2006){Livingstone}, {Kaspi}, {Gotthelf}, \&
  {Kuiper}}]{2006ApJ...647.1286L}
{Livingstone}, M.~A., {Kaspi}, V.~M., {Gotthelf}, E.~V., \& {Kuiper}, L. 2006,
  \apj, 647, 1286

\bibitem[{{Lopez} {et~al.}(2011){Lopez}, {Ramirez-Ruiz}, {Huppenkothen},
  {Badenes}, \& {Pooley}}]{2011ApJ...732..114L}
{Lopez}, L.~A., {Ramirez-Ruiz}, E., {Huppenkothen}, D., {Badenes}, C., \&
  {Pooley}, D.~A. 2011, \apj, 732, 114

\bibitem[{{Lorimer} {et~al.}(2006){Lorimer}, {Faulkner}, {Lyne}, {Manchester},
  {Kramer}, {McLaughlin}, {Hobbs}, {Possenti}, {Stairs}, {Camilo}, {Burgay},
  {D'Amico}, {Corongiu}, \& {Crawford}}]{Lorimer06-PSRpop}
{Lorimer}, D.~R., {Faulkner}, A.~J., {Lyne}, A.~G., {et~al.} 2006, \mnras, 372,
  777

\bibitem[{{Lozinskaya} {et~al.}(1993){Lozinskaya}, {Sitnik}, \&
  {Pravdikova}}]{1993ARep...37..240L}
{Lozinskaya}, T.~A., {Sitnik}, T.~G., \& {Pravdikova}, V.~V. 1993, Astronomy
  Reports, 37, 240

\bibitem[{{Maeda} {et~al.}(2009){Maeda}, {Uchiyama}, {Bamba}, {Kosugi},
  {Tsunemi}, {Helder}, {Vink}, {Kodaka}, {Terada}, {Fukazawa}, {Hiraga},
  {Hughes}, {Kokubun}, {Kouzu}, {Matsumoto}, {Miyata}, {Nakamura}, {Okada},
  {Someya}, {Tamagawa}, {Tamura}, {Totsuka}, {Tsuboi}, {Ezoe}, {Holt},
  {Ishida}, {Kamae}, {Petre}, \& {Takahashi}}]{2009PASJ...61.1217M}
{Maeda}, Y., {Uchiyama}, Y., {Bamba}, A., {et~al.} 2009, \pasj, 61, 1217

\bibitem[{{Mandelartz} \& {Becker Tjus}(2015)}]{Mandelartz13-Neutrino}
{Mandelartz}, M., \& {Becker Tjus}, J. 2015, Astroparticle Physics, 65, 80

\bibitem[{{Massaro} {et~al.}(2004){Massaro}, {Perri}, {Giommi}, \&
  {Nesci}}]{Massaro04-logp}
{Massaro}, E., {Perri}, M., {Giommi}, P., \& {Nesci}, R. 2004, \aap, 413, 489

\bibitem[{{Matheson} \& {Safi-Harb}(2010)}]{2010ApJ...724..572M}
{Matheson}, H., \& {Safi-Harb}, S. 2010, \apj, 724, 572

\bibitem[{{Mattox} {et~al.}(1996){Mattox}, {Bertsch}, {Chiang}, {Dingus},
  {Digel}, {Esposito}, {Fierro}, {Hartman}, {Hunter}, {Kanbach}, {Kniffen},
  {Lin}, {Macomb}, {Mayer-Hasselwander}, {Michelson}, {von Montigny},
  {Mukherjee}, {Nolan}, {Ramanamurthy}, {Schneid}, {Sreekumar}, {Thompson}, \&
  {Willis}}]{Mattox96-Likelihood}
{Mattox}, J.~R., {Bertsch}, D.~L., {Chiang}, J., {et~al.} 1996, \apj, 461, 396

\bibitem[{{McClure-Griffiths} {et~al.}(2001){McClure-Griffiths}, {Green},
  {Dickey}, {Gaensler}, {Haynes}, \& {Wieringa}}]{2001ApJ...551..394M}
{McClure-Griffiths}, N.~M., {Green}, A.~J., {Dickey}, J.~M., {et~al.} 2001,
  \apj, 551, 394

\bibitem[{{Miller} {et~al.}(2011){Miller}, {Reynolds}, {Maitra}, {Gultekin},
  {Gehrels}, {Kennea}, {Siegel}, {Gelbord}, \& {Kuin}}]{Miller11}
{Miller}, J.~M., {Reynolds}, M.~R., {Maitra}, D., {et~al.} 2011, The
  Astronomer's Telegram, 3415, 1

\bibitem[{{Milne} \& {Haynes}(1994)}]{1994MNRAS.270..106M}
{Milne}, D.~K., \& {Haynes}, R.~F. 1994, \mnras, 270, 106

\bibitem[{{Misanovic} {et~al.}(2010){Misanovic}, {Kargaltsev}, \&
  {Pavlov}}]{2010ApJ...725..931M}
{Misanovic}, Z., {Kargaltsev}, O., \& {Pavlov}, G.~G. 2010, \apj, 725, 931

\bibitem[{{Mori}(2009)}]{Mori09-EnhancementFactor}
{Mori}, M. 2009, Astroparticle Physics, 31, 341

\bibitem[{{Moriguchi} {et~al.}(2005){Moriguchi}, {Tamura}, {Tawara}, {Sasago},
  {Yamaoka}, {Onishi}, \& {Fukui}}]{2005ApJ...631..947M}
{Moriguchi}, Y., {Tamura}, K., {Tawara}, Y., {et~al.} 2005, \apj, 631, 947

\bibitem[{{Moriguchi} {et~al.}(2001){Moriguchi}, {Yamaguchi}, {Onishi},
  {Mizuno}, \& {Fukui}}]{2001PASJ...53.1025M}
{Moriguchi}, Y., {Yamaguchi}, N., {Onishi}, T., {Mizuno}, A., \& {Fukui}, Y.
  2001, \pasj, 53, 1025

\bibitem[{{Ng} {et~al.}(2007){Ng}, {Romani}, {Brisken}, {Chatterjee}, \&
  {Kramer}}]{2007ApJ...654..487N}
{Ng}, C.-Y., {Romani}, R.~W., {Brisken}, W.~F., {Chatterjee}, S., \& {Kramer},
  M. 2007, \apj, 654, 487

\bibitem[{{Nikoli{\'c}} {et~al.}(2013){Nikoli{\'c}}, {van de Ven}, {Heng},
  {Kupko}, {Husemann}, {Raymond}, {Hughes}, \&
  {Falc{\'o}n-Barroso}}]{2013Sci...340...45N}
{Nikoli{\'c}}, S., {van de Ven}, G., {Heng}, K., {et~al.} 2013, Science, 340,
  45

\bibitem[{{Nolan} {et~al.}(2012){Nolan}, {Abdo}, {Ackermann}, {Ajello},
  {Allafort}, {Antolini}, {Atwood}, {Axelsson}, {Baldini}, {Ballet}, \&
  et~al.}]{Nolan12-2FGL}
{Nolan}, P.~L., {Abdo}, A.~A., {Ackermann}, M., {et~al.} 2012, \apjs, 199, 31

\bibitem[{{Nynka} {et~al.}(2014){Nynka}, {Hailey}, {Reynolds}, {An},
  {Baganoff}, {Boggs}, {Christensen}, {Craig}, {Gotthelf}, {Grefenstette},
  {Harrison}, {Krivonos}, {Madsen}, {Mori}, {Perez}, {Stern}, {Wik}, {Zhang},
  \& {Zoglauer}}]{2014ApJ...789...72N}
{Nynka}, M., {Hailey}, C.~J., {Reynolds}, S.~P., {et~al.} 2014, \apj, 789, 72

\bibitem[{{Odegard}(1986)}]{1986ApJ...301..813O}
{Odegard}, N. 1986, \apj, 301, 813

\bibitem[{{Olive} \& {Particle Data Group}(2014)}]{Olive14-PDG}
{Olive}, K.~A., \& {Particle Data Group}. 2014, Chinese Physics C, 38, 090001

\bibitem[{{Pannuti} {et~al.}(2010){Pannuti}, {Allen}, {Filipovi{\'c}}, {De
  Horta}, {Stupar}, \& {Agrawal}}]{2010ApJ...721.1492P}
{Pannuti}, T.~G., {Allen}, G.~E., {Filipovi{\'c}}, M.~D., {et~al.} 2010, \apj,
  721, 1492

\bibitem[{{Pannuti} {et~al.}(2003){Pannuti}, {Allen}, {Houck}, \&
  {Sturner}}]{2003ApJ...593..377P}
{Pannuti}, T.~G., {Allen}, G.~E., {Houck}, J.~C., \& {Sturner}, S.~J. 2003,
  \apj, 593, 377

\bibitem[{{Paron} {et~al.}(2008){Paron}, {Dubner}, {Reynoso}, \&
  {Rubio}}]{2008AA...480..439P}
{Paron}, S., {Dubner}, G., {Reynoso}, E., \& {Rubio}, M. 2008, \aap, 480, 439

\bibitem[{{Paron} \& {Giacani}(2010)}]{Paron10-G35.6}
{Paron}, S., \& {Giacani}, E. 2010, \aap, 509, L4

\bibitem[{{Paron} {et~al.}(2009){Paron}, {Ortega}, {Rubio}, \&
  {Dubner}}]{2009AA...498..445P}
{Paron}, S., {Ortega}, M.~E., {Rubio}, M., \& {Dubner}, G. 2009, \aap, 498, 445

\bibitem[{{Paron} {et~al.}(2006){Paron}, {Reynoso}, {Purcell}, {Dubner}, \&
  {Green}}]{2006PASA...23...69P}
{Paron}, S.~A., {Reynoso}, E.~M., {Purcell}, C., {Dubner}, G.~M., \& {Green},
  A. 2006, \pasa, 23, 69

\bibitem[{{Pauls}(1977)}]{1977AA....59L..13P}
{Pauls}, T. 1977, \aap, 59, L13

\bibitem[{{Petriella} {et~al.}(2012){Petriella}, {Paron}, \&
  {Giacani}}]{2012A&A...538A..14P}
{Petriella}, A., {Paron}, S.~A., \& {Giacani}, E.~B. 2012, \aap, 538, A14

\bibitem[{{Pfeffermann} {et~al.}(1991){Pfeffermann}, {Aschenbach}, \&
  {Predehl}}]{1991AA...246L..28P}
{Pfeffermann}, E., {Aschenbach}, B., \& {Predehl}, P. 1991, \aap, 246, L28

\bibitem[{{Pineault} {et~al.}(1993){Pineault}, {Landecker}, {Madore}, \&
  {Gaumont-Guay}}]{1993AJ....105.1060P}
{Pineault}, S., {Landecker}, T.~L., {Madore}, B., \& {Gaumont-Guay}, S. 1993,
  \aj, 105, 1060

\bibitem[{{Planck and Fermi Collaborations} {et~al.}(2014){Planck and Fermi
  Collaborations}, {Ade}, {Aghanim}, {Aniano}, {Arnaud}, {Ashdown}, {Aumont},
  {Baccigalupi}, {Banday}, {Barreiro}, {Bartolo}, {Battaner}, {Benabed},
  {Benoit-Levy}, {Bernard}, {Bersanelli}, {Bielewicz}, {Bonaldi}, {Bonavera},
  {Bond}, {Borrill}, {Bouchet}, {Boulanger}, {Burigana}, {Butler}, {Calabrese},
  {Cardoso}, {Casandjian}, {Catalano}, {Chamballu}, {Chiang}, {Christensen},
  {Colombo}, {Combet}, {Couchot}, {Crill}, {Curto}, {Cuttaia}, {Danese},
  {Davies}, {Davis}, {de Bernardis}, {de Rosa}, {de Zotti}, {Delabrouille},
  {Desert}, {Dickinson}, {Diego}, {Digel}, {Dole}, {Donzelli}, {Dore},
  {Douspis}, {Ducout}, {Dupac}, {Efstathiou}, {Elsner}, {Ensslin}, {Eriksen},
  {Falgarone}, {Finelli}, {Forni}, {Frailis}, {Fraisse}, {Franceschi},
  {Frejsel}, {Fukui}, {Galeotta}, {Galli}, {Ganga}, {Ghosh}, {Giard},
  {Gjerlow}, {Gonzalez-Nuevo}, {Gorski}, {Gregorio}, {Grenier}, {Gruppuso},
  {Hansen}, {Hanson}, {Harrison}, {Henrot-Versille}, {Hernandez-Monteagudo},
  {Herranz}, {Hildebrandt}, {Hivon}, {Holmes}, {Hovest}, {Huffenberger},
  {Hurier}, {Jaffe}, {Jaffe}, {Jones}, {Keihanen}, {Keskitalo}, {Kisner},
  {Kneissl}, {Knoche}, {Kunz}, {Kurki-Suonio}, {Lagache}, {Lamarre}, {Lasenby},
  {Lattanzi}, {Lawrence}, {Leonardi}, {Levrier}, {Liguori}, {Lilje},
  {Linden-Vornle}, {Lopez-Caniego}, {Lubin}, {Macias-Perez}, {Maffei}, {Maino},
  {Mandolesi}, {Maris}, {Marshall}, {Martin}, {Martinez-Gonzalez}, {Masi},
  {Matarrese}, {Mazzotta}, {Melchiorri}, {Mendes}, {Mennella}, {Migliaccio},
  {Miville-Deschenes}, {Moneti}, {Montier}, {Morgante}, {Mortlock}, {Munshi},
  {Murphy}, {Naselsky}, {Natoli}, {Norgaard-Nielsen}, {Novikov}, {Novikov},
  {Oxborrow}, {Pagano}, {Pajot}, {Paladini}, {Paoletti}, {Pasian}, {Perdereau},
  {Perotto}, {Perrotta}, {Pettorino}, {Piacentini}, {Piat}, {Plaszczynski},
  {Pointecouteau}, {Polenta}, {Popa}, {Pratt}, {Prunet}, {Puget}, {Rachen},
  {Reach}, {Rebolo}, {Reinecke}, {Remazeilles}, {Renault}, {Ristorcelli},
  {Rocha}, {Roudier}, {Rusholme}, {Sandri}, {Santos}, {Scott}, {Spencer},
  {Stolyarov}, {Strong}, {Sudiwala}, {Sunyaev}, {Sutton}, {Suur-Uski},
  {Sygnet}, {Tauber}, {Terenzi}, {Tibaldo}, {Toffolatti}, {Tomasi}, {Tristram},
  {Tucci}, {Umana}, {Valenziano}, {Valiviita}, {Van Tent}, {Vielva}, {Villa},
  {Wade}, {Wandelt}, {Wehus}, {Yvon}, {Zacchei}, \&
  {Zonca}}]{2014arXiv1409.3268P}
{Planck and Fermi Collaborations}, {Ade}, P.~A.~R., {Aghanim}, N., {et~al.}
  2014, A\&A, accepted, arXiv:1409.3268

\bibitem[{{Plucinsky}(1998)}]{1998MmSAI..69..939P}
{Plucinsky}, P.~P. 1998, \memsai, 69, 939

\bibitem[{{Porter} {et~al.}(2008){Porter}, {Moskalenko}, {Strong}, {Orlando},
  \& {Bouchet}}]{Porter08-ISRF}
{Porter}, T.~A., {Moskalenko}, I.~V., {Strong}, A.~W., {Orlando}, E., \&
  {Bouchet}, L. 2008, \apj, 682, 400

\bibitem[{{Prinz} \& {Becker}(2012)}]{2012AA...544A...7P}
{Prinz}, T., \& {Becker}, W. 2012, \aap, 544, A7

\bibitem[{{Ptuskin} \& {Zirakashvili}(2003)}]{Ptuskin03-Emax}
{Ptuskin}, V.~S., \& {Zirakashvili}, V.~N. 2003, \aap, 403, 1

\bibitem[{{Ptuskin} \& {Zirakashvili}(2005)}]{Ptuskin05-Emax}
---. 2005, \aap, 429, 755

\bibitem[{{Radhakrishnan} {et~al.}(1972){Radhakrishnan}, {Goss}, {Murray}, \&
  {Brooks}}]{1972ApJS...24...49R}
{Radhakrishnan}, V., {Goss}, W.~M., {Murray}, J.~D., \& {Brooks}, J.~W. 1972,
  \apjs, 24, 49

\bibitem[{{Rakowski} {et~al.}(2006){Rakowski}, {Badenes}, {Gaensler},
  {Gelfand}, {Hughes}, \& {Slane}}]{2006ApJ...646..982R}
{Rakowski}, C.~E., {Badenes}, C., {Gaensler}, B.~M., {et~al.} 2006, \apj, 646,
  982

\bibitem[{{Rakowski} {et~al.}(2001){Rakowski}, {Hughes}, \&
  {Slane}}]{2001ApJ...548..258R}
{Rakowski}, C.~E., {Hughes}, J.~P., \& {Slane}, P. 2001, \apj, 548, 258

\bibitem[{{Reach} \& {Rho}(2000)}]{Reach00-IR-MC-SNRs}
{Reach}, W.~T., \& {Rho}, J. 2000, \apj, 544, 843

\bibitem[{{Reach} {et~al.}(2005){Reach}, {Rho}, \&
  {Jarrett}}]{2005ApJ...618..297R}
{Reach}, W.~T., {Rho}, J., \& {Jarrett}, T.~H. 2005, \apj, 618, 297

\bibitem[{{Reed} {et~al.}(1995){Reed}, {Hester}, {Fabian}, \&
  {Winkler}}]{1995ApJ...440..706R}
{Reed}, J.~E., {Hester}, J.~J., {Fabian}, A.~C., \& {Winkler}, P.~F. 1995,
  \apj, 440, 706

\bibitem[{{Reich} {et~al.}(1984){Reich}, {Furst}, \&
  {Sofue}}]{1984A&A...133L...4R}
{Reich}, W., {Furst}, E., \& {Sofue}, Y. 1984, \aap, 133, L4

\bibitem[{{Reich} {et~al.}(2003){Reich}, {Zhang}, \& {F{\"u}rst}}]{Reich03}
{Reich}, W., {Zhang}, X., \& {F{\"u}rst}, E. 2003, \aap, 408, 961

\bibitem[{{Renaud} \& {CTA Consortium}(2011)}]{Renaud11-SNRPWNcta}
{Renaud}, M., \& {CTA Consortium}. 2011, \memsai, 82, 726

\bibitem[{{Renaud} {et~al.}(2006){Renaud}, {Vink}, {Decourchelle}, {Lebrun},
  {den Hartog}, {Terrier}, {Couvreur}, {Kn{\"o}dlseder}, {Martin}, {Prantzos},
  {Bykov}, \& {Bloemen}}]{2006ApJ...647L..41R}
{Renaud}, M., {Vink}, J., {Decourchelle}, A., {et~al.} 2006, \apjl, 647, L41

\bibitem[{{Reynolds} {et~al.}(2012){Reynolds}, {Miller}, {Maitra}, {Gultekin},
  {Gehrels}, {Kennea}, {Siegel}, {Gelbord}, \& {Kuin}}]{Reynolds12}
{Reynolds}, M.~T., {Miller}, J.~M., {Maitra}, D., {et~al.} 2012, The
  Astronomer's Telegram, 3963, 1

\bibitem[{{Reynolds} {et~al.}(2013){Reynolds}, {Loi}, {Murphy}, {Miller},
  {Maitra}, {G{\"u}ltekin}, {Gehrels}, {Kennea}, {Siegel}, {Gelbord}, {Kuin},
  {Moss}, {Reeves}, {Robbins}, {Gaensler}, {Reis}, \& {Petre}}]{Reynolds13}
{Reynolds}, M.~T., {Loi}, S.~T., {Murphy}, T., {et~al.} 2013, \apj, 766, 112

\bibitem[{{Reynolds} {et~al.}(2008){Reynolds}, {Borkowski}, {Green}, {Hwang},
  {Harrus}, \& {Petre}}]{2008ApJ...680L..41R}
{Reynolds}, S.~P., {Borkowski}, K.~J., {Green}, D.~A., {et~al.} 2008, \apjl,
  680, L41

\bibitem[{{Reynoso} \& {Goss}(1999)}]{1999AJ....118..926R}
{Reynoso}, E.~M., \& {Goss}, W.~M. 1999, \aj, 118, 926

\bibitem[{{Reynoso} {et~al.}(2004){Reynoso}, {Green}, {Johnston}, {Goss},
  {Dubner}, \& {Giacani}}]{2004PASA...21...82R}
{Reynoso}, E.~M., {Green}, A.~J., {Johnston}, S., {et~al.} 2004, \pasa, 21, 82

\bibitem[{{Reynoso} {et~al.}(2006){Reynoso}, {Johnston}, {Green}, \&
  {Koribalski}}]{2006MNRAS.369..416R}
{Reynoso}, E.~M., {Johnston}, S., {Green}, A.~J., \& {Koribalski}, B.~S. 2006,
  \mnras, 369, 416

\bibitem[{{Reynoso} \& {Mangum}(2000)}]{2000ApJ...545..874R}
{Reynoso}, E.~M., \& {Mangum}, J.~G. 2000, \apj, 545, 874

\bibitem[{{Rosado} {et~al.}(1996){Rosado}, {Ambrocio-Cruz}, {Le Coarer}, \&
  {Marcelin}}]{1996AA...315..243R}
{Rosado}, M., {Ambrocio-Cruz}, P., {Le Coarer}, E., \& {Marcelin}, M. 1996,
  \aap, 315, 243

\bibitem[{{Routledge} {et~al.}(1991){Routledge}, {Dewdney}, {Landecker}, \&
  {Vaneldik}}]{1991AA...247..529R}
{Routledge}, D., {Dewdney}, P.~E., {Landecker}, T.~L., \& {Vaneldik}, J.~F.
  1991, \aap, 247, 529

\bibitem[{{Ruiz} \& {May}(1986)}]{1986ApJ...309..667R}
{Ruiz}, M.~T., \& {May}, J. 1986, \apj, 309, 667

\bibitem[{{Sallmen} \& {Welsh}(2004)}]{2004AA...426..555S}
{Sallmen}, S., \& {Welsh}, B.~Y. 2004, \aap, 426, 555

\bibitem[{{Sankrit} {et~al.}(2005){Sankrit}, {Blair}, {Delaney}, {Rudnick},
  {Harrus}, \& {Ennis}}]{2005AdSpR..35.1027S}
{Sankrit}, R., {Blair}, W.~P., {Delaney}, T., {et~al.} 2005, Advances in Space
  Research, 35, 1027

\bibitem[{{Schwentker}(1994)}]{1994AA...286L..47S}
{Schwentker}, O. 1994, \aap, 286, L47

\bibitem[{{Sedov}(1959)}]{1959sdmm.book.....S}
{Sedov}, L.~I. 1959, {Similarity and Dimensional Methods in Mechanics}

\bibitem[{{Seward}(1990)}]{Seward90}
{Seward}, F.~D. 1990, \apjs, 73, 781

\bibitem[{{Seward} {et~al.}(1995){Seward}, {Dame}, {Fesen}, \&
  {Aschenbach}}]{1995ApJ...449..681S}
{Seward}, F.~D., {Dame}, T.~M., {Fesen}, R.~A., \& {Aschenbach}, B. 1995, \apj,
  449, 681

\bibitem[{{Slane} {et~al.}(2014){Slane}, {Bykov}, {Ellison}, {Dubner}, \&
  {Castro}}]{Slane14-GeVSNRs}
{Slane}, P., {Bykov}, A., {Ellison}, D.~C., {Dubner}, G., \& {Castro}, D. 2014,
  Space Science Reviews, 1

\bibitem[{{Slane} {et~al.}(2002){Slane}, {Smith}, {Hughes}, \&
  {Petre}}]{2002ApJ...564..284S}
{Slane}, P., {Smith}, R.~K., {Hughes}, J.~P., \& {Petre}, R. 2002, \apj, 564,
  284

\bibitem[{{Sollerman} {et~al.}(2003){Sollerman}, {Ghavamian}, {Lundqvist}, \&
  {Smith}}]{2003AA...407..249S}
{Sollerman}, J., {Ghavamian}, P., {Lundqvist}, P., \& {Smith}, R.~C. 2003,
  \aap, 407, 249

\bibitem[{{Stecker}(1971)}]{Stecker71}
{Stecker}, F.~W. 1971, NASA Special Publication, 249

\bibitem[{{Stewart} {et~al.}(1993){Stewart}, {Caswell}, {Haynes}, \&
  {Nelson}}]{1993MNRAS.261..593S}
{Stewart}, R.~T., {Caswell}, J.~L., {Haynes}, R.~F., \& {Nelson}, G.~J. 1993,
  \mnras, 261, 593

\bibitem[{{Strong} {et~al.}(2010){Strong}, {Porter}, {Digel},
  {J{\'o}hannesson}, {Martin}, {Moskalenko}, {Murphy}, \&
  {Orlando}}]{Strong10-CREnergyBudget}
{Strong}, A.~W., {Porter}, T.~A., {Digel}, S.~W., {et~al.} 2010, \apjl, 722,
  L58

\bibitem[{{Stupar} \& {Parker}(2012)}]{Stupar12}
{Stupar}, M., \& {Parker}, Q.~A. 2012, \mnras, 419, 1413

\bibitem[{{Stupar} {et~al.}(2007){Stupar}, {Parker}, \&
  {Filipovi{\'c}}}]{2007MNRAS.374.1441S}
{Stupar}, M., {Parker}, Q.~A., \& {Filipovi{\'c}}, M.~D. 2007, \mnras, 374,
  1441

\bibitem[{{Sturner} \& {Dermer}(1995)}]{Sturner95}
{Sturner}, S.~J., \& {Dermer}, C.~D. 1995, \aap, 293, L17

\bibitem[{{Su} {et~al.}(2010){Su}, {Slatyer}, \&
  {Finkbeiner}}]{Su10-FermiBubbles}
{Su}, M., {Slatyer}, T.~R., \& {Finkbeiner}, D.~P. 2010, \apj, 724, 1044

\bibitem[{{Su} {et~al.}(2011){Su}, {Chen}, {Yang}, {Koo}, {Zhou}, {Lu},
  {Jeong}, \& {DeLaney}}]{2011ApJ...727...43S}
{Su}, Y., {Chen}, Y., {Yang}, J., {et~al.} 2011, \apj, 727, 43

\bibitem[{{Sun} {et~al.}(1999){Sun}, {Wang}, \& {Chen}}]{1999ApJ...511..274S}
{Sun}, M., {Wang}, Z.-r., \& {Chen}, Y. 1999, \apj, 511, 274

\bibitem[{{Sun} {et~al.}(2011){Sun}, {Reich}, {Reich}, {Xiao}, {Gao}, \&
  {Han}}]{2011AA...536A..83S}
{Sun}, X.~H., {Reich}, P., {Reich}, W., {et~al.} 2011, \aap, 536, A83

\bibitem[{{Tam} {et~al.}(2010){Tam}, {Wagner}, {Tibolla}, \&
  {Chaves}}]{Tam10_VHEforFermi}
{Tam}, P.~H.~T., {Wagner}, S.~J., {Tibolla}, O., \& {Chaves}, R.~C.~G. 2010,
  \aap, 518, A8

\bibitem[{{Taylor}(1950)}]{1950RSPSA.201..159T}
{Taylor}, G. 1950, Royal Society of London Proceedings Series A, 201, 159

\bibitem[{{Temim} {et~al.}(2013){Temim}, {Slane}, {Castro}, {Plucinsky},
  {Gelfand}, \& {Dickel}}]{2013ApJ...768...61T}
{Temim}, T., {Slane}, P., {Castro}, D., {et~al.} 2013, \apj, 768, 61

\bibitem[{{Thompson} {et~al.}(2012){Thompson}, {Baldini}, \&
  {Uchiyama}}]{Thompson12-FermiCRsReview}
{Thompson}, D.~J., {Baldini}, L., \& {Uchiyama}, Y. 2012, Astroparticle
  Physics, 39, 22

\bibitem[{{Tian} {et~al.}(2007{\natexlab{a}}){Tian}, {Haverkorn}, \&
  {Zhang}}]{2007MNRAS.378.1283T}
{Tian}, W.~W., {Haverkorn}, M., \& {Zhang}, H.~Y. 2007{\natexlab{a}}, \mnras,
  378, 1283

\bibitem[{{Tian} \& {Leahy}(2008{\natexlab{a}})}]{2008ApJ...677..292T}
{Tian}, W.~W., \& {Leahy}, D.~A. 2008{\natexlab{a}}, \apj, 677, 292

\bibitem[{{Tian} \& {Leahy}(2008{\natexlab{b}})}]{2008MNRAS.391L..54T}
---. 2008{\natexlab{b}}, \mnras, 391, L54

\bibitem[{{Tian} \& {Leahy}(2011)}]{2011ApJ...729L..15T}
---. 2011, \apjl, 729, L15

\bibitem[{{Tian} \& {Leahy}(2012)}]{2012MNRAS.421.2593T}
---. 2012, \mnras, 421, 2593

\bibitem[{{Tian} \& {Leahy}(2013)}]{2013ApJ...769L..17T}
---. 2013, \apjl, 769, L17

\bibitem[{{Tian} \& {Leahy}(2014)}]{2014ApJ...783L...2T}
---. 2014, \apjl, 783, L2

\bibitem[{{Tian} {et~al.}(2008){Tian}, {Leahy}, {Haverkorn}, \&
  {Jiang}}]{2008ApJ...679L..85T}
{Tian}, W.~W., {Leahy}, D.~A., {Haverkorn}, M., \& {Jiang}, B. 2008, \apjl,
  679, L85

\bibitem[{{Tian} {et~al.}(2007{\natexlab{b}}){Tian}, {Leahy}, \&
  {Wang}}]{2007AA...474..541T}
{Tian}, W.~W., {Leahy}, D.~A., \& {Wang}, Q.~D. 2007{\natexlab{b}}, \aap, 474,
  541

\bibitem[{{Tian} {et~al.}(2007{\natexlab{c}}){Tian}, {Li}, {Leahy}, \&
  {Wang}}]{2007ApJ...657L..25T}
{Tian}, W.~W., {Li}, Z., {Leahy}, D.~A., \& {Wang}, Q.~D. 2007{\natexlab{c}},
  \apjl, 657, L25

\bibitem[{{Tibolla}(2009)}]{Tibolla09_HESSUnIds}
{Tibolla}, O. 2009, in American Institute of Physics Conference Series, Vol.
  1112, American Institute of Physics Conference Series, ed. D.~{Bastieri} \&
  R.~{Rando}, 211--222

\bibitem[{{Torii} {et~al.}(2006){Torii}, {Uchida}, {Hasuike}, {Tsunemi},
  {Yamaguchi}, \& {Shibata}}]{2006PASJ...58L..11T}
{Torii}, K., {Uchida}, H., {Hasuike}, K., {et~al.} 2006, \pasj, 58, L11

\bibitem[{{Trimble}(1973)}]{1973PASP...85..579T}
{Trimble}, V. 1973, \pasp, 85, 579

\bibitem[{{Uchida} {et~al.}(2012){Uchida}, {Tsunemi}, {Katsuda}, {Mori},
  {Petre}, \& {Yamaguchi}}]{2012PASJ...64...61U}
{Uchida}, H., {Tsunemi}, H., {Katsuda}, S., {et~al.} 2012, \pasj, 64, 61

\bibitem[{{Uchiyama} {et~al.}(2010){Uchiyama}, {Blandford}, {Funk}, {Tajima},
  \& {Tanaka}}]{Uchiyama10-crushedClouds}
{Uchiyama}, Y., {Blandford}, R.~D., {Funk}, S., {Tajima}, H., \& {Tanaka}, T.
  2010, \apjl, 723, L122

\bibitem[{{Ueno} {et~al.}(2003){Ueno}, {Bamba}, {Koyama}, \&
  {Ebisawa}}]{2003ApJ...588..338U}
{Ueno}, M., {Bamba}, A., {Koyama}, K., \& {Ebisawa}, K. 2003, \apj, 588, 338

\bibitem[{{Vel{\'a}zquez} {et~al.}(2002){Vel{\'a}zquez}, {Dubner}, {Goss}, \&
  {Green}}]{2002AJ....124.2145V}
{Vel{\'a}zquez}, P.~F., {Dubner}, G.~M., {Goss}, W.~M., \& {Green}, A.~J. 2002,
  \aj, 124, 2145

\bibitem[{{Vink}(2004)}]{2004ApJ...604..693V}
{Vink}, J. 2004, \apj, 604, 693

\bibitem[{{Vink}(2012)}]{Vink12}
---. 2012, \aapr, 20, 49

\bibitem[{{Walker} \& {Zealey}(2001)}]{2001MNRAS.325..287W}
{Walker}, A.~J., \& {Zealey}, W.~J. 2001, \mnras, 325, 287

\bibitem[{{Welsh} {et~al.}(2003){Welsh}, {Sallmen}, {Jelinsky}, \&
  {Lallement}}]{2003AA...403..605W}
{Welsh}, B.~Y., {Sallmen}, S., {Jelinsky}, S., \& {Lallement}, R. 2003, \aap,
  403, 605

\bibitem[{{Westerhoff}(2014)}]{Westerhoff14_HAWC}
{Westerhoff}, S. 2014, Advances in Space Research, 53, 1492

\bibitem[{{White} {et~al.}(1987){White}, {Rainey}, {Hayashi}, \&
  {Kaifu}}]{White87}
{White}, G.~J., {Rainey}, R., {Hayashi}, S.~S., \& {Kaifu}, N. 1987, \aap, 173,
  337

\bibitem[{{Wolleben}(2007)}]{Wolleben07-LoopI}
{Wolleben}, M. 2007, \apj, 664, 349

\bibitem[{{Xiao} \& {Zhu}(2012)}]{2012AA...545A..86X}
{Xiao}, L., \& {Zhu}, M. 2012, \aap, 545, A86

\bibitem[{{Xilouris} {et~al.}(1993){Xilouris}, {Papamastorakis}, {Paleologou},
  {Andredakis}, \& {Haerendel}}]{1993AA...270..393X}
{Xilouris}, K.~M., {Papamastorakis}, J., {Paleologou}, E.~V., {Andredakis}, Y.,
  \& {Haerendel}, G. 1993, \aap, 270, 393

\bibitem[{{Yamaguchi} {et~al.}(2004){Yamaguchi}, {Ueno}, {Koyama}, {Bamba}, \&
  {Yamauchi}}]{2004PASJ...56.1059Y}
{Yamaguchi}, H., {Ueno}, M., {Koyama}, K., {Bamba}, A., \& {Yamauchi}, S. 2004,
  \pasj, 56, 1059

\bibitem[{{Yar-Uyaniker} {et~al.}(2004){Yar-Uyaniker}, {Uyaniker}, \&
  {Kothes}}]{2004ApJ...616..247Y}
{Yar-Uyaniker}, A., {Uyaniker}, B., \& {Kothes}, R. 2004, \apj, 616, 247

\bibitem[{{Yatsu} {et~al.}(2013){Yatsu}, {Asano}, {Kawai}, {Yano}, \&
  {Nakamori}}]{2013ApJ...773...25Y}
{Yatsu}, Y., {Asano}, K., {Kawai}, N., {Yano}, Y., \& {Nakamori}, T. 2013,
  \apj, 773, 25

\bibitem[{{Yuan} {et~al.}(2013){Yuan}, {Funk}, {J{\'o}hannesson}, {Lande},
  {Tibaldo}, \& {Uchiyama}}]{Yuan13-CasA}
{Yuan}, Y., {Funk}, S., {J{\'o}hannesson}, G., {et~al.} 2013, \apj, 779, 117

\bibitem[{{Yusef-Zadeh} {et~al.}(1999){Yusef-Zadeh}, {Goss}, {Roberts},
  {Robinson}, \& {Frail}}]{1999ApJ...527..172Y}
{Yusef-Zadeh}, F., {Goss}, W.~M., {Roberts}, D.~A., {Robinson}, B., \& {Frail},
  D.~A. 1999, \apj, 527, 172

\bibitem[{{Yusef-Zadeh} {et~al.}(1995){Yusef-Zadeh}, {Uchida}, \&
  {Roberts}}]{1995Sci...270.1801Y}
{Yusef-Zadeh}, F., {Uchida}, K.~I., \& {Roberts}, D. 1995, Science, 270, 1801

\bibitem[{{Yusef-Zadeh} {et~al.}(2007){Yusef-Zadeh}, {Arendt}, {Heinke},
  {Hinz}, {Hewitt}, {Pratap}, {Ramirez}, {Rieke}, {Roberts}, {Stolovy},
  {Wardle}, \& {Whitney}}]{2007IAUS..242..366Y}
{Yusef-Zadeh}, F., {Arendt}, R.~G., {Heinke}, C.~O., {et~al.} 2007, in IAU
  Symposium, Vol. 242, IAU Symposium, ed. J.~M. {Chapman} \& W.~A. {Baan},
  366--373

\bibitem[{{Zavlin} {et~al.}(2000){Zavlin}, {Pavlov}, {Sanwal}, \&
  {Tr{\"u}mper}}]{Zavlin00-PSR}
{Zavlin}, V.~E., {Pavlov}, G.~G., {Sanwal}, D., \& {Tr{\"u}mper}, J. 2000,
  \apjl, 540, L25

\bibitem[{{Zhou} \& {Chen}(2011)}]{2011ApJ...743....4Z}
{Zhou}, P., \& {Chen}, Y. 2011, \apj, 743, 4

\bibitem[{{Zhou} {et~al.}(2014){Zhou}, {Safi-Harb}, {Chen}, {Zhang}, {Jiang},
  \& {Ferrand}}]{2014ApJ...791...87Z}
{Zhou}, P., {Safi-Harb}, S., {Chen}, Y., {et~al.} 2014, \apj, 791, 87

\bibitem[{{Zhou} {et~al.}(2009){Zhou}, {Chen}, {Su}, \&
  {Yang}}]{2009ApJ...691..516Z}
{Zhou}, X., {Chen}, Y., {Su}, Y., \& {Yang}, J. 2009, \apj, 691, 516

\bibitem[{{Zhu} {et~al.}(2013){Zhu}, {Tian}, {Torres}, {Pedaletti}, \&
  {Su}}]{2013ApJ...775...95Z}
{Zhu}, H., {Tian}, W.~W., {Torres}, D.~F., {Pedaletti}, G., \& {Su}, H.~Q.
  2013, \apj, 775, 95

\bibitem[{{Zirakashvili} \& {Ptuskin}(2008)}]{Zirakashvili08-Emax}
{Zirakashvili}, V.~N., \& {Ptuskin}, V.~S. 2008, \apj, 678, 939

\end{thebibliography}

\end{document}